\documentclass[a4paper,12pt,twoside,openright]{book}
\pdfoutput=1
\pdfpagewidth=\paperwidth
\pdfpageheight=\paperheight
\usepackage[pass]{geometry}
\usepackage{float}

\usepackage{longtable}
\usepackage{booktabs}
\usepackage{array}

\usepackage{graphicx}
\usepackage[textfont={it}]{caption}
\usepackage{subcaption}

\usepackage{lscape}

\usepackage[numbers,sort&compress]{natbib}

\usepackage{psfrag}
\usepackage{soulutf8}
\sethlcolor{cyan}
\usepackage[utf8]{inputenc}

\PassOptionsToPackage{hyphens}{url}\usepackage[pdftex]{hyperref}
\usepackage{breakurl}
\hypersetup{pdftitle={Design, scale-up and characterization of the data acquisition system and software for the ANAIS dark matter experiment}}
\hypersetup{pdfsubject={Universidad de Zaragoza - PhD Thesis}}
\hypersetup{pdfauthor={Miguel Ángel Oliván Monge}}
\hypersetup{pdfkeywords={Dark matter; ANAIS; Electronics; VME; NIM; Nuclear Instrumentation Module; PC; Acquisition; Data acquisition; DAQ; Dead time; Linux; Software; Software Design; IRQ; Interruption; Poll; CFD; Constant fraction discriminator; PMT; photomultiplier; SER; Single electron response; Quantum efficiency; Dark counts; Low threshold experiment; Light Collection; Stability; Environmental parameters; Radon; Muon; Muon Flux; Underground muon flux; Pulse shape analysis; NTP; NTP in particle physics; Particle physics}}

\usepackage{fancyhdr}               
\usepackage{vmargin}
\usepackage{color}
\usepackage{titlesec}
\usepackage{listings}
\usepackage[nottoc,notlot,notlof]{tocbibind}
\usepackage[percent]{overpic}

\usepackage{amsmath}
\usepackage[ampersand]{easylist}
\usepackage{multirow}
\usepackage{mathtools}
\DeclarePairedDelimiter{\ceil}{\lceil}{\rceil}

\newcommand{\myblank}{\newpage
\thispagestyle{empty}
\cleardoublepage} 

\title{\textbf{Design, scale-up and characterization of the data acquisition system for the ANAIS dark matter experiment}}
\date{Diciembre 2015}
\author{Memoria presentada por \\\textbf{Miguel Ángel Oliván Monge}\\para optar al grado de \\Doctor en Física\\ \\ \\ \\Laboratorio de Física Nuclear y Astropartículas\\Área de Física Atómica, Molecular y Nuclear\\Departamento de Física Teórica\\\textbf{UNIVERSIDAD DE ZARAGOZA}\\ \\}

\makeatletter
  \def\cleardoublepage{\clearpage\if@twoside \ifodd\c@page\else%
    \hbox{}
    \thispagestyle{empty}               
       \newpage
    \if@twocolumn\hbox{}\newpage\fi\fi\fi}
\makeatother

\floatstyle{ruled}
\newfloat{Program}{htbp}{lop}[chapter]

\begin{document}

\frontmatter                    

\maketitle                      
\clearpage
\thispagestyle{empty}
\begin{table}[h]
\begin{tabular}{p{.9\textwidth}}
 \begin{center}
 \includegraphics[scale=1]{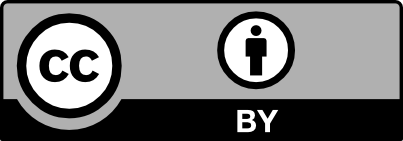}
 \end{center}
 \end{tabular}
 \end{table}
 This work is licensed under a Creative Commons Attribution 4.0 International License except the figures listed below:
\\ 
 \begin{easylist}[itemize]

 & Figure~\ref{fig:RotVec}. From K. Begeman, A. Broeils, and R. Sanders, “Extended rotation curves of spiral galaxies: Dark haloes and modified dynamics,” Monthly Notices of the Royal Astronomical Society, vol. 249, no. 3, pp. 523–537, 1991.
 & Figure~\ref{fig:BulletCluster}. X-ray: NASA/CXC/CfA/M.Markevitch et al.; Optical: NASA/STScI; Magellan/U.Arizona/D.Clowe et al.; Lensing Map: NASA/STScI; ESO WFI; Magellan/U.Arizona/D.Clowe et al.
 & Figure~\ref{fig:CMB_CABM}. K.A. Olive et al. (Particle Data Group), Chin. Phys. C, 38, 090001 (2014).
 & Figure~\ref{fig:PlankCMBAFit}. © ESA and the Planck Collaboration.
 & Figure~\ref{fig:DMExclusionExp}. P. Gondolo, from \href{https://indico.ibs.re.kr/event/7/contribution/11/material/slides/0.pdf}{\texttt{https://indico.ibs.re.kr/event/7/}\\\texttt{contribution/11/material/slides/0.pdf}}.
 & Figure~\ref{fig:DAMAModulation}. From R. Bernabei et al., “New results from DAMA/LIBRA,” The European Physical Journal C-Particles and Fields, vol. 67, no. 1, pp. 39–49, 2010.
 & Figure~\ref{fig:NaIWIMPRates}. From C. Cuesta ``ANAIS-0: Feasibility study for a 250 kg NaI(Tl) dark matter search experiment at the Canfranc Underground Laboratory'', Universidad de Zaragoza, 2013.
 & Figure~\ref{fig:Sensitivity}. From J. Amaré et al. ``Status of the ANAIS Dark Matter Project at the Canfranc Underground Laboratory''. Proceedings of 11th Patras Workshop on Axions, WIMPs and WISPs.
 & Figure~\ref{fig:PMT}. From ``Photomultiplier Tubes: Basics and Applications''. Hamamatsu Photonics KK, Iwata City, 2007. 
 & Figure~\ref{fig:MatacqChip}. From D. Breton, E. Delagnes, and M. Houry, “Very high dynamic range and high sampling rate VME digitizing boards for physics experiments,” Nuclear Science, IEEE Transactions on, vol. 52, no. 6, pp. 2853–2860, 2005. 
 & Figures~\ref{fig:NTPArchitecture} and~\ref{fig:NTPdiscipline}. D. L. Mills, from \url{https://www.eecis.udel.edu/~mills/ntp/html/warp.html}. 
 \end{easylist}

\myblank
\setcounter{page}{1}

\myblank
{\let\newpage\relax\vspace*{-1.5cm}\tableofcontents}                
\myblank



\mainmatter
\chapter{Introduction}\label{sec:intro}
The observation of our universe has been more and more precise and it has brought many data about its composition and dynamics. As a result, there are strong evidences supporting the existence of a non-luminous matter known as dark matter. This matter accounts for the majority of the material that forms galaxies and clusters of galaxies explaining observational phenomena and it can not be only composed by particles of the standard model of particle physics. The evidences of the existence of dark matter come from accurate measurements of galactic rotation curves, orbital galaxy velocities in clusters and cluster mass determination with gravitational lensing. In addition, the precise measurement of the Cosmic Microwave Background (CMB) combined with the standard cosmological model also gives a big dark matter proportion ($\sim$ 27\% of non-baryonic dark matter and $\sim$ 68\% of dark energy) after the Planck measurements~\cite{ade2013planck}.
\paragraph{}
The nature of the hypothetical dark matter remains unknown and the detection and characterization of this type of matter is an open field in particle physics that can solve many cosmological questions as well as give answers beyond the standard model of particle physics~\cite{beringer2012review}. The evidences supporting the dark matter existence, the possible candidates and the efforts to detect and characterize the dark matter are covered in this chapter.
\paragraph{}
DAMA/LIBRA stands out among these initiatives because of a positive signal of annual modulation at low energy attributed to dark matter~\cite{bernabei2008first,bernabei2010new,bernabei2013final}, being the only experiment with a positive result spanning a long period of time. The ANAIS project (see Chapter~\ref{sec:ANAIS}) has been a long effort to carry out an experiment with the same target and technique, which could confirm the DAMA/LIBRA result.
\section{Dark matter evidences}
The measurements of large astrophysical systems from galactic to cosmological scales show anomalies that can only be explained by assuming the existence of large amounts of unseen (dark) matter or by assuming deviations from the laws of gravitation and the theory of general relativity.
\paragraph{}
\subsection{The galactic scale}
The most direct and convincing evidences for dark matter existence on galactic scales are the galactic rotation curves first measured by Vera Rubin~\cite{rubin1983rotation}. The rotational velocities of visible stars or gas as a function of the distance to the galactic center exhibit a flat behavior contrasting with the expected Newtonian dynamics:
\begin{equation} 
	v(r) = \sqrt{\frac{GM(r)}{r}}\label{eq:th_rot}
\end{equation}

\begin{figure}[h!]
  \begin{center}
	  \includegraphics[width=.5\textwidth]{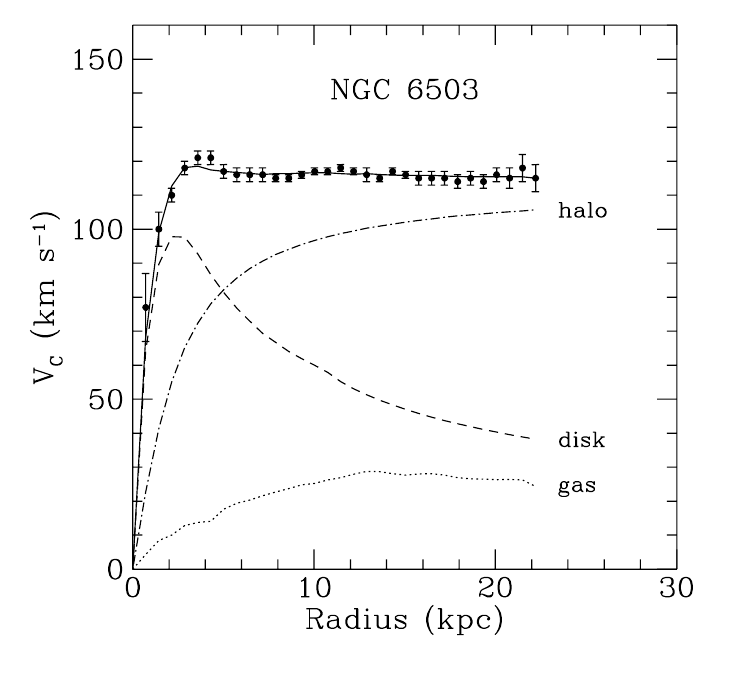}
    \caption[Galaxy rotation curve]{Example of galactic rotation curve. The dotted, dashed and dash-dotted lines are the contributions to rotational velocity of gas, disk and dark matter, respectively. From~\cite{begeman1991extended}\label{fig:RotVec}.}
  \end{center}
\end{figure}

This velocity should be falling $\propto 1/\sqrt{r}$ outside the optical disk. The fact that $v(r)$ is almost constant, as seen in Figure~\ref{fig:RotVec}, implies an halo with $M(r) \propto r $. This observations\footnote{The observed zones are inside and outside the optical disk revealing an important gas contribution in galaxy composition.}, usually measured by Doppler shift of 21 cm HI line combined with optical surface photometry, have been confirmed in many galaxies~\cite{begeman1991extended}.

There is a larger number of hints at galactic scale such as low surface brightness (LSB) galaxies, weak modulation of strong lensing around individual massive elliptical galaxies, velocity dispersion of dwarf spheroidal galaxies or velocity dispersion of spiral galaxy satellites (see~\cite{bertone2005particle} and references therein).
\paragraph{}
\subsection{The galaxy cluster scale}
\begin{figure}[h!]
  \begin{center}
	  \includegraphics[width=.9\textwidth]{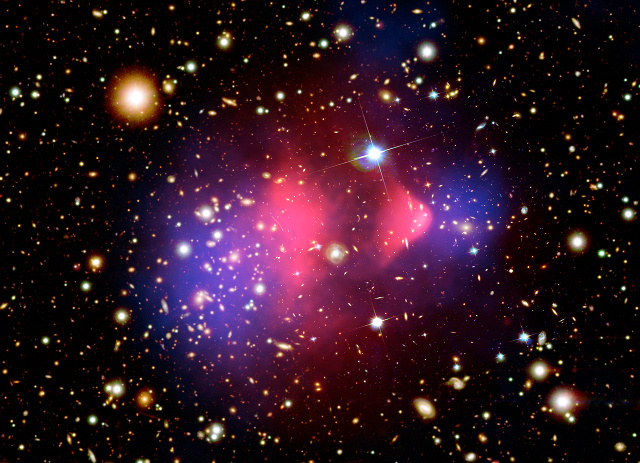}
	  \caption[Bullet cluster composite image]{Bullet cluster composite image. Hot gas detected by Chandra X-ray telescope is shown in pink. The optical image shows the galaxies in orange and white from Magellan and Hubble Space Telescope images. The mass distribution deduced by gravitational lensing is shown in blue. From~\cite{bulletclusterimage}\label{fig:BulletCluster}.}
  \end{center}
\end{figure}

Historically, the first hint about dark matter existence was at the cluster scale. Fritz Zwicky inferred the existence of unseen matter examining the Coma galaxy cluster in 1933~\cite{zwicky1933rotverschiebung} using the virial theorem and he was the first to refer the unseen matter as \emph{dunkle materie}.
\paragraph{}
Since then, other dark matter evidences such as gravitational lensing effects have been observed and used to determine cluster mass distribution. This gravitational lensing involves the distortion of a far galaxy by the gravitational field from a very massive cluster. The lensing can be used to estimate the mass profile of the foreground massive object. This is a especially valuable technique used in combination with other direct observational data using the visible and the X-ray spectra and it has been applied to colliding clusters~\cite{clowe2006direct, bradavc2008revealing, mahdavi2007dark}.
\paragraph{}
This analysis shows a matter distribution that does not fit with visible and X-ray patterns. This distribution can be seen in Figure~\ref{fig:BulletCluster} in a composite image showing the first cluster analyzed in this way, the 1E 0657-558 also known as Bullet Cluster. The hot gas detected by its electromagnetic interaction and shown in pink, contains the most part of the baryonic matter. Galaxies are shown in white and orange. The matter profile deduced by gravitational lensing is shown in blue. Most of the matter, shown in blue, does not interact which implies that this unseen matter only interacts gravitationally and/or very weakly. Therefore, this matter can only be compound of neutrinos or non-baryonic matter. The same result can be reported from other cluster collision, MACS J0025.4–1222 cluster~\cite{bradavc2008revealing} giving a similar distribution for every type of mass. However, not all collisions show the same profile and the Abell 520 galaxy cluster presents a mass distribution with the cluster center dominated by dark matter~\cite{mahdavi2007dark} being a challenge for theories and simulations about dark matter interaction properties~\cite{jee2012study}.
\subsection{The cosmological scale}
The dark matter is also a fundamental ingredient in the standard cosmological model. This model, based on general relativity with assumptions about space geometry and mass distribution and interaction is capable to offer falsifiable predictions some of them already measured experimentally with high accuracy like cosmic microwave background (CMB) and CMB anisotropy. This measurements also characterize the model including the amount of dark matter and dark energy in the universe. 
\paragraph{}
The first fact to take into account in a modern cosmological model is that the universe is expanding. Edwing Hubble found combining their own measurements with those of Vesto Slipher and Milton L. Humason that the galaxies velocity is proportional to their distance~\cite{hubble1931velocity} in the now known as \emph{Hubble's law}:
\begin{equation}
	v=H_0r
\end{equation}
\paragraph{}
Following this empirical measurements describing the average behavior of distant galaxies a set of solutions to Einstein equations were developed~\cite{lemaitre1927univers} giving a new era in cosmology: the big bang cosmology~\cite{beringer2012review}. In addition to the observed expansion, a cosmological model must explain other experimental evidences such as the accelerated expansion~\cite{riess1998observational,kowalski2008improved} and the CMB~\cite{penzias1965measurement} among others.
\subsubsection{The standard cosmological model}
The standard model of cosmology (see~\cite{goobar2006cosmology} for example) is based on the resolution of the Einstein equations with the isotropic and homogeneous Friedmann-Lemaître-Robertson-Walker model (or FLRW model) allowing vacuum energy (via cosmological constant), matter and radiation. It solves the equations using the perfect fluid model.
\paragraph{}
The FLRW metric emerges from the isotropic and homogeneous conditions. The FLRW line element should have the form:
\begin{equation}
	ds^2 = g_{\mu\nu}dx^\mu dx^\nu= dt^2 - a^2(t) \left(  \frac{dr^2}{1-kr^2} + r^2d\theta + r^2 sin^2 \theta d\phi^2 \right)
\end{equation}
being $k$ and $a(t)$ the curvature parameter and the time dependent scale factor respectively. The possible values of $k$ (being $r$ dimensionless) are +1, 0 or -1 depending on whether the constant curvature is positive, zero or negative. This model provides a metric $g_{\mu\nu}$ for the Einstein equations, with the cosmological constant term:
\begin{equation}
	R_{\mu\nu} - \frac{1}{2}g_{\mu\nu}R - \Lambda g_{\mu\nu} = 8\pi GT_{\mu\nu}
\end{equation}
being $R_{\mu\nu}$ the Ricci curvature tensor, $R$ ($g^{\mu\nu}R_{\mu\nu}$) the Ricci scalar, $\Lambda$ the cosmological constant, $G$ the gravitation constant and $T_{\mu\nu}$ the energy-momentum tensor. This tensor can be written in the form:
\begin{equation}
	T_{\mu\nu} = (p + \rho) u_\mu u_\nu - pg_{\mu\nu}
\end{equation}
for a perfect fluid, a fluid without viscosity or heat flux, with $u_\mu$ the four-velocity of the fluid element, $p$ the pressure and $\rho$ the matter density. All the above equations can be combined to give the Friedmann equations:
\begin{equation}
	\left( \frac{\dot a}{a} \right)^2 + \frac{k}{a^2} = \frac{8\pi G}{3}\rho_{tot} \qquad \qquad \frac{\ddot a}{a} = \frac{-4\pi G}{3} (\rho_{tot} + 3 p) \label{eq:Friedmann}
\end{equation}
being $\rho_{tot}=\rho_m+\rho_{rad} + \rho_{vac}$ and $\rho_{vac}=\rho_\Lambda=\frac{\Lambda}{8\pi G}$. This $\rho_{tot}$ takes into account the density of mass, radiation and vacuum energy.
\paragraph{}

The scale factor temporal behavior can be deduced using again the ideal fluid model with the state equation $p = \omega \rho$ for every density and assuming a flat universe $k=0$ (as pointed for WMAP/Planck measurements, see \emph{Cosmic microwave background experimental results} subsection). $\rho$ is dominated by rest mass for non-relativistic matter and the pressure is proportional to velocity ($v<<c$ for such a type of matter), so $\omega=0$ for non-relativistic matter. Considering only this matter:
\begin{equation}
	a(t) \propto t^{2/3} \qquad\text{(matter domination)}
\end {equation}
For radiation, $v=c=1$ and $p=\rho/3$ since pressure is averaged for three spatial directions, and $\omega=1/3$  gives:
\begin{equation}
	a(t) \propto \sqrt{t} \qquad\text{(radiation domination)}
\end {equation}
And for vacuum $p=-\rho$, $\omega=-1$:
\begin{equation}
	a(t) \propto e^{Ht} \qquad\text{(vacuum energy domination)}
\end {equation}
being $H \equiv \frac{\dot a}{a}$, the Hubble \emph{constant} or more properly the Hubble parameter.
\paragraph{}
These three particular solutions show a simplified set of solutions to Friedmann equations with an expanding universe. They are useful to understand the behavior of the universe expansion in special cases illustrating the three stages in an expanding flat universe: the initial radiation domination, the subsequent matter domination and the current vacuum energy domination.
\paragraph{}
This above described model can be extended to include perturbations explaining the universe deviation from homogeneity. In addition, adding a period of inflationary expansion solves the horizon, the flatness and the lack of magnetic-monopole problems~\cite{guth1981inflationary}. Next to this inflationary expansion, the use of the standard model of particle physics explains how the nuclei of the light elements were formed in a process called big bang nucleosynthesis~\cite{iocco2009primordial}. This cosmological model gives predictions for relative abundances of light elements and CMB that can be contrasted to measurements. The model is parametrized to perform such a verification as it can be seen in next section.
\subsubsection{$\Lambda CDM$ and the cosmological parameters}
One of the first attempts to describe dark matter from a cosmological point of view included the neutrino as candidate to dark matter in the so-called \emph{hot dark matter} models~\cite{primack2001hot}. However, these models were not able to explain the large scale structure (LSS) and galaxy formation giving more fragmented structures than the observed~\cite{blumenthal1984formation}. In contrast, the $\Lambda CDM$ (or lambda cold dark matter) proposes a non-relativistic matter as source of the non-luminous matter in addition with vacuum energy explaining the accelerating expansion.
\paragraph{}
In this section, the cosmological parameters that make up the so-called six-parameter model are described (see~\cite{beringer2012review} and references therein). These six parameters are used to contrast different cosmological models with measurements, as it can be seen in the next subsection. 
\paragraph{}
The (first) Friedmann equation~\ref{eq:Friedmann} can be used to define critical density setting $k=0$ and $\Lambda = 0$:
\begin{equation}
	\rho_c \equiv \frac{3H^2}{8\pi G}
\end {equation}
The scaled Hubble parameter, $h$, is defined as:
\begin{equation}
	H \equiv 100 \,h \,km \,s^{-1} \,Mpc^{-1}
\end {equation}
giving $\rho_c = 1.88 \times 10^{-26}\,h^2\,kg \,m ^{-3} = 1.05\times h^2 \,GeV \,cm ^{-3}$.
\paragraph{}

The critical density can be used to define adimensional cosmological density parameter as the energy density relative to critical density:
\begin{equation}
	\Omega_{tot} \equiv \frac{\rho_{tot}}{\rho_c}
\end {equation}
The (first) Friedmann equation~\ref{eq:Friedmann} can be written as:
\begin{equation}
	\frac{k}{H^2a^2} = \Omega_{tot} -1 \label{eq:FriedCurv}
\end {equation}
The overall geometry of the universe is determined by the total energy density parameter $\Omega_{tot}$: if $k>0$ and $\Omega_{tot}>1$ the universe is closed (and $a(t)$ can be chosen to be the radius), if $\Omega_{tot}<1$ and $k=-1$ the universe is open and when $\Omega_{tot}=1$ and k=0 the universe is spatially flat.
\paragraph{}
The $\Lambda CDM$ model breaks down the mass contribution, $\Omega_m$, in components depending of the type of matter $\Omega_m = \Omega_{b} + \Omega_{cdm}$ being $\Omega_{b}$ the baryon density and $\Omega_{cdm}$ the proposed cold dark matter, i.e. matter moving slowly compared with the speed of light (cold) that interacts weakly with electromagnetic radiation (dark). It is necessary to distinguish the different contributions to the density because of their different dynamics. For non-relativistic matter, $\Omega_m$ is defined as the \emph{present day} matter density and critical density ratio, and the same for relativistic matter $\Omega_{rad}$ using its \emph{present day} density. It is also possible to write $\Omega_{\Lambda} \equiv \frac{\Lambda}{3H_0^2}$ and thus $\Omega_{tot} = \Omega_m + \Omega_{rad} + \Omega_{\Lambda}$. The density of matter $\Omega_m$ includes visible and dark matter. 
\paragraph{}
The aforementioned primordial perturbations are usually modeled obeying a power law parametrized by $n_s$ as the density spectral index and $A_s$ as the perturbation amplitude. The last parameter used to the fit of the six-parameter model is $\tau$ being the Thomson scattering optical depth due to reionization that alters the pattern of CMB anisotropies as it can be seen in next subsection.
\paragraph{}
It is worth to mention that other cosmological parameters can be deduced from the above presented like curvature $k$ (or curvature ``density'' $\Omega_k \equiv \frac{k}{H^2a^2}$) from Equation~\ref{eq:FriedCurv}. A non-complete list of derived parameters includes the age of the universe, the present horizon distance, the baryon-to-photon ratio and the baryon to dark matter density ratio. 
\subsubsection{Cosmic microwave background}
The cosmic microwave background (CMB) radiation was first detected in 1964 by Penzias and Wilson~\cite{penzias1965measurement} and it was soon identified as relic radiation from the epoch of recombination or photon decoupling, once the protons and electrons formed neutral atoms. This background radiation measurement was a landmark for accepting Big Bang model and later it has been a benchmark to test any proposed cosmological model. The first measurements by Penzias and Wilson noted a high isotropy and the study and characterizations of the CMB signal has been increasingly precise revealing anisotropies at $10^{-5}$ level.
\paragraph{}
This anisotropy is usually expressed by using spherical harmonic expansion. The temperature can be expressed as:
\begin{equation}
	T(\theta, \phi) = \sum_{\ell m}{a_{\ell m}{Y_{\ell m}}(\theta,\phi)}
\end {equation}

\begin{figure}[h!]
  \begin{center}
	  \includegraphics[width=.9\textwidth]{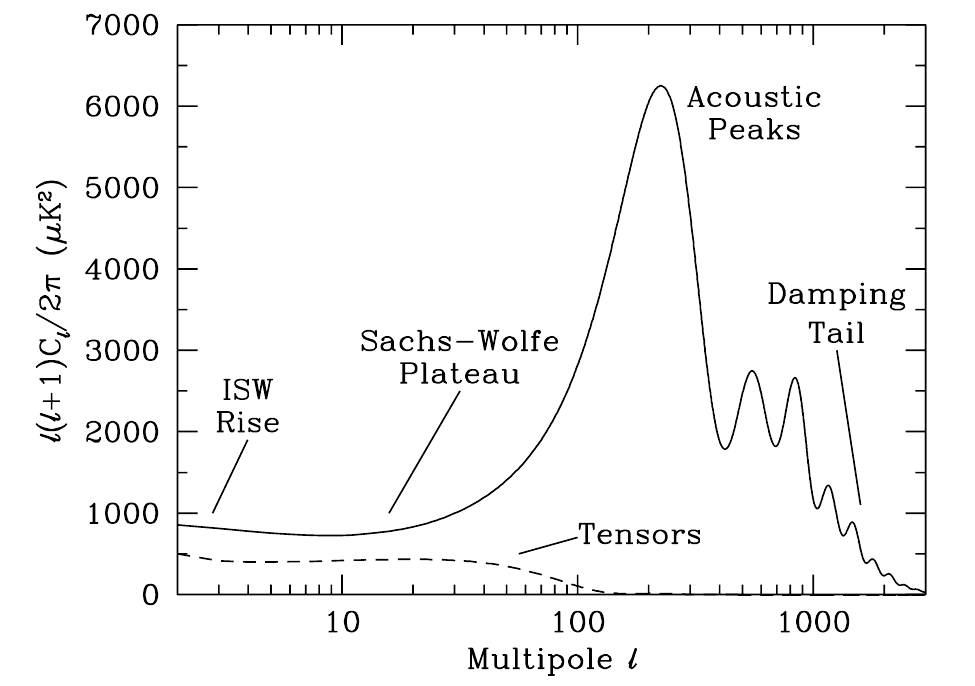}
	  \caption[Theoretical $\Lambda CMB$ temperature power spectrum]{Theoretical $\Lambda CMB$ temperature power spectrum\label{fig:CMB_CABM}. From~\cite{beringer2012review} using CAMB software~\cite{lewis2000efficient}.}
  \end{center}
\end{figure}

Given this expansion, it is easier to compare the features of the resultant power spectrum represented as $\ell (\ell +1) \frac{C_{\ell}}{(2\pi)}$ being $C_{\ell} = \frac{1}{2\ell+1}\sum_{m}{|a_{\ell m}|^2}$. The $\Lambda CDM$ predicted power spectrum, with the above described parameters as input parameters, can be calculated as it can be seen in Figure~\ref{fig:CMB_CABM}. It is worth to note that the monopole harmonic takes into account the black body spectrum of the universe giving the mean temperature and the dipole ($\ell = 1$) is interpreted as the result of the Doppler shift caused by the solar system motion relative to the nearly isotropic black-body field. The power spectrum has some peculiar features at higher harmonics such as Sachs-Wolfe plateau and the acoustic peaks~\cite{beringer2012review} also seen in Figure~\ref{fig:CMB_CABM}. 
\paragraph{}
In addition to temperature, different polarization modes can show their own anisotropies and power spectra $C_{\ell}^E$, $C_{\ell}^B$ and $C_{\ell}^{TE}$ for E-modes, B-modes and correlations between density and velocity perturbations at last scattering surface. The B-mode power spectrum can reveal signatures of gravitational waves~\cite{seljak1997signature}.
\paragraph{}
The acoustic oscillations arise from the tight coupling of baryons and photons in the early Universe: the propagation of sound waves through this medium gives rise to a characteristic scale in the distribution of perturbations corresponding to the distance traveled by the wave before recombination. This signal is imprinted in the distribution of both the matter and the radiation. The latter are seen as anisotropies in the CMB radiation, featuring the aforementioned acoustic peaks, while the former are the baryon acoustic oscillations, measured via astronomical surveys such as Sloan Digital Sky Survey III~\cite{anderson2012clustering}.

\subsubsection{Cosmic microwave background experimental results}

The COBE (Cosmic background explorer) NASA satellite, launched in 1989, was the first satellite dedicated to such measurements. It was capable to measure the universe black body spectrum with an unprecedented accuracy and it was the first experiment to report a faint anisotropy in CMB~\cite{smoot1992structure}, compatible with models of inflationary cosmology.
\paragraph{}
A new spacecraft, WMAP (Wilkinson Microwave Anisotropy Probe), was launched in 2001, again by the NASA, with the aim to characterize the COBE hinted anisotropy. WMAP has played a key role in establishing the $\Lambda CDM$ model as current standard model of cosmology. The predicted CMB anisotropies by this model fits with the WMAP experimental data in the previous presented six-parameter model. In addition with the values of the direct and derived parameters such as of baryonic, cold dark matter and dark energy~\cite{WMAP} the WMAP data was used to recalibrate the COBE FIRAS spectrometer giving the more accurate measurement for the black-body spectrum corresponding to a temperature of $2.72548 \pm 0.00057$ K~\cite{fixsen2009temperature}.
\paragraph{}

\begin{figure}[h!]
  \begin{center}
	  \includegraphics[width=.9\textwidth]{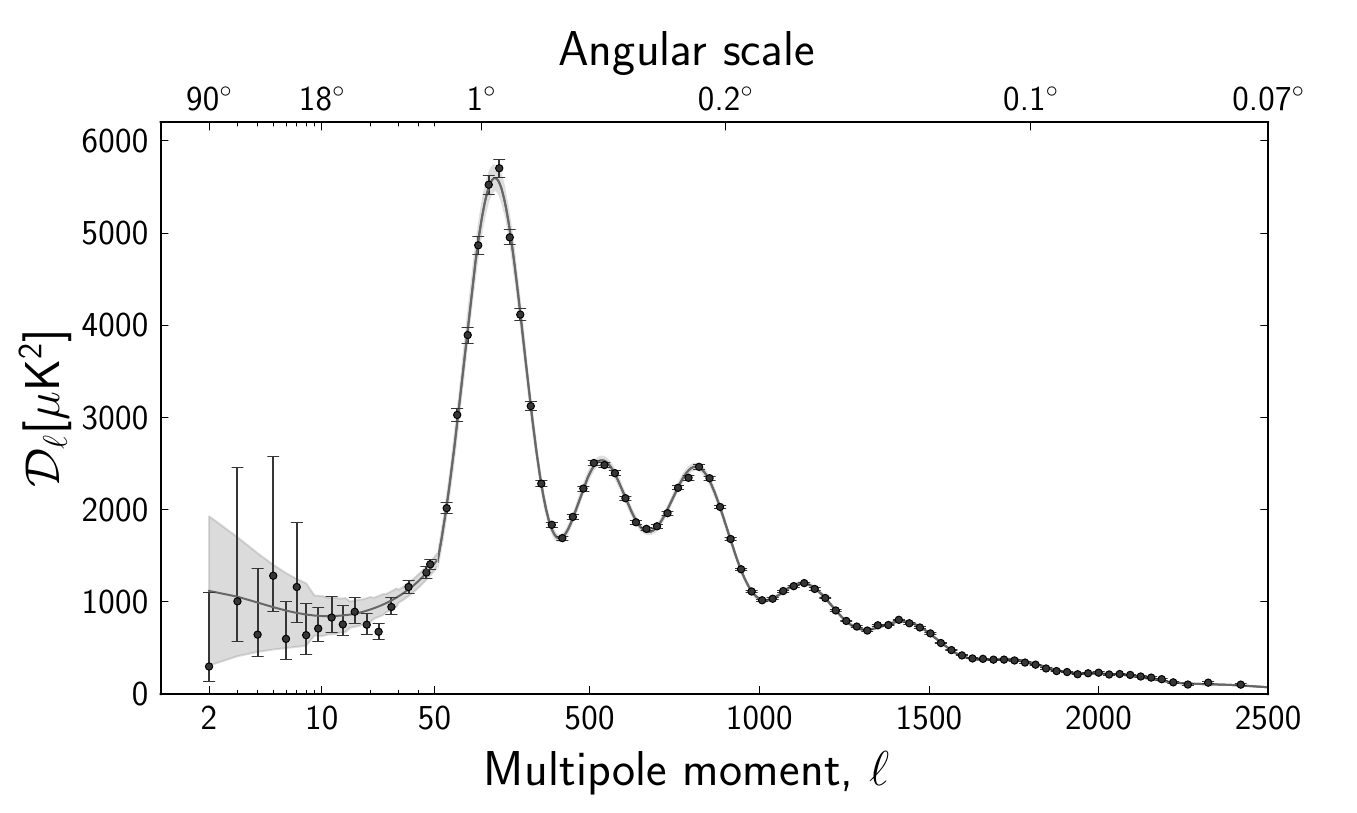}
	  \caption[Plank CMB temperature anisotropy data and fit]{Plank CMB temperature anisotropy data and six-parameter model fit\label{fig:PlankCMBAFit} with $\mathcal{D}_{\ell} \equiv \ell (\ell +1) \frac{C_{\ell}}{(2\pi)} $ (From~\cite{ade2013planckover}).}
  \end{center}
\end{figure}

The Planck satellite was launched in 2009 by the European Space Agency~\cite{ade2013planckover}. It was designed to improve the sensitivity and the angular resolution and the fitted cosmological parameter values are the more accurate at the present time. The CMB temperature anisotropy data and fit can be seen in Figure~\ref{fig:PlankCMBAFit}. It shows the temperature power spectrum in both real data and predicted by $\Lambda CDM$ model showing a very good fit at high multipoles and determining the acoustic peaks relative amplitude with an excellent accuracy.
\paragraph{}
The extracted values can be seen in Table~\ref{tab:PlanckResults} giving, as stated at the beginning, a $\sim$ 27\% of cold dark matter and a $\sim$ 68\% of dark energy~\cite{ade2013planck}. The Planck results, as seen in Table~\ref{tab:PlanckResults}, have been presented with two different combinations. The Planck data only using CMB data combined with lensing is labeled as \emph{CMB+lensing}. The Planck data, in combination with polarization measured by WMAP, high-$\ell$ anisotropies from experiments like ACT~\cite{sievers2013atacama} and SPT~\cite{keisler2011measurement} and data by baryon acoustic oscillations, described in previous section, is labeled as \emph{Planck+WP+highL+BAO}. The ACT and SPT data help to fix the foreground model at high $\ell$ and the low redshift measurements provided Baryon Oscillation Spectroscopic Survey (BOSS), which is part of the Sloan Digital Sky Survey III~\cite{anderson2012clustering}, allow to break some degenaracies still present in Planck data. The combination of these experiments provides the best constrains to the six-parameter model.
\paragraph{}
\begin{table}[h!]
	\begin{center}
		\begin{tabular}{ l l l} 
			\toprule
			 Symbol & Planck (CMB+lensing) & Planck+WP+highL+BAO\\
			\hline
			Fit Parameters: &&\\
			\hline
			$\Omega_b h^2$		& $0.02217\pm 0.00033$	& $0.02214\pm0.00024$ \\
			$\Omega_{cdm} h^2$	& $0.1186\pm 0.0031$	& $0.1187\pm0.0017$ \\
			$100\theta_{MC}$ 	& $1.04141\pm 0.00067$	& $1.04147 \pm 0.00056$ \\
			$ \tau$			& $ 0.089\pm 0.032$	& $0.0092\pm0.0013$ \\
			$ n_s$			& $0.9635 \pm 0.0094$	& $0.9608\pm0.0054$ \\
			$ ln(10^{10}A_s)$	& $3.085 \pm 0.057$	& $3.091\pm0.025$ \\
			\hline
			Derived Parameters: & &\\
			\hline
			$\Omega_{\Lambda}$	& $ 0.693\pm 0.019$	& $0.692\pm0.010$ \\
			$H_0$ (km s\textsuperscript{-1} Mpc\textsuperscript{-1})& $ 67.9\pm1.5$ &$67.80 \pm 0.77$ \\
			$t_0$ (Gyr)		& $13.796 \pm 0.058$	& $13.798 \pm 0.037$ \\
			$100\Omega_k$		& $-4.2^{+4.3}_{-4.8}$	& $-0.10^{+0.62}_{-0.65}$ \\
			\toprule
		\end{tabular}
		\begin{tabular}{ l l} 
			\toprule
			Parameter & Symbol \\
			\hline
			Baryon density today & $\Omega_b h^2$\\
			Cold dark matter density today&$\Omega_{cdm} h^2$ \\
			100 $\times$ approximation to $r_*/D_A$  &$100\theta_{MC}$ \\
			Scattering optical depth & $ \tau$\\
			Scalar spectrum power-law index & $ n_s$ \\
			Primordial perturbation amplitude & $ ln(10^{10}A_s)$ \\
			\hline
			Dark energy density today& $\Omega_{\Lambda}$\\
			Hubble parameter& $H_0$\\
			Age of the universe& $t_0$\\
			Curvature ``density'' & $100\Omega_k$ \\
			\toprule
		\end{tabular}

		\caption[Planck cosmological parameters]{Cosmological parameters derived from Planck measurements. From~\cite{ade2013planck}.} 
		\label{tab:PlanckResults} 
	\end{center}
\end{table}
A derived parameter from the fit parameters is the curvature ``density'', $\Omega_k$, fully compatible with a flat universe ($k=0$). The Planck results also establish other constrains for different aspects of the $\Lambda CDM$ like tensor-to-scalar ratio or inflation models~\cite{ade2013planckinf}. Tensor-to-scalar ratio is an important parameter for extensions to $\Lambda CDM$ including gravitational waves. This parameter has claimed to be measured by BICEP2 collaboration~\cite{ade2014bicep2} measuring B-mode polarization although an incorrect foreground estimation could jeopardize the result~\cite{flauger2014toward}.
\subsubsection{$\Lambda CDM$ simulations}
A set of simulations based on the $\Lambda CDM$ properties has been performed taking advantage of the models and data described earlier. These simulations can be very useful for trying to describe the dark matter distribution in the clusters and galaxies. At the galactic scale the results of the simulations hint to clumpsy haloes with streams components~\cite{diemand2008clumps}. Further gravitationally-only interacting N-body simulations such as Aquarius~\cite{springel2008aquarius} or Via Lactea II~\cite{kuhlen2009exploring} increased the resolution of Milky Way sized simulations and they give hints to indirect dark matter searches~\cite{pieri2011implications}.
\paragraph{}
These simulations can also be used as a crosscheck of the data of the observable universe through a post-processing procedure known as semi-analytic modeling. The modeling involves simple physical prescriptions over the N-body simulation results to estimate the distribution of galaxies~\cite{white1991galaxy,guo2011dwarf}.   
\paragraph{}
Recent simulations have been carried out using a more complete hydrodynamical approach taking into account baryonic matter from the beginning trying to offer self-consistent and predictive approach. The Illustris project has reproduced in such a way, with a large scale simulation, mixed populations of elliptical and spiral galaxies~\cite{vogelsberger2014properties}. The simulated spiral galaxies agree with many observational parameters exhibiting an almost flat rotational velocity profile~\cite{vogelsberger2014introducing} among other features of real galaxies. However, the simulation does not reproduce some observational parameters such as the steep slope and small scatter of the baryonic Tully-Fisher relation found by McGaugh~\cite{mcgaugh2012baryonic} but it gives more realistic results than previous simulations.
\subsubsection{$\Lambda CDM$ challenges}
$\Lambda CDM$ model appears to be very successful in predicting the CMB and large-scale structure but it is fair to note that there are still open questions. From the astrophysical point of view it is not able to explain the observed rotation curves of all dwarf and low surface brightness (LSB) galaxies~\cite{primack2012triumphs} .
\paragraph{}
Other remarkable issues are the weak agreement at low multipoles with the $\Lambda CDM$ predictions, the so-called \emph{cold-spot} (a region of significant temperature decrement, far from statistic fluctuation)~\cite{ade2013planckIso} and the primordial lithium problem~\cite{fields2012primordial}: the abundance of \textsuperscript{7}Li is far below (2-3 factor) of the predictions of big bang nucleosynthesis together with cosmic baryon density coming from CMB anisotropy fits. These anomalies, confirmed by WMAP and Planck measurements, need $\Lambda CDM$ refinement or physics beyond the standard models.
\subsubsection{Summary}
It can be concluded as a result of the cosmological observation and models that: 
\begin{easylist}[itemize]
& The universe is almost homogeneous and it is expanding at an accelerating rate.
& The accelerated expansion is ``explained'' via cosmological constant and a consequent dark energy ($\sim$ 68\%).
& The universe has a large amount of non-baryonic, beyond the standard model, cold dark matter ($\sim$ 27\%).
\end{easylist}

\subsection{Alternatives to dark matter}\label{sec:MOND}
The need of dark matter has been questioned by models that have tried to explain the observed anomalies in an alternative way by modifying gravity laws. The most promising models have been MOND (modified Newtonian dynamics) and its relativistic version TeVeS (tensor–vector–scalar gravity)~\cite{famaey2012modified}. MOND was proposed as a correction to Newtonian gravity by modifying the predicted acceleration at small values of gravitational force giving acceptable fits for rotational velocity of galaxies~\cite{milgrom1983modification} and predicting some observational parameters such as the baryonic Tully-Fisher relation~\cite{mcgaugh2012baryonic}.
\paragraph{}
The theory was challenged by the dark matter hints at all scales. TeVeS has been capable to explain effects such as strong gravitational lensing and some of the features of the CMB anisotropy but in an incomplete way giving inaccurate predictions about the weak gravitational lensing effect and the ratio between acoustic peaks in CMB temperature power spectrum~\cite{famaey2012modified,bertone2010particle} without the addition of cold dark matter. For these reasons and for its consistency with general relativity the $\Lambda CDM$ model is the most widely accepted cosmological model.
\section{Dark matter candidates}
Dark matter candidates have to exhibit several properties in order to match with the models and data described earlier. Candidates for non-baryonic dark matter have to be weakly interacting with ordinary matter (otherwise they wouldn't qualify as \emph{dark} matter), non-relativistic at the onset of galaxy formation, massive and stable or very long lived on cosmological time scales (otherwise they would have decayed by now)~\cite{bertone2010particle}. 

\paragraph{}
The most obvious candidate (partially) fulfilling these conditions is the neutrino, having ``the undisputed virtue of being known to exist''~\cite{bergstrom2000non} but cosmological observations and simulations are incompatible with neutrinos as unique dark matter component, as seen in the previous section. These facts force the need of candidates beyond the standard model of particle physics (BSM).  
\paragraph{}
Candidates fulfilling the expected properties include axions, sterile neutrinos and weakly interacting massive particles (WIMPs). These are \emph{well motivated candidates} because they are proposed to solve problems in principle unrelated to dark matter and whose properties can be computed within a well-defined particle physics model. Primordial black holes, a frequently cited candidate, have been ruled out as primary dark matter constituent in an extremely wide range of masses by a variety of searches and constrains~\cite{carr2010new}.  
\paragraph{}
Axions were proposed by Peccei and Quinn to resolve the strong CP problem in quantum chromodynamics explaining the lack of measured broken CP-symmetry in strong interactions~\cite{peccei1977constraints}. Several constrains from experiments and astrophysical observations give an axion mass less than $3 \times10^{-3}$ eV still enough to be a suitable cold dark matter candidate~\cite{axionsbertone}.
\paragraph{}
Sterile neutrinos are proposed neutrinos without any interaction with other particles except through gravity solving the lack of observed right handed neutrinos~\cite{drewes2013phenomenology}. Due to their predicted collisionless and long lived nature, they are candidates to dark matter. The Plank data also establish constrains to their masses~\cite{ade2013planck}.
\paragraph{}
Weakly interacting massive particles (WIMPs) are hypothetical particles that interact with standard model particles via weak nuclear force (or another interaction similar in strength). The fact that annihilation cross-section at electroweak scales gives the appropriate relic abundance~\cite{gondolo1991cosmic} has been seen as a major reason to highly consider WIMPs as dark matter candidate. Another strong reason to consider WIMPs is the natural emergence of this kind of particles in supersymmetric theories. Supersymmetry (or SUSY) is a hypothetical space-time symmetry that allows to link matter particles and force carriers and also link gravity with the other fundamental interaction solving the hierarchy problem~\cite{nilles1984supersymmetry}. The introduction of such a symmetry to extend the particle standard model (SM) predicts the existence of new particles. The simplest supersymmetry extension to SM, called minimal supersymmetric standard model (MSSM), gives naturally \emph{superpartners} for every SM particle: every fermion has a boson partner and every boson has a fermion partner. The lightest \emph{superparters} have properties to be the non-baryonic dark matter particle such as the neutralino or, including gravity in the model, the gravitino~\cite{ellisolivebertone}.  
\paragraph{}
These WIMPs beyond the standard model (BSM) could be produced at colliders~\cite{SUSYColiders} such Large Hadron Collider (LHC). Other alternative theories, like extra dimension theories, also predict new particles BSM with distinguishable features. These alternatives gives different spin or number of \emph{superpartners}~\cite{ExtraDimLHC} to the expected new particles. These searches have been performed at accelerators with negative results for the moment. No hints of physics beyond the standard model have been observed and cross section limits have been established by Tevatron CDF~\cite{PhysRevLett.108.211804} with $p\overline{p}$ collisions and LHC CMS~\cite{chatrchyan2012searchmono,chatrchyan2012searchphoton} and ATLAS~\cite{aad2013search,PhysRevLett.110.011802,aad2014search} experiments with $pp \rightarrow \chi \tilde{\chi} + X$ being $X$ a hadronic jet, photon or $W$ or $Z$ bosons. 
\paragraph{}
There are many efforts to detect and characterize dark matter. Direct detection aims to recover information about the WIMPs measuring their scattering with ordinary matter. Other complementary information about dark matter can be obtained by detecting secondary products of self-annihilation or decay of dark matter particles. These by-products are searched by indirect detection experiments. In the next subsections both indirect and direct detection techniques are covered.
\section{Indirect dark matter detection}
Despite being stable or very long lived particles, dark matter constituents could decay and despite their weak interacting nature they could self-annihilate originating ordinary matter (see~\cite{silk1984cosmic} for an early reference and~\cite{cirelli2012indirect} for a recent review). Indirect searches try to detect an excess of ordinary particles not explained by other known physical mechanism.
\paragraph{}
The case of gamma-ray products is particularly interesting because of their directionality: it can give hints about dark matter decays if diffuse emission is detected where background is less intense and annihilation should come from the center of the galaxies, where dark matter density must be higher~\cite{gammabertone}. These type of products could be detected by Cherenkov telescopes or by space telescopes. Current data taking Cherenkov telescopes like HESS~\cite{abramowski2011hess}, MAGIC~\cite{aleksic2011searches} or VERITAS~\cite{aliu2012veritas} have not found any excess attributable to self-annihilation coming from analyzed galaxies. The future CTA (Cherenkov Telescope Array) could greatly improve the sensitivity in this search~\cite{actis2011design}. Fermi-LAT space telescope data have been claimed to contain a 130 GeV line that could be interpreted as a result of WIMP annihilation~\cite{weniger2012tentative} but further analyses by Fermi-LAT collaboration disfavor this interpretation~\cite{2013PhRvD..88h2002A}.
\paragraph{}
Neutrinos can also be used as signature from dark matter annihilation. The source of this annihilation can be the center of the galaxy or the Sun. The center of the galaxy has to have a higher density of dark matter and the Sun could scatter WIMPs and bound them gravitationally giving a signature of high energy neutrinos~\cite{neutrinosunbertone}. None of these two effects has been detected by IceCube neutrino telescope~\cite{abbasi2011search, abbasi2012multiyear}.
\paragraph{}
The presence of cosmic antimatter could also be a signature of dark matter annihilation~\cite{salati2010indirect}. Unexplained excess in positrons have been observed by several experiments: PAMELA~\cite{PhysRevLett.111.081102}, Fermi-LAT~\cite{PhysRevLett.108.011103}, and AMS-01~\cite{alcaraz2000leptons} among others. The new results for AMS-02 are inconclusive about the end of the positron spectrum that could be attributed to the total mass of the dark matter particle in its self-annihilation~\cite{PhysRevLett.113.121102}.
\section{Direct dark matter detection}
\label{sec:DirectDetection}
As it has be seen before, the dark matter candidates more suitable for their direct detection are axions and WIMPs. The detection of the axion has been proposed to be through the inverse Primakoff effect~\cite{sikivie1983experimental}, in which the axion is converted into a photon in presence of macroscopic magnetic fields. There have been experiments searching axions with photon regeneration technique (“shining light through a wall”)~\cite{arik2014search}, from galactic origin~\cite{asztalos2010squid}, and produced in the Sun~\cite{arik2014search}, all with negative results. A more sensitive helioscope (IAXO) has been proposed in order to continue the axion searches~\cite{armengaud2014conceptual}.
\paragraph{}
The Weakly Interacting Massive Particle (WIMP) has been regarded as a very well-motivated candidate because its existence would give the appropriate relic abundance as well as its natural emergence from SUSY models as seen above. The idea of their direct detection relies on the elastic scattering of the WIMPs with the detector nuclei giving a tiny energy deposition with low rate of interactions (see~\cite{baudis2012direct} and references therein). For this reason, a detector devoted to these searches has to have very low energy threshold and low radioactive background.
\subsection{WIMP detection rate}
\label{sec:WIMP_rate}
The rate for WIMP elastic scattering off detector nuclei can be expressed as~\cite{cerdeno2010direct}:
\begin{equation}
	\frac{dR}{dE_R} = N_N \frac{\rho_0}{m_W} \int_{v_{min}}^{v_{max}}  f(\textbf{v}) \, v \, \frac{d\sigma_{WN}}{dE_R} d\textbf{v} 
\end {equation}
being $N_N$ the number of target nuclei, $m_W$ the WIMP mass, $\rho_0$ the local WIMP density in the galactic halo, $\textbf{v}$ and $f(\textbf{v})$ the WIMP velocity and velocity distribution in Earth reference frame and $d\sigma_{WN}/dE_R$ the WIMP-nucleus differential cross-section. The energy transferred to the recoiling nucleus is:
\begin{equation}
	E_R = \frac{p^2}{2m_N} = \frac{m_r^2 v^2}{m_N}(1-cos\theta)\label{eq:recoil_energy}
\end {equation}
where $p$ is the momentum transfer, $\theta$ is the scattering angle in WIMP-nucleus center-of-mass reference frame, $m_N$ is the nuclear mass and $m_r$ is the WIMP-nucleus reduced mass ($m_r = \frac{m_N m_W}{m_N+ m_W}$). The minimum velocity needed to produce a recoil of $E_R$ energy is:
\begin{equation}
	v_{min} = \sqrt{\frac{m_N E_R}{2m_r^2}}
\end{equation}
and the maximum velocity is given by the escape WIMP velocity in the Earth's reference frame. The rate of detected interactions depends on several parameters from particle physics and nuclear physics to astrophysics. These parameters must be investigated in order to recognize patterns for a possible positive signature of detection.
\paragraph{}
Particle and nuclear physics can give inputs and constrains to some of the aforementioned parameters. In particular, WIMP mass and interaction strength can be loosely constrained given the largely different values predicted by different \emph{beyond the standard model} theories. The WIMP-nucleus differential cross-section encodes particle physics inputs (with many uncertainties) including WIMP interaction properties. It depends on the WIMP-quark interaction strength, which is calculated from the microscopic description of the model, and it is later promoted to WIMP-nucleon cross-section. The calculation of WIMP-nucleus cross-section needs the use of the hadronic matrix elements, which describe the nucleon content in quarks and gluons, introducing large uncertainties. In the case of spin-1/2 or spin-1 WIMP field, the WIMP-nucleus cross-section can be separated as a sum of a spin independent (SI) and a spin dependent (SD) terms:
\begin{equation}
	\frac{d\sigma_{WN}}{dE_R} = \left( \frac{d\sigma_{WN}}{dE_R} \right)_{SI} + \left( \frac{d\sigma_{WN}}{dE_R} \right)_{SD} = \frac{m_N}{2m^2_rv^2}\left(\sigma^{SI}_0 F^2_{SI}(E_R) + \sigma^{SD}_0 F^2_{SD}(E_R)\right)
\end {equation}
where $F(E_R)$ are the nuclear form factors that encode the momentum transfer and depend on the recoil energy. $\sigma_0^{SI,SD}$ are the WIMP-nucleon cross-sections in zero momentum transfer limit. They can be expressed as:
\begin{align}
\begin{split}
	\sigma^{SI}_{0} = \frac{4 m_{rn}^2}{\pi} \left( Zf_p + (A-Z)f_n\right)^2 \\
	\sigma^{SD}_{0} = \frac{32 m_{rn}^2}{\pi} G^2_F \frac{J+1}{J}\left(a_p \langle S_p\rangle + a_n \langle S_n\rangle \right)
\end{split}
\end{align}
with $G_F$ being the Fermi coupling constant, $m_{rn}$ the nucleon-WIMP reduced mass, $f_{p,n}$ and $a_{p,n}$ the effective WIMP coupling to protons and neutrons in spin-independent and spin dependent, respectively. $\langle S_{p,n}\rangle$ are the expectation values of proton and neutrons spin operators in the limit of zero momentum transfer, and it can be determined using detailed nuclear model calculations. 
\paragraph{}
The astrophysics input is also subject to uncertainties~\cite{green2012astrophysical,read2014local}. In particular, WIMP velocity distribution $f(\textbf{v})$ and local dark matter density $\rho_0$ will translate their uncertainties into the predicted event rate. The velocity distribution is usually modeled with the so-called standard halo model describing an isotropic, isothermal sphere of collissionless particles with density profile $\rho_r \propto r^{-2}$ and a Maxwellian velocity distribution. Other halo models use a multivariate Gaussian velocity distribution that is in better agreement with simulations at galaxy scales~\cite{vogelsberger2009phase} but due to the spatial resolution of this simulations the structure at local scale is still an open issue. From the experimental point of view, the GAIA satellite~\cite{perryman2001gaia} data will largely improve the knowledge of the dark matter distribution in the Milky Way. For a recent review of the local dark matter density from simulations to current experimental data and future prospects see~\cite{read2014local}.
\paragraph{}
Some signatures in the detected signal can be predicted given the previous input parameters. In particular, the shape of the recoil energy spectrum greatly differs with the WIMP and nucleus mass ratio. Taking into account the Equation~\ref{eq:recoil_energy} some extreme scenarios can be explored: $m_W << m_N$ gives $E_R \propto m_W^2 $ but $m_W >> m_N$ shows an energy spectrum independent of the WIMP mass. The WIMP mass can be more accurately determined when its mass is comparable with the target nucleus and using different targets with different nucleus mass can help in providing better constrains on $m_W$~\cite{pato2011complementarity}. Anyway, the expected recoil spectrum will have a shape $\propto e^{-E_R}$ for the standard halo model~\cite{lewin1996review}.
\paragraph{}
Other very important expected signature is the so-called annual modulation, i.e. a seasonal variation of the total event rate~\cite{PhysRevD.33.3495,freese2012annual}. This signal arises because of the Earth's motion in the galactic rest frame. This motion is a superposition of Earth's rotation around the Sun and the Sun's rotation around the galactic center. Considering the standard halo model, the amplitude of the modulation will be small taking into account the ratio between velocities (of the order of $v_{orb}/v_c \simeq 0.07$ being $v_{orb}$ the Earth's orbital speed and $v_c$ the Sun's circular speed). Following this model the differential rate can be approximated to:
\begin{equation}
	\frac{dR}{dE_R}(E_R,t) \simeq \frac{dR}{dE_R}(E_R)\left[ 1 + \Delta(E_R) \mathrm{cos}\frac{2\pi(t-t_0)}{T}\right]
\end {equation}
where the period $T$ = 1 year and the phase is $t_0$ = 150 days. $\Delta(E_R)$ is positive for high energy recoils and it becomes negative for small values. This feature implies a modulation phase change: the peak rate is in winter for small recoil energies and in summer for larger recoils~\cite{primack1988detection}. The energy at which the phase changes is commonly referred as crossing energy and it can be used to determine the WIMP mass~\cite{PhysRevD.70.043501}, although this technique will require an ultra-low energy threshold. The previous analyses have been performed with the standard halo model. The effect in the annual modulation for non-standard haloes has been also explored. Introducing the effect of a dark disk gives a second low energy maximum in the amplitude of the modulation and a significant phase change~\cite{green2010dependence}. The effect of a dark matter stream can also affect the shape of the modulation introducing some non-cosine effects at some energies~\cite{bernabei2006investigating,savage2006annual}.
\paragraph{}
A stronger signature would come from the ability to detect the axis and direction of the recoiling nucleus. The direction of the dark matter should vary in the laboratory reference frame with the motion of the Earth through the Galaxy and pointing in the direction of the motion of the Sun, to the Cygnus galaxy. Therefore the recoils should peak in the opposite direction~\cite{spooner2010direct}. The experimental challenge is to build massive enough detectors sensitive to the direction of the incoming WIMP.
\subsection{Background sources}\label{sec:Backgrounds}
Minimizing and characterizing the background noise is a necessary and very challenging task for a WIMP detector due to the expected signal detection rates~\cite{baudis2012direct}. Low background techniques are used in order to achieve as much sensitivity as possible to the previously described signatures~\cite{heusser1995low}. The main contributions to this background come from the environmental radioactivity, the cosmic rays and their secondaries and activation at Earth surface of the detector materials. Other more subtle but potentially harmful sources are intrinsic activity of the detector materials and muon induced neutrons. The ultimate limit in the background will come from the neutrino flux and will start to play an increasing important role in future experiments because of its irreducibility.
\paragraph{}
The cosmic rays background can be highly reduced by locating the detector underground. The hadronic component becomes negligible by tens of meters water equivalent overburden, but muon flux is more difficult to attenuate producing high energy fast neutrons. These neutrons can produce keV nuclear recoils in elastic scatter with detector nuclei~\cite{PhysRevD.73.053004}. Thus, active veto detectors are necessary to reduce background by tagging the original muon or its associated cascade. The muon flux underground also has seasonal variations of amplitude~\cite{bellini2012cosmic}. To avoid the induced muon signal mimicking a possible WIMP signature is an additional reason to keep muon flux under control. Other measures such as passive gamma and neutron shieldings are needed to isolate the detector from the environmental background. This shielding is typically flushed by boiled-off $N_2$ in order to suppress airborne radon decays.
\paragraph{}
The internal background of the detector materials is a very dangerous contribution to the sensitivity. The neutron spectra coming from natural chains can be calculated by measuring the amount of $^{238}U$ and $^{232}Th$. The signal of these neutrons may be difficult to distinguish from a potential WIMP signal~\cite{mei2009evaluation}. Other sources of intrinsic internal background of some WIMP targets include $^{39}Ar$, $^{85}Kr$ and $^{40}K$. In addition to these contributions, the decay of long lived isotopes produced by cosmic rays in target and detector materials during their exposure at Earth's surface, known as cosmogenic radionuclides, can be a relevant and harmful fraction of the total background.
\paragraph{}
Additionally to the use of shielding and radiopure detectors the use of particle discrimination can be very useful to reduce the background. In particular, the ability of perfectly distinguish nuclear from electronic recoils could separate all $\gamma/\beta$ background from neutron and WIMPs contribution, with a great background improvement.
\paragraph{}
\subsection{Direct detection techniques, current status and future prospects}
\begin{figure}[h!]
  \begin{center}
	  \includegraphics[width=1\textwidth]{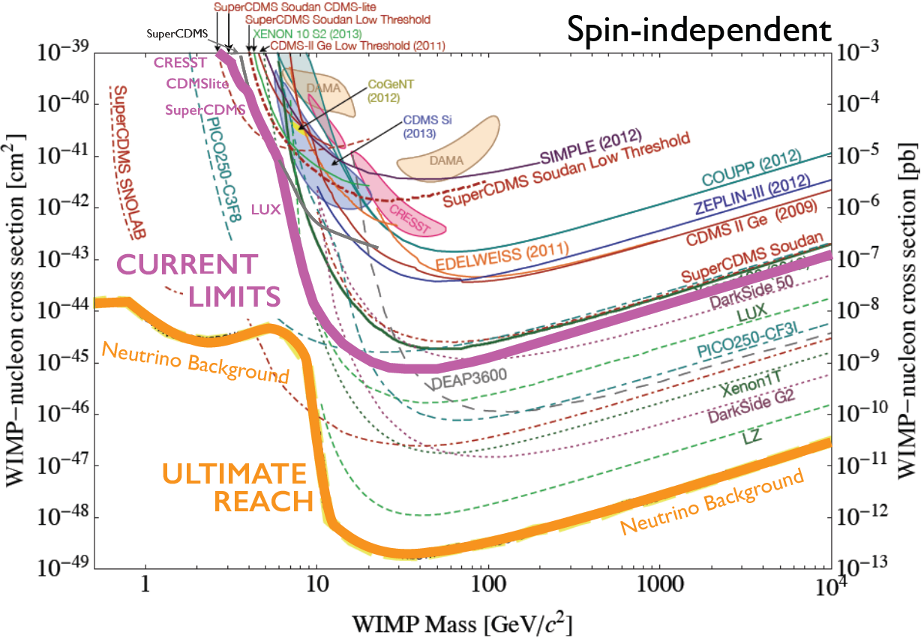}
	  \caption[Spin independent exclusion plot ]{Spin independent exclusion plot with current measured better limits (in pink), current experimental individual results (solid lines) and prospect projections (dashed lines)\label{fig:DMExclusionExp}. From~\cite{gondolo2014talk}.}
  \end{center}
\end{figure}

\begin{figure}[h!]
  \begin{center}
	  \includegraphics[width=1\textwidth]{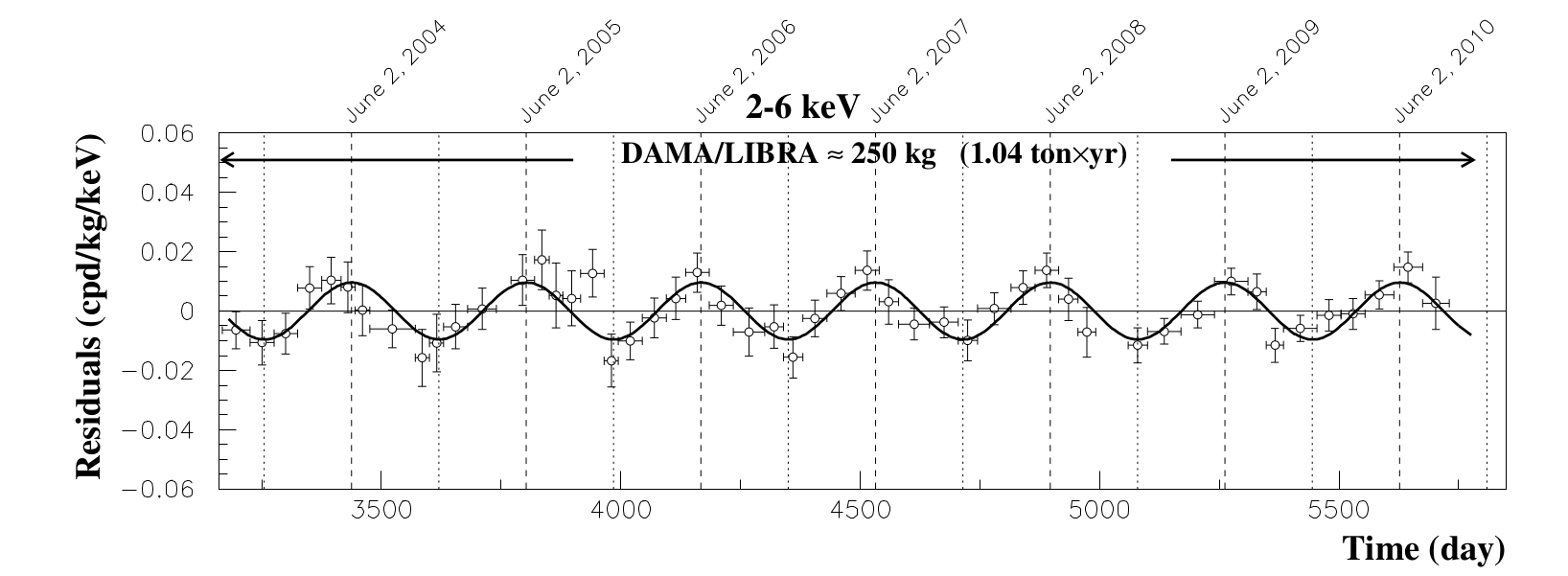}
	  \caption[DAMA/LIBRA residual rate and fit]{DAMA/LIBRA residual rate of \emph{single hit} scintillation event in the 2-6 keV energy window. From~\cite{bernabei2013final}\label{fig:DAMAModulation}.}
  \end{center}
\end{figure}

To observe a WIMP-induced spectrum a low energy threshold, an ultra-low background noise and a large target mass are mandatory. A detector must convert the scattered nucleus kinetic energy to a mensurable signal such as ionization, scintillation light or heat. The simultaneous detection of two different channels of energy conversion gives a powerful discrimination against background events by determining the nature of the incident particle.
\paragraph{ }
The experimental search for the dark matter started in the 1980s profiting the underground facilities and the low background techniques developed for the search of the neutrinoless double beta decay. The first relevant experimental result was to discard heavy standard Dirac neutrinos as candidate to cold dark matter using $Ge$ detectors~\cite{ahlen1987limits}. In the 1990s, NaI scintillators started to be used given its ability to be sensitive to spin-dependent interactions and the feasibility of growing big mass crystals. More recently, the solid-state and noble gases detectors have been the most massive and sensitive experiments, as we will see later. The comparison of the main recent experimental results in the latest years (solid lines) and prospects (dashed lines) can be observed in Figure~\ref{fig:DMExclusionExp} in terms of exclusion plots in the cross-section/WIMP mass space for spin-independent interaction, including the regions singled out by the DAMA/LIBRA positive signal.
\paragraph{}
Solid-state cryogenic detectors operated at sub-Kelvin temperatures have been the most traditional detectors devoted to the direct dark matter detection due to their low threshold ($<$ 10 keV), excellent resolution ($<$ 1\% at 10 keV) and the ability to discriminate nuclear from electronic recoils~\cite{booth1996low}. CDMS collaboration has run several experiments consisting of semiconductors operated at 40 mK. During 2003-2008 CDMS-II consisted of 4.6 kg of $Ge$ detectors and 1.2 kg of $Si$ detectors reporting an excess of three events in the $Si$ detectors~\cite{agnese2013silicon} and no excesses nor modulation in the $Ge$ detectors~\cite{ahmed2012search}. The following experiment upgrade at SOUDAN underground laboratory with 9 kg of new designed $Ge$ detectors, SuperCDMS, has presented data analysis lowering the previous exclusion limits~\cite{agnese2014search}. The same collaboration has also run another experiment CDMSLite. It takes advantage of the Luke-Neganov amplification of the phonon signal but without discrimination between electronic and nuclear recoils, however it lowers the threshold excluding more low-mass WIMP parameter space~\cite{cdmslite2014}. The next upgrade of SuperCDMS at SNOLAB will have around 100 Kg of germanium and 10 Kg of silicon detectors~\cite{cerdeno2014talk}.
\paragraph{ }
The EDELWEISS experiment is also a cryogenic germanium detector measuring ionization and phonon signal. The EDELWEISS-II upgrade consisted of 4 kg of $Ge$ detectors finding an excess of five events with an estimate of three~\cite{ armengaud2011final}. The same set-up was used to explore the low-mass WIMP parameter space~\cite{armengaud2012search}. The next upgrade called EDELWEISS-III~\cite{armengaud2013background} will consist of forty detectors with improved technology and radiopurity accounting 24 kg of fiducial mass. 
\paragraph{ }
The CRESST collaboration has designed and operated several cryogenic detectors with simultaneous phonon and photon detection. The CRESST-II experiment consist of eight detector modules of 300 g each reporting sixty-seven events in the nuclear recoil region~\cite{angloher2012results} not explained by known backgrounds. The last results of CRESST-II including a new detector design have not confirmed the aforementioned excess~\cite{angloher2014results}. This collaboration has planned to merge with EDELWEISS collaboration to perform a high scale cryogenic experiment called EURECA~\cite{kraus2010eureca}.
\paragraph{ }
Germanium ionization detectors operated at 77 K can reach sub-keV energy threshold and low backgrounds~\cite{barbeau2007large} but without the ability of distinguish electronic from nuclear recoils. This type of experiments includes CoGeNT~\cite{aalseth2011search}, TEXONO~\cite{li2013limits}, MALBEK~\cite{aalseth2011astroparticle} and CEDEX~\cite{zhao2013first}. Unlike the other p-type point-contact germanium experiments, CoGeNT has reported the presence of an annual modulation in the event rate~\cite{aalseth2011search,aalseth2014search} but it is worth to mention that recent analyses result in contradictory conclusions~\cite{aalseth2014maximum,davis2014quantifying,kelso2014talk}.
\paragraph{}
Liquid noble elements can be used for building self-shielding detectors. Liquid Xenon (LXe) and liquid Argon (LAr) are good scintillators and ionizers in response to radiation. The simultaneous scintillation and ionization detection allows to identify the primary interacting particle in the liquid. In addition, the 3D position of an interaction can be determined in a time projection chamber (TPC). These features, together with the relative ease of scale-up to large masses, have contributed to make LXe and LAr powerful targets for WIMP searches~\cite{aprilebaudisbertone} giving the most stringent limits to spin-independent couplings as it can be seen in Figure~\ref{fig:DMExclusionExp}. The LUX experiment, a double phase liquid/gas $Xe$ TPC experiment at Sanford Underground Research Facility, has currently set the best limits in the high WIMP masses zone with data corresponding to 85.3 live days with a fiducial volume of 118 kg~\cite{akerib2014first}. Other $Xe$ TPC with a good experimental limit is XENON100~\cite{aprile2012dark}. Some ton or multiton-scale Xe TPC experiments have been proposed~\cite{baudis2012direct} some of them such as XENON1T~\cite{aprile2013xenon1t} in commissioning stage. In addition to $Xe$ experiments, some LAr experiment are taking data. DarkSide-50 has presented the first results~\cite{agnes2014first} reporting the presence of cosmogenic $^{39}Ar$ and considering the use of underground $Ar$ with a factor 150 of $^{39}Ar$ below atmospheric argon. ArDM, a double phase $Ar$ TPC experiment is currently taking data at the Canfranc Underground Laboratory (LSC)~\cite{badertscher2013ardm}.
\paragraph{ }
Studying spin-dependent channel requires target nuclei with uneven total angular momentum. A very favorable candidate is $^{19}F$ exhibiting a cross-section a factor ten above the other used nuclei because of its unpaired proton. This target is present in WIMP detectors using superheated liquids composed of $C_4F_{12}$, $CF_3I$ or $C_2ClF_5$. An energy deposition can move the system from its metastable state forming bubbles that can be detected both acoustically and optically. The temperature and pressure of the detector can be adjusted to lead to the formation of bubbles only for nuclear recoils given the fact that the threshold to produce bubbles depends on a \emph{critical energy} deposited within a \emph{critical radius}, both dependent on pressure and temperature~\cite{boukhira2000suitability}. Additional discrimination of nuclear recoils from alpha particles has been developed using the acoustic signal~\cite{aubin2008discrimination}. Several low target mass experiments have been carried out. PICASSO~\cite{archambault2012constraints} consisted of $C_4F_{10}$ droplets in 4.5 liters of polymerized emulsion with a total mass 0.72 kg of $^{19}F$ and an exposure of 114 kg-day without positive result. COUPP-4~\cite{behnke2012first} consisted of 4 kg of $CF_3$ for an effective exposure of 437.4 kg-day lowering the exclusion limits except in the low mass region. Both collaborations joined efforts in a new PICO collaboration. Results from COUPP-60, a 40 liter $CF_3I_4$ update of COUP-4, and PICO2L, a 2-liter $C_3F_8$ chamber currently running at SNOLAB, are expected in the short term~\cite{bou2014talk}. PICO-250L~\cite{bou2014talk} will be an experiment in the ton-scale range of target mass. Another technique being explored for this type of detectors is the geyser technique~\cite{bertoni2014new} that could eventually be applied in the MOSCAB experiment.
\paragraph{}
Scintillation radiation detectors have been also used in the search of dark matter from the 1990s as seen earlier. DAMA/LIBRA, a 250 kg Thallium doped Sodium Iodine experiment, has observed an annual variation of \emph{single-hit} events with a total exposure of 1.33 ton$\times$year (combining the results with the previous phase of the experiment, DAMA/NaI), corresponding to 14 annual cycles and a statistical significance of 9.3 $\sigma$ in the 2-6 keV energy window~\cite{bernabei2008first,bernabei2010new,bernabei2013final}. In Figure~\ref{fig:DAMAModulation} it can be seen the seven cycles corresponding to the DAMA/LIBRA-phase1 set-up. The modulation is not present at higher energy regions or in coincidence (\emph{multiple-hit}) events.

\paragraph{}
The difficulty to find dark matter candidates explaining all experimental results~\cite{frandsen2012resolving}, the dependence on WIMP and halo models considered to compare different targets~\cite{cerdeno2013complementarity} along with the uncertainties in nuclear recoil calibration (see Section~\ref{sec:NaINuclRecoils}) and the partial understanding of experimental backgrounds at low energy~\cite{kudryavtsev2010expected,bernabei2010technical,nygren2011testable,bernabei2012no,kuzniak2012surface} make highly interesting confirming the DAMA/LIBRA annual modulation in a model independent way. The need for this confirmation, with the same target and technique, lead to several groups to perform large mass NaI(Tl) experiments: ANAIS~\cite{amare2014anais}, DM-Ice~\cite{cherwinka2014first} and KIMS~\cite{kim2014tests}. KIMS group already carried out an experiment with 103.4 kg of CsI(Tl) scintillators with 25 ton$\times$day exposure and without any signal of annual modulation~\cite{kim2012new}.
\section{NaI(Tl) as target for WIMP searches}
\label{sec:NaINuclRecoils}
\begin{figure}[h!]
  \begin{subfigure}[b]{0.5\textwidth}
  \includegraphics[width=1\textwidth]{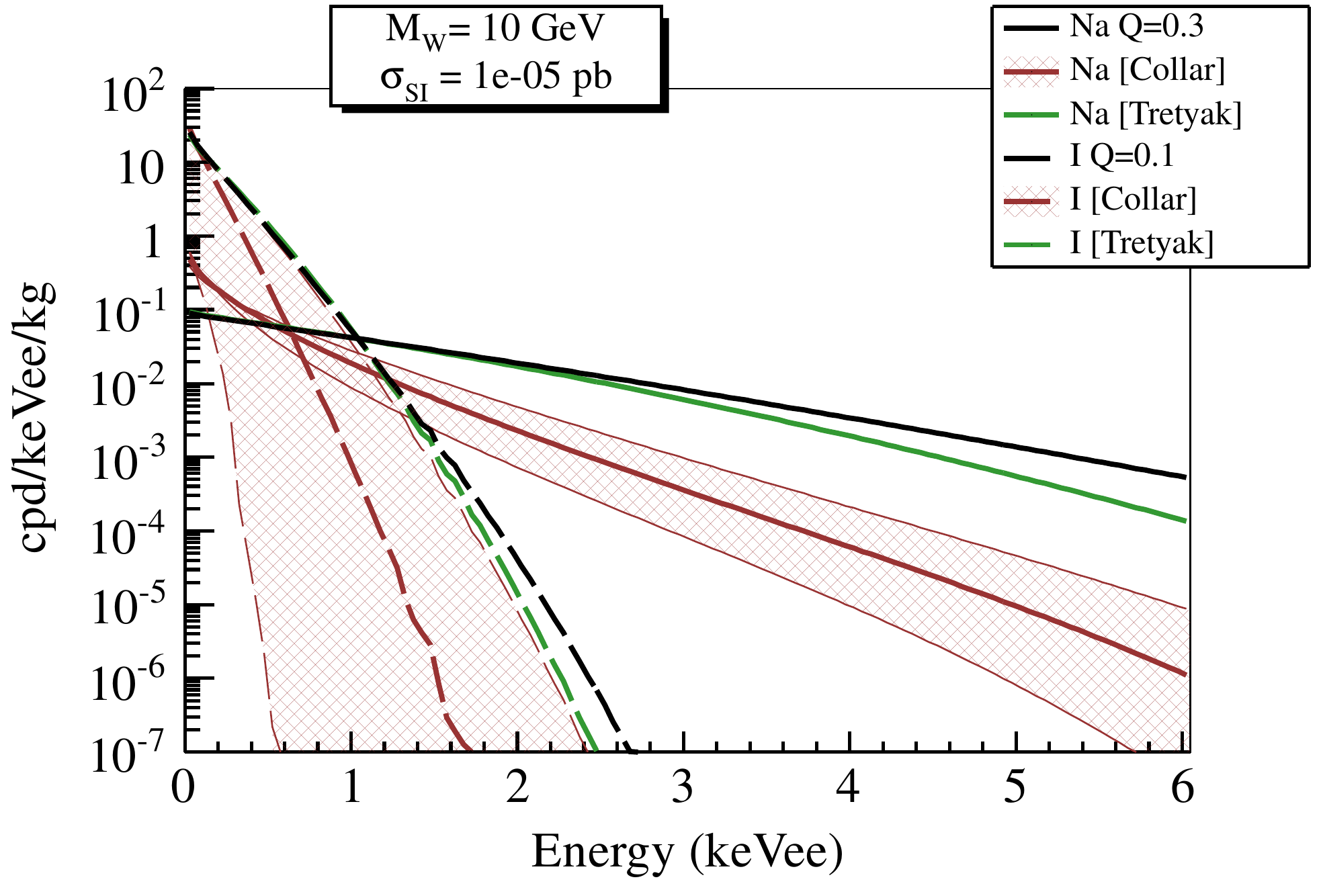}
  \caption[]{}
  \end{subfigure}
        ~ 
  \begin{subfigure}[b]{0.5\textwidth}
  \includegraphics[width=1\textwidth]{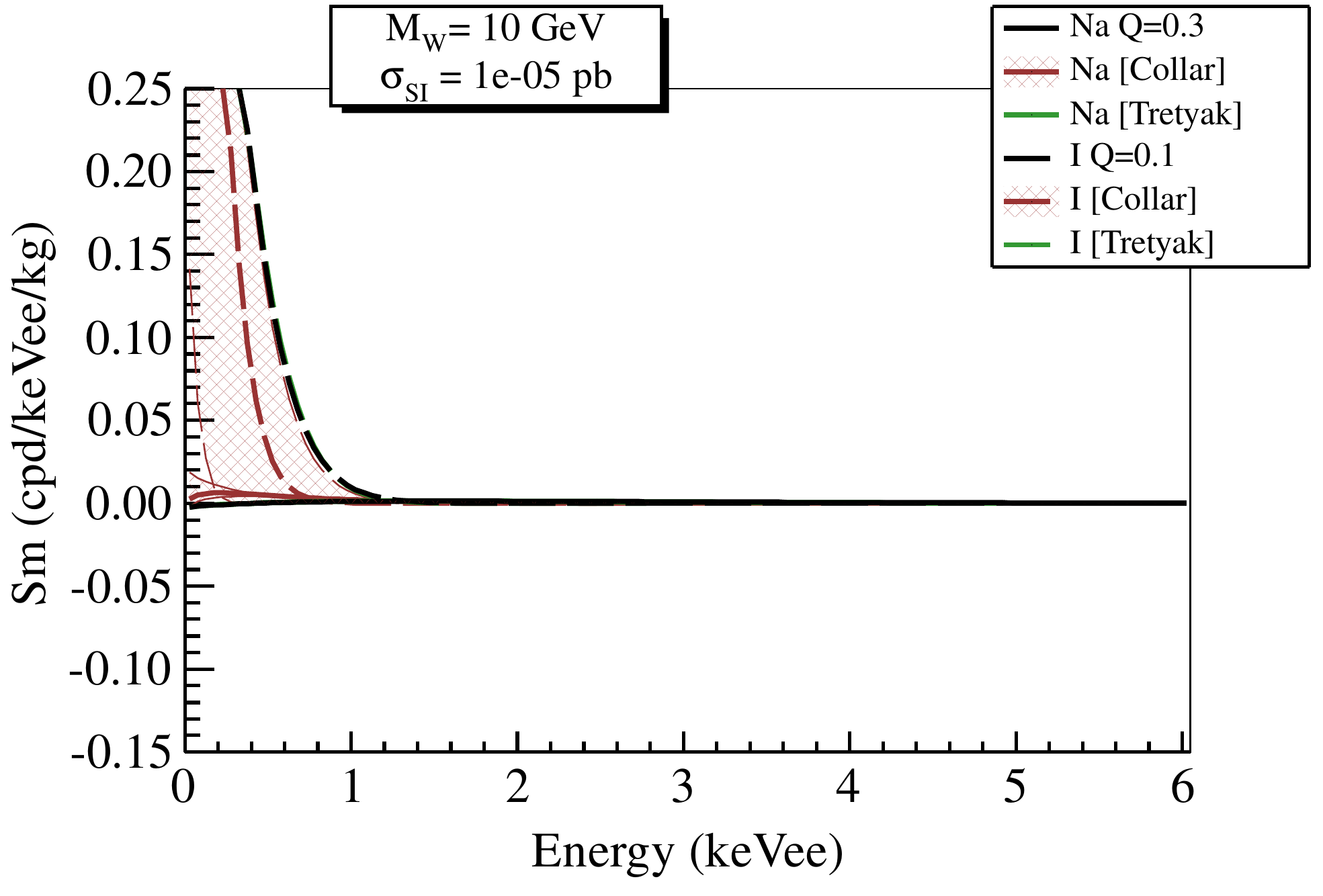}
  \caption[]{}
  \end{subfigure}
  
  \begin{subfigure}[b]{0.5\textwidth}
  \includegraphics[width=1\textwidth]{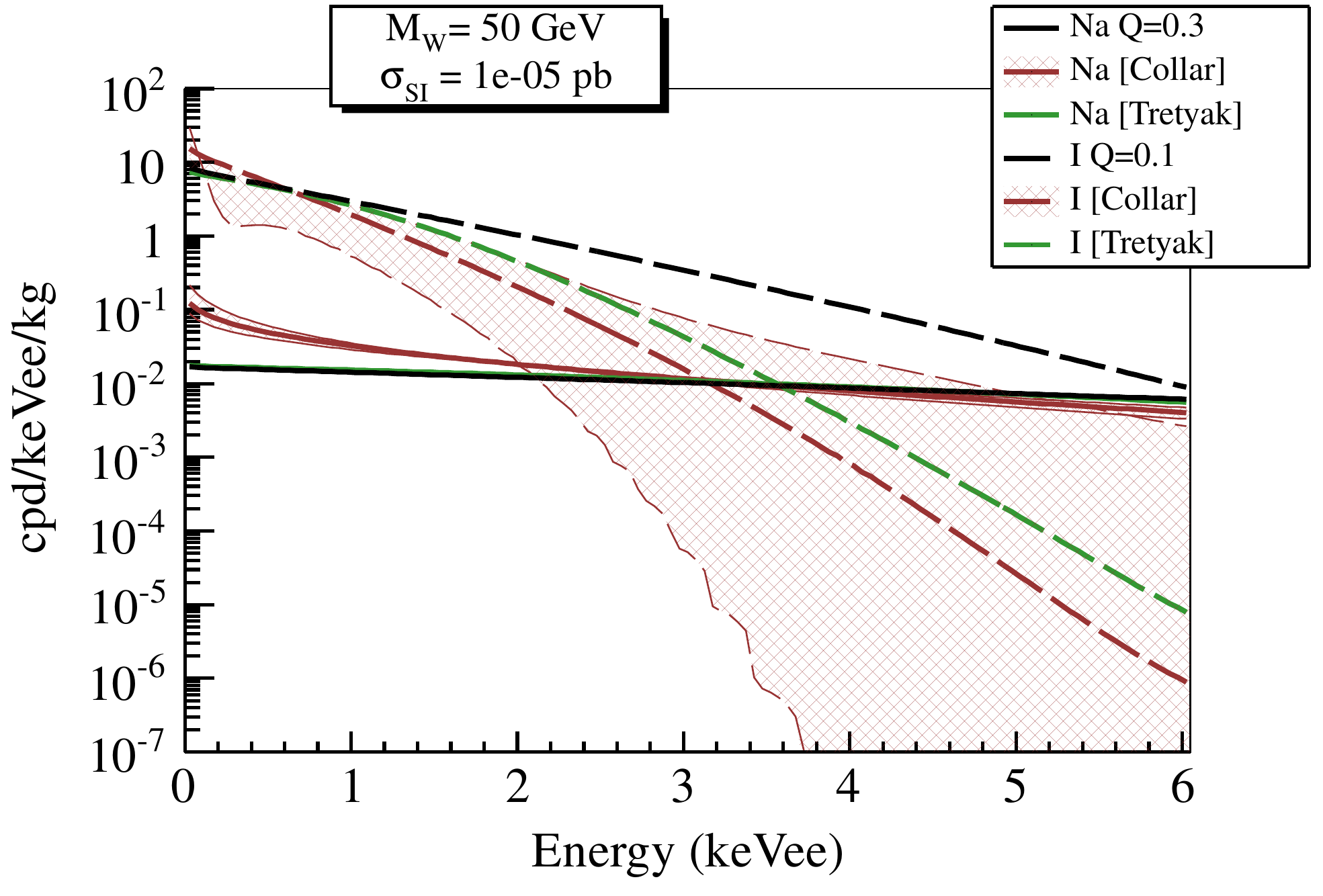}
  \caption[]{\label{fig:ANAIS25RnBox}}
  \end{subfigure}
  ~ 
  \begin{subfigure}[b]{0.5\textwidth}
  \includegraphics[width=1\textwidth]{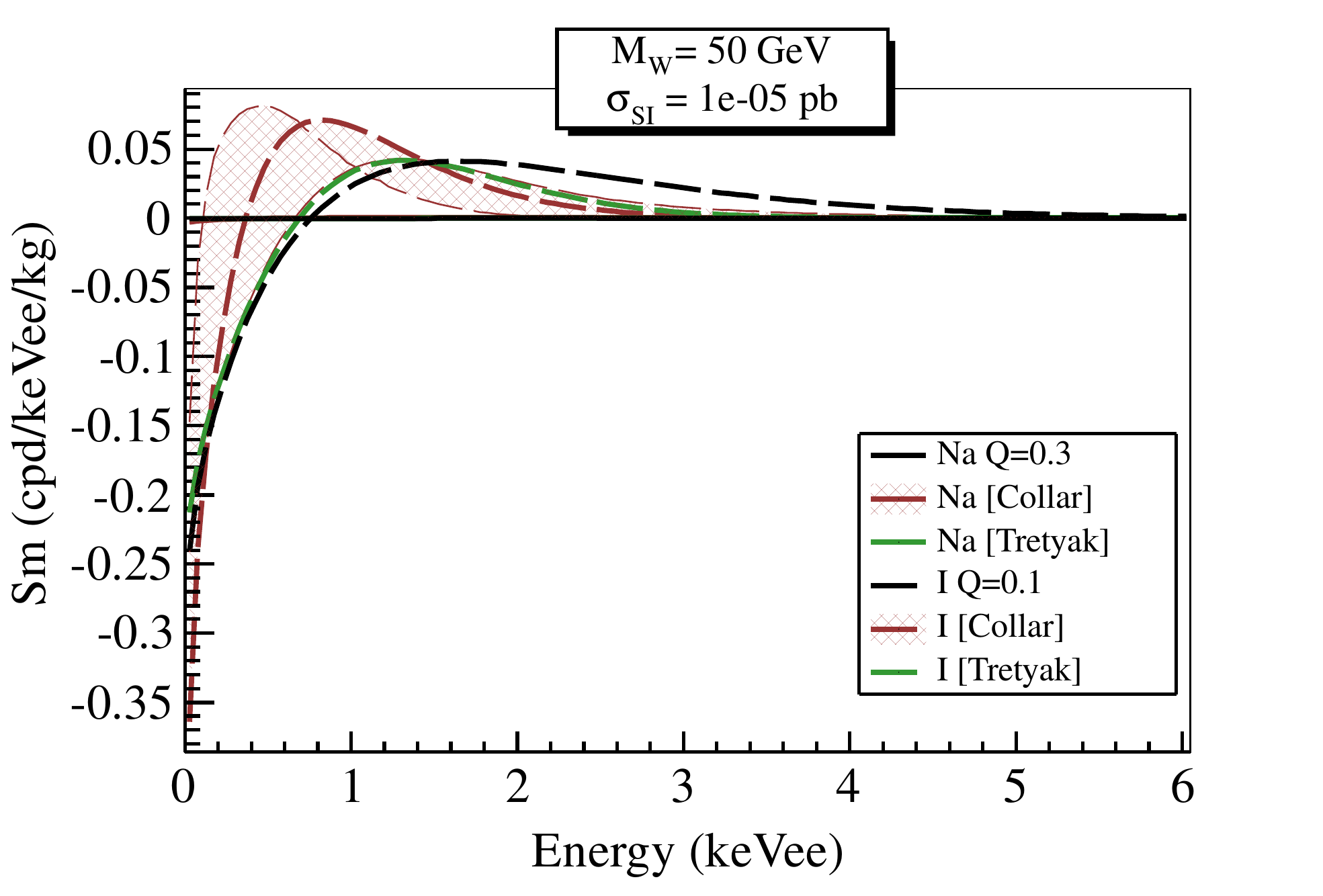}
  \caption[]{}
  \end{subfigure}
  \begin{center}
\caption[Expected NaI WIMP Rates]{Expected WIMP rates (left) and modulation amplitude (right) with NaI target using 10 GeV (top) and 50 GeV (bottom) as WIMP mass and $\sigma_{SI} = 10^{-5} pb$. From~\cite{CCUESTA}.}
\label{fig:NaIWIMPRates}       
  \end{center}
\end{figure}

The first feature to be aware for the use of NaI(Tl) for measuring nuclear recoils is the relative efficiency factor or quenching factor, usually denoted by Q. The quenching factor is defined as the relative light yield of nuclear recoils (the expected WIMP interaction, see Section~\ref{sec:WIMP_rate}) and electronic recoils. It is well known that different interacting particles with different interaction mechanism produce different amount of light for the same recoil energy. The Q value has been measured giving $\sim$ 0.3 for Na recoils and $\sim$ 0.1 for I recoils~\cite{spooner1994scintillation,gerbier1999pulse}. New measurements pointed Q energy dependence and a slight increase at low energies~\cite{chagani2008measurement, tretyak2010semi}, however recent measurements have been presented showing lower values for Na recoils in the low energy range and a strong energy dependence~\cite{collar2013quenching}.
\paragraph{ }
The NaI(Tl) detectors are usually calibrated with X-ray and gamma sources using their photopeaks, or peaks in the spectrum created by the total absorption of the photon energy. In order to not include the aforementioned Q uncertainties, these calibrations are usually expressed in electron equivalent energies (keVee or sometimes only keV).
\paragraph{}
These uncertainties have a major impact in the interpretation of a NaI(Tl) experiment because the energy and the energy threshold are expressed in electron equivalent energy. The impact of such uncertainties in the expected rate of WIMPs has been studied~\cite{CCUESTA} as can be seen in Figure~\ref{fig:NaIWIMPRates}. The quenching factor also has a key role in the comparison of NaI results with other experiments with different target and technique.
\paragraph{}
As a result of the study seen in Figure~\ref{fig:NaIWIMPRates} the expected rate for the amplitude of the annual modulation is lower than 0.05 count/(keVee kg day) above 1 keVee. It also illustrates the importance of a threshold as low as possible and the need for a clean background that does not mask such a subtle effect. 
\paragraph{}
Other NaI(Tl) feature to take into account is the pulse shape discrimination. Different interacting particles have different pulse shape (see~\cite{kudryavtsev1999characteristics} for example). However, the different shape can not be used as discrimination technique at low energies~\cite{gerbier1999pulse,miramonti2002study}. Without discrimination, the background at low energy must be as clean as possible because of their indistinguishability from nuclear recoils. Hence, K-shell electron binding energy following electron capture in \textsuperscript{40}K (3.2 keV) and \textsuperscript{22}Na (0.9 keV) adding counts to low energy are harmful for a NaI(Tl) experiment.
\paragraph{}
Once reviewed the context of the ANAIS experiment, the experiment itself is presented in the next chapter.
\chapter{The ANAIS experiment}\label{sec:ANAIS}
The ANAIS (Annual modulation with NaI Scintillators) experiment is intended, as stated in the previous chapter, to search for dark matter annual modulation with ultrapure NaI(Tl) scintillators in order to provide a model-independent confirmation of the signal reported by the DAMA/LIBRA collaboration using the same target and technique. It will consist of more than one hundred kilograms of NaI(Tl) and it will be conducted at the Canfranc Underground Laboratory (LSC) in Spain.  
\paragraph{}
The ANAIS experiment will be the conclusion of a large series of previous efforts dating back to the mid-nineties. The NaI32~\cite{sarsa1994dark, sarsa1996searching, sarsa1997results} experiment searched for dark matter at the LSC with NaI detectors. It used three BICRON hexagonal detectors operated and stored in underground since the late eighties. This experiment, amounting 32.1 kg of target mass, accumulated two years of data taking (1993-94). Bounds of WIMP masses and cross-section were established given the absence of positive signal, neither in total rate nor in a pioneer modulation analysis.
\paragraph{}
One of these detectors was modified and it was used to build the ANAIS prototype I~\cite{TesisSusana}, and, after decoupling the photomultiplier and removing the original encapsulation it was used in ANAIS II~\cite{TesisMaria} and III prototypes. The main result of the study of these prototypes was the measurements of the potassium content~\cite{cuesta2014analysis} of the bulk crystal being too high for a dark matter experiment. A new crystal was tested: a 9.6 kg NaI(Tl) parallelepipedic module bought to Saint Gobain, with the same shape and size of DAMA/LIBRA crystals~\cite{bernabei2008dama}, named ANAIS-0. Again, the result of the potassium content was too high to use it in an experiment devoted to search dark matter annual modulation~\cite{CCUESTA,cuesta2014analysis}. Nevertheless, many technical aspects of the set-up and the study of these prototypes were used in the set-up, taking data and data analysis of the next prototypes.
\paragraph{}
Several providers of purified NaI powder were contacted and several samples were analyzed in order to have a clean starting material~\cite{CCUESTA}. Alpha Spectra company~\cite{AlphaSpectra} provided a sample with low potassium content compatible with the background of the GIFNA HPGe test bench at LSC. Two modules were purchased for quality assessment in both radiopurity and light collection quality. These detectors formed the ANAIS-25 set-up described in Section~\ref{sec:ANAIS25} featuring an acceptable content in $^{40}K$ but a high $^{210}Pb$ contamination. The origin of such a contamination was identified at purification and growing processes and a new module was incorporated later to form ANAIS-37 set-up (see Section~\ref{sec:ANAIS37}). The reduction of the Pb content was checked in addition to the reproducibility of the previous properties such as the excellent light collection and potassium content.
\paragraph{}
Once successfully tested the last module and having good sensitivity prospects (yet conservative, see Section~\ref{sec:SensitProsp}) based on the measured backgrounds, the steps towards the ANAIS full experiment have being started. The phase-I will consist of nine modules accounting more than 100 kg of ultrapure NaI(Tl) and it is expected to be fully commissioned along 2016. The ANAIS experimental set-up is described in Section~\ref{sec:ANAISSetup}.
\section{ANAIS experimental requirements}
\label{sec:ExperimentalRequirements}
The expected signal of an annual modulation is subtle and the region of interest is at very low energies as described in Section~\ref{sec:NaINuclRecoils}. The rates are low and the annual modulation is expected to be less than a 10\% of the total rate (of the order of 0.05 count/(keVee kg day) above 1 keVee).
\paragraph{}
For these reasons, a threshold as low as possible is desirable, any noise near this threshold must be avoided and enough exposure to have statistical significance is required. Another very important requirement is stability. Since the experiment must take data for several annual cycles, the conditions under which these data has been acquired have to be as stable as possible.
\paragraph{}
These requirements, especially a low energy threshold, high exposure and stability are the framework in which the present work has been developed, having to fulfill them as a primary goal.
\subsection{Low energy threshold}
The energy threshold depends on some experimental aspects like the light collection efficiency and the hardware triggering strategy. The light collection efficiency is marked by the optical quality of the crystal and other optical components such as quartz windows, optical couplings and photomultiplier (PMT) quantum efficiency. The low energy events and their conversion into electrical signal throughout PMTs are covered in Chapter~\ref{sec:OpticalToElectrical} and light collection of the Alpha Spectra modules is analyzed in Section~\ref{sec:LY}. The hardware trigger efficiency at low energy has been studied in Section~\ref{sec:HWTriggEff}.
\paragraph{}
On the other hand, the triggering strategy must ensure the storage of all relevant events but without a high rate that could increase the dead time. This is achieved by coupling two PMTs to each crystal and triggering with the coincidence of the two PMT signals. Due to the nature of the low energy events, consisting on few separate photons, the trigger must be performed at photoelectron level. The trigger modules tests are presented in Section~\ref{sec:TestTriggerStrat} and the trigger at photoelectron level was evaluated in Section~\ref{sec:SERTrEff}. The trigger can be affected by electric noises at such low level and for this reason a baseline characterization and improvements were performed and are described in Section~\ref{sec:Baseline}.
\subsection{Low background at low energy}
The region of interest for the WIMP annual modulation in our detectors is below 6~keV. The known background must be kept as low as possible in this region. Hence, an ultrapure NaI powder must be used, a careful crystal growing and detector manufacturing and a very thorough selection of external radiopure materials must be done~\cite{CCUESTA}. In particular, the contamination of $^{40}$K and $^{22}$Na can be especially relevant because the K-shell electron binding following an electron capture are 3.2 keV and 0.9 keV respectively.
\paragraph{}
In addition to internal background, it is important to shield and control the possible influence of the external background. The passive shielding will consist of lead, boron-loaded water and polyethylene. An anti-radon box will keep the system tightly closed and flushed with boil-off Nitrogen. These techniques to avoid external background are described in Section~\ref{sec:ANAISSetup}.
\paragraph{}
Another relevant external background is the contribution of cosmic rays. For this reason, an underground experiment is mandatory and ANAIS will be carried out at the LSC. Anyway, a residual muon flux still survives the rock shielding. This flux can produce other particles, such as neutrons, in the shielding or the detector contributing to the low energy region. An active muon veto system is used to tag muons using plastic scintillators to reject these potentially harmful events. This system is described and characterized in Chapter~\ref{sec:Veto}.
\subsection{Stability}
A very important experimental aspect in an annual modulation search is the stability of the environmental and experimental parameters. The data will be taken for several annual cycles and the conditions under which they are taken must be as stable as possible. It is mandatory to check any seasonal variation that could affect to any key experimental aspect: gain, threshold or efficiency for example. Therefore, all relevant environmental and external parameters must be monitored in order to detect and report any anomaly and store data to allow correlation with detector data. This slow control system and the stability of the environmental parameters are described in Chapter~\ref{sec:Stability}. This chapter also covers the test of the stability of the data acquisition parameters such as the gain, the trigger level and the coincidence window. The stability of the real time clock and comparisons with the system time via Internet (NTP system) is reviewed in Section~\ref{sec:RealTimeTest}.
\subsection{Exposure maximization}
Another reason to require stability and robustness to the system is to maximize the exposure (see Section~\ref{sec:SensitProsp}). When the target mass is fixed, the exposure is marked by the effective measurement time, or \emph{live time}. Hence, the system has to maintain the dead time, the time after each event during which the system is busy and it is not able to record another event due to the acquisition process; the down time, maintenance or system malfunction time; or bad experimental conditions, due to ambient radon inside the shielding, electric noise, etc., as minimum as possible. A study about the dead time of the experiment and the strategies to keep it controlled can be seen in Section~\ref{sec:DeadTimeMeas} and collected statistics of live time, dead time and down time from the latest prototypes are presented in Section~\ref{sec:LiveTimeMeas}.
\section{ANAIS experimental set-up}
\label{sec:ANAISSetup}
The experimental set-up for the full ANAIS experiment has been designed taking into account the aforementioned experimental requirements and the experience of the previous prototypes.
\paragraph{}
The target mass will consist of nine ultrapure 12.5 kg NaI(Tl) detectors arranged in a $3\times3$ matrix. These detectors will be enclosed in several layers of shielding as represented in Figure~\ref{fig:ANAISSetup}, from inner to outer layer:
\begin{figure}[h!]
  \begin{center}
	  \includegraphics[width=1\textwidth]{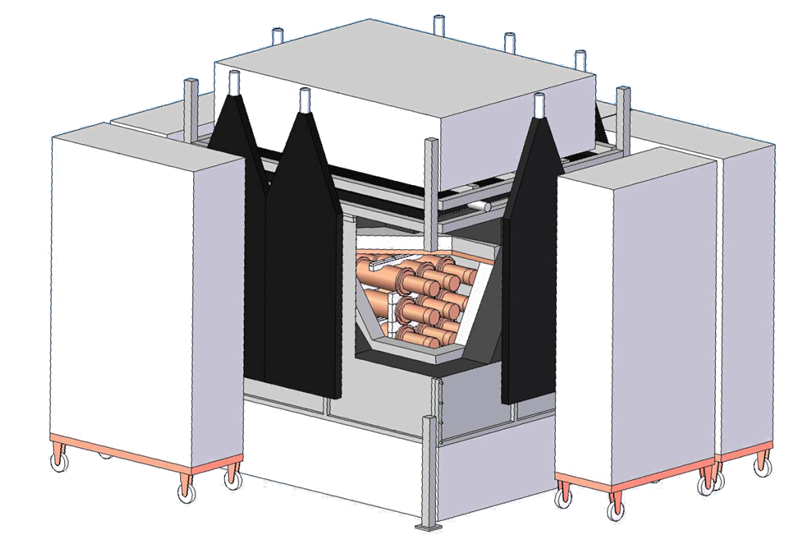}
	  \caption[Full ANAIS experiment set-up]{ANAIS experimental set-up artistic view.\label{fig:ANAISSetup}}
  \end{center}
\end{figure}

\begin{easylist}[itemize]
& 10 cm of ancient very low activity lead plus 20 cm of low activity lead to attenuate the $\gamma$ background. All lead required for the full experiment is already available underground at the LSC.
& A stainless steel anti-radon box that will tightly close the shielding to avoid the entrance of airborne radon inside the lead shielding. The box will be continuously flushed with radon free boil-off nitrogen.
& Plastic scintillators covering the top and the lateral faces of the anti-radon box in order to act as active shielding against muons and muon-related events as described earlier.
& The most external layer will consist of a combination of boron-loaded water and polyethylene blocks acting as neutron shielding.
\end{easylist}
\paragraph{}
The experimental set-up is completed by the electronic front-end, the data acquisition and the slow control systems. All of them are the evolution of the previous prototypes and their scale-up to the full experiment is a nuclear part of the work described in this thesis.
\section{ANAIS-25 set-up}
\label{sec:ANAIS25}
\begin{figure}[h!]
\begin{subfigure}[b]{0.5\textwidth}
  \begin{center}
  \includegraphics[width=1\textwidth]{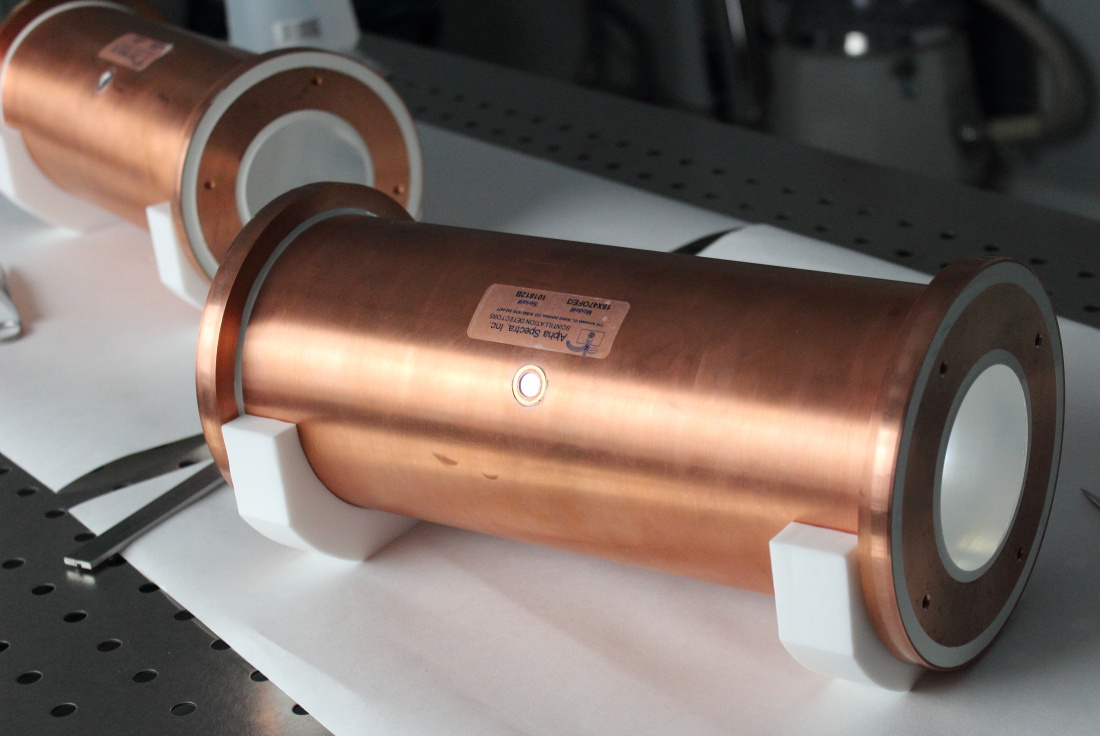}
  \caption[]{\label{fig:ANAIS25Det}}
  \end{center}
  \end{subfigure}
        ~ 
  \begin{subfigure}[b]{0.5\textwidth}
  \begin{center}
  \includegraphics[width=1\textwidth]{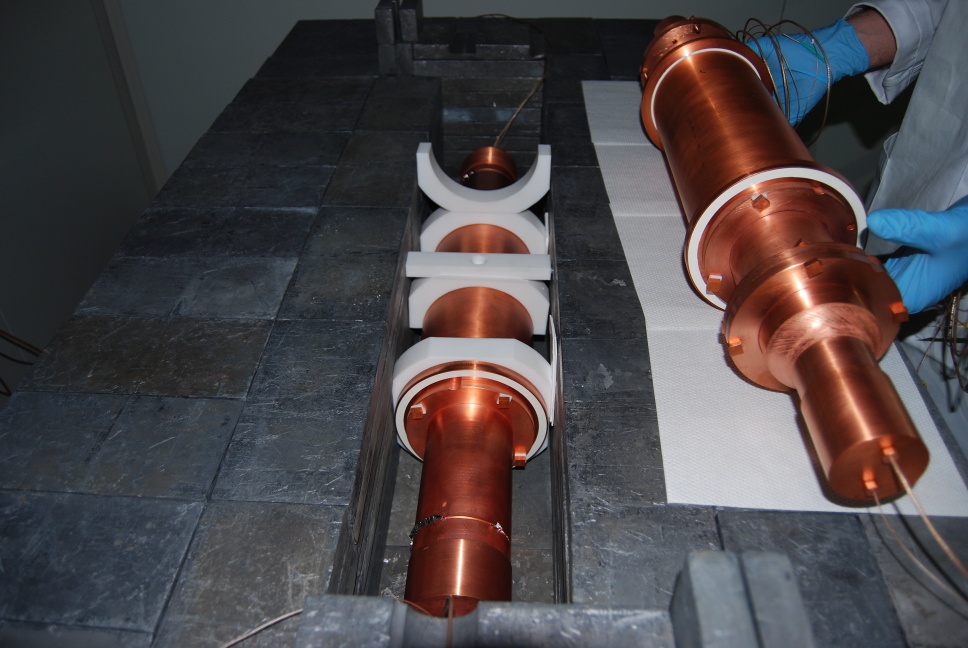}
  \caption[]{\label{fig:ANAIS25Lead}}
  \end{center}
  \end{subfigure}
  
  \begin{subfigure}[b]{\textwidth}
  \begin{center}
  \includegraphics[width=.5\textwidth]{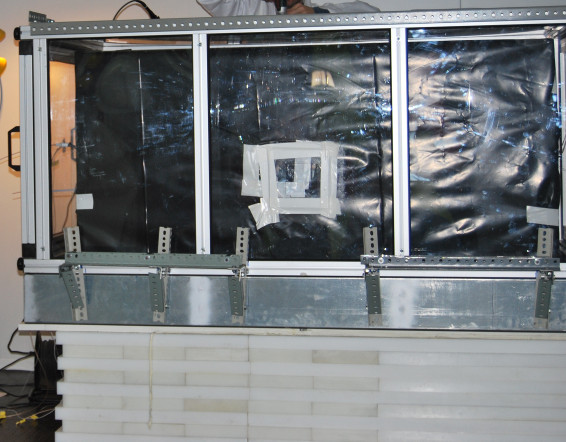}
  \caption[]{\label{fig:ANAIS25RnBox}}
  \end{center}
  \end{subfigure}
  \caption[ANAIS-25 set-up]{ANAIS-25 detectors without PMTs (a) and being introduced inside the shielding (b). The anti-radon box (c).}
\end{figure}

ANAIS-25~\cite{amare2014preliminary,CCUESTA} set-up consisted of two modules of 12.5 kg each (named D0 and D1) provided by Alpha Spectra. The main goals for this set-up have been to measure the crystal internal contamination, determine light collection efficiency, fine tune the data acquisition and test the filtering and analysis protocols. The modules are cylindrical, 4.75” diameter and 11.75” length, with quartz windows for PMTs coupling (see Figure~\ref{fig:ANAIS25Det}). A Mylar window in the lateral face allows for low energy calibrations. Two types of photomultipliers have been tested (Hamamatsu R12669SEL2 and Hamamatsu R11065SEL) but the model R12669SEL2 was finally selected. The studies referred as ANAIS-25 in this work correspond to data taken with four PMTs of this model.
\paragraph{}
The modules have been surrounded by a reduced version of the aforementioned shielding without boron-loaded water: 10+20 cm of lead, anti-radon box and plastic scintillators. The introduction of the modules inside the lead shielding can be observed in Figure~\ref{fig:ANAIS25Lead} and the anti-radon box in Figure~\ref{fig:ANAIS25RnBox}.

\begin{figure}[h!]
\begin{subfigure}[b]{0.5\textwidth}
  \begin{center}
  \includegraphics[width=1\textwidth]{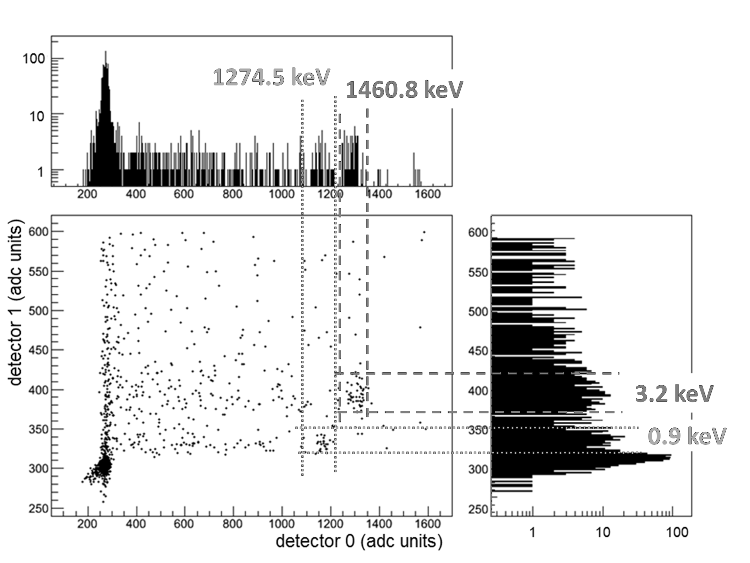}
  \caption[]{\label{fig:CoincThr}}
  \end{center}
  \end{subfigure}
        ~ 
\begin{subfigure}[b]{0.5\textwidth}
  \begin{center}
  \includegraphics[width=1\textwidth]{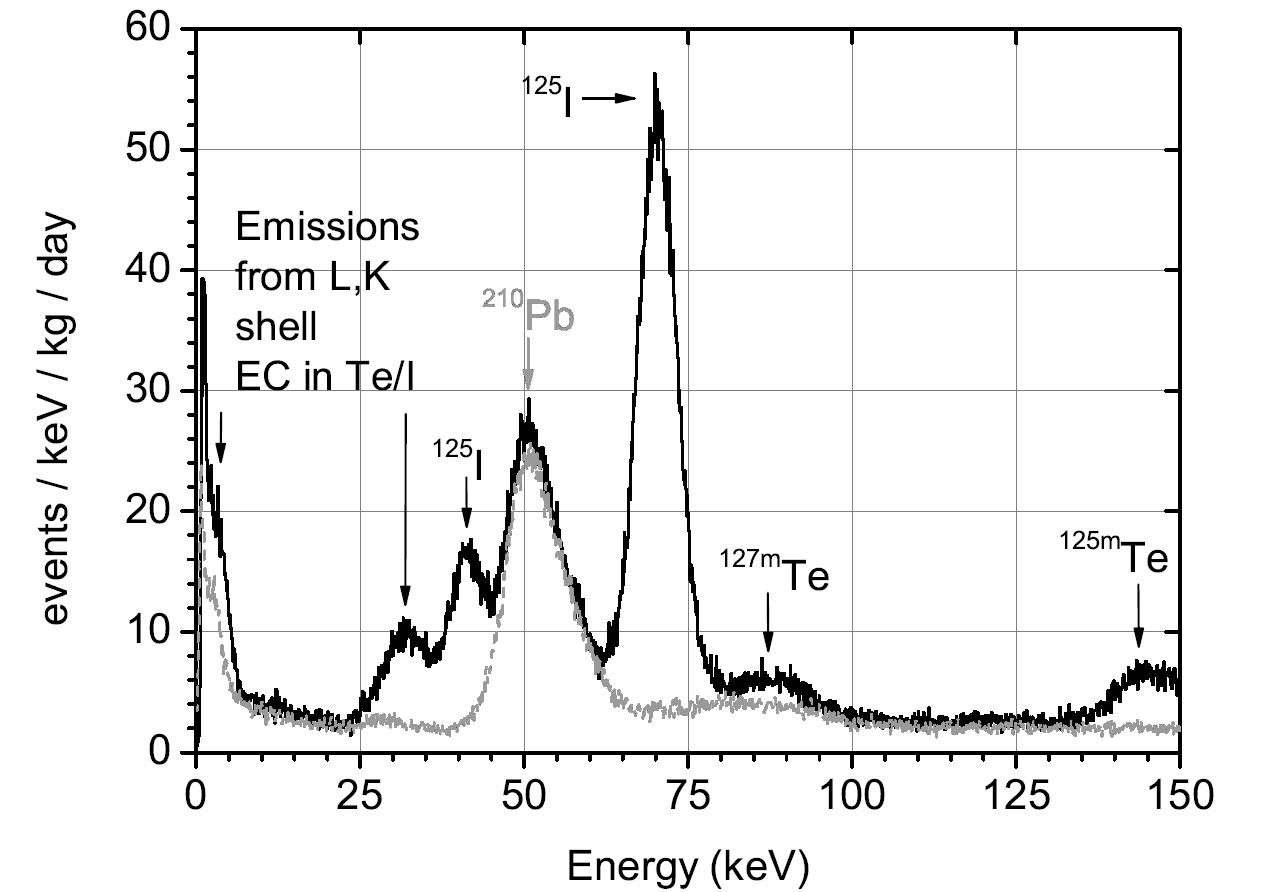}
    \caption[]{\label{fig:CosmoActiv}}
  \end{center}
  \end{subfigure}

\begin{subfigure}[b]{1\textwidth}
\begin{center}
  \includegraphics[width=0.5\textwidth]{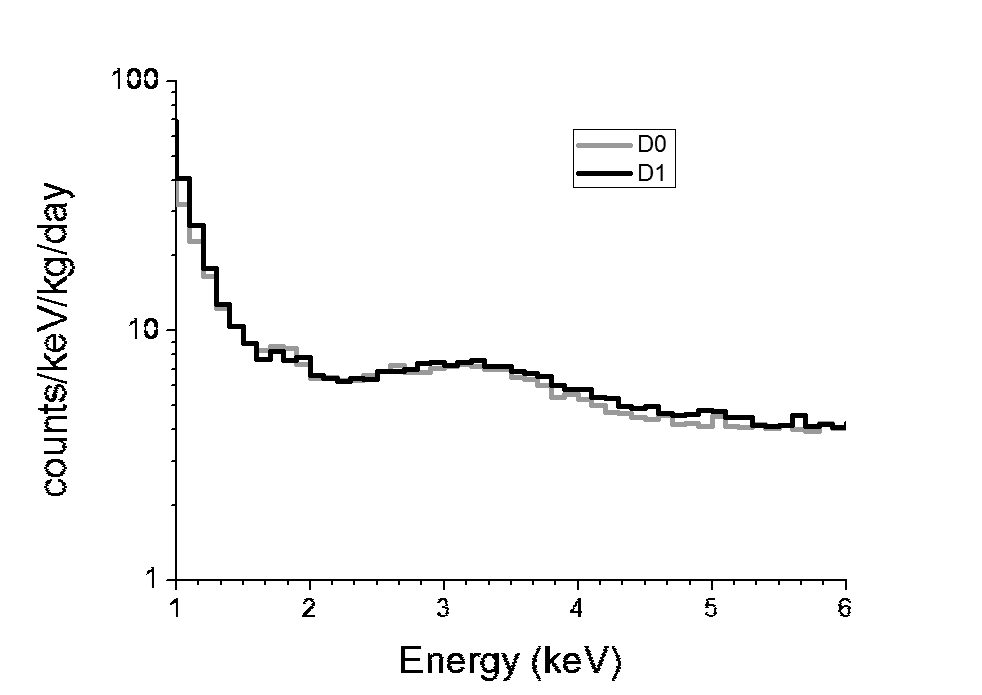}
  \caption[]{\label{fig:A25Bck}}
  \end{center}
  \end{subfigure}
  \caption[ANAIS-25 coincidence events and cosmogenic radionuclide decay]{ANAIS-25 Background. Coincidence scatter plot showing events from \textsuperscript{40}K and \textsuperscript{22}Na (a), cosmogenic radionuclide decay at low energy (b) showing the spectrum for the first month of data taking (black) and fifteen months later (gray) and filtered and efficiency corrected background  1-6 keV for D0 and D1  modules (c).}
\end{figure}

\begin{table}[h!]
\centering

\begin{tabular}{cccc}
\hline
\textsuperscript{40}K (mBq/kg) & \textsuperscript{238}U (mBq/kg) & \textsuperscript{210}Pb (mBq/kg)& \textsuperscript{232}Th (mBq/kg)\\\hline
1.25 $\pm$ 0.11 (41 ppb K) &  0.010 $\pm$ 0.002& $\sim $3.15 & 0.002$\pm$ 0.001 \\
\hline
\end{tabular}
\caption{NaI(Tl) crystals internal contamination measured in ANAIS-25 prototype.\label{tab:InternalContamination}}
\end{table}
\paragraph{}
This set-up was taking data from December 2012 to March 2015. The first feature to be remarked is the excellent light collection~\cite{CCUESTA}, tested again in this work as it can be seen in Section~\ref{sec:LY}. This light collection has a good impact in both resolution and energy threshold. A preliminary study has been done with the coincident events in both prototype modules showing two low energy populations that have been attributed to internal \textsuperscript{40}K and cosmogenic \textsuperscript{22}Na as it is shown in Figure~\ref{fig:CoincThr}. The K-shell electron binding energy following electron capture in \textsuperscript{40}K (3.2 keV) and \textsuperscript{22}Na (0.9 keV) can be tagged by the coincidence with a high energy $\gamma$ ray (1461 keV and 1274 keV respectively). Hence, a threshold of the order of 1 keVee seems to be achievable. The trigger and filtering efficiencies at the threshold level are currently under study.

\paragraph{}
On the other hand, background contributions have been thoroughly analyzed. Figure~\ref{fig:CosmoActiv} shows the low energy spectrum at the beginning of the data taking and fifteen months later, showing a high suppression of most of the lines except the corresponding to \textsuperscript{210}Pb, highlighting their cosmogenic origin; a more detailed study of radionuclide production in NaI(Tl) derived from this data can be found at reference~\cite{amare2014cosmogenic}. Table~\ref{tab:InternalContamination} shows the results of the activities determined for the main crystal contaminations. \textsuperscript{40}K content has been measured performing coincidence analysis between 1461 keV and 3.2 keV lines~\cite{CCUESTA, cuesta2014analysis}. The activities from \textsuperscript{210}Pb and \textsuperscript{232}Th and \textsuperscript{238}U chains have been determined by quantifying Bi/Po sequences and by comparing the total alpha rate with the low energy depositions attributable to \textsuperscript{210}Pb, which are fully compatible.
\paragraph{ }
These results give a moderate contamination of \textsuperscript{40}K, above the initial goal of ANAIS (20 ppb of K), but acceptable (see Section~\ref{sec:ANAISProspects}); a high suppression of \textsuperscript{232}Th and \textsuperscript{238}U chains, but a high activity of \textsuperscript{210}Pb at the mBq/kg level. The origin of such contamination was identified as radon contamination at powder purifying and crystal growing procedures and it was addressed by Alpha Spectra. A new module, grown following a new protocol, was available for radiopurity checks at the beginning of 2015. These measurements confirmed the better background as it can be seen in the next section.
\paragraph{}
Other goals of this set-up were fine tuning the data acquisition and testing the filtering and analysis protocols. The filtered background of ANAIS-25 modules is shown in Figure~\ref{fig:A25Bck}. The filtering process removes non-bulk scintillation based on reference~\cite{cuesta2014bulk}. It was adapted to ANAIS-25 and it is being improved with the knowledge of PMT only events. This process is described in Section~\ref{sec:EventSelection}.
\section{ANAIS-37 set-up}\label{sec:ANAIS37}
\begin{figure}[h!]
  \begin{center}
  \includegraphics[width=0.7\textwidth]{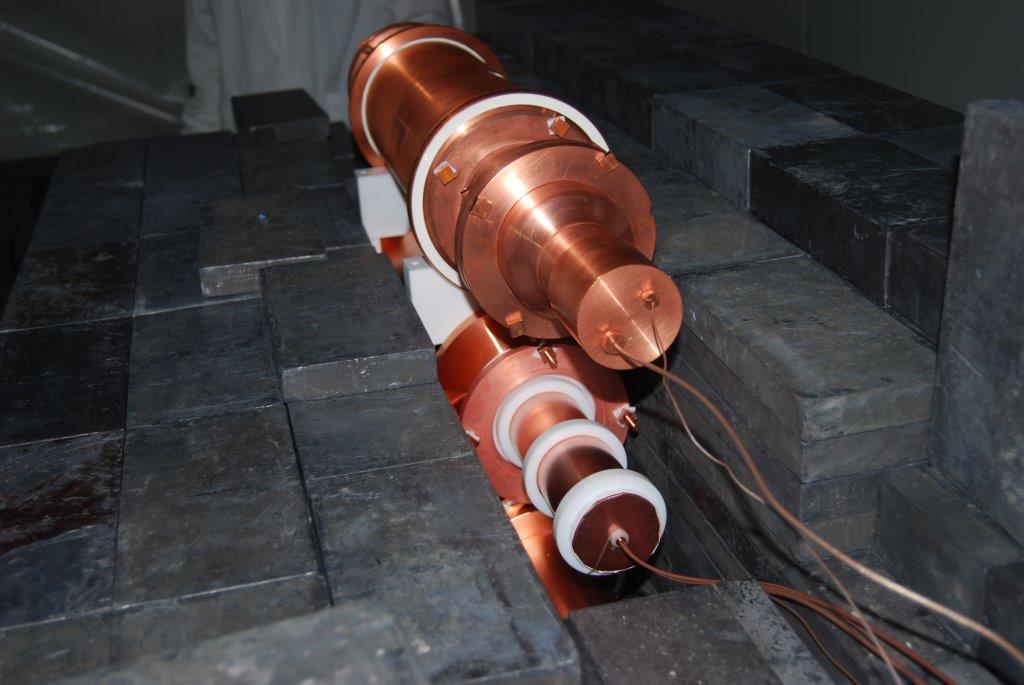}
  \end{center}
%
  \caption[ANAIS-37 set-up]{The three ANAIS-37 modules.\label{fig:A37Setup}}
\end{figure}

\begin{figure}[h!]
  \begin{center}
  \includegraphics[width=0.7\textwidth]{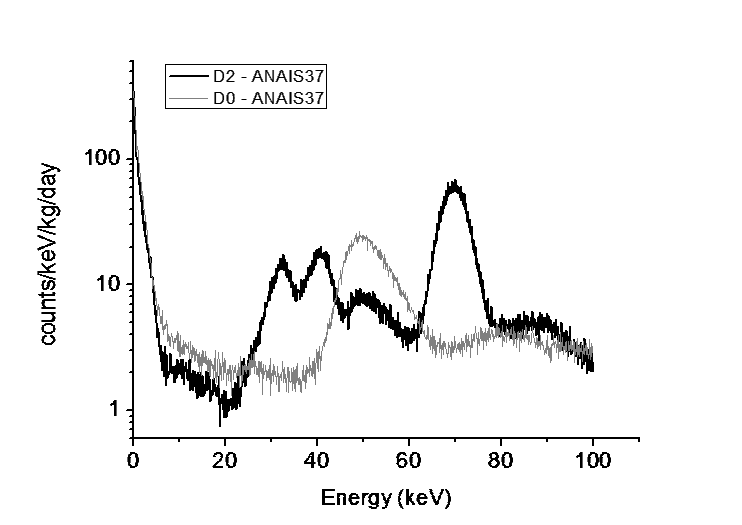}
  \end{center}
  \caption[ANAIS-37 Background]{Raw background for D0 and D2.\label{fig:A37RawBkg}}
\end{figure}

The origin of $^{210}Pb$ contamination was identified and was addressed by Alpha Spectra. A new 12.5 kg crystal was grown and a new module was constructed using new improved protocols. This detector, named D2, was integrated in the ANAIS-25 set-up to form the new ANAIS-37 set-up as it can be seen in Figure~\ref{fig:A37Setup}. 
\paragraph{ }
A preliminary determination of the $\alpha$ rate gave a value of $0.58 \pm 0.01$ mBq/kg which is five times lower than ANAIS-25 values, concluding that effective reduction of Rn entrance in the growing or/and purification processes at Alpha Spectra has been achieved. This reduction can also be appreciated at low energy (see Figure~\ref{fig:A37RawBkg}) in both peak and continuum in spite of the still decaying cosmogenic background.

\paragraph{ }
\sloppypar{The potassium content was also determined with the same technique earlier mentioned obtaining $44\pm4$ ppb, compatible with the content of D0 and D1 (see Table~\ref{tab:InternalContamination}). The contamination measurements were introduced in the aforementioned background model in order to explain the measured raw background down to 5~keV~\cite{amare2015background}}.
\paragraph{ }
In addition to the measurement of the internal contamination and the understanding of the low energy background, the ANAIS-37 set-up was also used to test some other improvements for the complete experiment such as the complete plastic scintillator coverage, the new calibration system and was also used to the test new approaches to scale up the experiment.
\subsubsection{Complete plastic scintillator coverage}
An almost total plastic scintillator coverage was mounted and put into operation at the time of the ANAIS-37 set-up with eleven scintillators. This system and its final extension to sixteen scintillators in order to cover the total ANAIS experiment are presented in Chapter~\ref{sec:Veto}.
\begin{figure}[h!]
     \begin{center}
	\begin{subfigure}[b]{0.50\textwidth}
                \centering
                \includegraphics[width=\textwidth]{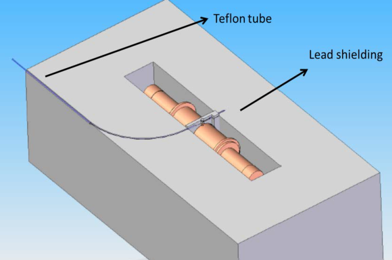}
                \caption{}
                \label{fig:Calib0}
        \end{subfigure}%
        ~
	\begin{subfigure}[b]{0.50\textwidth}
                \centering
                \includegraphics[width=\textwidth]{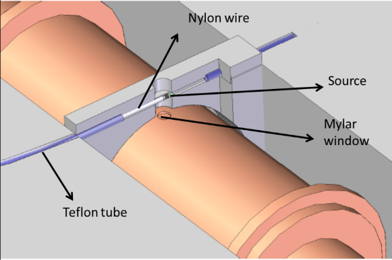}
		\caption{}
                \label{fig:Calib1}
        \end{subfigure}
	
	\begin{subfigure}[b]{0.50\textwidth}
                \centering
                \includegraphics[width=\textwidth]{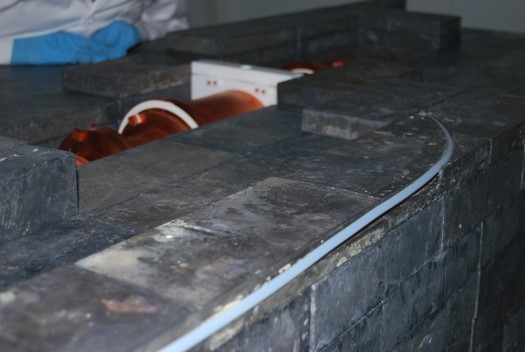}
		\caption{}
                \label{fig:Calib2}
        \end{subfigure}
	\caption{Calibration system design (a), design detail (b) and its implementation (c).\label{fig:CalibSys}}
\end{center}
\end{figure}
\begin{figure}[h!]
     \begin{center}
	\begin{subfigure}[b]{0.45\textwidth}
                \centering
                \includegraphics[width=\textwidth]{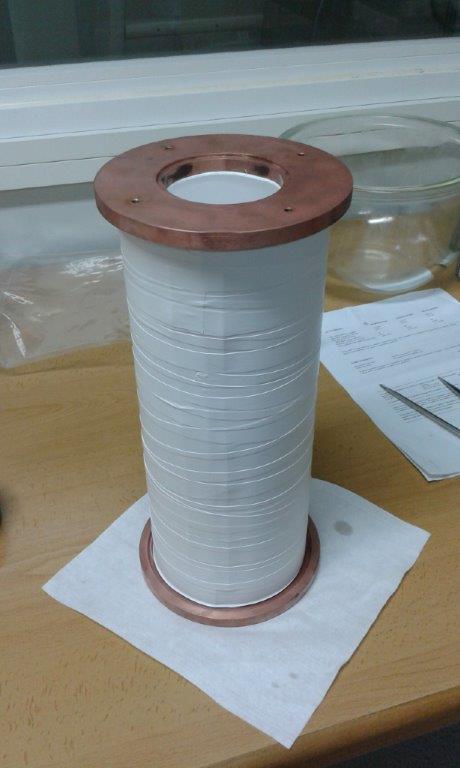}
        \end{subfigure}%
        ~
	\begin{subfigure}[b]{0.45\textwidth}
                \centering
                \includegraphics[width=\textwidth]{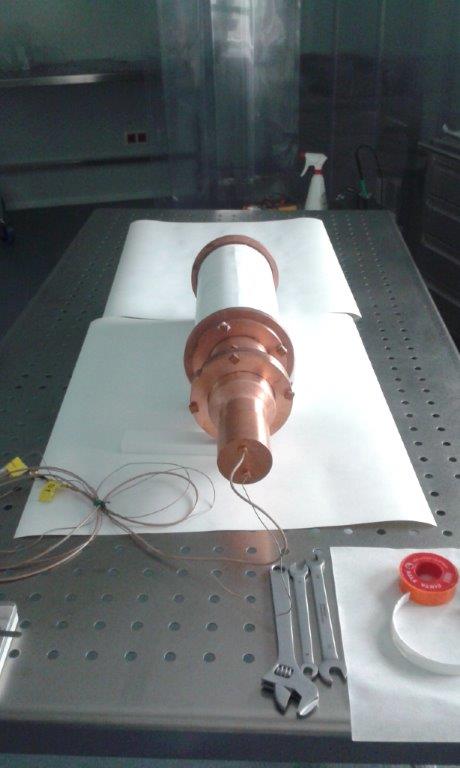}
        \end{subfigure}
	\caption{Blank module assembly.\label{fig:Blank}}
\end{center}
\end{figure}
\subsubsection{New calibration system}
The ANAIS modules have been designed with a Mylar window to allow low energy calibration as seen previously. The calibration is performed by introducing a $\gamma$ or X-ray sources inside the shielding and in front of the thin Mylar window in order to measure the response to known energy depositions. The introduction of these sources inside the shielding was previously done by means of an aperture big enough to introduce a tray with a disk shaped source holder and opening the anti-radon box in the operation. This calibration system was changed at the end of the ANAIS-25 set-up in order to keep the system free of radon. 
\paragraph{ }
The new system consists of flexible wires with $^{57}Co$ and $^{109}Cd$ customized sources along them. The sources are outside the shielding and they are introduced inside the shielding facing the Mylar windows during calibration keeping the system closed and radon free. Mylar windows of adjacent detectors face each other so that one wire can calibrate two detectors at a time. The design and implementation of this system can be seen in Figure~\ref{fig:CalibSys}. This system has been tested at the end of the ANAIS-25 set-up with one wire and a new wire was installed to calibrate the three ANAIS-37 detectors.
\subsubsection{Blank module testing}
A blank module consisting of the D1 PMTs and a copper housing without NaI crystal or quartz windows was mounted in place of the D1 detector in September 2015. This set-up was conceived to fully understand, characterize and try to filter and reject all PMT noise. This better filtering will help to understand the raw background below 1.5 keVee and to lower the energy threshold as much as possible.
\section{ANAIS sensitivity prospects}
\label{sec:ANAISProspects}\label{sec:SensitProsp}
\begin{figure}[h!]
\begin{center}
\includegraphics[width=1\textwidth]{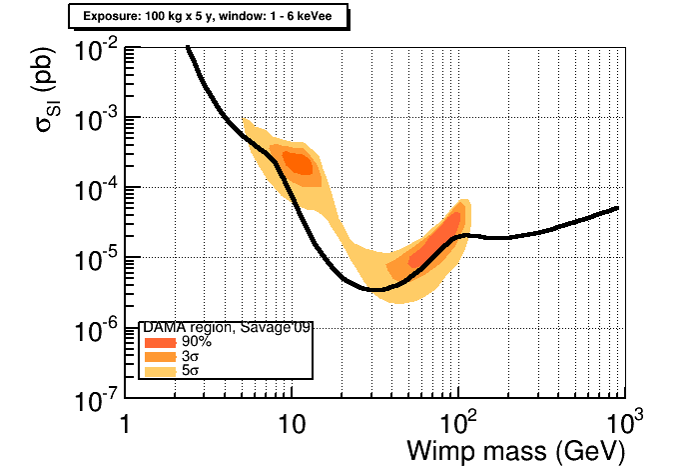}
\caption[ANAIS Sensitivity projection]{Sensitivity projection for 1-6 keVee energy window and five years of data taking (without considering anticoincidence rejection). From~\cite{amare2015status}.}
\label{fig:Sensitivity}       
\end{center}
\end{figure}

The expected sensitivity to annual modulation of the ANAIS full experiment has been evaluated with the measured background in ANAIS-25 but using the \textsuperscript{210}Pb measured in ANAIS-37 D2. The prospects of the sensitivity to the annual modulation in the WIMP mass-cross-section parameter space are shown in Figure~\ref{fig:Sensitivity} using a configuration of 100 kg and 5 years of data taking. The energy window considered was 1 to 6 keVee. These prospects correspond to a detection limit at 90\% CL with a critical limit at 90\% CL, following reference~\cite{cebrian2001sensitivity}. It has to be noted that further background reduction from the anticoincidence pattern is expected. Even with this conservative approach, there is a considerable discovery potential of dark matter particles within the DAMA/LIBRA singled-out parameter space regions~\cite{savage2009compatibility}.

\section{This work}
This chapter has been dedicated to give a complete picture of the ANAIS experiment, its goals, history and status. The complete information and results have been published in the above referenced articles. 
\paragraph{ }
This work is devoted to the design and the scale-up of a suitable data acquisition system for the ANAIS experiment following the experimental requirements described in Section ~\ref{sec:ExperimentalRequirements}. Additionally, the data analysis tools of previous prototypes have been adapted to the new data structures and an automation framework has been developed. Finally, the whole system has been characterized and its stability reviewed.
\paragraph{ }
Chapter~\ref{sec:OpticalToElectrical} shows the development of PMTs characterization protocol and the testing results of the first units. Chapter~\ref{sec:FrontEnd} describes the electronic front-end and the characterization of the key modules. Chapter~\ref{sec:Veto} presents the design and first results for the muon detection system. The data acquisition software design and implementation are covered in Chapter~\ref{sec:DAQSW} whereas the analysis software and its use is seen in Chapter~\ref{sec:AnalysisSW}. Finally, Chapter~\ref{sec:FullCharac} reviews the global tests, the performance of the detectors and Chapter~\ref{sec:Stability} the stability of both the environmental parameters and the full data acquisition system.

\chapter{Optical to electrical conversion}
\label{sec:OpticalToElectrical}
The scintillation light produced in certain materials is one of the oldest techniques to detect ionizing radiation and it remains as one of the most useful available methods for detection and spectroscopy of a wide range of radiations. This effect is thoroughly covered in references~\cite{birks1964theory} and~\cite{knoll2010radiation}. The scintillation light is usually detected by photomultiplier tubes (PMTs) converting photons into electric signal.
\paragraph{ }
The processes involved in the light generation and its conversion to electrical signal are described in this chapter. The characterization of the ANAIS PMTs is also covered focusing in their critical aspects to ANAIS: radiopurity, dark counts, gain and quantum efficiency.
\section{Scintillation process in NaI(Tl)}\label{sec:NaIScint}
The ideal scintillator material should have some desired properties: convert kinetic energy of the incident particle into detectable light with high scintillation efficiency; a linear light yield with deposited energy; the medium should be transparent to the emitted wavelength; the decay time should be as short as possible; it has to have good optical quality to produce practical large size detectors; and its refraction index should be near that of the glass ($\sim$1.5) to allow efficient photomultiplier coupling.
\paragraph{}
There are two main groups of scintillation materials: inorganic and organic-based liquids and plastics. The most widely used scintillators include the inorganic alkali halide crystals (such as NaI), exhibiting the best light output and good linearity but with a response relatively slow compared to organic scintillators, which are much faster but have a lower light yield.
\paragraph{}
The scintillation process in organic scintillators arises from transitions in the energy level structure of a single molecule. A large category of organic scintillators are based on organic molecules with $\pi$-electron structure which gives a principal emission path with a decay time of the order of a few nanoseconds. For example, plastic scintillators are used in ANAIS as a veto system as described in Chapter~\ref{sec:Veto}.
\paragraph{}
\begin{figure}[h!]
  \begin{center}
    \includegraphics[width=.7\textwidth]{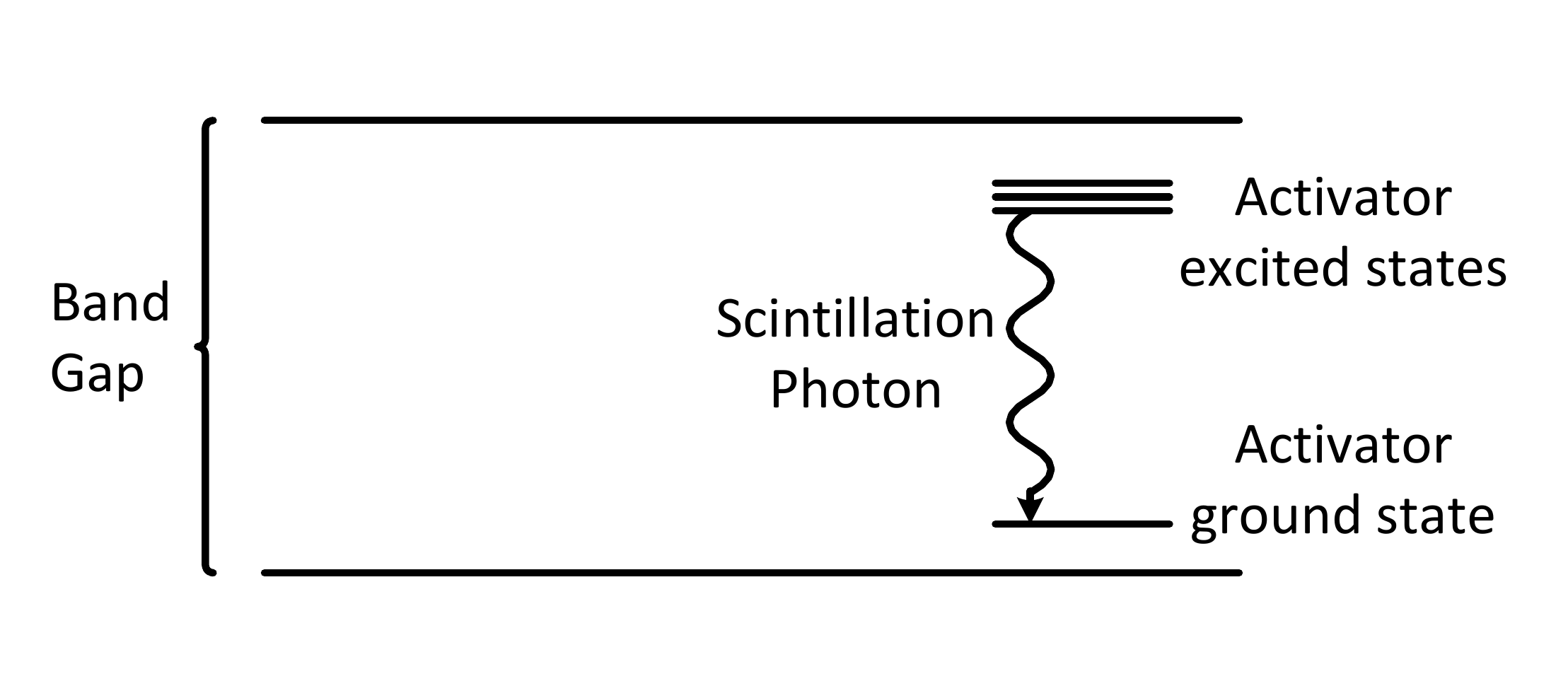}
    \caption{\label{fig:CrystalBandStructure}Energy band structure of an activated crystalline scintillator.}
  \end{center}
\end{figure}

The scintillation mechanism in inorganic materials depends on the energy states determined by the crystal lattice of the material. As shown in Figure~\ref{fig:CrystalBandStructure}, electrons only have available discrete bands of energy in insulators or semiconductor materials. Small amounts of an impurity are commonly added to inorganic scintillators in order to enhance the probability of visible photon emission during the de-excitation process. These impurities, called activators, create special sites in the lattice modifying the normal band structure of the pure crystal. As a result, new energy states are created within the forbidden gap through which the electron can de-excite back to the valence band with the emission of a photon. This de-excitation will occur with typical half-lives of $\sim$10\textsuperscript{-7} s and, with a careful selection of the activator, with a photon in the visible energy range. 
\paragraph{}
Competing to the aforementioned scintillation process in inorganic materials there are other processes:
\begin{easylist}[itemize]
& \textbf{Slow components}: the electron can create an excited configuration at the impurity site whose transition to the ground state is forbidden. Thermal excitation can give an increment of energy to raise it to a higher state from which the de-excitation is possible resulting in a slower component of light called \emph{phosphorescence}.
& \textbf{Quenching}: Certain radiationless transitions are possible between some excited states and the ground state via electron capture, with no visible photons emitted. These processes are called \emph{quenching} and account for the loss in the conversion of particle energy into scintillation light.
\end{easylist}
These effects have an impact in the efficiency of the scintillation process. Various experimental determinations have shown that the absolute scintillation efficiency of NaI(Tl) is about 13\%.
\paragraph{}
Thallium doped sodium iodide scintillators have been widely used in the gamma spectroscopy field. The material allows the growing of large mass detectors, exhibits an excellent light yield and has a close to linear response. For these reasons, it has been accepted as standard scintillation material for routine gamma spectroscopy. The disadvantages for some applications are the long decay time (230 ns), the crystal fragility and its hygroscopic nature that requires a tight encapsulation. In addition, the phosphorescent effect can be more evident at high activity creating an ``afterglow'' effect~\cite{cuesta2013slow}.
\paragraph{}
NaI(Tl) exhibits other features like differences in shape and light yield for different interaction mechanism. These features are very relevant to the use of NaI as WIMP-nucleus recoil detector, as described in Section~\ref{sec:NaINuclRecoils}.
\subsection{NaI(Tl) scintillation parameters}
\label{sec:NaIScintParameters}
Even taking into account the shape difference between particle interactions, the time constant of the main scintillation channel only ranges from 200 to 320 ns~\cite{kudryavtsev1999characteristics,cuesta2013slow}. This fact, alongside with the response of the light detector device, will mark the features needed by the front-end electronics in order to process the signal properly (see Section~\ref{sec:LESignal}). Additionally, the spectral distribution of the scintillation photons has its maximum around 420 nm~\cite{knoll2010radiation}.
\paragraph{ }
Another parameter to take into account is the scintillation yield which is tens ($\sim$40) of photons per keV for $\gamma$ rays~\cite{murray1975energy}. Given this fact and considering the light collection efficiency factor, the typical scintillation constant and the energy range of interest (below 6 keV), a device sensitive to individual photons of the order of 420 nm is needed in order to have a threshold as low as possible. High quantum efficiency bialkali photomultipliers fulfill all the aforementioned requirements and they are described in the next section.
\section{Photomultipliers}
\label{sec:PMTs}
\begin{figure}[h!]
  \begin{center}
	\begin{subfigure}[b]{0.5\textwidth}
        \centering
	\includegraphics[width=\textwidth]{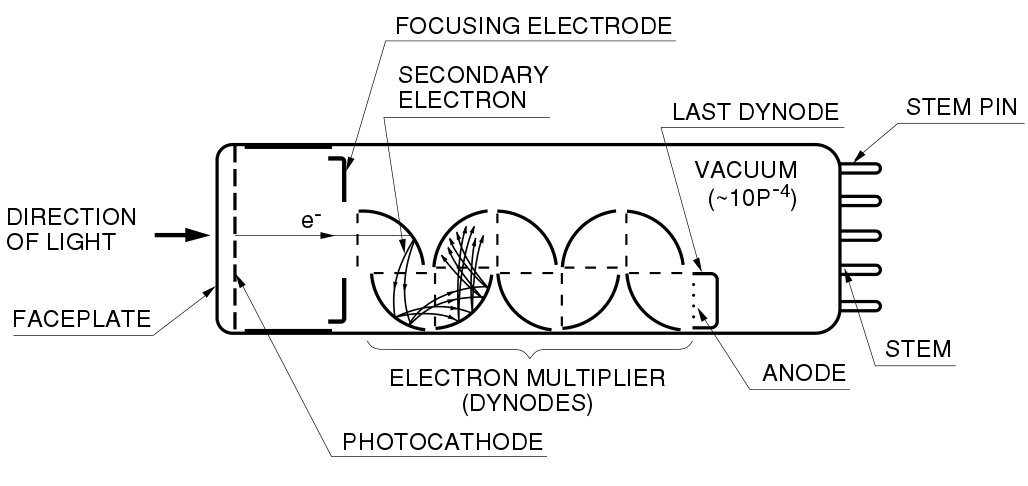}
	\caption{\label{fig:PMTScheme}}
	\end{subfigure}%
	~ 
	\begin{subfigure}[b]{0.5\textwidth}
        \centering
	\includegraphics[width=\textwidth]{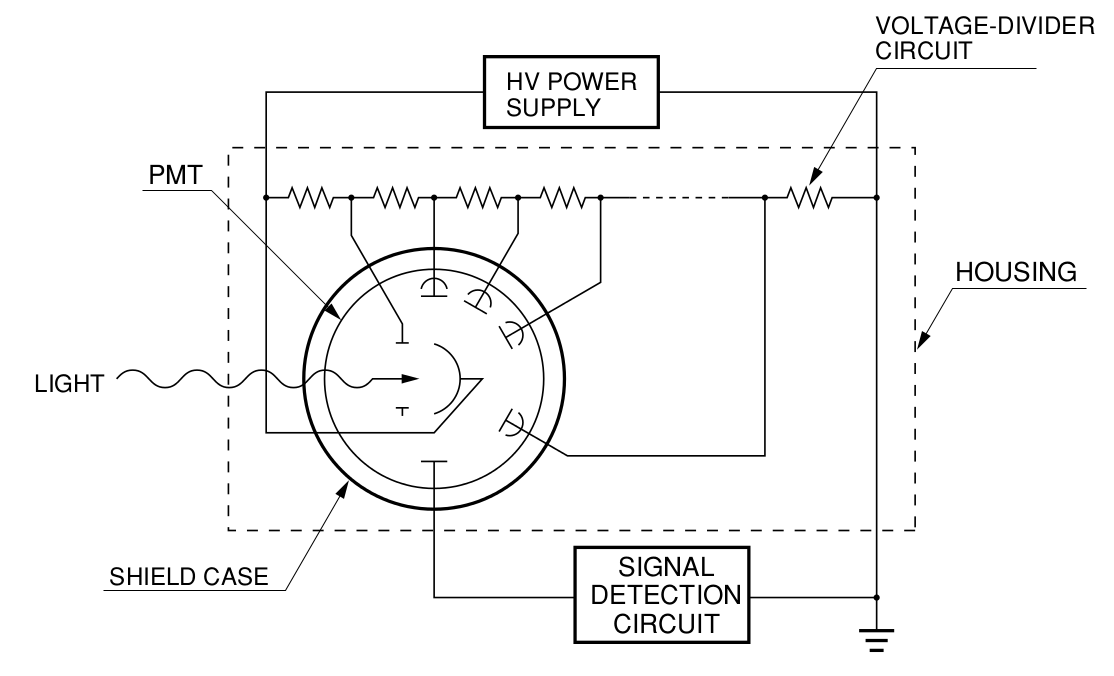}
	\caption{\label{fig:PMTOper}}
	\end{subfigure}%
    \caption[Schematic of a photomultiplier tube]{Schematic of a photomultiplier tube (a) and a usual operating method (b). From~\cite{tubes2006basics}.\label{fig:PMT}.}
  \end{center}
\end{figure}

A photomultiplier tube (PMT) is a vacuum tube consisting of an input window, a photocathode with a photosensitive layer, focusing electrodes, electron multipliers called dynodes and an anode~\cite{knoll2010radiation,tubes2006basics}. The process of converting light into a measurable electric signal (see Figure~\ref{fig:PMTScheme}) consists of several steps involving photoelectric and secondary electron emission effects. First, the light passes through the input window and excites the photocathode so that photoelectrons are emitted into the vacuum via photoelectric effect. The photoelectrons are accelerated in the vacuum and focused by the focusing electrode onto the first dynode where they are multiplied taking advantage of the secondary emission effect. This step is repeated at each successive dynodes. Finally, the electrons emitted by the last dynode are collected by the anode. 

\paragraph{}
A PMT needs some peripheral devices for proper operation as seen in Figure~\ref{fig:PMTOper}: a stable high voltage power supply, a voltage divider circuit, a housing system to shield the external light and a signal processing set-up. The high voltage supply is used to provide potential difference to the dynodes in order to allow the secondary emission effect. The main voltage is distributed among the dynodes by the voltage divider circuit. The typical high voltage of a photomultiplier is of the order of one thousand volts.
\paragraph{ }
There are several parameters for the photomultiplier characterization:
\begin{easylist}[itemize]
& The \textbf{spectral response} is the ratio between detected and incident photons as a function of their wavelength. It shows the wavelength range to which the PMT is sensitive. It has to match the spectrum of the light to be detected as much as possible.
& The \textbf{quantum efficiency} is the spectral response at a given wavelength. It is usually expressed as percentage and given at the peak of spectral response or at a fixed wavelength.
& The \textbf{gain} is the number of electrons coming from the anode for each photoelectron produced in the photocathode.
& The \textbf{rise time} of the photomultiplier temporal response to a single photon quantifies the time to rise from 10\% to 90\% of the signal peak. The \textbf{fall time} is defined as the time to fall from the 90\% to the 10\% of the signal peak.  
& The \textbf{transit time} is the time interval from the arrival of the light to the photocathode and the appearance of the output pulse at the anode.   
& The \textbf{dark current} is the current in the anode not produced by light. The processes that could be sources of dark current are thermionic emission in the photocathode and the dynodes, leakage current between anode and other electrodes, photocurrent produced by scintillation in the glass envelope or electrode supports, field emission current, ionization current from residual gases and noise current caused by cosmic rays, radiation from radioisotopes contained in the glass envelope and environmental gamma rays.
& The \textbf{dark counts} are the rate of dark current events that could be accounted as external photon events.
& The \textbf{single electron response} (sometimes referred as SER) is the charge spectrum of the photomultiplier to single photons. It takes account of the response dispersion of the dynode system and the portion of light events producing significantly less signal for hitting the second dynode or for inelastic scattering of the photoelectron~\cite{de2011methods}. The distribution of these two type of events is often characterized by the \textbf{peak-to-valley ratio} (see Section~\ref{sec:PTVRatio}).
\end{easylist}
\paragraph{ }
There are several types of PMTs depending on the photocathode type and the geometric disposition and number of dynodes. The material of the photocathode determines the spectral response of the photomultiplier in terms of wavelength and quantum efficiency. The type of the photocathode is selected taking into account the spectrum of the light to be detected. The type and number of dynodes will determine the gain of the PMT.
\paragraph{}
In this section the PMTs chosen for the ANAIS experiment and the peripheral devices are described.
\subsection{ANAIS requirements}
A suitable photomultiplier for the ANAIS experiment has to meet certain requisites coming from the experimental requirements of the ANAIS experiment (see Section~\ref{sec:ExperimentalRequirements}) in addition to the features of the NaI(Tl) scintillation (see Section~\ref{sec:NaIScintParameters}):
\\
\begin{easylist}[itemize]
& \textbf{High quantum efficiency:} The requirement of low energy threshold, the light yield of the NaI(Tl) and its main scintillation constant demand a high efficiency in the detection of the light.
& \textbf{Low background:} The materials of the photomultiplier must be as radiopure as possible in order to minimize their contribution to the ANAIS background.
& \textbf{Low dark counts:} The dark counts must be as low as possible because they will trigger the acquisition given the low energy threshold needed by ANAIS producing undesirable high triggering rate.
\end{easylist}
\paragraph{ }
All these features were considered in the PMT selection and testing~\cite{CCUESTA}. The model Hamamatsu R12669SEL2 PMT (previously named as R6956MOD) was chosen fulfilling all requirements. This model is similar to the installed in the upgrade of the DAMA/LIBRA experiment (DAMA/LIBRA–phase2)~\cite{bernabei2012performances}.
\subsection{Hamamatsu R12669SEL2 PMT}\label{sec:PMTs}
	Hamamatsu R12669SEL2 (see Figure~\ref{fig:PMT}) photomultipliers tubes have bialkali photocathode with a high quantum efficiency ($>$ 33\%) with a maximal response at 420~nm and dark counts below 500 Hz. They have ten stages and a gain factor around $10^6$. The full technical information can be seen in Table~\ref{tab:HamR12699}.

	\begin{figure}[h!]
	  \begin{center}
	    \includegraphics[width=.7\textwidth]{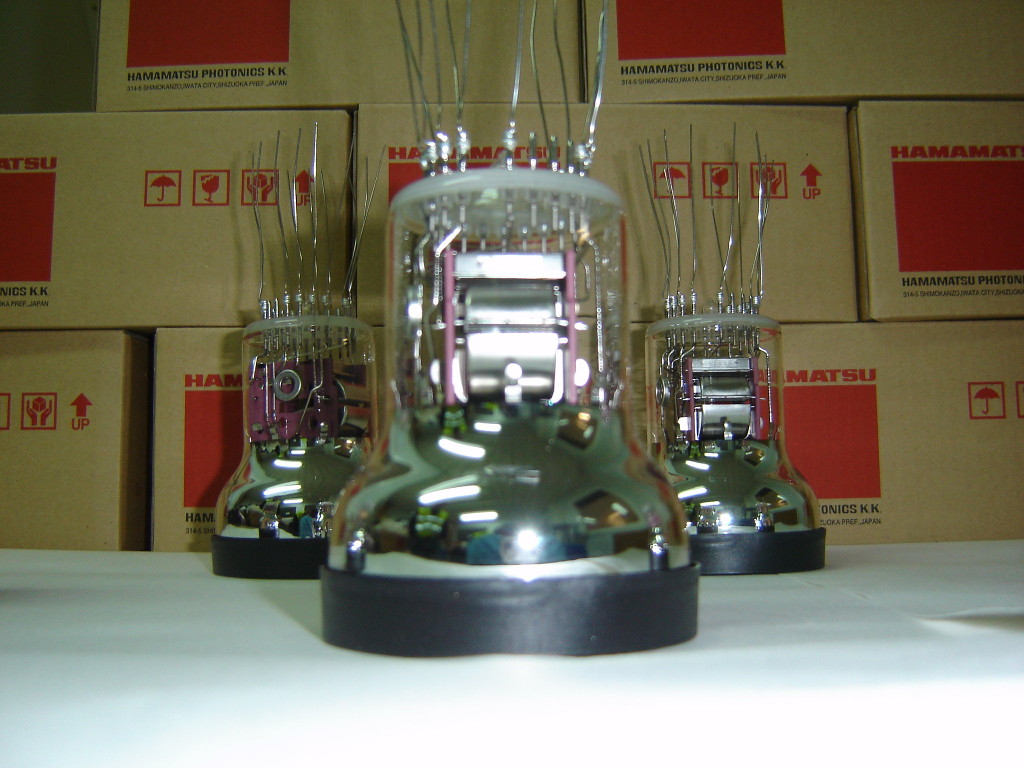}
	    \caption{Hamamatsu R12669SEL2 photomultipliers\label{fig:PMT}.}
	  \end{center}
	\end{figure}

	\begin{table}[h!]
	\centering
	\begin{tabular}{ll}
	\toprule
	Characteristics& \\
	\toprule
	Diameter & 3'' \\
	Spectral response & 300-650 nm \\
	$\lambda$ of max. response & 420 nm\\
	Photocathode material & Bialkali \\
	Photocathode min. effective & 70 mm \\
	Dynode structure & Box-and-grid + Linear-focused \\
	Number of stages & 10  \\
	Operating ambient T & -30/+50 ºC \\
	Max. supply voltage & 1500 V \\
	Max. avg anode current  & 0.1 mA \\
	\toprule
	Characteristics at 25 º C & \\
	\toprule
	Cathode luminous sens.  & 100 $\mu$A/lm \\
	Quantum efficiency at peak & $>$ 33\% \\
	Anode luminous sens & 100 A/lm \\
	Gain & 10\textsuperscript{6} \\
	Anode dark current (max.) & 6 (60) nA \\
	Dark current rate & $<500$ Hz \\
	Anode pulse rise time  & 9.5 ns \\
	Electron transit time  & 60 ns \\
	Transit time spread & 13 ns \\
	Pulse linearity & 30 ($\pm$ 2\%) mA \\
	\toprule
	\end{tabular}
	\caption{Hamamatsu R12669SEL2 technical data.}
	\label{tab:HamR12699}       
	\end{table}
	\paragraph{ }
	Forty two high quantum efficiency low background Hamamatsu R12669SEL2 PMTs were received at the LSC and tested. High purity germanium spectroscopy were performed at LSC verifying the required radiopurity levels and the homogeneity among them (see Section~\ref{sec:PMTHPGe}). Operational parameters such as SER, gain, relative quantum efficiency and dark rate were tested at the University of Zaragoza test bench. This process is described in the rest of this chapter.
	\paragraph{ }
	First, a set-up to expose PMTs to controlled ultraviolet LED light, data acquisition chain and algorithms to extract SER parameters and dark rate were developed and tested as described in Section~\ref{sec:PMTSignal}. A protocol to characterize all units was implemented and the results are presented in Section~\ref{sec:PMTResults}.

	\subsection{Voltage divider}\label{sec:VoltageDivider}
\begin{figure}[h!]
	  \begin{center}
		\begin{subfigure}[b]{1\textwidth}
		\centering
		\includegraphics[width=1\textwidth]{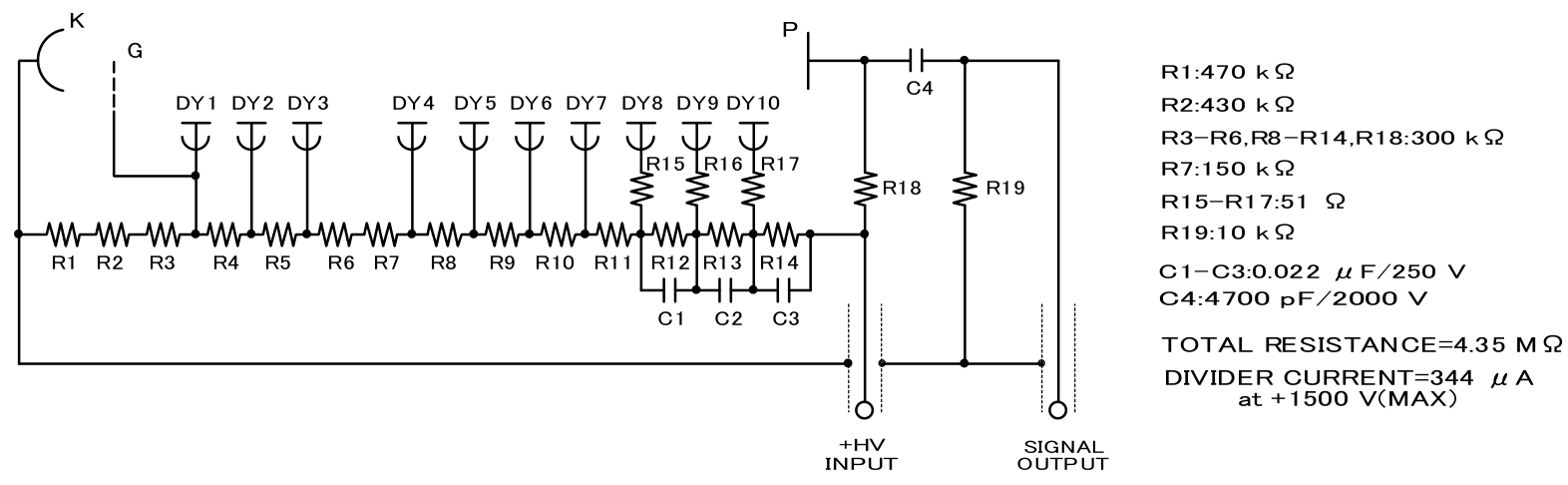}
		\caption{\label{fig:PMTDivider}}
		\end{subfigure}%
		
		\begin{subfigure}[b]{0.5\textwidth}
		\centering
		\includegraphics[width=\textwidth]{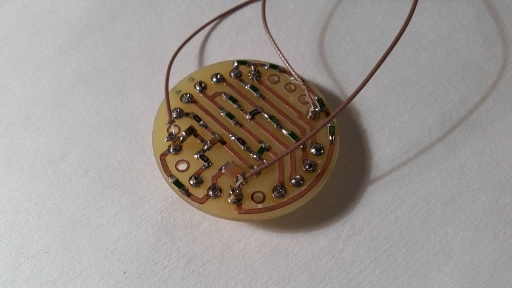}
		\caption{\label{fig:PMTDivCuFlon}}
		\end{subfigure}%
		
		\caption[R12669SEL2 voltage divider]{R12669SEL2 voltage divider design recommended by Hamamatsu (a) and its implementation in CuFlon PCB (b).\label{fig:PMTDiv}}
	  \end{center}
	\end{figure}

	The voltage divider scheme used with these PMTs is the recommended by Hamamatsu (see Figure~\ref{fig:PMTDivider}). The construction of such a circuit must be as radiopure as possible because of the ANAIS experimental requirements (see Section~\ref{sec:ExperimentalRequirements}). For this reason, CuFlon PCB was designed and manufactured as can be seen in Figure~\ref{fig:PMTDivCuFlon}. CuFlon is a tested radiopure material~\cite{aznar2013assessment}. All components in the PCBs are in SMD format to minimize their mass and consequently their activity and were measured in the HPGe low background test bench at LSC in order to check their low contribution to the ANAIS background.

	\subsection{High voltage power supply}
	A high voltage power supply with a high density of channels is needed for the full experiment. For this reason a CAEN SY2527 Universal Multichannel Power Supply System was acquired. This system offers a high scalability thanks to its six slots to house High Voltage boards. Two boards were acquired: an A1833BP with 28 independent channels and an A1535P with 24 independent channels, accounting a total of 52 channels.
	\paragraph{}
	The ripple specifications of the aforementioned boards ($<$50 mV pp for A1833BP and $<$30 mV for A1535P) was not enough for the ANAIS requirements and a filtering of the output signal was needed. In addition, a crosstalk between channels was detected risking the coincidence analysis. These effects were prevented using a proper filter as described in Section~\ref{sec:Baseline}.
	\paragraph{}
	In addition to the high density of channels, the SY2527 system has monitoring and communication features, very convenient for the ANAIS purposes. The mainframe can be monitored and controlled, giving the values of current and voltage and allowing the remote power on and off of the channels. The results of the monitoring can be seen in Section~\ref{sec:HVPSData}.

	\section{Photomultiplier radiopurity}\label{sec:PMTHPGe}
\begin{figure}[h!]
	  \begin{center}
	    \includegraphics[width=.7\textwidth]{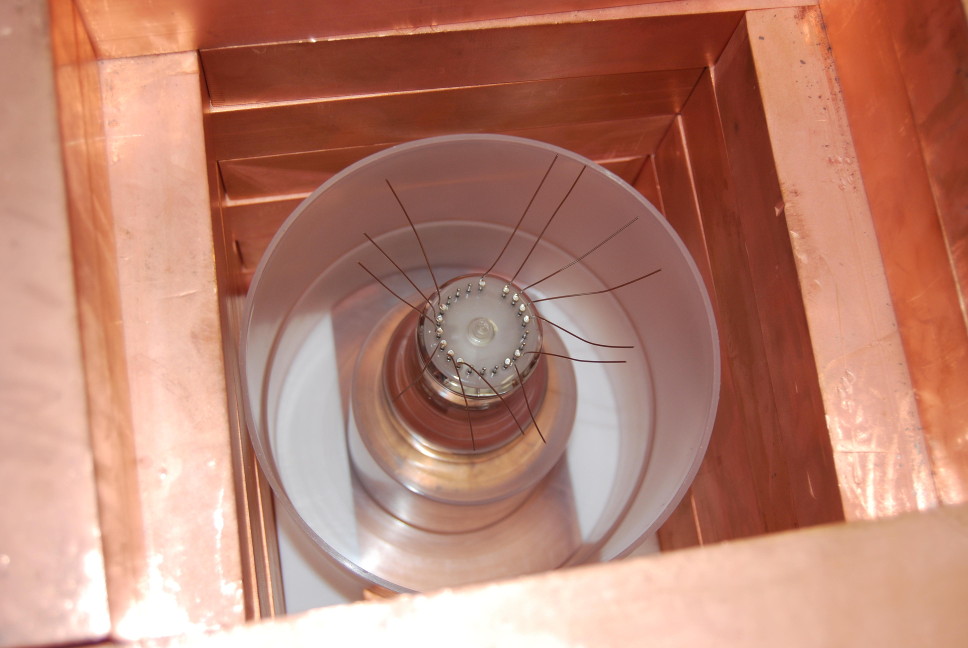}
	    \caption{Hamamatsu R12669SEL2 radiopurity screening with a HPGe detector\label{fig:PMTHPGe}.}
	  \end{center}
	\end{figure}

	The measurement of the radioactive isotopes contained in Hamamatsu PMTs was performed using a of HPGe low background test bench at the LSC (see Figure~\ref{fig:PMTHPGe}).

	\paragraph{}
	The test of all PMTs is still ongoing. The result of the already measured units can be seen in Table~\ref{tab:HamR12699HPGe}. The first two units, coupled to D0 in ANAIS-25, are R6956MOD. The rest of the PMTs have a new model reference, R12669SEL2, due to a Hamamatsu catalog change. The data show a good homogeneity in contamination levels among units and they are in agreement with the specifications requested to Hamamatsu. These data have been used as input for the background model of the ANAIS-0 prototype~\cite{cebrian2012background} and the subsequent refinements for the following prototypes.
	\begin{table}[h!]
	\begin{center}
	\begin{tabular}{ccccc}
	\toprule
	PMT& $^{232}Th$ &$^{238}U$&$^{262}Ra$&$^{40}K$\\ 
	&(mBq/PMT)&(mBq/PMT)&(mBq/PMT)&(mBq/PMT)\\
	\toprule
	ZK5902 & 20 $\pm$ 2 & 128 $\pm$ 38 &  84 $\pm$ 3 & 97 $\pm$ 19\\
	\hline
	ZK5908 & 20 $\pm$ 2 & 150 $\pm$ 34 & 88 $\pm$ 3 & 133 $\pm$ 13\\
	\hline
	FA0010 &18 $\pm$ 2 & 77 $\pm$ 42 & 78 $\pm$ 4 & 113 $\pm$ 18\\
	\hline
	FA0016 &18 $\pm$ 2 & 179 $\pm$ 51 & 83 $\pm$ 4 & 123 $\pm$ 20\\
	\hline
	FA0018 &21 $\pm$ 3 & 161 $\pm$ 58 & 79 $\pm$ 5 & 108 $\pm$ 29\\
	\hline
	FA0020 & 25 $\pm$ 4 & 260 $\pm$ 84 & 77 $\pm$ 6 & 95 $\pm$ 36\\
	\hline
	FA0022 &18 $\pm$ 2 & 158 $\pm$ 53 & 77 $\pm$ 4 & 93 $\pm$ 19\\
	\hline
	FA0034 &20 $\pm$ 3 & 144 $\pm$ 33 & 89 $\pm$ 5 & 155 $\pm$ 36\\
	\hline
	FA0035 &22 $\pm$ 2 & 148 $\pm$ 29 & 79 $\pm$ 4 & 79 $\pm$ 14\\
	\hline
	FA0036 &16 $\pm$ 2 & 139 $\pm$ 31 & 71 $\pm$ 4 & 95 $\pm$ 27\\
	\hline
	FA0037 &19 $\pm$ 2 & 170 $\pm$ 37 & 78 $\pm$ 5 & 123 $\pm$ 32\\
	\hline
	FA0051 &19 $\pm$ 2 & 133 $\pm$ 52 & 89 $\pm$ 4 & 94 $\pm$ 16\\
	\hline
	FA0053 &25 $\pm$ 2 & 114 $\pm$ 29 & 87 $\pm$ 4 & 132 $\pm$ 19\\
	\hline
	FA0057 &19 $\pm$ 2 & 153 $\pm$ 25 & 82 $\pm$ 3 & 128 $\pm$ 23\\
	\hline
	FA0058 &22 $\pm$ 2 & 168 $\pm$ 31 & 84 $\pm$ 4 & 118 $\pm$ 27 \\ 
	\hline
	FA0059 &26 $\pm$ 2 & 171 $\pm$ 32 & 85 $\pm$ 4 & 104 $\pm$ 24\\
	\hline
	FA0060 &22 $\pm$ 2 & 145 $\pm$ 29 & 88 $\pm$ 4 & 95 $\pm$ 24\\
	\hline
	FA0064 &21 $\pm$ 2 & 180 $\pm$ 34 & 85 $\pm$ 4 & 131 $\pm$ 31\\
	\hline
	FA0066 &19 $\pm$ 2 & 156 $\pm$ 31 & 82 $\pm$ 4 & 117 $\pm$ 29\\
	\hline
	FA0068 &18 $\pm$ 2 & 185 $\pm$ 29 & 81 $\pm$ 4 & 138 $\pm$ 25\\
	\hline
	FA0069 &18 $\pm$ 2 & 159 $\pm$ 29 & 79 $\pm$ 3 & 105 $\pm$ 15\\
	\hline
	FA0070 &21 $\pm$ 2 & 127 $\pm$ 30 & 82 $\pm$ 4 &  103 $\pm$ 20\\
	\hline
	FA0072 &27 $\pm$ 3 & 172 $\pm$ 35 & 91 $\pm$ 4 & 91 $\pm$ 27\\
	\hline
	FA0073 &22 $\pm$ 2 & 181 $\pm$ 31 & 78 $\pm$ 4 & 112 $\pm$ 25\\
	\toprule
	Mean value & 20.7 $\pm$ 0.5 & 157 $\pm$ 8 & 82.5 $\pm$ 0.8 & 111 $\pm$ 5\\
	\hline
	Standard deviation&  2.8 &  32.3 &  4.9 &  18.2\\

	\toprule
	\end{tabular}
	\caption{ Contamination levels of PMT units from Hamamatsu (R6956MOD and R12669SEL2) tested at low background HPGe test bench in Canfranc.}
	\label{tab:HamR12699HPGe}       
	\end{center}
	\end{table}

	\section{Low energy signals}\label{sec:LESignal}
\begin{figure}[h!]
	  \begin{center}
	    \includegraphics[width=1\textwidth]{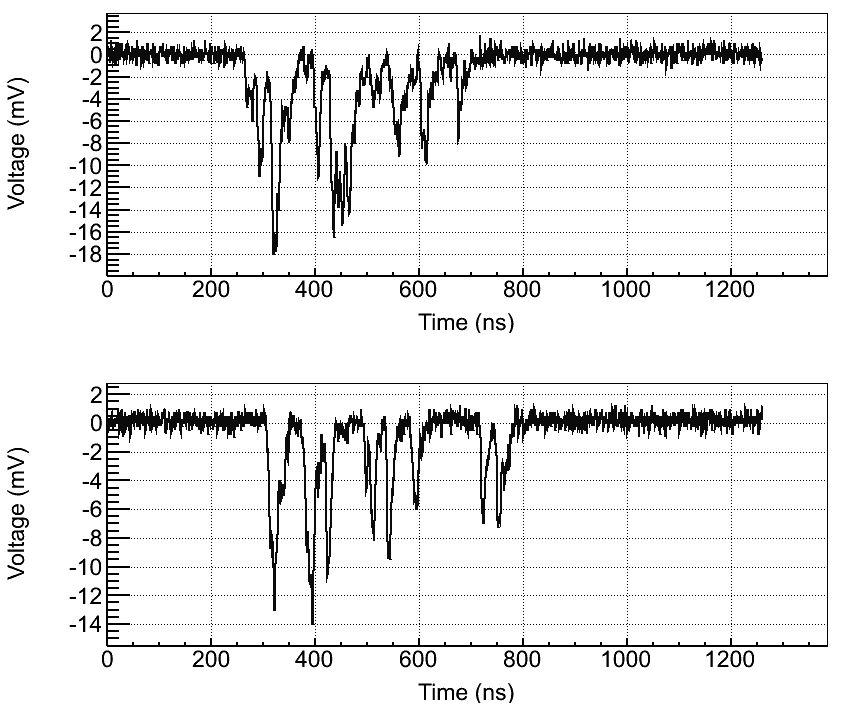}
	    \caption[Low energy NaI(Tl) scintillation]{\label{fig:LESignal}Low energy NaI(Tl) scintillation: event of 3.2 keV.}
	  \end{center}
	\end{figure}

	The scintillation light convoluted with the photomultiplier response gives rise to the electrical signal that will be processed by the ANAIS electronic front-end (see \mbox{Chapter~\ref{sec:FrontEnd}}). The signal features will determine the front-end requirements, especially in the lower energy region, below 6 keV, the ANAIS region of interest. The signal of a PMT of one of the ANAIS-25 detectors from the aforementioned energy region is shown in Figure~\ref{fig:LESignal} in order to illustrate the typical shape of the waveform. Taking into account the light yield and the scintillation constants, an event of few keV consists of some discrete photons (photoelectrons) following an exponential temporal distribution with a constant of few hundreds of nanoseconds.
	\paragraph{ }
	This waveform requires a hardware trigger set at the photoelectron level. Therefore, the study of the response of the photomultiplier to a single photon, the single electron response or SER, is crucial. This study can be seen in Section~\ref{sec:PMTSignal}. 

	\section{Photomultiplier response characterization}\label{sec:PMTSignal}
	The theory and techniques to characterize the PMTs are described in this section. A good knowledge of the PMT response at low light intensity is essential to understand the scintillation signal at very low energies. Particularly, a characterization of the SER allows to understand the most elementary PMT signal.
	\paragraph{}
	In this section an introduction to the Single Electron Response (SER) deconvolution from PMT spectra is first presented. Next, the actual set-up, algorithms and procedures are discussed. Finally, in the next section a summary of the results of the parameter extraction is shown.
	\subsection{Single electron response deconvolution}\label{sec:SERDeconv} 
	The deconvolution of the PMT charge distribution using a pulsed light as source is a well-established technique~\cite{bellamy1994absolute,dossi2000methods,gaug2006calibration,de2011methods}. It is based on modeling the photomultiplier in two basic steps: photodetection and amplification.
	\paragraph{}
	A pulsed source of light produces a flux of photons with a Poisson distribution in the number of photons per pulse. Photons are converted into electrons at the photocathode via photoelectric effect as described earlier. The conversion of photons into electrons and the collection by the dynode system is a random binary process. So, the photodetection process gives again a Poisson distribution in the number of detected photons:
	\begin{equation} 
	P(n;\mu) = \frac{\mu^n e^{-\mu}}{n!}\label{eq:phe_poisson}
	\end{equation}
	where $m$ is the mean photons hitting the photocathode and $\mu$ is defined as:
	\begin{equation}
	\mu = m q
	\label{eq:qe}
	\end{equation}
	where $\mu$ is the mean number of photoelectrons collected by the dynode system, $P(n;\mu)$ the probability that $n$ photoelectrons will be observed and $q$ the quantum efficiency.
	\paragraph{}
	Given a photoelectron coming from the cathode, the amplification process of a multiplicative dynode system can be approximated by a sum of a Gaussian distribution (parametrized by $Q_0$ and $\sigma_0$) and an exponential component (parametrized by the $\alpha$ constant). The Gaussian part describes the normal second emission effect in all dynodes and the exponential models the photoelectron being inelastically back-scattered by the first dynode (with a probability $w$). The PMT output charge (denoted by $x$) will be:
	\begin{equation}
		SER_{ideal}(x) = \frac{1-w}{\sigma_0 \sqrt{2\pi}} e^{-\frac{(x-Q_0)^2}{2\sigma_0^2}} + w \alpha e^{-\alpha x}\label{eq:ser_ideal}
	\end{equation}
	\paragraph{}
	In addition to the response for a photon, other background processes can have influence in the photomultiplier response. The possible background processes consist of noise in absence of light signal such as thermoelectron emission in the photocathode or in the dynodes, electron autoemission by electrodes, photon and ion feedback and external or internal radioactivity. Additionally, photon flux can also contribute to this background with photoemission from focusing electrodes and dynodes or other sources. All these discrete processes can appear in the measured signal with non-zero probability. The set-up and analysis described later are designed to avoid such effects as much as possible. The contribution of these effects, described earlier as dark counts and usually modeled as an exponential contribution, are considered negligible for this SER response. This assumption is reviewed later (see Section~\ref{sec:PMTTesting}). The only contribution to non-photon produced signal to be considered is the electrical noise, modeled as a Gaussian distribution:
	\begin{equation}
	B(x) = \frac{1}{\sigma_p\sqrt{2\pi}} e^{- \frac{(x-Q_p)^2}{2 \sigma _p^2} }\label{eq:pmt_bkg}
	\end{equation}
	where $Q_p$ and $\sigma_p$ determines the electric noise pedestal.
	\paragraph{ }
	A realistic PMT SER response is the convolution of the ideal response (\ref{eq:ser_ideal}) and the noise distribution function (\ref{eq:pmt_bkg}):
	\begin{align}
	\begin{split}
		SER(x) &= SER_{ideal}\otimes B(x) =\\ \label{eq:ser_real}\\
		&= \frac{1-w}{\sqrt{2\pi (\sigma_p^2+\sigma_0^2)}} e^{- \frac{(x-Q_p-Q_0)^2}{2 (\sigma_p^2+\sigma_0^2)} } \\
		&+ \frac{w \alpha}{2} e^{\frac{\alpha}{2}(2 Q_p+ \alpha \sigma_p^2 -2x)}\left( 1+ Erf\left( \frac{Q_p + \alpha \sigma_p^2 -x }{\sqrt{2} \sigma_p} \right)\right)
	\end{split}
	\end{align}
	$Q_1$ is defined as the average charge when an electron is collected in the first dynode and $\sigma_1$ the standard deviation. So, the gain can be obtained from $Q_1 = eg$, being $g$ the actual gain and $e$ the electron charge. These values can be deduced from the SER parameters using an approximation to low light:
	\begin{align}
	\begin{split}
	Q_1 &\approx (1-w)Q_0 + w/\alpha \\
	\sigma_1^2 &\approx (1-w)(\sigma_0^2 + Q_0^2)+ 2 w\alpha^2-Q_1^2 \label{eq:q1_aprox}
	\end{split}
	\end{align}
	\paragraph{}
	For a number of photoelectrons $n \geq 2$, the response will be approximately Gaussian~\cite{dossi2000methods} (This approximation is also reviewed later, in Section~\ref{sec:SERDecFit}). Assuming a linear response, the Gaussian response to $n$ photoelectrons ($G_n$) will be:
	\begin{align}
	\begin{split}
	G_n(x) &=\frac{1}{\sigma_n \sqrt{2\pi}} e^{-\frac{(x-Q_n)^2}{2\sigma_n^2}}\\ 
	Q_n &= Q_p + nQ_1\\
	\sigma_n &= \sqrt{\sigma_p + n\sigma_1^2} \label{eq:pmt_n_gauss}
	\end{split}
	\end{align}
	Having all contributions in mind (equations~\ref{eq:phe_poisson},~\ref{eq:pmt_bkg},~\ref{eq:ser_real} and~\ref{eq:pmt_n_gauss}), the function to deconvolute the PMT response will be:
	\begin{flalign}
	\begin{split}
	\begin{aligned}
		f(x) &= N\left(\vphantom{\int}\ e^{-\mu}B(x) \right.&\text{(0 photons)} \\
		&+ \left.\mu e^{-\mu} SER(x) \right.& \text{(1 photon)} \\
		&+ \left.\vphantom{\int} \sum_{n=2}^{\infty}{\frac{\mu^n e^{-\mu}}{n!} G_n(x)}\right)&\text{(n photons)}\label{eq:pmt_real}\\ 
	\end{aligned}
	\end{split}
	\end{flalign}

	\paragraph{ }
	where the function has been split in several parts: the pedestal or the electrical noise contribution to the charge spectrum, the SER and the components due to higher number of photons.
	\paragraph{ }
	This response function has seven free parameters: $Q_p$ and $\sigma_p$ define the pedestal, $w$ and $\alpha$ the exponential part of the SER,  $Q_0$ and $\sigma_0$ the Gaussian part and $\mu$ the intensity of the detected light. 
	\subsection{PMT test bench}\label{sec:PMTTesting}
\begin{figure}[h!]
	  \begin{center}
	    \begin{subfigure}[b]{1\textwidth}
	    \includegraphics[width=1\textwidth]{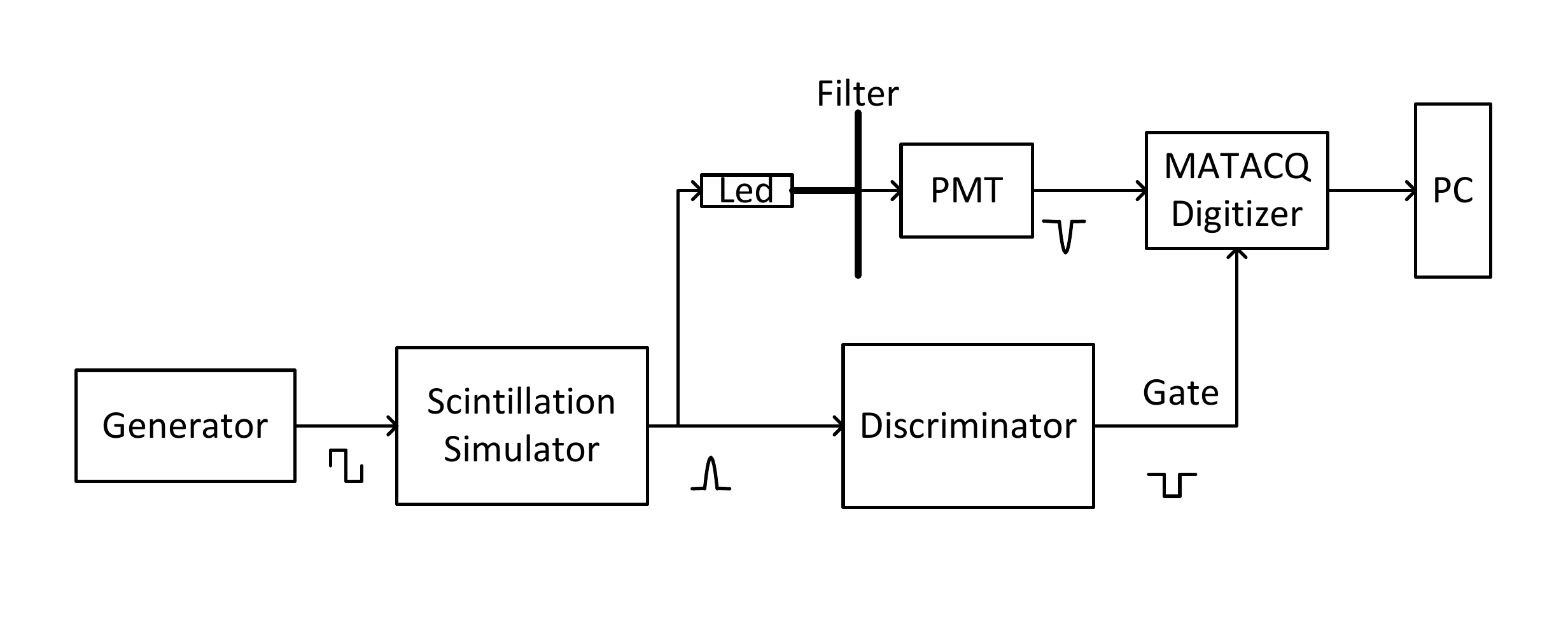}
	    \caption[Set-up scheme for PMT characterization]{\label{fig:SERSetupScheme}}
	    \end{subfigure}
	    
	    \begin{subfigure}[b]{0.5\textwidth}
	    \includegraphics[width=1\textwidth]{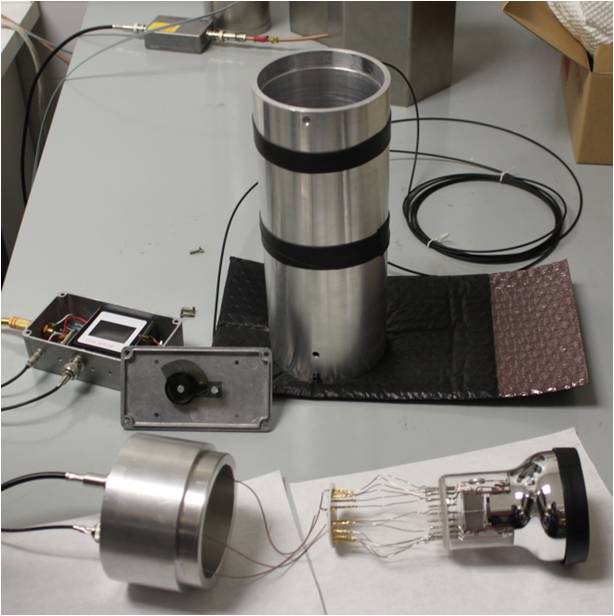}
	    \caption[Set-up for PMT characterization]{\label{fig:PMTTestBench}}
	    \end{subfigure}

	    \caption[Set-up for PMT characterization]{Set-up for PMT characterization: scheme (a) and actual set-up (b).\label{fig:SERSetup}}
	  \end{center}
	\end{figure}

	A dedicated set-up was designed and mounted in order to study and characterize the response of all PMTs using the aforementioned deconvolution. A periodically pulsed ultraviolet LED~\cite{CPOBES} was used as light source and as the trigger for a MATACQ digitizer (see sections~\ref{sec:VMEModules} and~\ref{sec:MatacqCharac}) as it can be seen in Figure~\ref{fig:SERSetup}.

\begin{figure}[h!]
	  \begin{center}
	    \includegraphics[width=.7\textwidth]{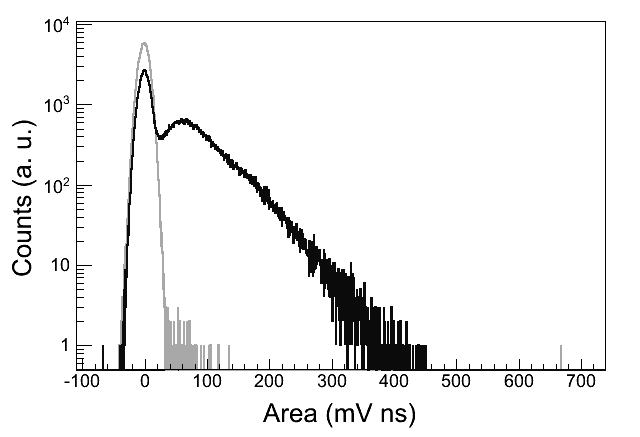}
	    \caption[PMT Background Distribution]{\label{fig:PMTBckg}PMT background distribution (in gray) compared with light spectrum (in black, $\mu\simeq0.8$).}
	  \end{center}
	\end{figure}

	\paragraph{}
	 This set-up was used to:
	\begin{easylist}[itemize]
	& Deconvolute PMT spectra in order to obtain fundamental PMT information: gain and single electron response (SER) parameters~\cite{bellamy1994absolute} and, with reproducible source LED conditions, quantum efficiency.
	& Study the single and multiple photoelectron distributions and test the peak detection algorithms with expected distributions.
	& Test all PMTs and draw their curve gain versus high voltage.
	& Test the dark counts behavior at different voltages.
	\end{easylist}
	\paragraph{ }
	The light spectrum has been expressed in terms of collected charge while the described set-up uses digitized waveform. A specific software analysis configuration was developed to integrate voltage in a fixed window (see Section~\ref{sec:AnalysisConfig}) with the same width that the original pulsed signal. The obtained integrated area is therefore proportional to the collected charge in the aforementioned time window.
	\paragraph{}
	A testing of the assumption of a negligible PMT background in this set-up was performed. The frequency of random coincidences is:
	\begin{equation}
		f_{random} = f_{dark} f_{trigg} \tau_{gate} \label{eq:f_rand}
	\end{equation}
	being $f_{dark}$ the dark rate,  $f_{trigg}$ the trigger frequency and $\tau_{gate}$ the integration time. The frequency of real light events at low light intensity (the worst scenario for the background influence) is:
	\begin{equation}
		f_{events} = (1-P(0)) f_{trigg} \simeq \mu f_{trigg} \label{eq:f_evt}
	\end{equation}
	$\tau_{gate}$ is 200 ns, the maximum time of LED excitation signal and the dark count is below 500 Hz (see Section~\ref{sec:DarkCounts}). The minimum amount of light to have a dark count contribution below 1\% is $\mu \geq \frac{f_{dark} \tau_{gate}}{0.01} = 0.01$. 
	\paragraph{ }
	A fully opaque layer was used in order to test the above assumptions. The result can be seen in gray in Figure~\ref{fig:PMTBckg}. Very few events were in coincidence with the integration window (with probability $5 \sim \times 10^{-5}$) giving a dark counts result (using equations~\ref{eq:f_rand} and~\ref{eq:f_evt}) of 260 Hz, fully compatible with the evolution of the dark current seen later (see Section~\ref{sec:DarkCounts}). A typical spectrum for the same number of triggers and with light ($\mu \simeq 0.8$) is also shown as reference in black in Figure~\ref{fig:PMTBckg}.

	\subsection{Deconvolution fit}\label{sec:SERDecFit}
\begin{figure}[h!]
	  \begin{center}
	    \begin{subfigure}[b]{0.7\textwidth}
	    \centering    
	    \includegraphics[width=\textwidth]{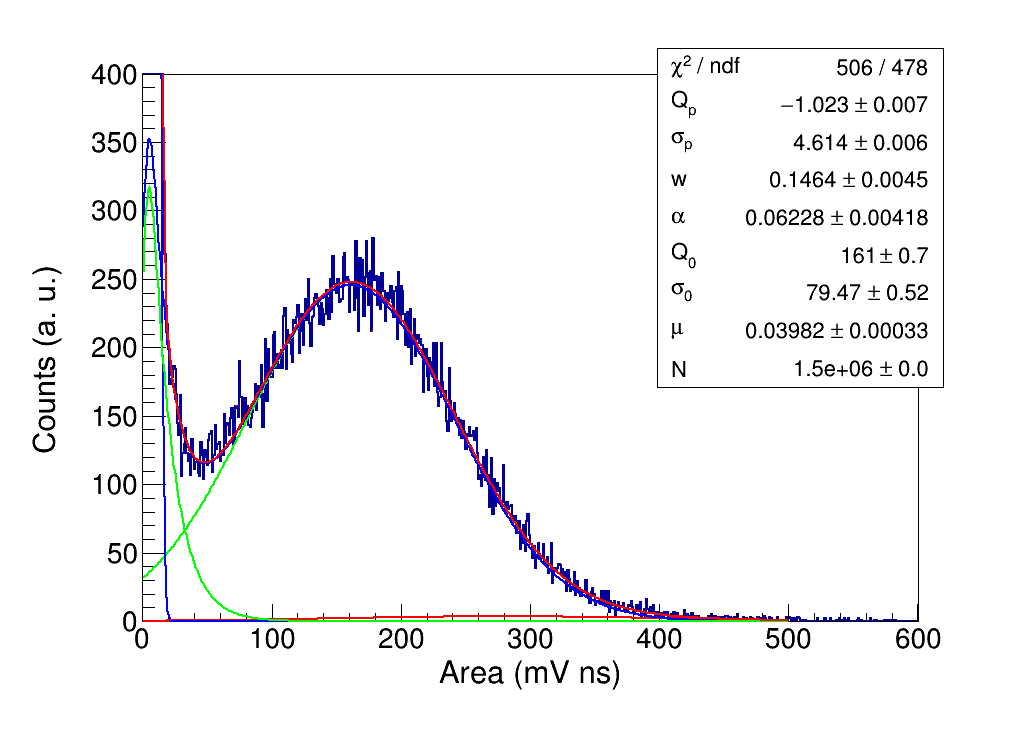}
	    \caption{\label{fig:SERFit}}
	    \end{subfigure}
	  \end{center}

	    \begin{subfigure}[b]{0.5\textwidth}
	    \centering    
	    \includegraphics[width=\textwidth]{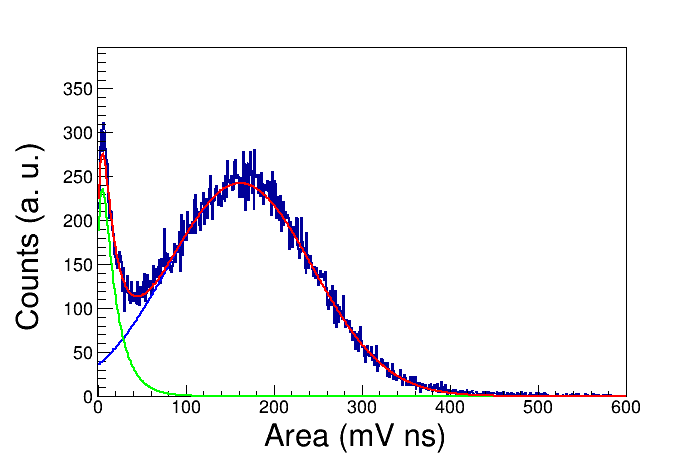}
	    \caption{\label{fig:SERFitSER}}
	    \end{subfigure}
	    ~
	    \begin{subfigure}[b]{0.5\textwidth}
	    \centering    
	    \includegraphics[width=.9\textwidth]{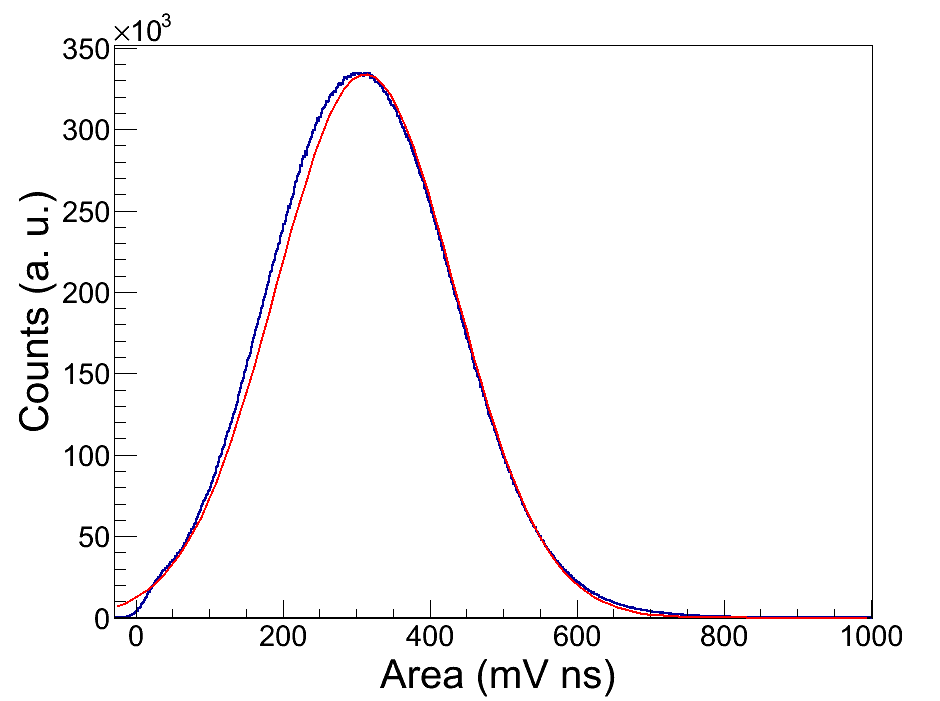}
	    \caption{\label{fig:SERFitPHE2}}
	    \end{subfigure}

	  \begin{center}
	    \caption[Photoelectron distribution fit]{Photoelectron distribution fit (a), only SER fit (b) and two photoelectron Gaussian component (red) compared with SER autoconvolution (blue) (c) (see text for details).}
	  \end{center}
	\end{figure}

	The reproducibility of the deconvolution fit and the light intensity stability were first studied before establishing a PMT characterization procedure.
	\paragraph{}
	The result of the data taking with a high density filter (low light intensity, $\mu<1$) and the data fit with the Equation~\ref{eq:pmt_real} can be observed in Figure~\ref{fig:SERFit}. The fit can be seen in red in the figure, the blue left Gaussian line corresponds to Gaussian pedestal (Gaussian baseline noise), the green line to the exponential SER component part and the green line to SER Gaussian component. The terms corresponding to higher number of photons is barely observed because of their little contribution to the fit. 

\begin{figure}[ht!]
		\begin{subfigure}[b]{0.5\textwidth}
		\centering
		\includegraphics[width=\textwidth]{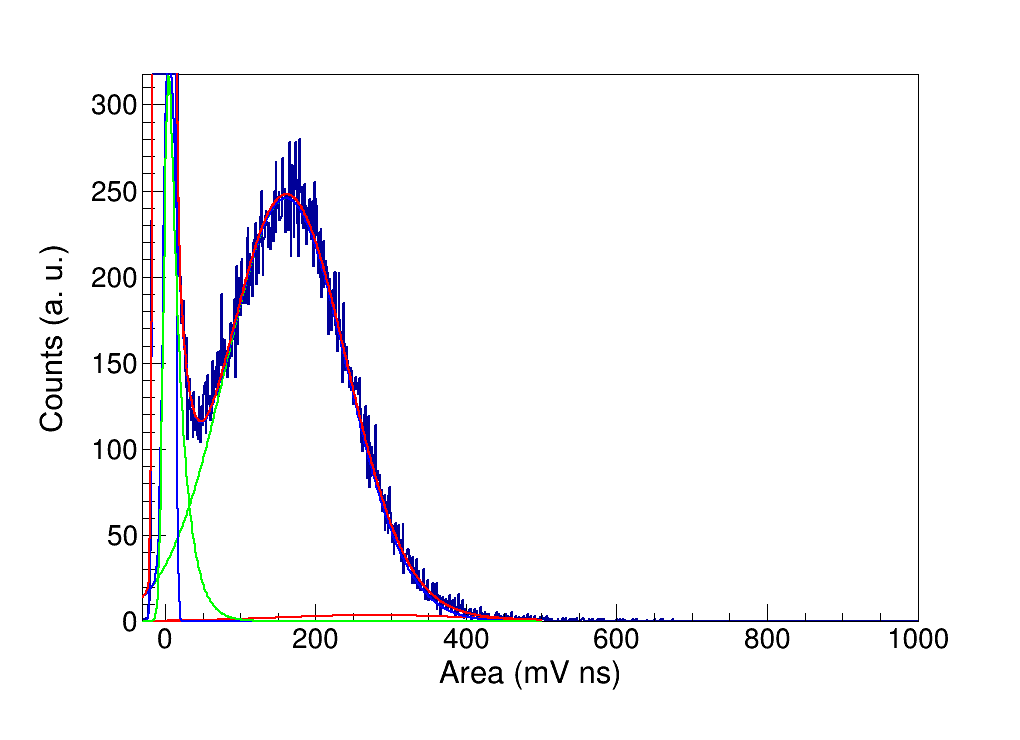}
		\end{subfigure}%
		~ 
		\begin{subfigure}[b]{0.5\textwidth}
		\centering
		\includegraphics[width=\textwidth]{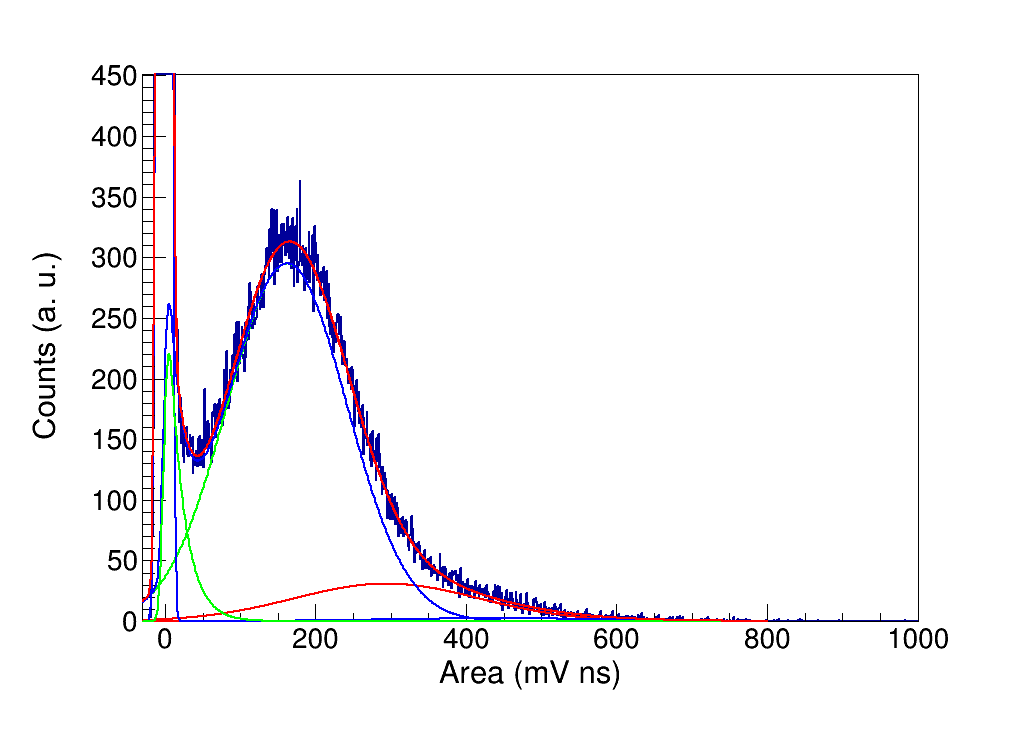}
		\end{subfigure}%
		
		\begin{subfigure}[b]{0.5\textwidth}
		\centering
		\includegraphics[width=\textwidth]{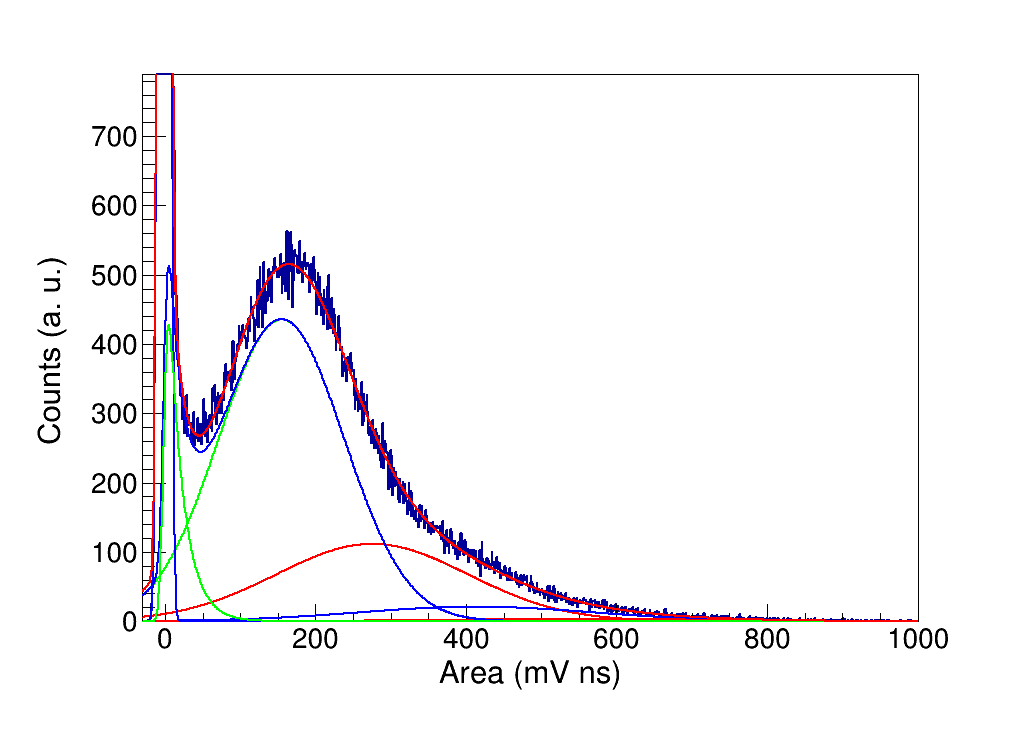}
		\end{subfigure}%
		~
		\begin{subfigure}[b]{0.5\textwidth}
		\centering
		\includegraphics[width=\textwidth]{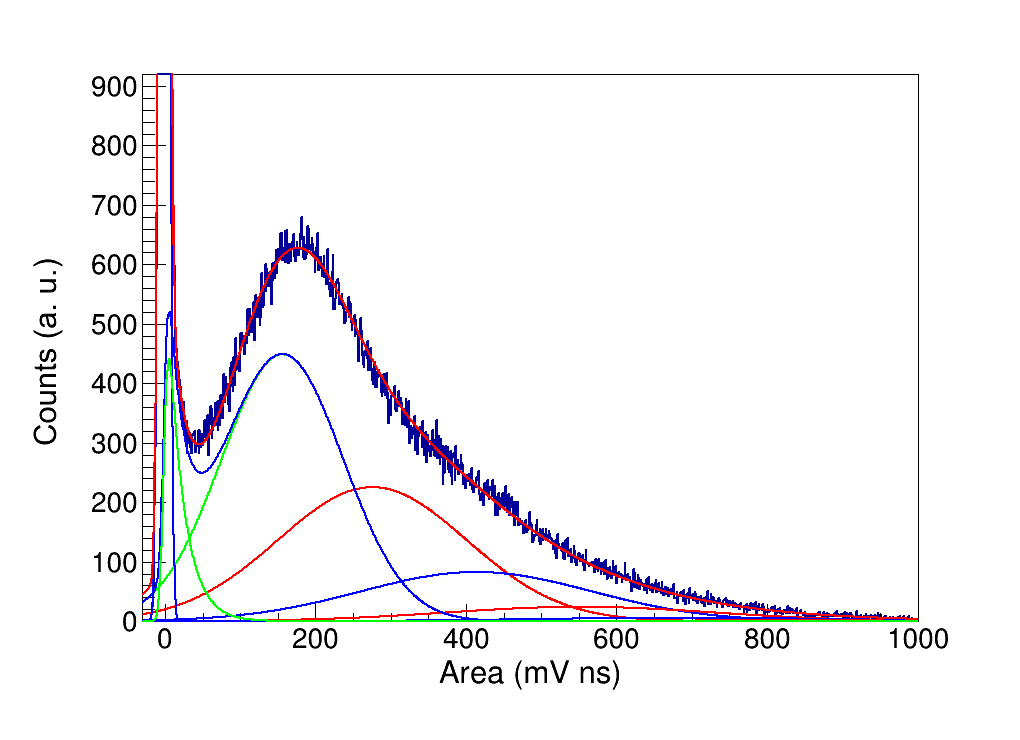}
		\end{subfigure}%
		
		\begin{subfigure}[b]{0.5\textwidth}
		\centering
		\includegraphics[width=\textwidth]{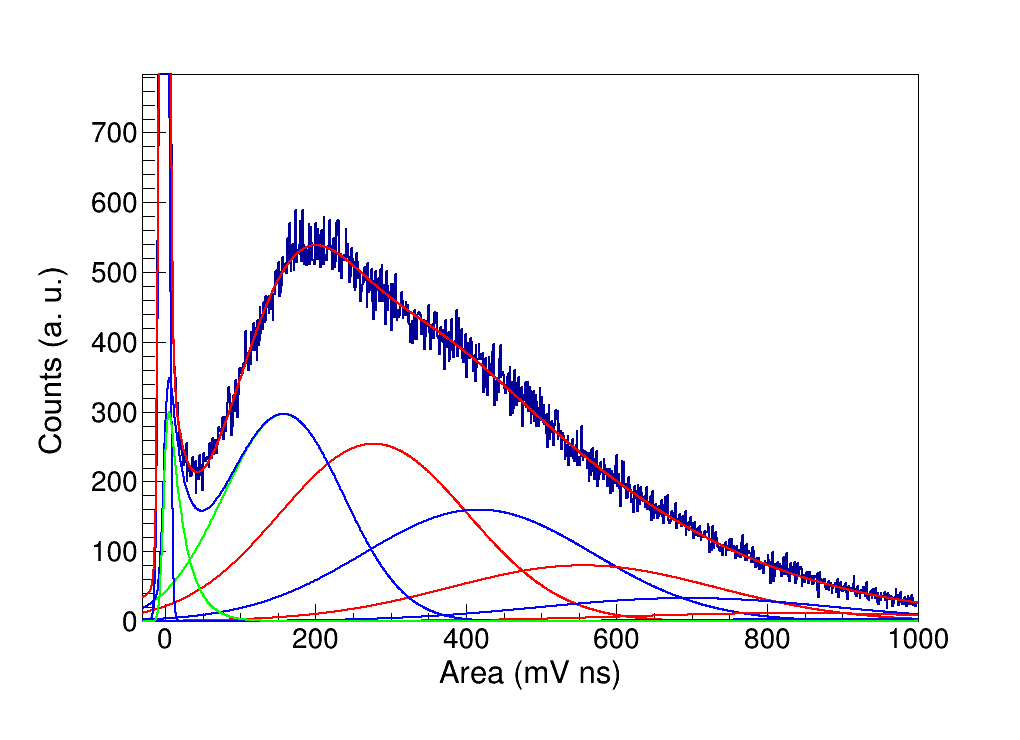}
		\end{subfigure}
		~
		\begin{subfigure}[b]{0.5\textwidth}
		\centering
		\includegraphics[width=\textwidth]{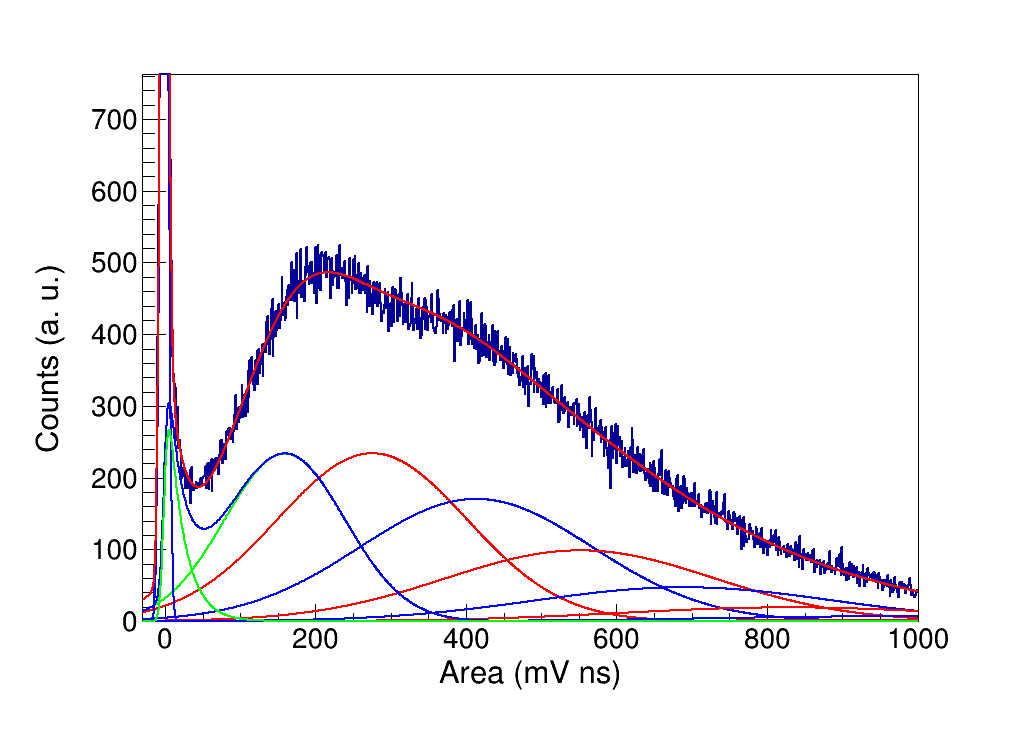}
		\end{subfigure}

	  \begin{center}
		\caption[Deconvolution fit test with increasing light intensity]{Deconvolution test with increasing light intensity at same voltage\label{fig:SERFitTest} showing the contribution of more than one photoelectron.}
	  \end{center}
	\end{figure}

	\paragraph{}
	Two crosschecks were performed in order to study the fit. First, a simplified fit was performed using the \emph{signal only} spectrum. This spectrum can be composed by events with one or more detected peaks using the peak detection algorithm (see Section~\ref{sec:PeakDetection}). The fit uses the SER equation only (\ref{eq:ser_real}) assuming two or more photoelectron components negligible ($\mu \simeq$ 0.04). This fit can be seen in Figure~\ref{fig:SERFitSER} showing both Gaussian and exponential components, confirming the SER shape and giving hints to $\alpha$ and $w$ values. 
	\paragraph{ }
	The second crosscheck was to test the Gaussian shape for terms corresponding to more than two photoelectrons. The autoconvolution of the SER extracted with the peak detection algorithm was performed and compared with the shape of the Gaussian predicted using Equation~\ref{eq:pmt_n_gauss}. The result can be seen in Figure~\ref{fig:SERFitPHE2} exhibiting a good general agreement with a little discrepancy at the beginning of the spectrum.
	\paragraph{ }
	The deconvolution of the PMT spectra at different light intensities was done at the same high voltage. This test must give a similar value of gain and standard deviation of the single electron response (SER). The results can be seen in Figure~\ref{fig:SERFitTest} showing the fit with the seven free parameters in the SER parameters. The contribution of the components of more than one photoelectron can be observed in the fits (red even number of photoelectrons, blue odd number of photoelectrons) as described in Equation~\ref{eq:pmt_real}.

	\paragraph{}
	The parameters extracted from the fits seen in Figure~\ref{fig:SERFitTest} exhibit similar gain results with discrepancies of the order of 5\% in $Q_0$ for $\mu>0.1$ due to the use of the approximative $Q_1$ expression seen in Equation~\ref{eq:q1_aprox}).
	\paragraph{}
	As a conclusion, a deconvolution fit was developed to quantify the gain of a photomultiplier and the light detected from a pulsed light source. In the next sections the quality, limitations and range of application are studied.
	\subsection{Light intensity detection and LED stability}\label{sec:LEDUnstab}
\begin{figure}[h!]
	  \begin{center}
		\includegraphics[width=0.7\textwidth]{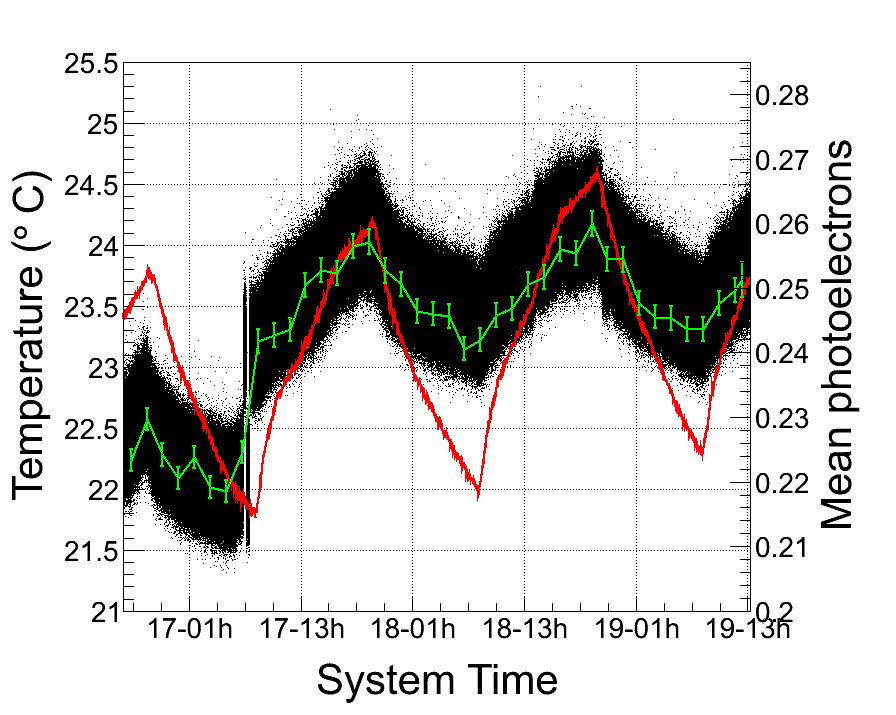}
		\caption[Led Instabilities]{LED stability study: LED current (black points), extracted light (green line) and temperature (red)\label{fig:LEDUnstabCh}.}
	  \end{center}
	\end{figure}

	The deconvolution seen in the previous section was studied by trying to extract the same light intensity at different working PMT voltages, assuming the same LED emission. This test was not successful and it gave reasons to examine the LED stability.
	\paragraph{ }
	A very long run of four days at low trigger frequency was performed in order to see its long term behavior. Some additional parameters were measured in this run: the voltage in the serial resistor (proportional to LED intensity) and the temperature in the LED box. The results can be seen in Figure~\ref{fig:LEDUnstabCh} showing the LED intensity in black points, the temperature in red and the light detected ($\mu$ and error) in green. It can be seen a good correlation among the three magnitudes and a LED instability (\mbox{$> 20$\%}) correlated with temperature. The LED stability is crucial for measuring parameters such as quantum efficiency and a more stable set-up is intended to be designed and tested. In the meantime, the light characterization runs have been as short as possible.

	\subsection{Mean photoelectrons algorithms and relative quantum efficiency}\label{sec:MuAlgor}
\begin{figure}[h!]
	  \begin{center}
		\begin{subfigure}[b]{0.5\textwidth}
		\centering
		\includegraphics[width=\textwidth]{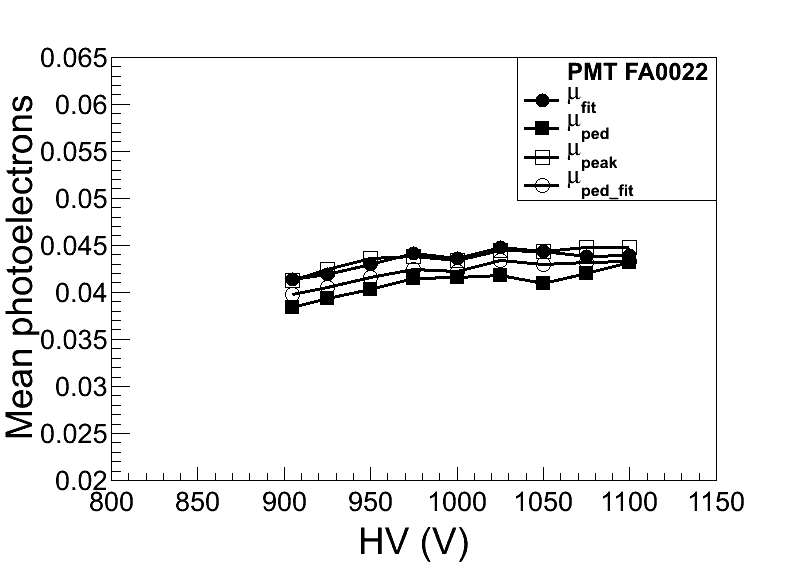}
		\caption[]{\label{fig:Mu_22}}
		\end{subfigure}%
		~ 
		\begin{subfigure}[b]{0.5\textwidth}
		\centering
		\includegraphics[width=\textwidth]{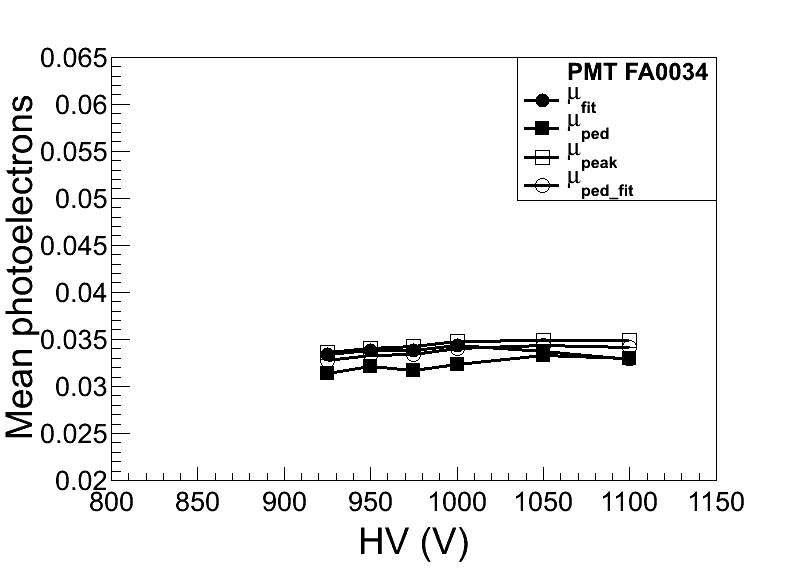}
		\caption[]{\label{fig:Mu_34}}
		\end{subfigure}%

		\begin{subfigure}[b]{0.5\textwidth}
		\centering
		\includegraphics[width=\textwidth]{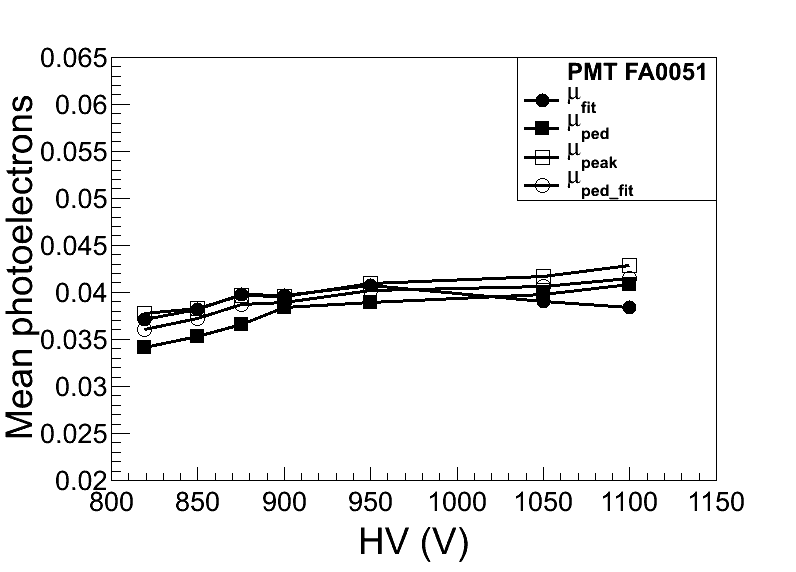}
		\caption[]{\label{fig:Mu_51}}
		\end{subfigure}%
		~ 
		\begin{subfigure}[b]{0.5\textwidth}
		\centering
		\includegraphics[width=\textwidth]{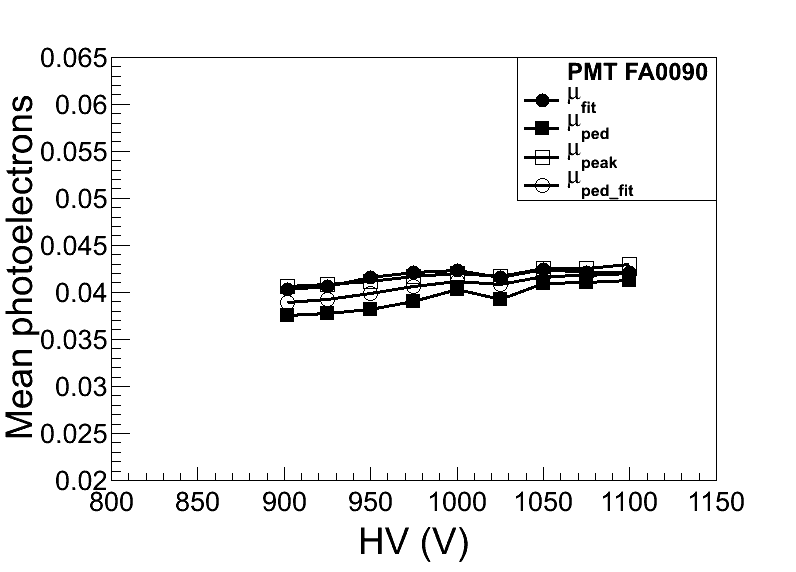}
		\caption[]{\label{fig:Mu_90}}
		\end{subfigure}%
		
		\caption[Mean photoelectron extraction methods]{Mean photoelectron different extraction methods\label{fig:PMTMuAlgor}.}
	  \end{center}
	\end{figure}

The quantum efficiency of the PMTs is a crucial parameter for their use with scintillators at very low energy threshold and checking this feature is important in the ANAIS PMT testing procedure. A direct measurement requires a well calibrated source of light. However, an easier way to crosscheck the quantum efficiency is to measure the relative quantum efficiency by detecting light from a stable source. Once determined $\mu$, the relative quantum efficiency can be deduced following Equation~\ref{eq:qe}:
	\begin{equation}
		 \frac{q_1}{q_2} = \frac{\mu_1}{\mu_2}
	\end{equation}
\paragraph{ }
A description of the light measurement procedure and the relative quantum efficiency determination can be seen in Section~\ref{sec:PMTResults}.
	\paragraph{ }	A study of several algorithms for the extraction of the pulsed LED light intensity was done in order to compare them and perform PMT relative quantum efficiency measurements. 
\paragraph{}
	The mean number of photoelectrons can be extracted from the signal spectral distribution with several complementary methods~\cite{dossi2000methods} in addition to the deconvolution fit seen earlier. The simplest way is to determine the ``non-photoelectron'' part of the Poissonian SER distribution with a low light intensity. Given this component, the mean photoelectron can be determined with the expression: 
	\begin{equation} 
		P(0) = \frac{N_{0}}{N_{trigg}} = e^{-\mu}
	\end{equation}
	\paragraph{}
where $N_{0}$ is the number of events without signal and $N_{trigg}$ is the total of events triggered with the external LED trigger. $N_{0}$ can be determined in several ways: by adjusting the Gaussian pedestal and determining its area ($\mu_{ped}$) or by counting events with ``no signal''. This second method can be done by subtracting events with signal to the total $N_{trigg}$ events and the events with signal can be quantified with the detection peak algorithm (see Section~\ref{sec:PeakDetection}) obtaining an estimation $\mu_{peak}$.
	\paragraph{}
	Two other mean photoelectron estimators were used. $\mu_{fit}$ is obtained in the deconvolution fit described in the previous subsection. Additionally, a mixed method correcting the $\mu_{ped}$ by calculating the SER inside the pedestal from the fit function gives a new estimator: $\mu_{ped\_fit}$.
	\paragraph{ }
	A comparison of these algorithms was performed at different HV testing $\mu$ stability with the aforementioned estimators. This comparison can be seen in Figure~\ref{fig:PMTMuAlgor}. All the PMT data were taken at $\mu \simeq 0.04$. There are some similar behaviors in the four estimators among PMTs. $\mu_{ped}$ and $\mu_{ped\_fit}$ underestimate the light because they do not take into account the signal inside the pedestal. This effect is corrected for $\mu_{fit}$ and $\mu_{peak}$. These last estimators have a very good agreement at low and medium voltage. It has to be noted an increase in all the estimators except $\mu_{fit}$ at high voltage. This effect may be due to the increase of the dark counts that affects only at high gain and low light (see Figures~\ref{fig:Mu_34} and~\ref{fig:Mu_51}). Every PMT unit has a different dependence of the dark counts increase as a function of HV as we will see in Section~\ref{sec:DarkCountsMeas} and the unit 51 has a very strong increase (see Figure~\ref{fig:DarkCountsvsHV}).
	\paragraph{}
	As a conclusion, the preferred estimators for $\mu$ are $\mu_{fit}$ and $\mu_{peak}$ for their agreement. Consequently, the recommended protocol for measuring light includes low light intensity and a low/medium voltage as we will see in Section~\ref{sec:PMTProtocol}.

	\subsection{Peak-to-valley ratio}\label{sec:PTVRatio}
\begin{figure}[h!]
	  \begin{center}
	    \includegraphics[width=.7\textwidth]{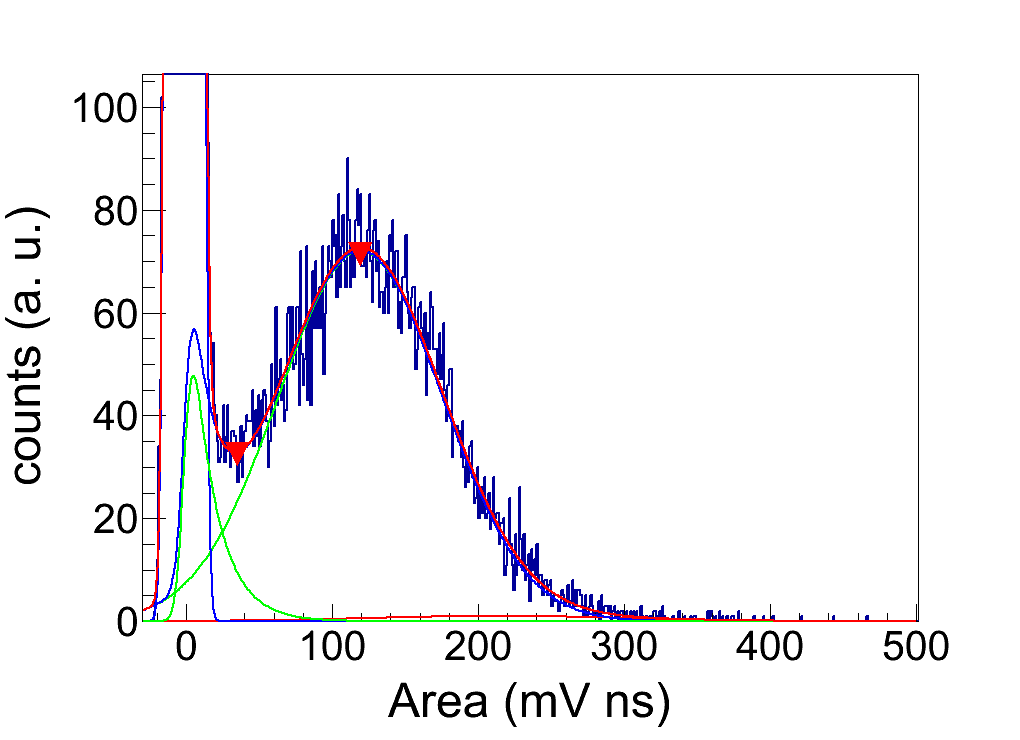}
	    \caption[Peak and valley SER extraction]{\label{fig:PV}Peak and valley SER extraction (red triangles).}
	  \end{center}
	\end{figure}

	The peak-to-valley ratio is defined as the ratio between the peak and the valley of the SER charge distribution. It is useful to characterize the discrimination of high charge and low charge peaks. The low charge responses are caused by the portion of light events producing significantly less signal, hitting the second dynode or being inelastic scattered in addition to dark current events.
	\paragraph{ }
	A high peak-to-valley ratio allows to better reject the baseline and dark noises consisting of low charge events (see Section~\ref{sec:PMTSetHVTrigg}). The extraction of the location and value of the peak and the valley can be seen in Figure~\ref{fig:PV}. The results of the peak-to-valley ratio of all tested PMTs can be seen in Section~\ref{sec:PMTResults}.

	\subsection{Dark counts}\label{sec:DarkCounts}
\begin{figure}[h!]
	  \begin{center}
		\begin{subfigure}[b]{0.5\textwidth}
		\centering
		\includegraphics[width=\textwidth]{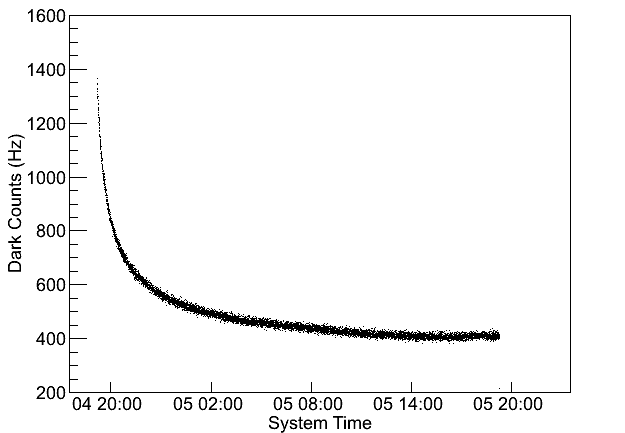}
		\caption[PMT dark counts after light exposure]{\label{fig:DarkCountsvsTime}}
		\end{subfigure}%
		~ 
		\begin{subfigure}[b]{0.5\textwidth}
		\centering
		\includegraphics[width=\textwidth]{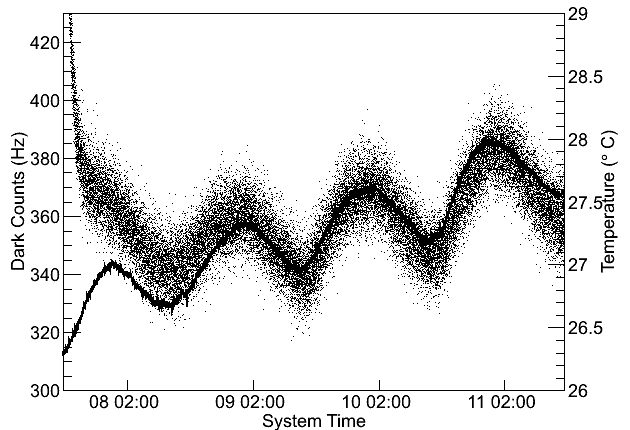}
		\caption[PMT dark vs. Time and temp]{\label{fig:DarkCountsvsTimeTemp} }
		\end{subfigure}%
		\caption[Dark count vs. time]{PMT dark counts decay after light exposure (a) and dark counts (points) and temperature (line) correlation (b).\label{fig:DarkCountsvsTimeTempandHV}}
	  \end{center}
	\end{figure}

	The dark counts (see Section~\ref{sec:PMTs}) are a natural source of noise using PMTs. Therefore, the characterization of the PMT dark counts is needed. It will give information about the expected rate of dark events and its influence in the trigger rate and the data.
	\paragraph{}
	A careful light isolation is required for the PMT in order to avoid external photon counting. For this reason, the PMT is introduced in a closed cylinder (as it can be seen in Figure~\ref{fig:PMTTestBench}) in a dark environment. The dark counts were measured counting the triggers with a scaler at regular time intervals. The trigger level was set at 2~mV, a value above the electric noise and near the typical value of the ANAIS trigger level. Two different measurements were performed: dark counts versus time in order to see the de-activation after the PMT light exposure and dark counts over HV once these counts were stable.
	\paragraph{}
	The dark counts over time after light exposure can be observed in Figure~\ref{fig:DarkCountsvsTime}. It can be seen a fast decay, but the counts become stable only after twenty hours. This is the temporal window needed for proper dark counts characterization, i.e. the time windows used in the systematic measurements (see Section~\ref{sec:DarkCountsMeas}). In addition to the decay, a more subtle effect can be observed. A clear correlation with the temperature can be seen in the long run when the activation effect becomes negligible. An example of such a correlation can be seen in Figure~\ref{fig:DarkCountsvsTimeTemp} showing the end of the activation and the temperature correlation. The dark counts increase 10\% with a temperature rise of 1 ºC. This is why a good stability in the temperature is required and its monitoring and control is needed for a long term experiment such as the ANAIS experiment.
	\paragraph{}
	The characterization of dark current as a function of high voltage and the comparison with the vendor data can be seen in Section~\ref{sec:DarkCountsMeas}. The protocol for subsequent measurements can be seen in Section~\ref{sec:PMTProtocol}.

	\section{Characterization protocol and results}\label{sec:PMTResults}
	A characterization protocol is needed due to the high number of PMTs to test. This protocol and associated tools must provide reproducibility and ease of use in order to be able to extract the parameters of all tested PMTs. In this section the automation of the characterization process and the obtained preliminary data for few PMTs are presented.
	\subsection{Parameter extraction automation}\label{sec:PMTExtraction}
	The first step to define a characterization protocol was to design and implement a set of tools for the automatic parametrization of the algorithms described in Section~\ref{sec:PMTSignal}. This is a critical step taking into account the difficulty of \emph{a priory} knowing the initial value of some parameters such as $Q_0$, $\alpha$ or $w$ in the deconvolution fit for all gain range. The quality of the fit improves with accurate initial values. All this automation assumes very low light.
	\paragraph {}
	The steps followed for an appropriate parametrization of the fit are:
	\begin{easylist}[itemize]
	& Fit the pedestal to a Gaussian, obtaining $Q_p$ and $\sigma_p$.
	& Fit the distribution of the events with one or more peak to a SER distribution, obtaining  $\alpha$ and $w$ (as it can be seen in Figure~\ref{fig:SERFitSER}).
	& Obtain $\mu_{ped}$ from the integral of the pedestal. This will be the initial value to $\mu$.
	& Obtain peak and valley values. The position of the peak is used as initial value of $Q_0$. 
	& Fit the SER fixing  $Q_p$, $\sigma_p$, $\alpha$ and $w$, obtaining $Q_0$, $\sigma_0$, $\mu_{fit}$.
	& Calculate $\mu_{peak}$, $\mu_{ped\_fit}$ and $Q_1$ and $\sigma_1$ with Equation~\ref{eq:q1_aprox}.
\end{easylist}
\paragraph{ }
This extraction procedure was used in the PMT characterization seen in the following subsections.
	\subsection{Gain curves}
\begin{figure}[h!]
	  \begin{center}
		\begin{subfigure}[b]{0.7\textwidth}
		\centering
		\includegraphics[width=\textwidth]{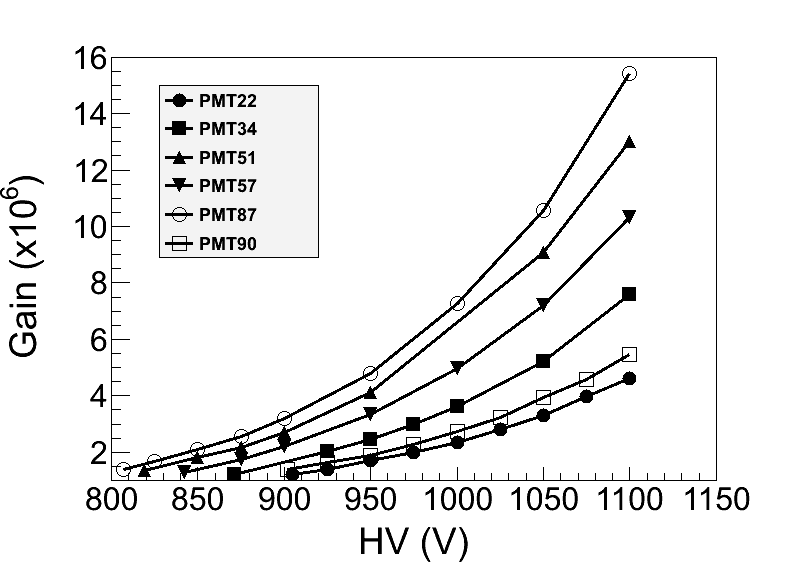}
		\caption[]{\label{fig:PMTsGainCurve}}
		\end{subfigure}%

		\begin{subfigure}[b]{0.7\textwidth}
		\centering
		\includegraphics[width=\textwidth]{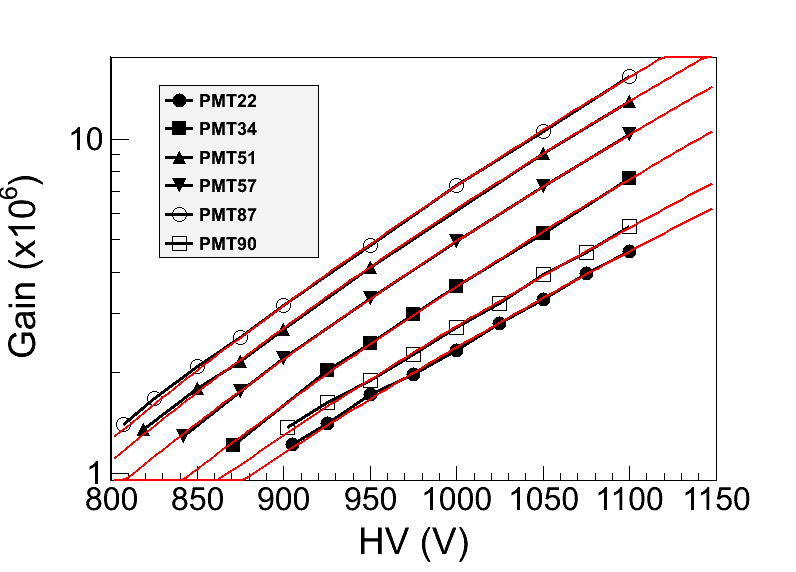}
		\caption[PMT Gain vs. HV]{\label{fig:PMTsGainCurveFit}}
		\end{subfigure}%
		
	    \caption[PMT gain curve]{\label{fig:PMTGainCurve}PMTs gain curve (a) and power law fit (b).}
	  \end{center}
	\end{figure}

	The extracted values for $Q_1$ were used to plot the gain curve of every PMT as it can be seen in Figure~\ref{fig:PMTsGainCurve}. The gain is calculated as the charge collected compared with the elemental incident charge (the electron produced by photoelectric effect). The charge collected is:
	\begin{align}
	\begin{split}
		Q = \int_{t0}^{t1} \frac{V(t)}{R}dt
	\end{split}
	\end{align}
and for a correct comparison with the electron elementary charge, the input impedance of the digitizer ($R$) has to be taken into account once the voltage signal is integrated.
	\paragraph{ }
	It is well known that the behavior of the gain with the voltage approximates to a power expression \hbox{$G = A V^K\label{eq:GainHV}$}~\cite{tubes2006basics}. The gain curves were fitted to the previous equation extracting $A$ and $K$ parameters as seen in Table~\ref{tab:HamR12699GainFit}. The value is $K=kn$, being $n$ the number of dynodes (9 for the R12699) and $k$ a constant depending on the dynode structure and the material having a value from 0.7 to 0.8~\cite{tubes2006basics}. Additionally, $A$ depends on the dynode collection efficiency. The result table is fully compatible with these facts.

	\paragraph{ }

	\begin{table}[h!]
	\begin{center}
	\begin{tabular}{ccc}
	\toprule
		PMT&A  & K\\ 
	\toprule
	FA0022 & $(1.62 \pm 0.08) \times 10^{-20}$  & $6.72\pm 0.01$\\
	\hline
	FA0034 & $(2.04 \pm 0.07) \times 10^{-23}$  & $7.75\pm0.01$\\
	\hline
	FA0051 & $(3.73 \pm 0.10) \times 10^{-23}$  & $7.74\pm0.01$\\
	\hline
	FA0057 & $(3.24 \pm 0.04) \times 10^{-23}$ & $7.73\pm0.02$\\
	\hline
	FA0087 & $(2.61 \pm 0.09) \times 10^{-23}$ & $7.81\pm0.01$\\
	\hline
	FA0090 & $(7.67 \pm 0.43) \times 10^{-22}$ & $7.18\pm0.01$\\

	\hline
	\end{tabular}
	\caption{Hamamatsu R12669SEL2 Gain fit.}
	\label{tab:HamR12699GainFit}       
	\end{center}
	\end{table}

	Once extracted the gain curves, they are used help the operating voltage selection (See Section~\ref{sec:PMTSetHVTrigg}). The curves describe the dependence of gain and current with the voltage and they can give the setting for the PMT once selected the desired gain or the current for the maximum of the energy range of interest avoiding the non-linear ranges.

	\subsection{Peak-to-valley}
	The data used to calculate the peak-to-valley ratio were the same that the used to characterize the gain curve and the light detection algorithms (see Section~\ref{sec:MuAlgor}) and consequently at low light intensity. For this reason, the statistical significance of the peak and valley zones is limited and the errors increase. In any case, the results (see Table~\ref{tab:HamR12699PVRatio}) show a good peak-to-valley ratio, always above 2.0. 
	\begin{table}[h!]
	\begin{center}
	\begin{tabular}{cc}
	\toprule
		PMT& Peak-to-valley ratio\\ 
	\toprule
	FA0022 & $2.0\pm 0.1$\\
	\hline
	FA0034 & $2.5\pm0.2$\\
	\hline
	FA0051 & $2.5\pm0.1$\\
	\hline
	FA0057 & $2.2\pm0.1$\\
	\hline
	FA0087 & $2.1\pm0.2$\\
	\hline
	FA0090 & $2.2\pm0.2$\\

	\hline
	\end{tabular}
	\caption{Hamamatsu R12669SEL2 peak-to-valley ratio.}
	\label{tab:HamR12699PVRatio}       
	\end{center}
	\end{table}
	\paragraph{ }
	These peak-to-valley values show the suitability of the photoelectron triggering (see Section~\ref{sec:PMTSetHVTrigg}).
	\subsection{Dark counts}\label{sec:DarkCountsMeas}
\begin{figure}[h!]
	  \begin{center}
		\begin{subfigure}[b]{0.7\textwidth}
		\centering
		\includegraphics[width=\textwidth]{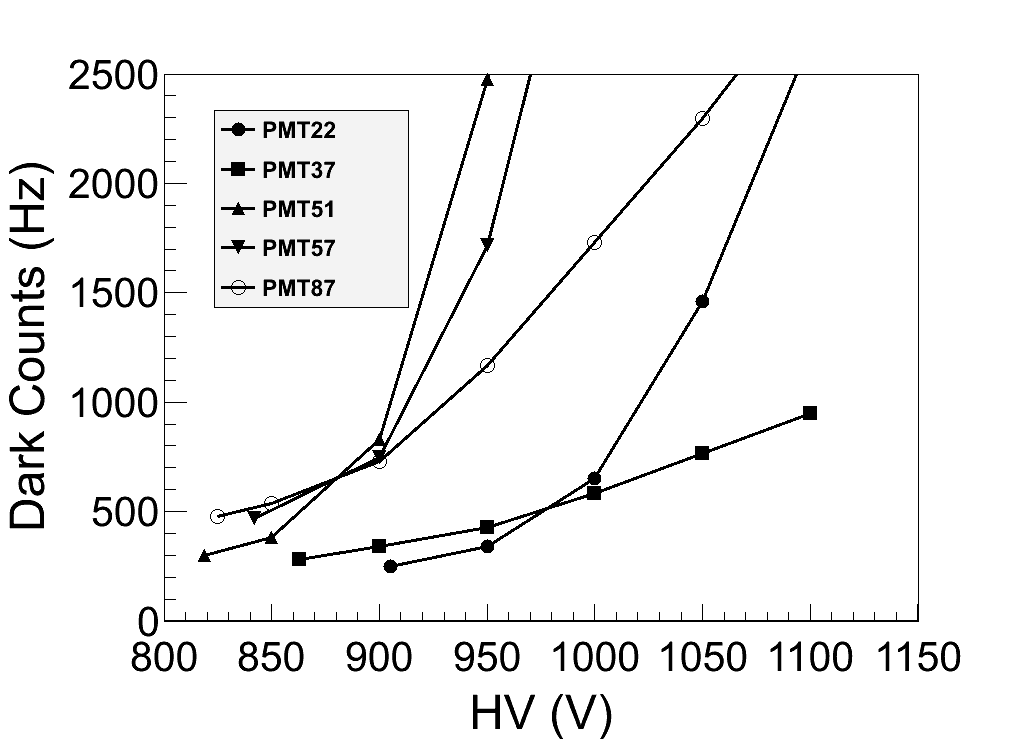}
		\end{subfigure}%

		\caption[PMT dark counts vs. HV]{\label{fig:DarkCountsvsHV}PMT dark counts vs. HV.}
	  \end{center}
	\end{figure}

	The dark counts were measured twenty hours after their last light exposition in order to wait for their de-excitation following the conclusion seen in Section~\ref{sec:DarkCounts}. The trigger level was 2 mV.
	\paragraph{ }
	The dark counts behavior with high voltage for five photomultipliers can be seen in Figure~\ref{fig:DarkCountsvsHV} showing an exponential growth with a very intrinsic behavior for every PMT. The comparison of stable dark current at voltage corresponding to gain $\times 10^{6}$ with the Hamamatsu data-sheet can be seen in Table~\ref{tab:DarkCurrent}. It can be noted that the measured values are higher than the Hamamatsu data. The environmental and triggering conditions for the Hamamatsu test are unknown. In any case, the obtained values are below the 500 Hz upper limit, fulfilling the initial requirements.
	\paragraph{}

	\begin{table}[h!]
		\begin{center}
			\begin{tabular}{ c c c }
				\toprule
				 PMT &  Dark current (Hz) & Dark current Hamamatsu (Hz) \\
				\toprule
			 FA0022 &  $248.5\pm 5.8 $ &  135.00\\
			 \hline
			 FA0037 &  $283.1\pm 7.2 $ &  71.10\\
			 \hline
			 FA0051 &  $298.8\pm 6.2 $ &  43.60\\
			 \hline
			 FA0057 &  $467.5\pm 9.2 $ &  84.70\\
			 \hline
			 FA0087 &  $479.5\pm 9.6 $ &  297.00\\
			 \hline

		\end{tabular}
		\caption[Dark counts at EBBV]{Dark counts at gain $1 \cdot 10^6$.} 
		\label{tab:DarkCurrent} 
	\end{center}
\end{table}

	\subsection{Relative quantum efficiency}
	Relative quantum efficiency measurements require high stability in the light source and in the physical set-up in order to ensure the same incident light to the photomultiplier. For this reason, a careful and fast PMT change must be performed taking into account the intrinsic drift of the set-up (see Section~\ref{sec:LEDUnstab}) and the harmful effects of an undesired change in the optical fiber, filter and diffuser positions.
	\paragraph{}
	The data were collected at very low light ($\mu \simeq 0.05$) to take advantage of the better behavior of the detection light algorithms seen in Section~\ref{sec:MuAlgor}. The measured $\mu$ is compared with the $\mu$ from the reference PMT (FA0087). The results of the measurements can be seen in Table~\ref{tab:RQETest}. It can be noted that the discrepancies between the expected and measured relative quantum efficiency are above the statistical error possibly due to the instabilities in light emission and collection. A more stable set-up is in preparation as noted earlier. Anyway, this set-up is useful as crosscheck in order to detect units with significant deviations from the Hamamatsu specification (see Section~\ref{sec:PMTs}).     
	\begin{table}[h!]
		\begin{center}
			\begin{tabular}{ c c c }
				\toprule
				FA0051 $\mu$  & FA0087 $\mu$ & Relative QE \\
				\toprule
				$0.03975 \pm 0.00058$ & $0.04593 \pm 0.00061$ & $ 0.865\pm 0.017$\\
				\hline
				FA0051 QE  & FA0087 QE & Relative QE (Hamamatsu)  \\
				\hline
				33.19 & 40.37 & 0.822\\
				\hline
				& & \\
				& & \\
				\toprule
				FA0057 $\mu$  & FA0087 $\mu$ & Relative QE \\
				\toprule
				$0.05665 \pm 0.00072$ & $0.06432 \pm 0.00072$ & $ 0.818\pm 0.016$\\
				\hline
				FA0057 QE  & FA0087 QE & Relative QE (Hamamatsu)  \\
				\hline
				33.95 & 40.37 & 0.841\\
				\hline
				& & \\
				& & \\
				\toprule
				FA0090 $\mu$  & FA0087 $\mu$ & Relative QE \\
				\toprule
				$0.04266 \pm 0.00063$ & $0.04493 \pm 0.00060$ & $ 0.949\pm 0.019$\\
				\hline
				FA0090 QE  & FA0087 QE & Relative QE (Hamamatsu)  \\
				\hline
				38.97 & 40.37 & 0.965\\

			\end{tabular}
			\caption[Relative quantum efficiency results]{Relative quantum efficiency results.} 
			\label{tab:RQETest} 
		\end{center}
	\end{table}
	\subsection{Characterization protocol}\label{sec:PMTProtocol}
	The proposed protocol to test all the ANAIS PMTs is based on the set-up and algorithms previously described in Section~\ref{sec:PMTSignal} taking advantage of accumulated experience in the analysis of the PMT data presented in this section. The steps of such a protocol will be:
\begin{easylist}
	& Data taking at different voltages in order to draw the gain curve. The starting point will be the voltage of gain $\times 10^6$ with steps of 25 V. The data will be deconvoluted obtaining the SER parameters, including gain and peak-to-valley ratio with the procedure described in Section~\ref{sec:PMTExtraction}.
	& Taking dark counts data at least twenty hours after the PMT installation in order to get a stable dark rate. The starting point will be the $10{^6}$ gain.
	& Taking a run at low voltage and change the PMT for the reference PMT (FA0087) in order to compare the measured light and calculate the relative quantum efficiency. This PMT was chosen for its high quantum efficiency.  
\end{easylist}

\section{Operating voltage and trigger level selection}\label{sec:PMTSetHVTrigg}
\begin{figure}[h!]
	  \begin{center}
		\begin{subfigure}[b]{0.5\textwidth}
		\centering
		\includegraphics[width=\textwidth]{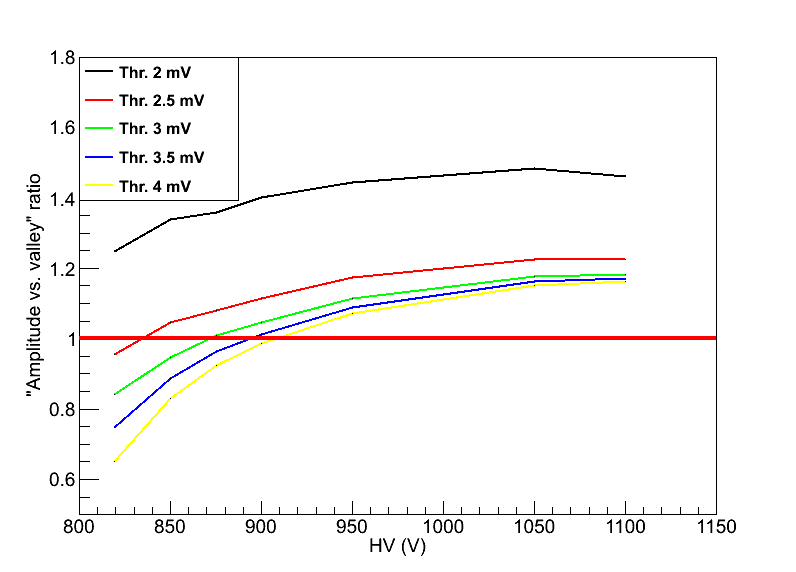}
		\caption{\label{fig:AmpvsValley}}
		\end{subfigure}%
		~ 
		\begin{subfigure}[b]{0.5\textwidth}
		\centering
		\includegraphics[width=\textwidth]{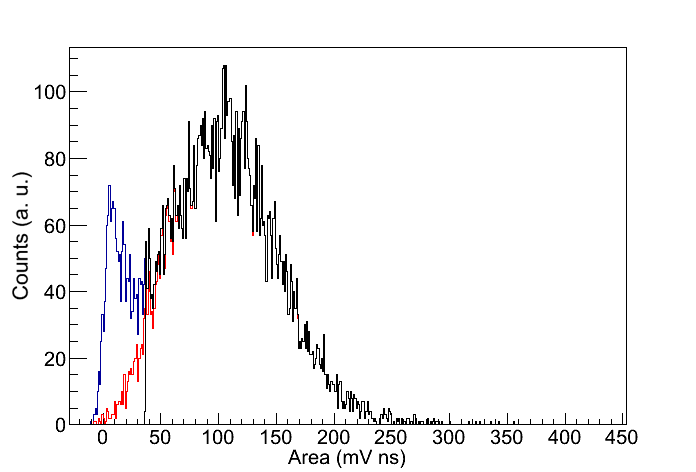}
		\caption{\label{fig:AmpvsValley_spc}}
		\end{subfigure}%
		\caption[Voltage and trigger level selection]{\label{fig:HVTriggSelection}Voltage and trigger level selection. Ratio between amplitude triggered events and \emph{above the SER valley} events at different amplitude trigger level (see text for details) (a). The SER at selected voltage (blue) and triggered events at selected trigger level (red) compared with valley cut (black) (b).}
	  \end{center}
	\end{figure}

The operating high voltage of every PMT can be selected once it is fully characterized. Several factors must be taken into account in order to get the proper selection: signal-to-noise ratio, the PMT linearity range and the photoelectron signal. 
	\paragraph{}
	Triggering at photoelectron level has a direct impact on the high voltage settings: the SER distribution has to be above the baseline RMS. The higher the voltage, the better the signal to noise ratio. Nevertheless, the operating voltage rise can produce other unwanted effects such as PMT saturation, reflections and dark counts. The first would have a negative impact in the linearity and the latter two in the signal-to-noise ratio.
	\paragraph{}
	The PMT linearity depends on the cathode and anode current linearities with light intensity and both depends only on the current~\cite{tubes2006basics}. For this reason, the energy range of interest imposes restrictions to the high voltage. The maximum of the energy range of interest should create a current below the saturation range. This is a major restriction when characterizing the detectors because the alpha rate plays a key role in the background understanding (see Sections~\ref{sec:ANAIS25} and~\ref{sec:ANAIS37}). On the other hand, a significant increase on the operating voltage can cause signal reflections which could be mistaken for photoelectrons by the peak identification algorithm and it can also increase the dark count rate. These two effects limit the range of the voltage increase.
	\paragraph{}
	Once selected the operation voltage with the energy region of interest in mind, the trigger level must be naturally selected to trigger all the SER distribution. Here, a method for finding the associated trigger level to a voltage is presented. The method is to locate the SER valley in the pulsed LED data used in the previous sections.
	\paragraph{}
	It has to be noted that the valley is better algorithmically found in charge than in amplitude distribution but the ANAIS trigger is performed in amplitude. A software comparison between amplitude and area distributions was performed in order to help to set trigger level. Figure~\ref{fig:AmpvsValley} shows the ratio between the events above the charge SER valley and the events above the amplitude trigger level at different trigger levels. The horizontal red line at a ratio of 1 marks the equal rate of both amplitude and area triggers and it is useful as reference. Following this reference, the trigger level would be 3 mV for an operating voltage of 875 V. Figure~\ref{fig:AmpvsValley_spc} shows the SER charge distribution at the selected voltage (875 V, blue line) and the comparison of the events above the SER valley (black) and the events above the selected trigger (red) showing a good agreement and the suitability as trigger selection method.

\section{Coincidence Trigger}\label{sec:CoincTrigg}
The low energy threshold required for the experiment makes triggering a very critical step. Dark current, as described and measured in Section~\ref{sec:DarkCounts}, has a high rate of events produced in the PMTs without the presence of light with a signal level compatible with light. Triggering at that rate would saturate the data acquisition with a dead time unacceptable for the experiment (see Section ~\ref{sec:ExperimentalRequirements}).
\paragraph{ }
The coincidence trigger technique uses two (or more) PMTs in a crystal detector in order to ensure having signal in all detector PMTs. This technique drastically reduces the trigger rate allowing to collect more light and discard other non-bulk scintillation events. Coincidence must be conceived from the very beginning of the experiment because the crystals must allow to couple two (or more) PMTs. The ANAIS modules uses two PMTs as seen in Section~\ref{sec:ANAIS25} for example.
\paragraph{}
The coincidence trigger is defined by the \emph{coincidence window}. This window is the time in which two signals are considered coincident. It must be set carefully because a big window may trigger many random coincidences but a small one can lose some genuine events, depending on light collection and dark rate. The random coincidence rate as a function of the two rates ($r_1$ and $r_2$) and the coincidence window ($\tau$) is~\cite{knoll2010radiation}:  
\begin{equation} 
	R_{random\_coinc} = 2 \tau r_1 r_2 \label{eq:random_coincidence}
\end{equation}
The dark rate for the R12669SEL2 has its upper limit at 500 Hz, so the upper limit random rate for a 200 ns coincidence window is 0.1 Hz. The resulting upper limits of the total random rate in the full experiment, with nine detectors, is of the order of the hertz, being acceptable for the acquisition system (see Section~\ref{sec:DeadTimeMeas}).    

\chapter{Data acquisition electronic front-end}\label{sec:FrontEnd}
This chapter covers the design, implementation and the tests of the electronic front-end modules for the ANAIS experiment. The design presented here is the evolution of the front-end of the previous prototypes~\cite{CPOBES,TesisMaria} adapted to VME electronics. Some parts have been developed specifically in this work such as the preamplifier (see Section~\ref{sec:Preamp}), the real time/live time system (see Section~\ref{sec:TimeScalers}) and the NIM delay module (see Section~\ref{sec:NIMDelay}).
\section{Front-end requirements}\label{sec:HWReq}
The electronic front-end must be designed with the scintillation signal described in the previous chapter in mind. It has to be able to get information in a wide range of energies in order to understand the experiment background while optimizing the low energy range. The system must have the lowest possible hardware threshold level in order to achieve the lowest energy threshold. Hence, a baseline as clean as possible is needed.
\subsubsection{Low threshold} An energy threshold below 2 keV must be achieved as described in Section~\ref{sec:ExperimentalRequirements}. Being able to lower the energy threshold will allow to better explore the DAMA/LIBRA region and it could enable the observation of expected features for the WIMP interactions such as the phase inversion.
\paragraph{}
This requirement implies a hardware trigger at the photoelectron level with a good signal to noise ratio, as it can be seen in Section~\ref{sec:LESignal}.
\subsubsection{Stability} The requirement of stable operation conditions described in Section~\ref{sec:ExperimentalRequirements} is particularly relevant from electronic front-end point of view. The system must ensure a valid set of stable enough energy estimators during all the data taking phase. Taking into account the nature of the experiment, the stability must be ensured during several annual cycles. These estimators must be accurately calibrated and effects like gain drift or fitful noises must be avoided as far as possible.
\paragraph{}
The above conditions are translated to the front-end design:
 
\begin{easylist}[itemize]
& Trigger level must be at the photoelectron level and it must be as stable as possible.
& Electronic noise must be minimized and controlled.
& The system must allow energy calibrations to control the stability.
& The whole system must be robust, stable and scalable.
\end{easylist}

\paragraph{}
The design of the front-end and the fulfillment of these requirements will be discussed in the following. The tests for the modules are reported in Section~\ref{sec:FrontEndTest}, the hardware trigger level is discussed in Chapter~\ref{sec:FullCharac} and the stability of the data acquisition parameters is covered in Chapter~\ref{sec:Stability}.

\section{High Level Description}

\begin{figure}[h!]
  \begin{center}
    \includegraphics[width=1\textwidth]{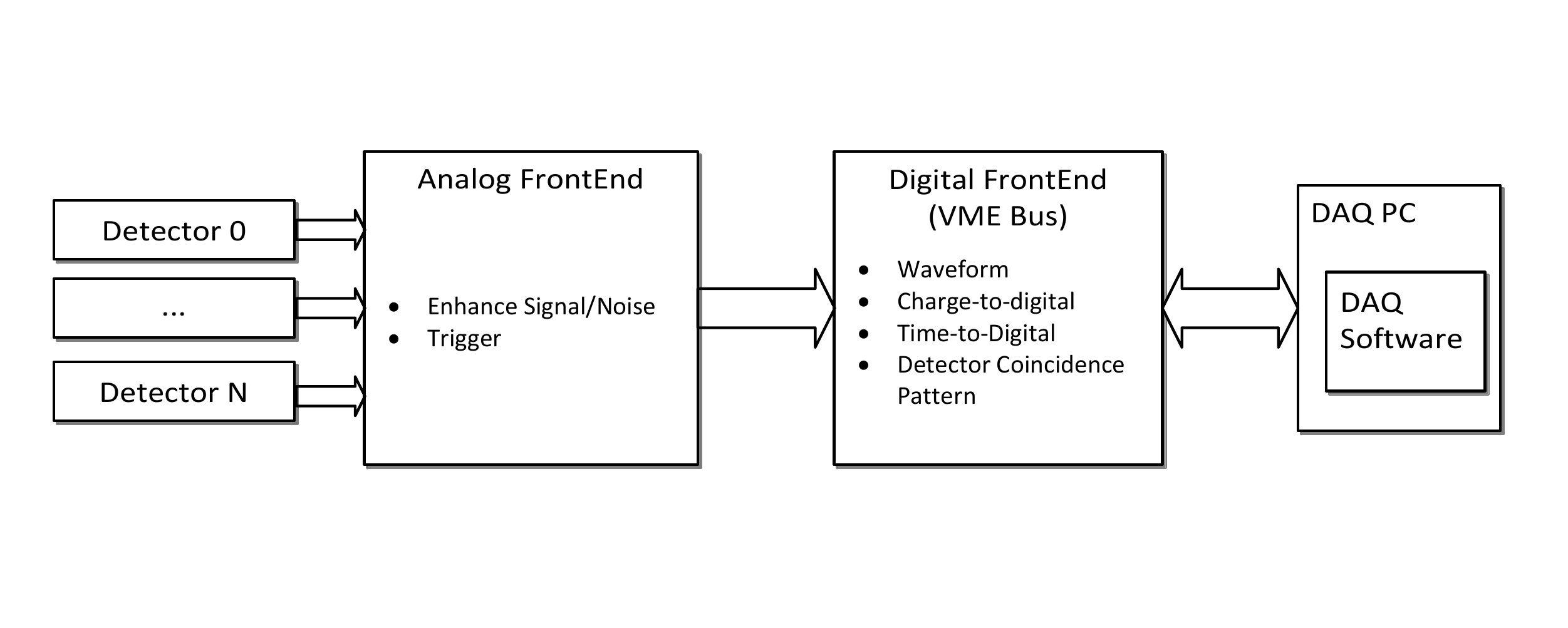}
    \caption[DAQ System Block Diagram]{\label{fig:DAQBlockDiagram}DAQ System Block Diagram.}
  \end{center}
\end{figure}

The electronic front-end needs several stages in order to achieve the above described requirements as it can be seen in Figure~\ref{fig:DAQBlockDiagram}. The basic stages are: 

\begin{easylist}[itemize]
& Analog front-end: It is responsible for processing PMT signals in order to achieve:
&& Trigger 
&& Improve Signal/Noise ratio
& Digital front-end: It is responsible for digitizing all the relevant signal parameters:
&& PMT waveforms
&& Charge generated at the PMTs
&& Pattern of triggering for all detectors
&& Triggering Time for all the PMT signals
&& Real time of the trigger and accumulated live time.
\end{easylist}

\section{Analog front-end}\label{sec:AnalogFrontend}
\begin{figure}[h!]
  \begin{center}
    \includegraphics[width=1\textwidth]{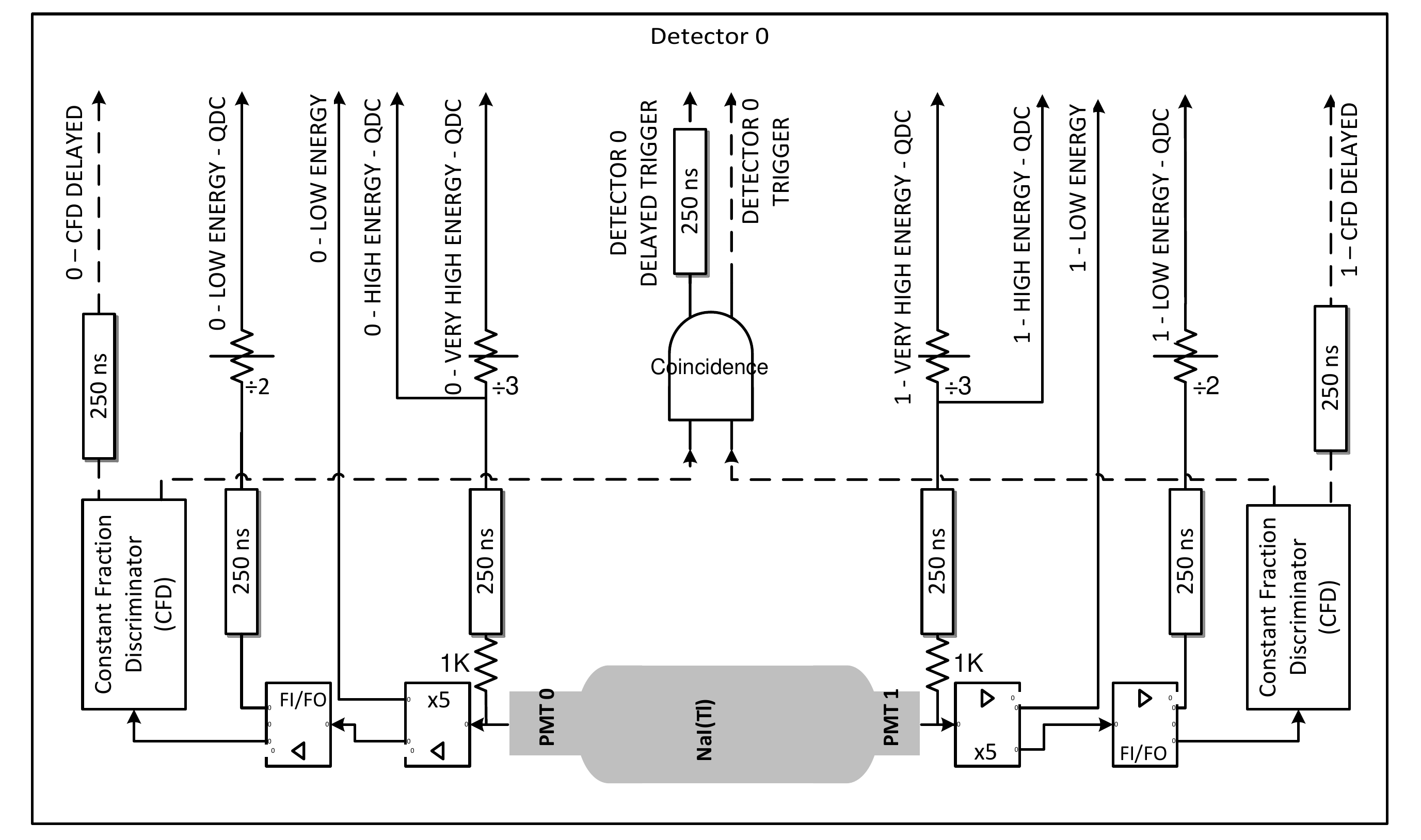}
    \caption[Detector Analog front-end]{\label{fig:AnalogFrontend}Detector Analog front-end.}
  \end{center}
\end{figure}

The analog front-end consists of a home-made preamplifier and several standard NIM~\cite{NIM} modules. The design of this block can be seen in Figure~\ref{fig:AnalogFrontend}.
The first step consists of an AD8009 based preamplifier in order to strengthen the signal as close to the PMT as possible. This preamplifier has two outputs: one for digitizing and one for the rest of analog front-end. This last signal is duplicated using a Fan-In/Fan-Out from which the trigger signal is obtained.
\paragraph{}
The triggering signal is used as input of a Constant Fraction Discriminator (CFD). The detector trigger is the logical AND of the two detector CFDs. The global trigger is the logical OR of the detector triggers. More information about triggering strategies in ANAIS can be seen in Section~\ref{sec:TriggMod}.
\paragraph{}
Figure~\ref{fig:AnalogFrontend} also shows the high energy signal. There is an optional resistive divider of 1~k$\Omega$ inside the preamplifier in order to get an attenuated signal. This signal do not suffer limitations in dynamic range previous to the digitization step and it can be used for the study of the high energy ranges. 
\paragraph{}
All modules are listed and described in following subsections: preamplifier in Section~\ref{sec:Preamp}, trigger modules in Section~\ref{sec:TriggMod} and NIM modules in Section~\ref{sec:NIMModules}.
\paragraph{}
\subsection{Preamplifier}\label{sec:Preamp}
\begin{figure}[h!]
  \begin{center}
    \includegraphics[width=1\textwidth]{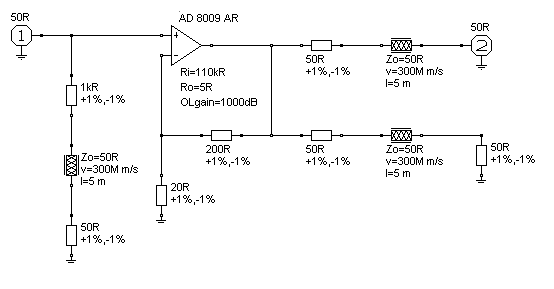}
    \caption[Preamplifier schematics]{\label{fig:preamp_scheme}Simplified schematics for the AD8009 based preamplifier.}
  \end{center}
\end{figure}

The particular restrictions of radiopurity and low background seen in Section~\ref{sec:ExperimentalRequirements} difficult the signal amplification near its production. The preamplifiers are located outside the shielding for low background reasons, but close to it to minimize signal loss and noise issues and to improve the signal/noise ratio.
\paragraph{}
The preamplifier design is based on the AD8009 low distortion amplifier. This amplifier is an ultrahigh speed current feedback amplifier with a good slew rate, enabling the resulting preamplifier to have a good bandwidth. It has been designed having ANAIS requirements in mind, providing two main outputs, one for triggering and one for digitization. It also provides an attenuated output in order to explore the high energy range without limitation in dynamic range. Figure~\ref{fig:preamp_scheme} shows a simplified version of the schematics without the power supply filters.
\paragraph{}

\begin{figure}[h!]
\begin{subfigure}[b]{0.5\textwidth}
  \begin{center}
    \includegraphics[width=\textwidth]{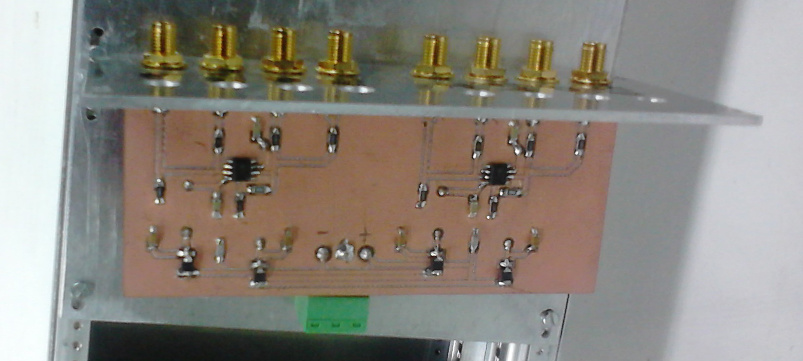}
    \caption[Preamplifier prototype]{\label{fig:preamp_pyhs_rack}PCB for two preamplifiers.}
  \end{center}
  \end{subfigure}
        ~ 
\begin{subfigure}[b]{0.5\textwidth}
  \begin{center}
    \includegraphics[width=\textwidth]{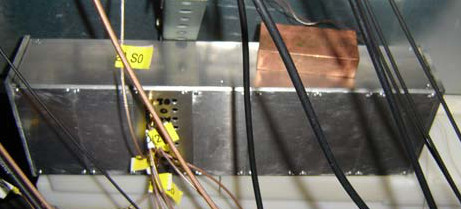}
    \caption[Preamplifier prototype]{\label{fig:preamp_shield_rack}Preamplifiers rack.}
  \end{center}
  \end{subfigure}
        \caption{AD8009 based preamplifier design for rack.}\label{fig:preamp_rack}
\end{figure}

The preamplifier was designed taking into account the nominal bandwidth of the MATACQ digitizer board (300 MHz). In order to achieve such a bandwidth, a minimal physical layout was designed and SMA connectors were used. The first prototypes were designed as a proof of concept, testing basic functionality, linear behavior and noise issues. The next step was to scale up the design in order to achieve modularity, ease of installation and maintainability of the whole set of needed amplifiers. The basic part of this new design was a PCB with two preamplifiers on it (see Figure~\ref{fig:preamp_pyhs_rack}). This PCB can be mounted in a rack located next to the shielding as can be seen in Figure~\ref{fig:preamp_shield_rack}.
\paragraph{}
The development of this amplifier was extensively tested as it can be seen in Section~\ref{fig:PreampTest} showing a good linear response and an increase of the signal/noise ratio.
\subsection{Trigger Modules}\label{sec:TriggMod}
This section describes the possible strategies to perform the PMT signal trigger and the needed modules for these strategies. Finally, ANAIS global trigger strategy is presented and discussed.
\\
\begin{easylist}[itemize]
& \textbf{Low threshold discriminators.} The simplest strategy is to detect signals with a discriminator; a signal triggers when a voltage threshold is exceeded. This is the low threshold discriminator (LTD) mechanism. LTD can be used for triggering low energy events if the signal distribution can be separated from the noise in amplitude.
& \textbf{Constant Fraction Discriminators.} Constant fraction discriminator (CFD) technique is similar to LTD but it gives a better trigger in temporal terms~\cite{gedcke1968design} avoiding jitter in the output discriminator signal, whereas a LTD triggers when the signal passes the trigger level. A CFD triggers when the signal passes a constant fraction of the maximum of the pulse. This feature is obtained by subtracting a delayed and attenuated copy of the signal to the original one and detecting the zero crossing for signals greater than the threshold level. This kind of discriminators have two configurable parameters: threshold and delay.
\end{easylist}
\paragraph{}
ANAIS has a very low energy threshold as target so the detectors are designed to couple with two PMTs and trigger in coincidence. This coincidence, logical AND, can be performed with a coincidence module (see Section~\ref{sec:NIMModules}).
\paragraph{}
Some tests were performed for choosing discrimination strategy (see Section~\ref{sec:TestTriggerStrat}) and as a result the CFD option was selected. This election has some advantages:
\begin{easylist}[itemize]
& CFD trigger is a better temporal reference.
& this particular CFD (see next section) saves a window generator because the maximum width exceeds 200 ns. LTD has a maximum window of 95 ns.
\end{easylist}

It has some disadvantages such as requiring some careful delay choosing as it can be seen in Section~\ref{sec:TestTriggerStrat}.
\paragraph{}
The discriminator digital output width is a critical parameter because, being wide enough, it can determine the coincidence window as input of the coincide module. ANAIS-25 and ANAIS-37 set-ups were performed with 200 ns coincidence window. A discussion for the impact in the reduction of such window can be found in Sections~\ref{sec:HWTriggEff} and~\ref{sec:100nsNoise}. Additionally, ANAIS does not use shaping because a photoelectron triggering can be achieved (see Section~\ref{sec:SERTrEff}) and the baseline is clean enough to avoid the need of an additional shaping stage.
\subsection{NIM Modules }\label{sec:NIMModules}
All NIM modules used in the ANAIS analog front-end are listed and described in this section explaining their most important features and their use in ANAIS.
\paragraph{Linear Fan-In/Fan-Out:} CAEN N625 (Quad Linear FAN-IN FAN-OUT) has four sections of four inputs and four outputs. Each section produces in all its output connectors the analog sum of all section inputs. The sum can be inverted optionally. The module bandwidth is 100 MHz. It is used in ANAIS to obtain signal copies for all modules except for the MATACQ VME input to preserve the signal bandwidth.
\paragraph{Constant fraction discriminator:} CAEN N843 (16 Channel Constant Fraction Discriminator) accepts 16 negative inputs and it produces 2 NIM signals and 1 inverted NIM signal for every channel when it exceeds the threshold level that ranges from 1-255 mV and it is expressed as a ``constant fraction" of the input signal. The constant fractions of this CFD are fixed at 20\%. It is used in ANAIS to detect a significant signal in one PMT and generate 200 ns gates. It uses two ordinary outputs, one for coincidence as input of the Coincidence Unit and the other, conveniently delayed, as input of the TDC VME module in order to save a precise timestamp of the arrival of every signal. 
\paragraph{Coincidence unit:}CAEN N455 (Quad Coincidence Logic Unit) has four sections and can perform selectable AND/OR logical operations. It has three NIM outputs and one inverted NIM output with selectable width via trimmer. In addition, it has an overlap output with a width equal to the interval in which the logical function is true. In ANAIS this module is used in a straightforward way as AND logical coincidence. It uses two normal outputs, one for the global trigger to the OR module and another as input of the pattern unit VME module.
\paragraph{OR Module:} CAEN N113 (Dual OR 12 IN - 2 OUT). This module has two sections with twelve inputs and two outputs every section. The output is the logical OR of every logical NIM input signal. This module has in the ANAIS set-up the role of joining all detector triggers in a global trigger. The output goes to the VME I/O Register that prevents a new trigger when boards are busy.
\paragraph{Gate generator:} N93B (Dual Timer) has two pulser generators from 50 ns to 10~s. Both input and output are logic NIM signals. This module is used in ANAIS to generate a precise 1 $\mu$s window that must be propagated to all digital VME boards (see Section~\ref{sec:DigitalFrontEnd} and Figure~\ref{fig:DigitalFrontEnd}) as trigger or acquisition window. The input comes from the VME I/O Register and the output goes to a logical Fan-In/Fan-Out.

\paragraph{Logical Fan-In/Fan-Out:}CAEN N454 (4-8 Logic FAN-IN/FAN-OUT) has four sections with four inputs and six outputs. The module performs a logical OR. Its outputs are four normal NIM and two complementary (inverted) NIM signals. ANAIS uses a logical Fan-In/Fan-Out as trigger propagator making copies of the output of the Gate Generator described in the previous paragraph and triggering all VME modules.
\paragraph{Delay Module:} A home-made high channel density NIM module. It takes NIM signals as input and produces a delayed gate as output. The delay is adjustable via external trimmer. The design and testing can be seen in Section~\ref{sec:NIMDelay}.
\section{Digital front-end}\label{sec:DigitalFrontEnd}
\begin{figure}[h!]
  \begin{center}
    \includegraphics[width=1\textwidth]{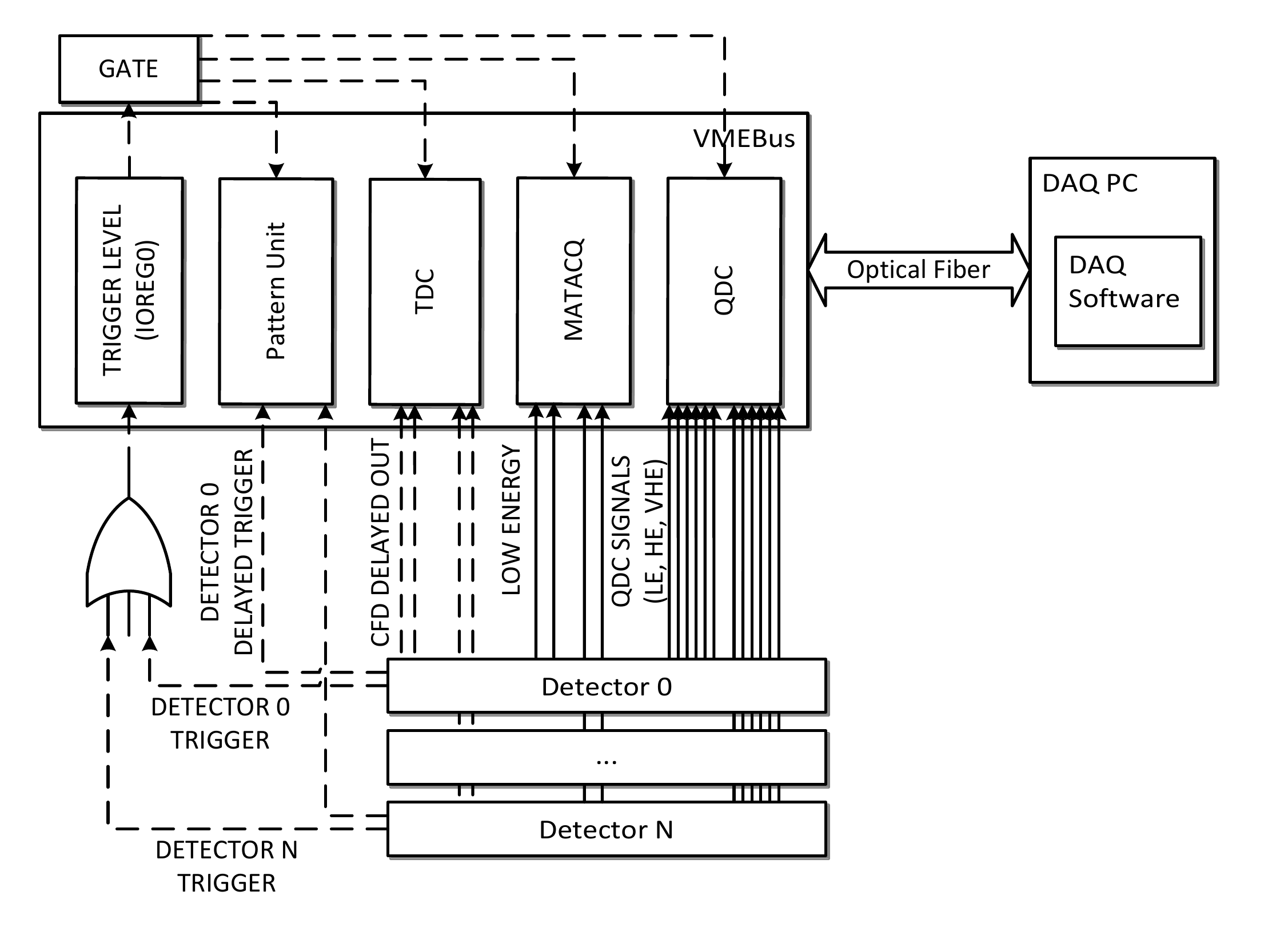}
    \caption[Digital front-end]{\label{fig:DigitalFrontEnd}Digital front-end.}
  \end{center}
\end{figure}

The ANAIS digital front-end consists of several VME~\cite{VME} modules. These modules convert the relevant signal parameters to digital values. They are controlled by the trigger of the analog front-end and store digital information in their buffers. The DAQ program reads these buffers and must ensure their validity.
\paragraph{}
The design of this section can be seen in Figure~\ref{fig:DigitalFrontEnd}. The global trigger is the logical OR of the detector trigger but it is filtered by the I/O Register. This module has the key role of marking the busy state of the system. The trigger is distributed throughout the system if the I/O register is in idle state (OFF). This distribution is performed by a gate generator with a trigger window of 1~$\mu$s. The I/O Register also marks the start of the DAQ Software acquisition and stores the busy system state (ON). The busy state must be reset by the DAQ program once the buffers are read, all information is saved and all boards are ready for a new acquisition. An OR trigger is discarded when the I/O register is on busy ON state in order to preserve the buffer integrity. The I/O register ON state is also used to measure the real time and the live time of the experiment as described in Section~\ref{sec:TimeScalers}. The VME boards receive both trigger and signals and convert signal and signal parameters into digital. 

\paragraph{}
The digitized parameters are listed and discussed here:
\paragraph{Signal Waveform:} The 230 ns decay constant of NaI(Tl) scintillators (see Section~\ref{sec:NaIScint}) and the expected signal from low energy events (see Section~\ref{sec:LESignal}) determines the digitization window and minimal bandwidth in order to get all relevant information. The digitization window has a minimal width of 1 $\mu$s and the minimal bandwidth must be 200 MHz and a sampling rate of the order of 1 GS/s.
\paragraph{Charge:} The energy yield of the interaction is proportional to the charge obtained from a PMT as described in Section~\ref{sec:PMTSignal}, so it is a very important parameter to digitize. It can be obtained by integrating the area in the previous waveform in the dynamic range of the digitizer or using a Charge-to-Digital converter (QDC).
\paragraph{Trigger time:} The relative trigger time of PMT signals for the same detector can be a useful information in order to discriminate random coincidence from bulk scintillation. This information can be digitized directly with a Time-to-Digital Converter (TDC) with the CFD outputs as inputs or it can be deduced analyzing the waveform.
\paragraph{Coincidence pattern:} It is very important to know which detectors were triggered in an event. This information can be saved in a structure called coincidence pattern: every bit of that structure has a related detector. This bit marks if the detector has triggered in that event. This information could be reconstructed at analysis time seeing what detector has signal or it can be done with a specific module.

\subsection{Time scalers}\label{sec:TimeScalers}
\begin{figure}[h!]
  \begin{center}
    \includegraphics[width=1\textwidth]{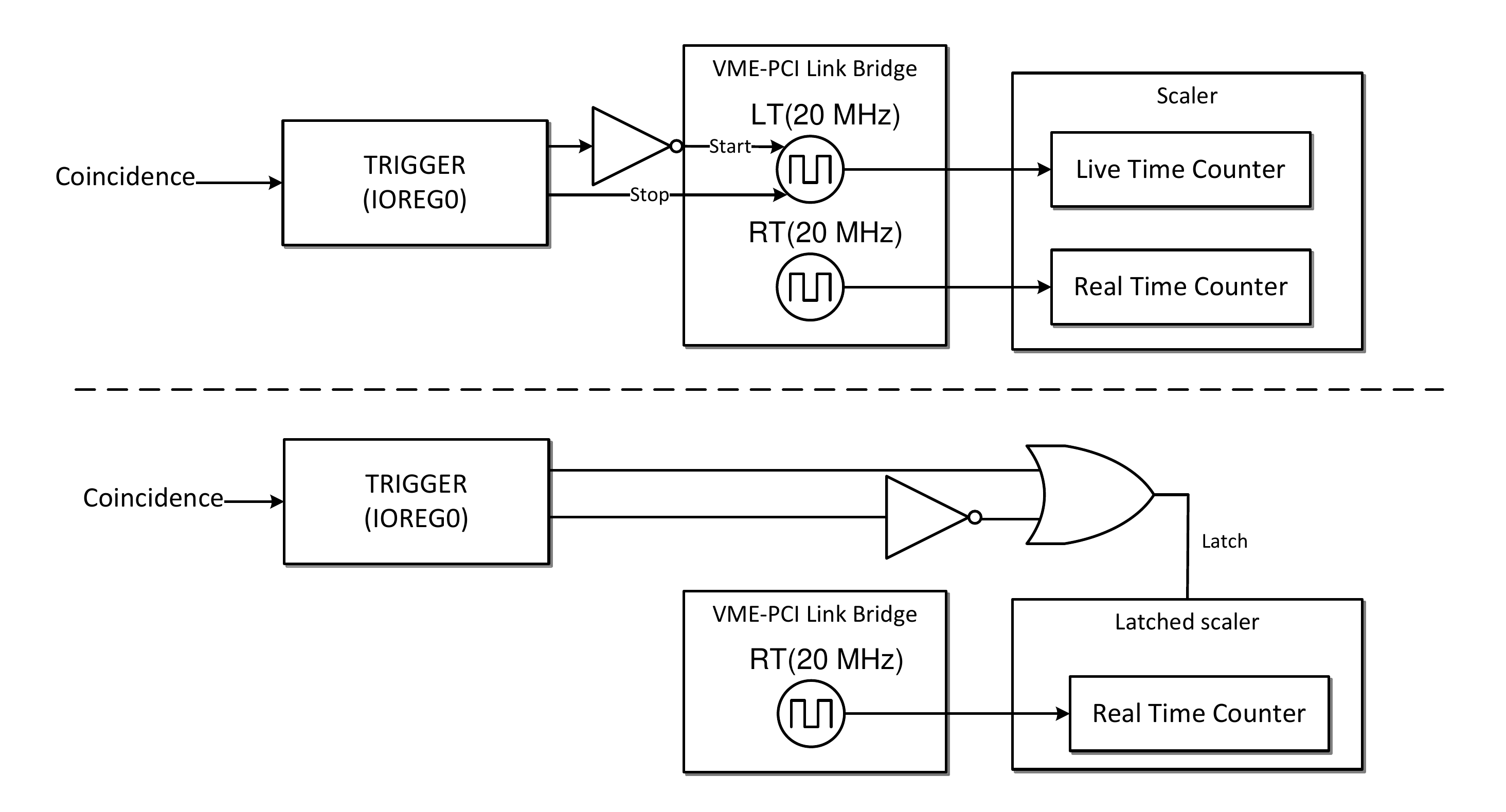}
    \caption[Time scalers]{\label{fig:Counters}Time scalers in Digital front-end. 64 bit scaler (top) and 32 bit latched scaler (bottom).}
  \end{center}
\end{figure}

The digital front-end must also measure the trigger real time and the accumulated live time. The real time of every trigger should be stored to temporal event tagging and to perform temporal studies such as time distance between consecutive events (see Section~\ref{sec:DeltaRT}). The accumulated live time must be measured in order to properly calculate the exposure of the measurements (as seen in Section~\ref{sec:ExperimentalRequirements}). Therefore, the acquisition dead time, the time in which the system is not able to detect a new event, should be discounted from the real time to measure the live time.
\paragraph{}
All this information is stored in scalers counting clock ticks coming from the VME/PCI Bridge clocks. This module is programmed to generate a 50 ns clock signals, used as real time clock master. This clock is distributed and logically processed in order to get the correct information. Two different modules are used: the first test were performed with a 64 bits scaler and a 32 bit latched scaler was acquired later for a more precise temporal tagging. The design of the scaler system can be seen in Figure~\ref{fig:Counters}.
\paragraph{}
The real time master clock, generated by the VME/PCI Bridge and started at the beginning of the run, is stored in the 64 bit scaler channel. Dead time is determined by the time that the I/O Register is in busy state as described at the beginning of this section. In addition, live time is the time that the I/O Register is in idle state. Another VME/PCI Bridge controller clock is used to measure live time. This clock is stopped when the I/O Register enters in busy state and it is started when the state is reset. This clock is stored in a 64 bit scaler. 
\paragraph{}
The 32 bit latched scaler can record start and stop timestamps of DAQ acquisition performing the latch process in any change of I/O Register state. The accurate tagging of the trigger event was the reason to acquire and install this new counter. It uses only the real time clock master and it requires some postprocessing in order to compute the event dead time: the difference between start and stop times (see Section~\ref{sec:TemporalParams}). The crosscheck for both (64-bit and 32-bit latched) measurements of real time and dead/live time can be seen in Sections~\ref{sec:RealTimeTest} and~\ref{sec:DeadTimeMeas} respectively.
\subsection{VME modules}\label{sec:VMEModules}
All VME modules used in the ANAIS digital front-end are listed and described in this section explaining their more important features and their use in ANAIS.
\paragraph{I/O Register}CAEN V977 (16 Channel Input/Output Register) can work either as general purpose I/O Register or as Multihit Pattern Unit. Every channel has its own status storing a \emph{true} logic value if input was true for some time. This state is stored until the channel is reset via VME command or a CLEAR signal. It is used as I/O register in ANAIS storing the trigger DAQ busy state. This state is reset via VME command by the DAQ program. The busy state is used for measuring dead time and its positive change triggers all VME boards via gate generator. In addition, it is configured to generate an interruption when the 0 output changes to a \emph{true} logical value and it is the board that starts the DAQ software process. This configuration allows saving dead time as can it be seen in Section~\ref{sec:DeadTimeMeas}.
\paragraph{MATACQ} CAEN V1729 (4 Channel 12 bit 2 GS/s Switched-Capacitor Digitizer) and CAEN V1729A (4 Channel 14 bit 2 GS/s Switched-Capacitor Digitizer) have four analog channels of bandwidth up to 300 MHz with sampling frequency up to 2 GS/s and a depth of 2520 points. Every sample is coded into 12 bit (14 bit in V1729A) in 650 $\mu$s giving a total dead time near 1ms. The configuration for ANAIS is 2GS/s (1260 $\mu$s digitization window), pretrigger 300 ns and external trigger.
\paragraph{}
The default dynamic range for this module is (-0.5V,0.5V) but it allows shifting by modifying some SMD components over the PCB. Given the mostly negative signal of PMTs as described in Section~\ref{sec:PMTSignal} the dynamic range was modified to be (0.1V, -0.9V) in V1729 model. A complete description and characterization of this module can be seen in Section~\ref{sec:MatacqCharac}.

\paragraph{QDC, Charge-to-Digital Converter}CAEN V792 (32 Channel Multievent QDC) integrates negative pulses. It converts the values into 12 bits and has a buffer of 32 events. This module is used in ANAIS with an integration window of 1 $\mu$s. The high density of channels allows the use of this module in several range of energies in order to get a reliable background information in these ranges.
\paragraph{TDC, Time-to-Digital Converter}CAEN V775 (32 Channel Multievent TDC) converts time intervals between ECL inputs and common gate into digital. Full Scale Range can be selected via VME from 140 ns to 1.2 $\mu$s. The board can operate in both \emph{common start} and \emph{common stop} modes. The configuration used in ANAIS is \emph{common start}. 
\paragraph{Pattern Unit}CAEN V259 (16 Bit Strobed Multihit Pattern Unit) module is able to store internally a 16 bit input pattern. The board has 16 inputs and an input coincidence circuit, controlled by a gate signal, selects the hits which are stored in a pattern register. It has also a multiplicity register. This module is used in ANAIS in a straightforward way without using the multiplicity register.
\paragraph{Scaler}CAEN V560 is a 16 Channel Scaler that allows internal cascading to produce eight 64 bit counters. In ANAIS it is used as Real Time and Live Time counters. 64 bits can store 50 ns tick counters in a typical run period (one or two weeks).
\paragraph{Latched Scaler}CAEN V830 is a 32 Channel Latching Scaler with ECL inputs. The latching feature allows to obtain a precise time measurements: the scaler stores counters when latch input triggers. It is used in ANAIS to store Real Time and Dead Time counters. The counters have only 32 bit and this width is not enough to store run counters of weeks at 50 ns clock tick. For this reason ANAIS uses both scalers, latched for high precision less significant bits and V560 for 64 bit counter, higher bits.
\paragraph{Clock generator and VME-PCI Optical Link Bridge} ANAIS uses VME controller (V2718 VME-PCI Optical Link Bridge) as clock generator. It uses all two pulse generator hosted in as described in Section~\ref{sec:TimeScalers}. Real Time clock is configured in \emph{manual} mode starting it at the beginning of the acquisition. The start and stop signals of Live Time clock are configured as input driven. This board is also used as VME-PCI bridge allowing high speed communication between VME and DAQ program.
\section{Front-end characterization}\label{sec:FrontEndTest}
\subsection{Preamplifier}\label{fig:PreampTest}
This section covers the tests performed to the AD8009 based preamplifier (see Section~\ref{sec:Preamp}). First, amplitude test to check linearity and dynamic range is presented. Next, a more detailed area test is shown and finally, the desired baseline improvement is checked.
\subsubsection{Basic amplitude tests}

\begin{figure}[h!]
     \begin{center}
\begin{subfigure}[b]{0.5\textwidth}
                \centering
                \includegraphics[width=\textwidth]{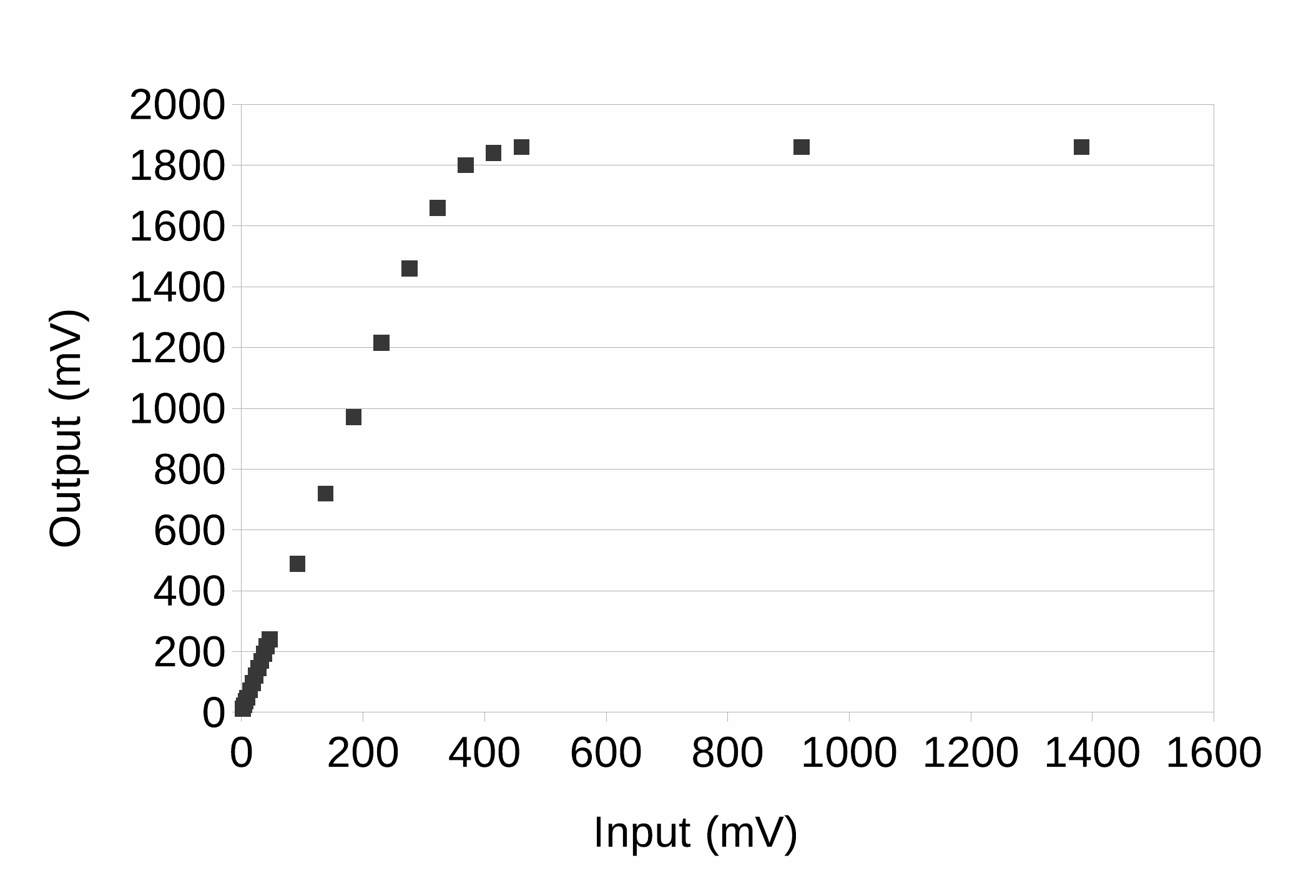}
                \caption{Gain.}
                \label{fig:preamp_gain}
        \end{subfigure}%
        ~ 
        \begin{subfigure}[b]{0.5\textwidth}
                \centering
                \includegraphics[width=\textwidth]{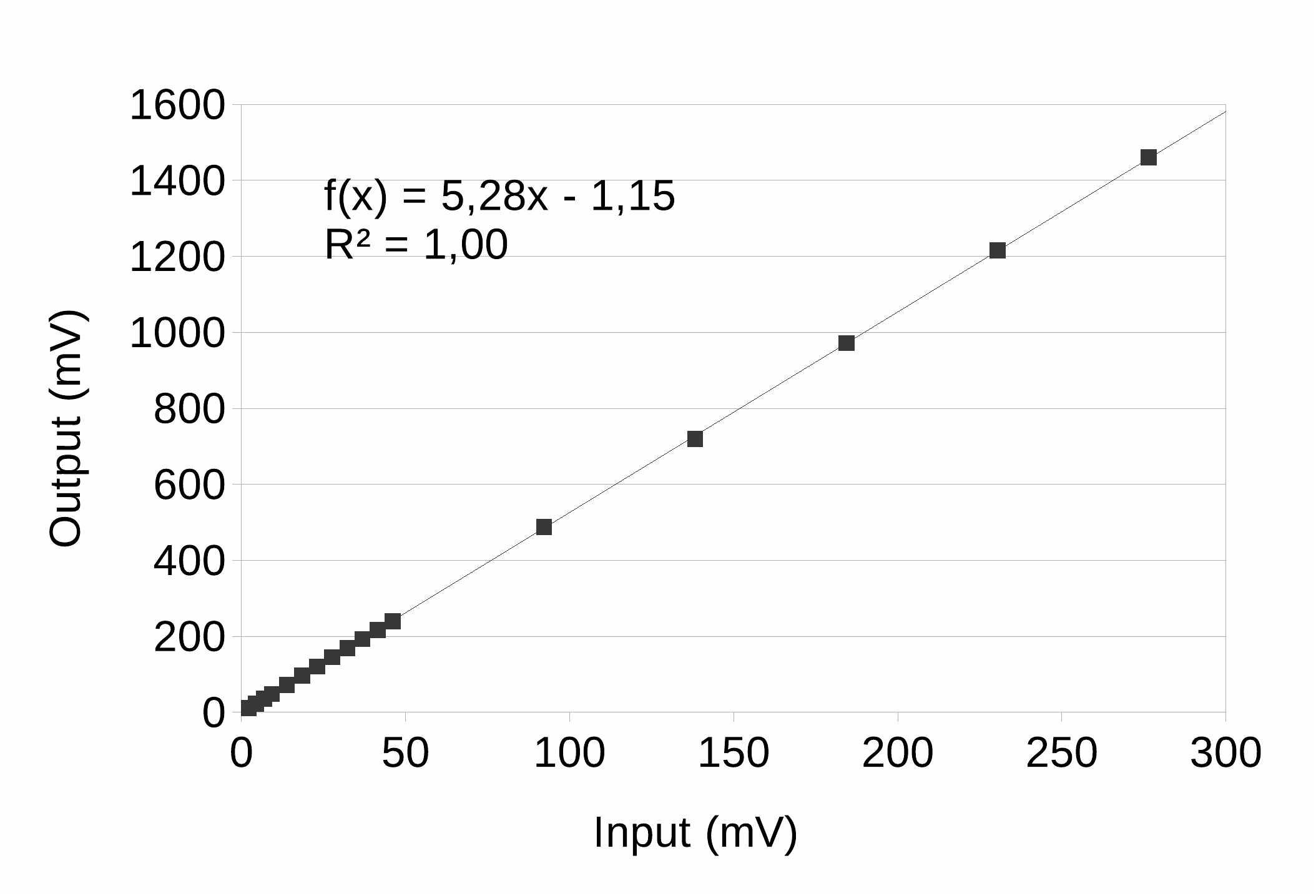}
                \caption{Gain in linear range (Fit).}
                \label{fig:preamp_gain_fit}
        \end{subfigure}

\begin{subfigure}[b]{0.5\textwidth}
                \centering
                \includegraphics[width=\textwidth]{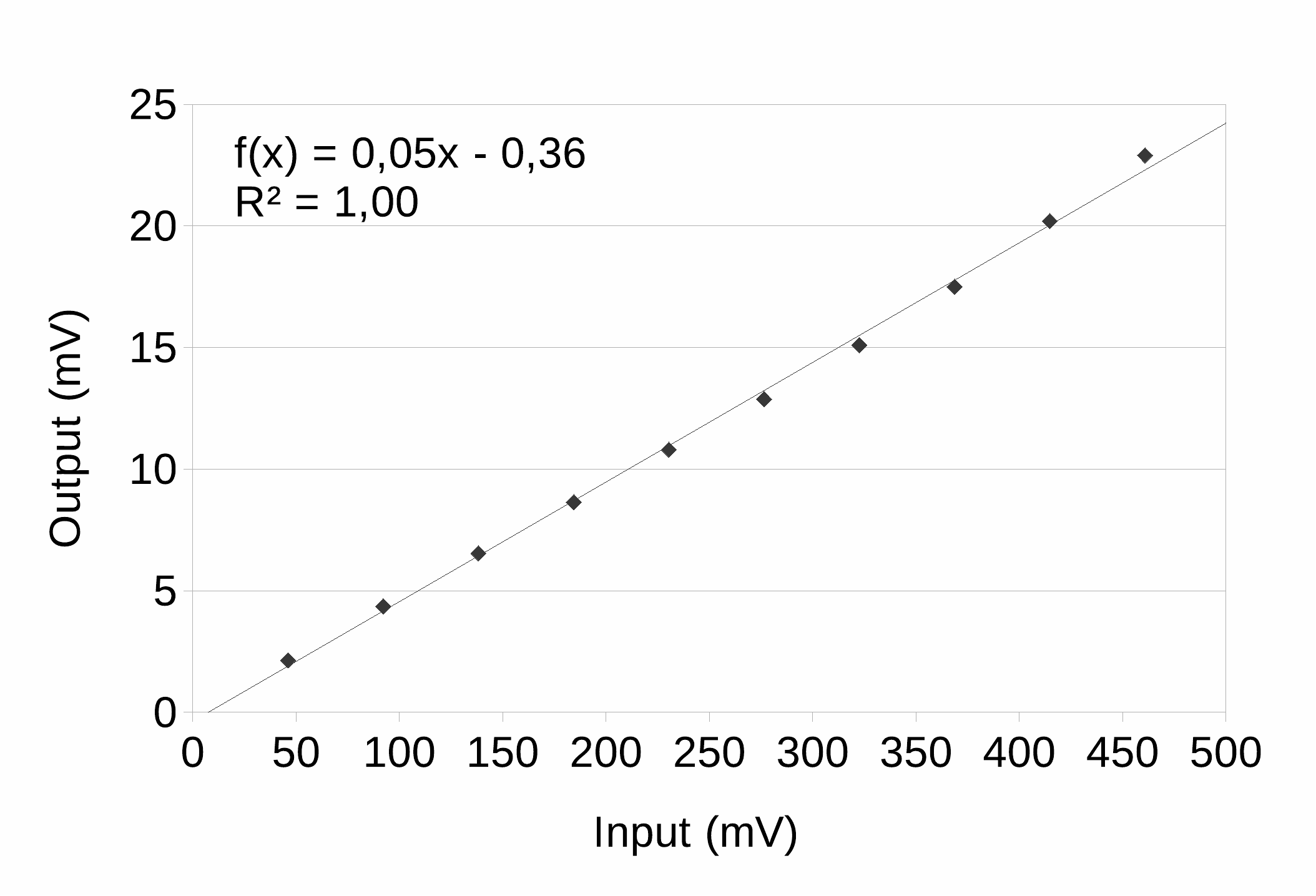}
                \caption{Attenuated output in amplified output linear range.}
                \label{fig:preamp_att_LE}
        \end{subfigure}%
        ~ 
        \begin{subfigure}[b]{0.5\textwidth}
                \centering
                \includegraphics[width=\textwidth]{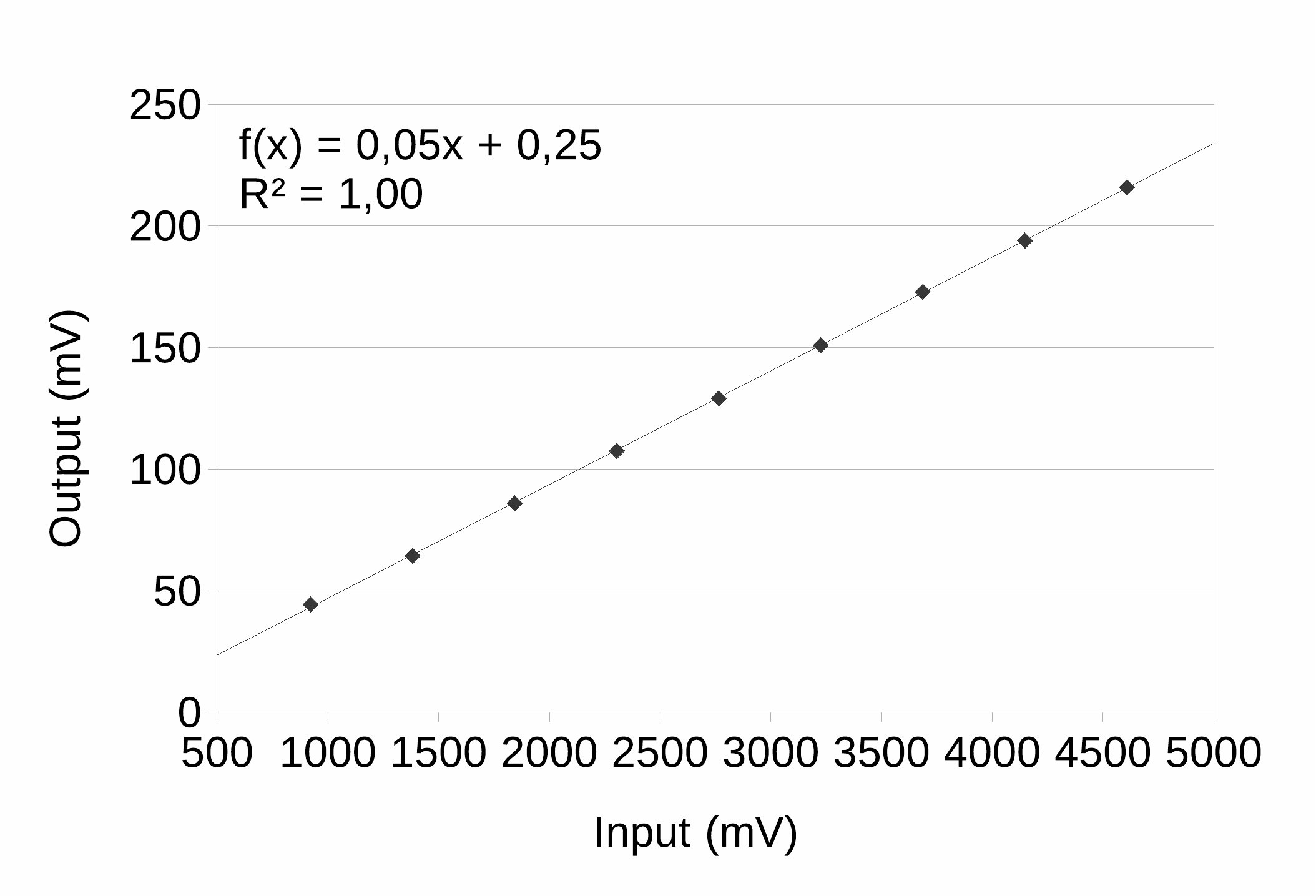}
                \caption{Attenuated output in amplified output saturation range .}
                \label{fig:preamp_att_HE}
        \end{subfigure}
 
        \caption{Amplitude linearity test from the preamplifier.}\label{fig:preamp_amp_tests}
\end{center}
\end{figure}

Several tests were performed in order to check the behavior of the AD8009 based preamplifier. The first test was done using negative pulses generated with a TTi-TGA12101 arbitrary pulse generator. The input waveform was designed to be similar to a typical photoelectron. The amplitude was variable to explore all requirements: low distortion and linear response at both low and high energy range.
\paragraph{}
The input amplitude was ranged from a few millivolt to several volts. Figure~\ref{fig:preamp_amp_tests} shows the response in both low and high amplitude ranges. It can be observed a good linearity in amplified and attenuated outputs. The preamplifier gain can be seen in Figure~\ref{fig:preamp_gain} from the linear zone to saturation. The linear zone is fitted in Figure~\ref{fig:preamp_gain_fit}. The attenuated signals in both low and high input amplitude are shown in Figures~\ref{fig:preamp_att_LE} and~\ref{fig:preamp_att_HE}. The latter shows the most important range to this signal because it is used to explore the high energy range when the preamplifier saturates.

The dynamic range of the preamplifier is (300mV,-300mV) in input amplitude, so the output amplitude range is (1.5V,-1.5V) and it is higher than the (0.1,-0.9V) MATACQ digitizer modified dynamic range (see Section~\ref{sec:VMEModules}). This means that the preamplifier dynamic range does not limit the global dynamic range, determined by the MATACQ range.
%
%
\subsubsection{Area test}\label{sec:PreampAreaTest}
The next step was to test the linearity between input and output areas. A new software was developed to acquire three signals and analyze them using a Tektronix TDS5034B oscilloscope. This test was performed using a triangular signal generated by the aforementioned pulse generator. The duration of the signal was configured to be twice NaI(Tl) scintillation constant, so we obtained an input signal similar to a scintillation signal in area and timing. Linearity and dynamic range of the amplifier can be quantified by the digitized signals analysis.
\begin{figure}[h!]
\begin{subfigure}[b]{0.5\textwidth}
                \centering
                \includegraphics[width=\textwidth]{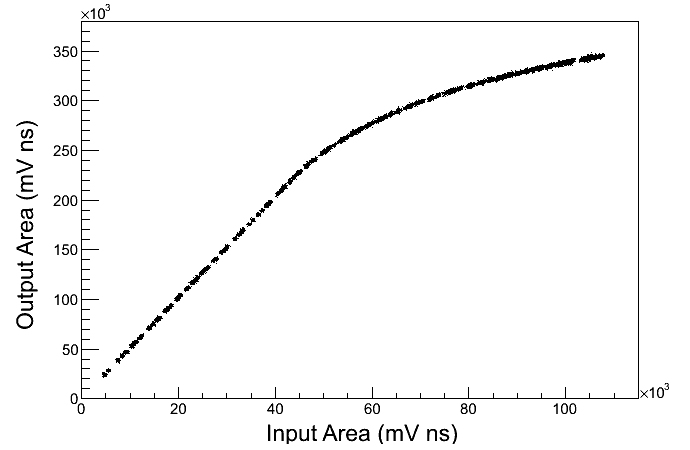}
                \caption{Input vs. amplified output.}
                \label{fig:preamp_area_amp}
        \end{subfigure}%
        ~ 
        \begin{subfigure}[b]{0.5\textwidth}
                \centering
                \includegraphics[width=\textwidth]{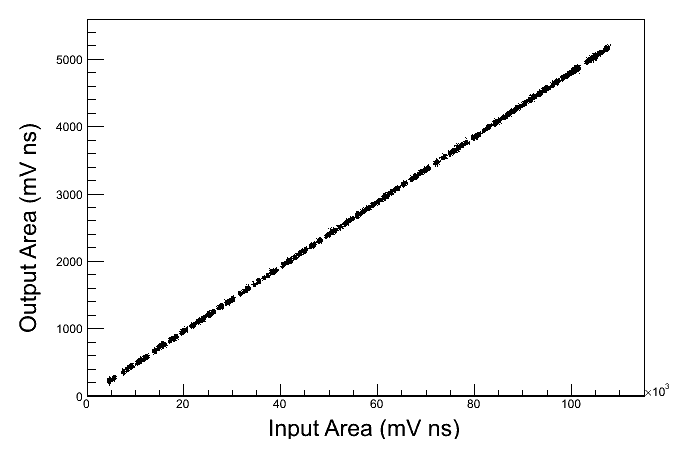}
                \caption{Input vs. attenuated output.}
                \label{fig:preamp_area_att}
        \end{subfigure}

\begin{subfigure}[b]{0.5\textwidth}
                \centering
                \includegraphics[width=\textwidth]{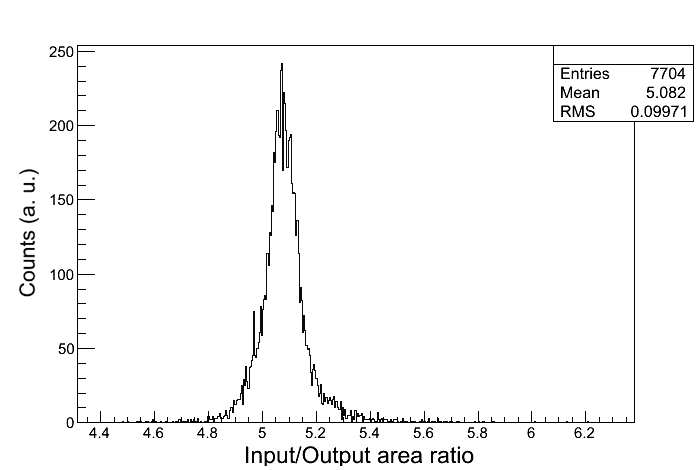}
                \caption{Ratio between amplified and input area in linear range.}
                \label{fig:preamp_area_gain}
        \end{subfigure}
        ~ 
        \begin{subfigure}[b]{0.5\textwidth}
                \centering
                \includegraphics[width=\textwidth]{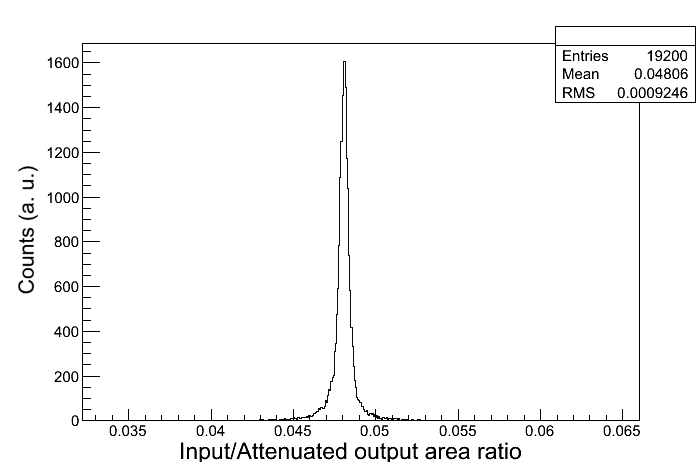}
                \caption{Ratio between attenuated and input areas.\\\hspace{\textwidth}}
                \label{fig:preamp_area_gain_att}
        \end{subfigure}%

     \begin{center}
        \caption{Area test from the preamplifier.}\label{fig:preamp_area_tests}
\end{center}
\end{figure}
\paragraph{}
The linearity of the attenuated signal and the amplified signal out of the saturation zone can be observed in Figure~\ref{fig:preamp_area_tests} (area is represented in mV$\times$ns units). Figure~\ref{fig:preamp_area_amp} shows the scatter plot between input and amplified output with linear and saturation zones. Figure~\ref{fig:preamp_area_att} is a similar scatter but with attenuated output in \emph{y} axis showing the linearity in all the range as expected. The bottom pictures quantify these linearities: Figure~\ref{fig:preamp_area_gain} shows the distribution of the ratio between input and amplified output in the linear range (input amplitude less than 400 mV) and Figure~\ref{fig:preamp_area_gain_att} ratio between input and attenuated output. Both show a FWHM smaller than 2\% and the same behavior for two distributions. This width can be attributed to systematic effects like sampling or area computing: the voltage divider nature of the attenuated output must give the same area ratio as the divider resistances ratio (as can be seen in Figure~\ref{fig:preamp_scheme}, resistors nominal values $49.9 \Omega/ 1k\Omega = 0.0499$ theoretical, 0.048 mean in Figure~\ref{fig:preamp_area_gain_att}, compatible taking into account the resistor tolerances and measurement errors). 
\begin{figure}[h!]
  \begin{center}
    \includegraphics[width=.8\textwidth]{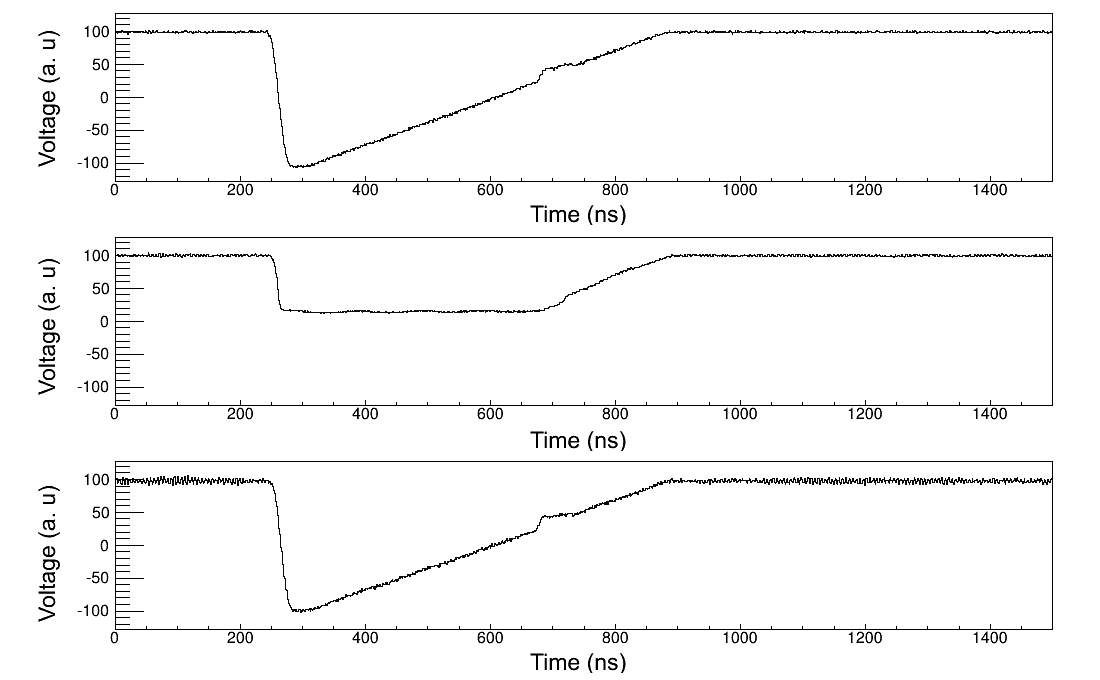}
    \caption[Saturation effect in a triangular signal]{\label{fig:preamp_triang}Effect of the saturation of the input (top), amplified (middle) and attenuated (down) signals.}
  \end{center}
\end{figure}

\begin{figure}[h!]
     \begin{center}
	\begin{subfigure}[b]{0.5\textwidth}
                \centering
                \includegraphics[width=\textwidth]{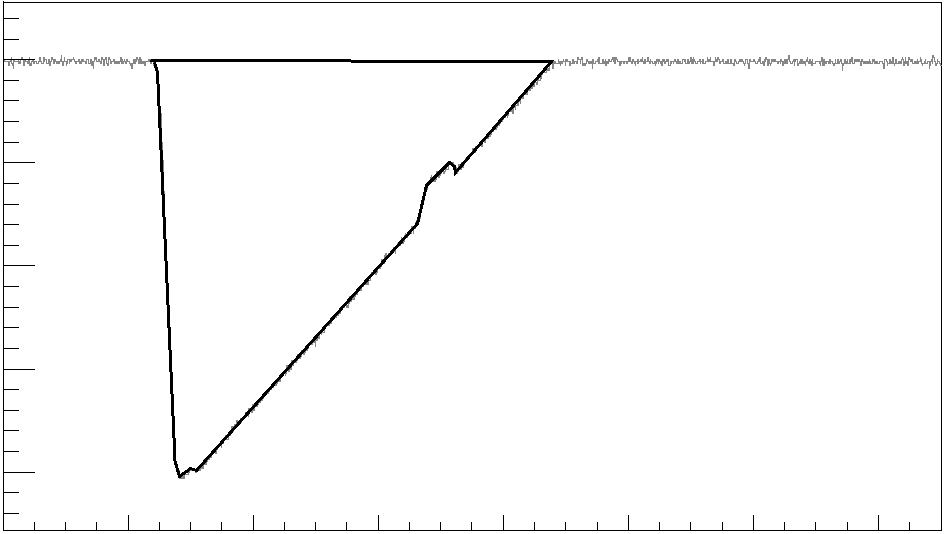}
                \caption{Area Estimation.}
                \label{fig:PreampAreaPoly}
        \end{subfigure}%
        ~ 
        \begin{subfigure}[b]{0.5\textwidth}
                \centering
                \includegraphics[width=\textwidth]{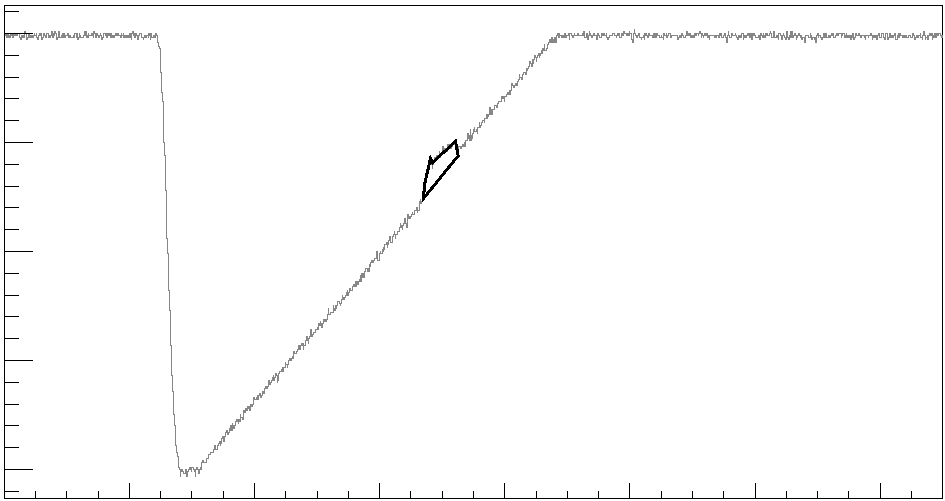}
                \caption{Glitch Area Estimation.}
                \label{fig:PreampAreaPolyGlitch}
        \end{subfigure}
        \caption{Polygon Area Estimation.}\label{fig:PreampPoly}
\end{center}
\end{figure}

\paragraph{}
As we can see in this test, linearity was achieved but an undesired effect was found. A \emph{glitch} can be observed in Figure~\ref{fig:preamp_triang} in the three digitized signals: input, amplified and attenuated outputs. This effect can be explained by a change in the input impedance of the amplifier when it leaves the saturation zone. An area calculation was done in order to quantify the effect. The total area and the \emph{glitch} area were computed taking polygonal areas as it can be seen in Figures~\ref{fig:PreampAreaPoly} and \ref{fig:PreampAreaPolyGlitch} respectively. Next, areas were computed by \texttt{TGraph::Integral} from the ROOT framework for different input sizes and all of them gave a \emph{glitch} area smaller than 1\% of total area.  

\paragraph{}
It is worth noting that events at low energy are free from this effect and all saturated signals are affected. It is also relevant to note that this effect is always smaller than 1\% in area, and consequently in energy, at high energy ranges. 
\subsubsection{Baseline Test}
The most important feature expected from the preamplifier was the improvement of signal/noise ratio. This baseline test can be seen in Section~\ref{sec:PreampBaselineTest}, in the context of baseline characterization.


\subsection{Delay module}\label{sec:NIMDelay}
\begin{figure}[h!]
     \begin{center}
	\begin{subfigure}[b]{1\textwidth}
            \centering
	    \includegraphics[width=1\textwidth]{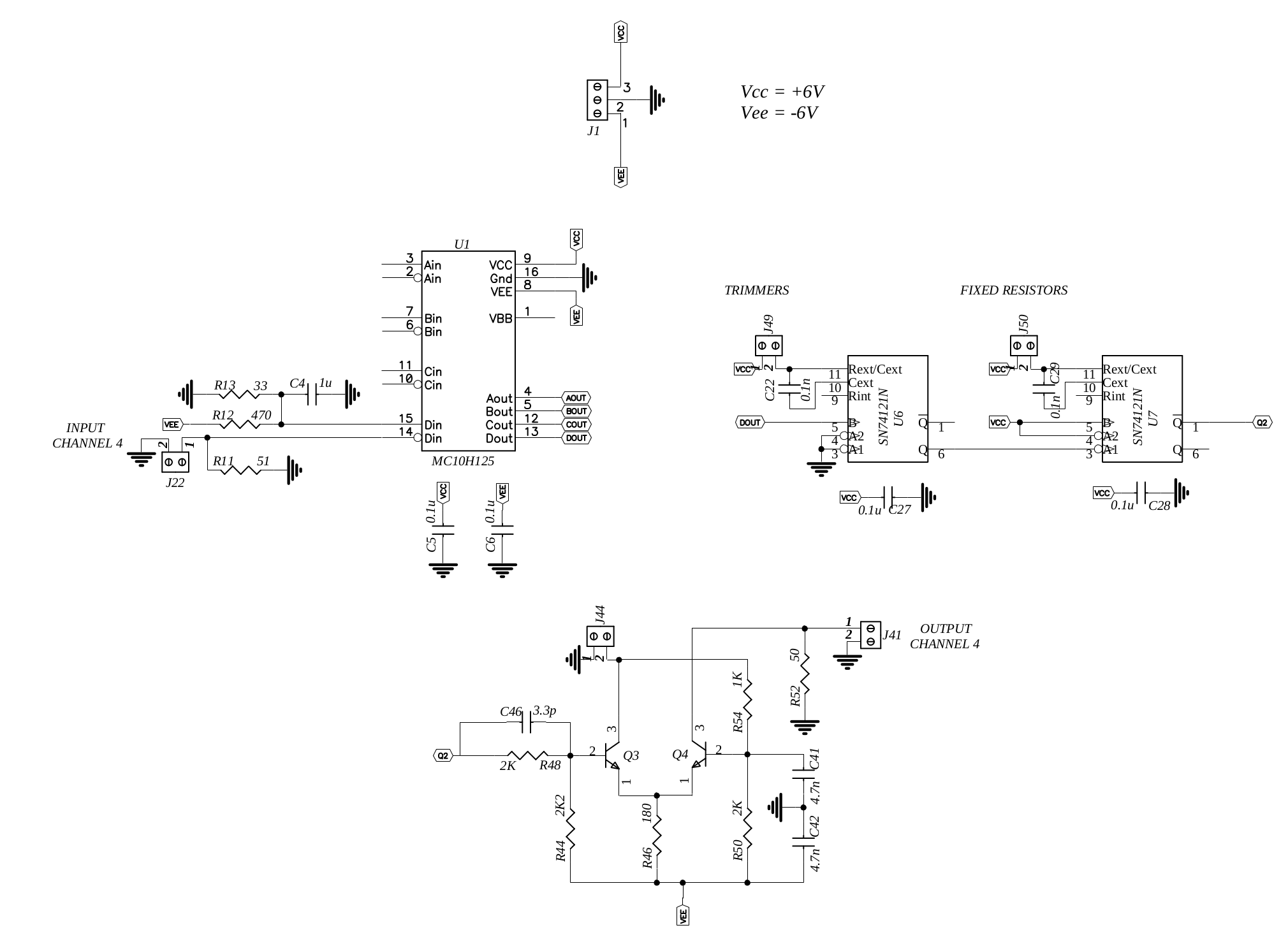}
	  \end{subfigure}

        \begin{subfigure}[b]{.5\textwidth}
                \centering
                \includegraphics[width=\textwidth, angle=90]{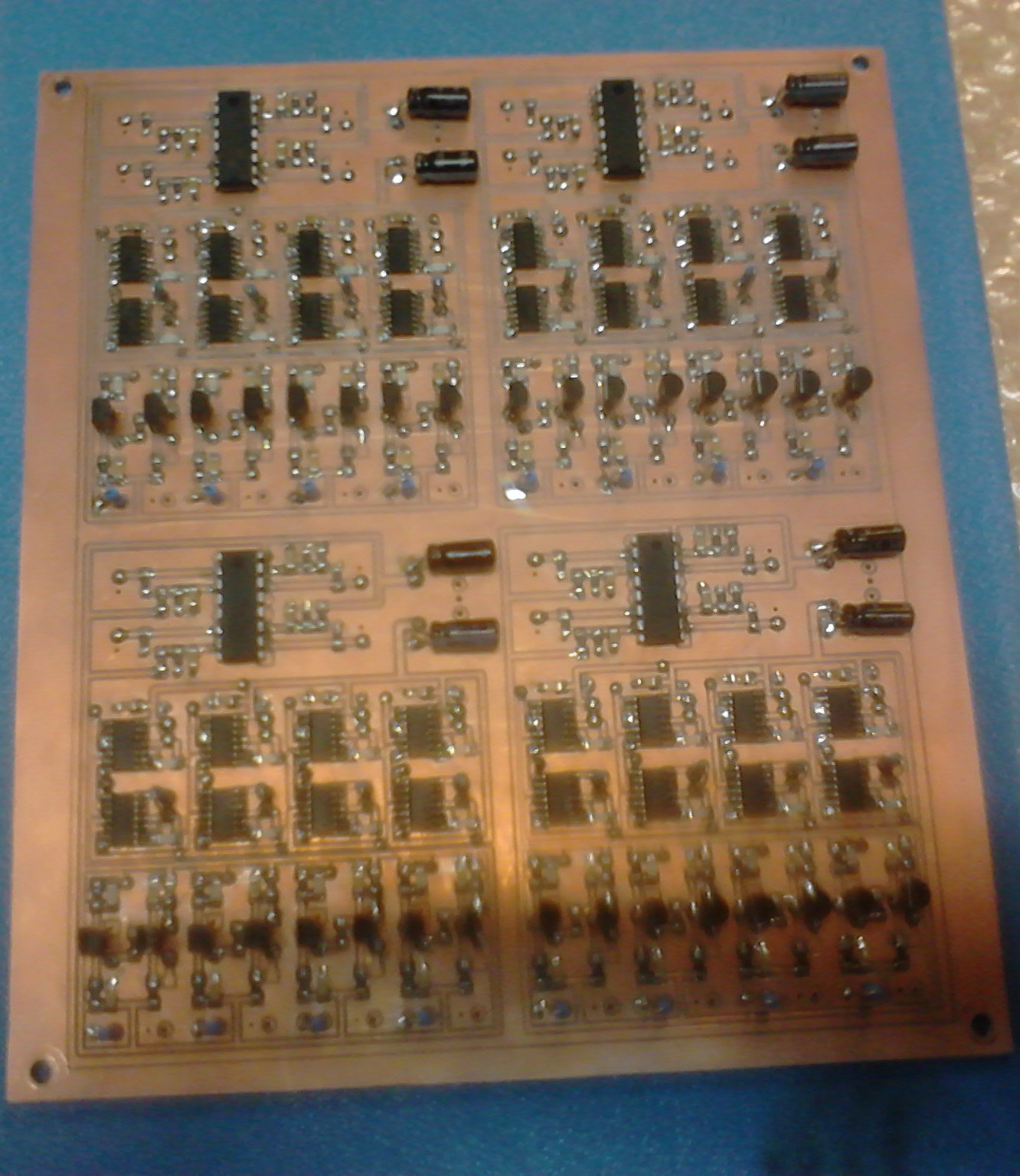}
        \end{subfigure}%

        \caption[Delay Module design and development]{Delay Module design and development.}\label{fig:DelayModuleScaleUp}
\end{center}
\end{figure}

\begin{figure}[h!]
  \begin{center}
    \includegraphics[width=.7\textwidth]{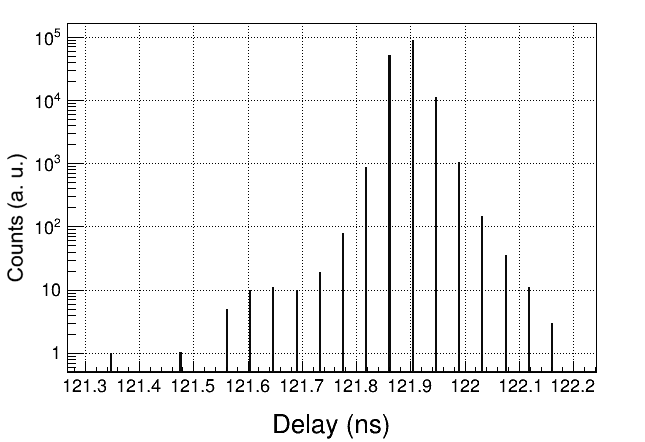}
    \caption[Delay module dispersion test]{\label{fig:DelayModuleDispersion}Delay module dispersion test.}
  \end{center}
\end{figure}

Every detector needs three delayed NIM signals according to the design seen in Section~\ref{sec:AnalogFrontend} (see Figure~\ref{fig:AnalogFrontend}). Hence, a high density of channels was needed. Since the commercial modules for this task do not have the required channel density, a high density NIM delay module was designed.
\paragraph{}
It is based on a Delay \& Gate Generator~\cite{DFORTUNO} previously implemented making it less versatile but multiplying $\times 4$ the number of channels. It uses the 74121 monostable multivibrator as TTL triggered gate generator with low jitter. In addition the module has NIM-to-TTL and TTL-to-NIM level adapters in order to match with all other NIM modules. The one channel design and the process of developing and testing from the test board to the whole module can be seen in Figure~\ref{fig:DelayModuleScaleUp}.
\paragraph{}
The basic delay feature was tested with a periodic NIM signal duplicated with a Logical Fan-In/Fan-Out and used as input of the Delay Module and as trigger of a CAEN V775 TDC (see Section~\ref{sec:VMEModules}). The delayed output was used as input of the TDC obtaining the delay time of a channel. The test of one channel can be seen in Figure~\ref{fig:DelayModuleDispersion}. 
\paragraph{}
The data of all channels can be seen in Table~\ref{tab:DelayModule} based on 150,000 samples of every channel. This test shows a very good delay time accuracy in every channel showing a typical standard deviation of tens of picoseconds and a worst case jitter of the order of 1 ns. This test also found an unstable channel with more jitter and it was excluded from the working channels set.
\begin{table}[h!]
	\begin{center}
		\begin{tabular}{ c c c c c }
			\toprule
			Channel & Mean (ns) & Standard deviation (ns) & Minimum (ns) & Maximum (ns) \\
			\toprule
1 & 128.52 & 0.033 & 127.88 & 128.83 \\
2 & 122.33 & 0.033 & 121.99 & 122.65 \\
3 & 122.68 & 0.455 & 121.34 & 123.33 \\
4 & 124.03 & 0.046 & 123.70 & 124.39 \\
5 & 128.67 & 0.041 & 128.29 & 129.05 \\
6 & 128.88 & 0.042 & 128.49 & 129.19 \\
7 & 128.57 & 0.038 & 128.22 & 128.91 \\
8 & 115.10 & 0.055 & 114.59 & 115.78 \\
9 & 125.80 & 0.038 & 125.02 & 126.06 \\
10 & 126.29 & 0.040 & 125.78 & 126.62 \\
11 & 121.81 & 0.042 & 121.33 & 122.17 \\
12 & 124.82 & 0.038 & 124.46 & 125.09 \\
13 & 120.04 & 0.041 & 119.66 & 120.36 \\
14 & 119.01 & 0.042 & 118.34 & 119.32 \\
15 & 121.44 & 0.043 & 120.98 & 121.7 \\
16 & 124.40 & 0.039 & 123.84 & 124.7 \\
			\toprule
		\end{tabular}
		\caption[Delay Module Test]{Delay Module channels test.} 
		\label{tab:DelayModule} 
	\end{center}
\end{table}

\subsection{MATACQ Module}\label{sec:MatacqCharac}
\begin{figure}[h!]
  \begin{center}
    \includegraphics[width=.7\textwidth]{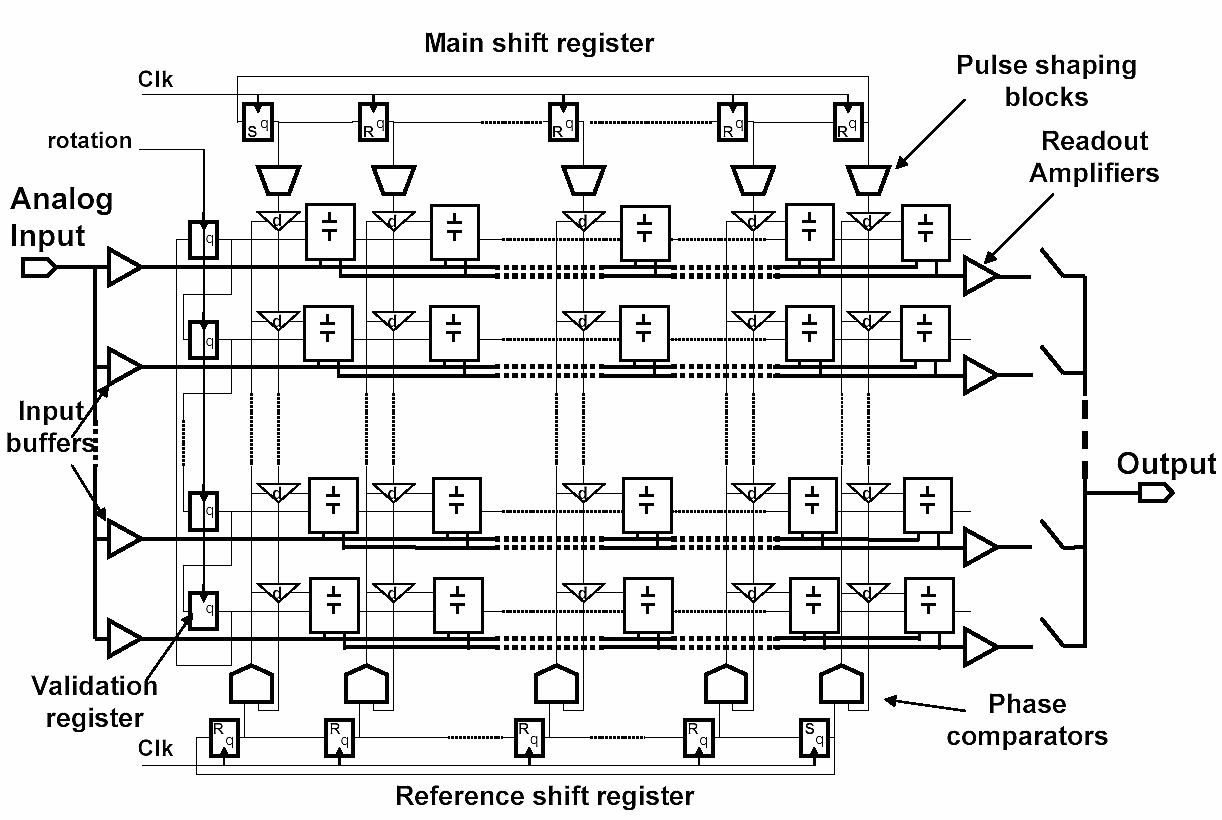}
    \caption[MATACQ chip design]{\label{fig:MatacqChip}MATACQ chip design showing switched capacitors matrix and associated Delay Locked Loops.}
  \end{center}
\end{figure}

The MATACQ boards are based on the MATACQ (MATrix ACQuision) digitization chip. The digitizer functional block of the chip is a matrix of switched capacitors associated with Delay Locked Loops~\cite{breton2005very, breton200614} configuring a sample matrix as it can be seen in Figure~\ref{fig:MatacqChip}. The matrix has 128 columns of 20 switched capacitor arrays giving a pulse depth of 2560 points (2520 valid points). This matrix is mapped to a circular memory and it must be unfolded in order to obtain the desired pulse. The pedestal of every sample cell must be calibrated in order to subtract it from the sampled value. The pedestal calibration process is covered later.
\paragraph{}
The MATACQ digitizer was extensively tested due its key role in the data acquisition system and the pervasive use of digitized pulse signal along the analysis process (see Chapter~\ref{sec:AnalysisSW} for more information). Many functional tests were performed and some peculiar characteristics were found.
\subsubsection{Pattern in mean signal and pedestal calibration}
\begin{figure}[h!]
     \begin{center}
	\begin{subfigure}[b]{0.5\textwidth}
                \centering
                \includegraphics[width=\textwidth]{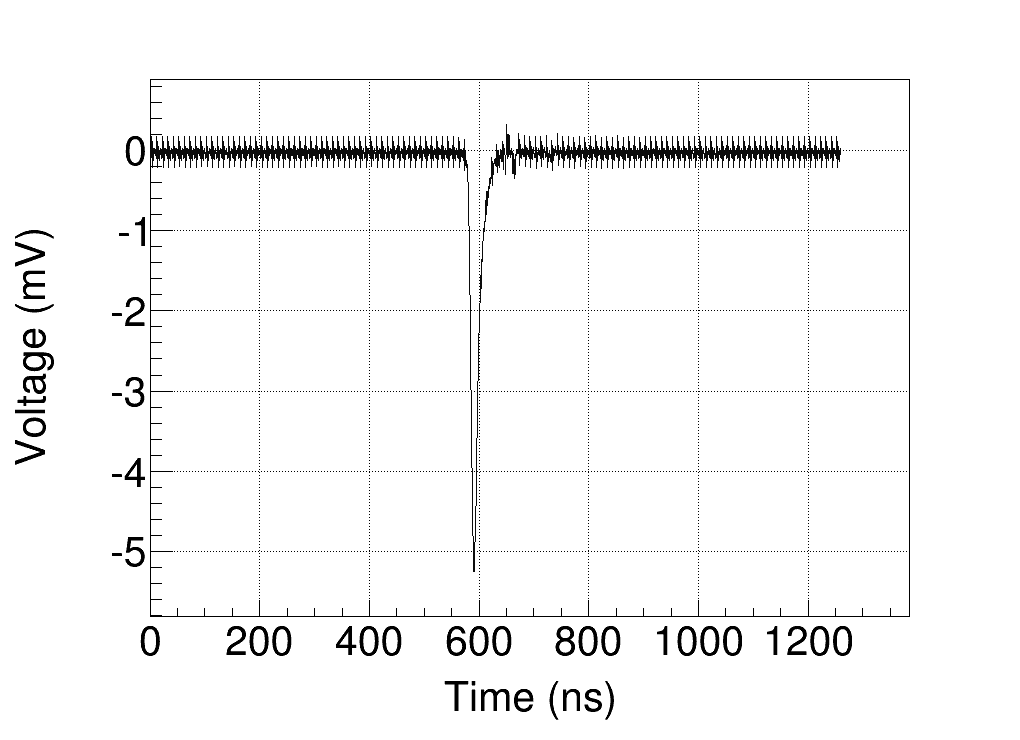}
        \end{subfigure}%
        ~ 
        \begin{subfigure}[b]{0.5\textwidth}
                \centering
                \includegraphics[width=\textwidth]{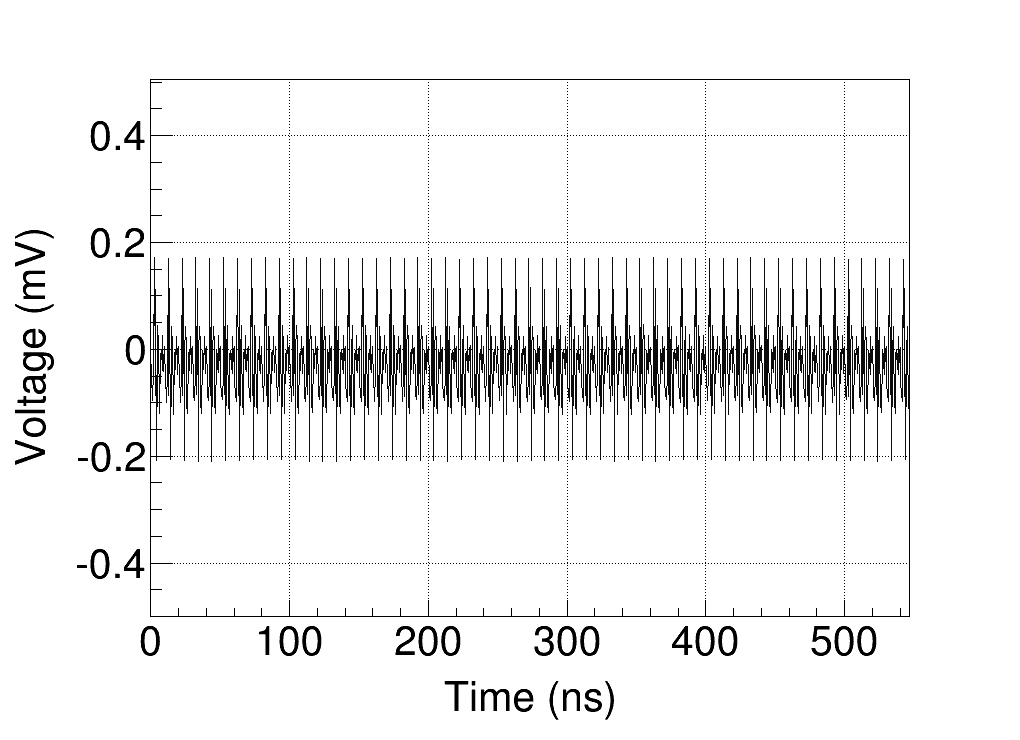}
        \end{subfigure}
        \caption[Baseline pattern in mean pulses]{Baseline pattern in mean pulses.\label{fig:Pedestal50Mean}}
\end{center}
\end{figure}

\begin{figure}[h!]
  \begin{center}
    \includegraphics[width=.7\textwidth]{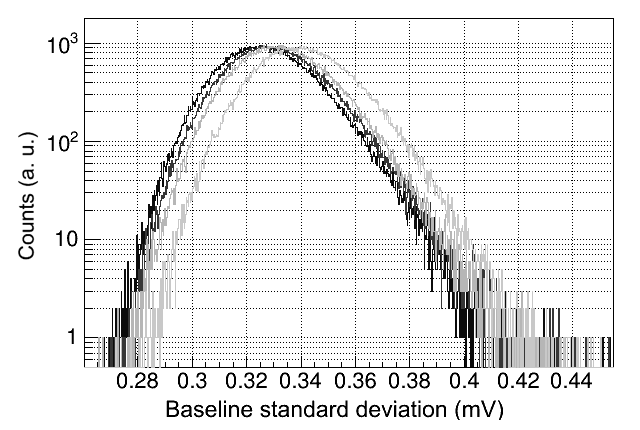}
    \caption[Baseline RMS with different pedestal calibrations]{\label{fig:RMSPedestals}Baseline RMS with different pedestal calibrations (right to left: 50, 500, 5,000 and 50,000 acquisitions).}
  \end{center}
\end{figure}

The more demanding scenario for MATACQ board was the SER characterization (see Section~\ref{sec:PMTSignal}) due to the signal size and the adverse signal/noise ratio. In these conditions, a subtle MATACQ behavior was observed: an unexpected pattern in mean pulses was discovered. This effect can be seen in Figure~\ref{fig:Pedestal50Mean}: a regular pattern in the baseline of the mean pulse. This effect was causing in the analysis program a bogus pulse onset detection with a low software threshold giving a strange periodicity.
\paragraph{}
The pattern seen in the baseline seems to be related with the MATACQ chip physical memory matrix seen before. For this reason, improving the pedestal calibration process has a positive impact in this effect. The plots of Figure~\ref{fig:Pedestal50Mean} were computed with fifty acquisitions. The V1729 MATACQ board manual states that for pedestal calibration it is enough with ``few tens of raw acquisitions of the baselines for all of the cells (disconnected or grounded inputs). The trigger must then be either automatic, or external'' and pedestal is computed as the mean of taken values. Several tests were performed in order to take pedestals without signal: switching off the preamplifier or the high voltage from the PMT but no improvement was observed in the mean pulse. The next test was to increase the number of acquisitions when the signal is present as suggested in the aforementioned manual. The baseline of a mean pulse acquired with a pedestal calibration of fifty acquisitions was compared with a pedestal of 50,000 acquisitions. The effect was reduced from 0.4 to 0.15 mV peak-to-peak.
\paragraph{}
A comparison of the baseline root mean square was carried out in order to check the pedestal calibration impact. Several runs were performed with different number of pedestal acquisitions and the RMS of the pretrigger points was computed. An improvement of the baseline correlated to the number of acquisitions was observed as it can be seen in Figure~\ref{fig:RMSPedestals}.
\paragraph{}

\begin{figure}[h!]
     \begin{center}
	\begin{subfigure}[b]{0.5\textwidth}
                \centering
		\begin{overpic}[width=\textwidth]{Pedestal50Mean.png}
			\put(50,28){\includegraphics[width=0.4\textwidth]{Pedestal50Meanbaseline.png}}
		\end{overpic}
        \end{subfigure}%
        ~ 
        \begin{subfigure}[b]{0.5\textwidth}
                \centering
		\begin{overpic}[width=\textwidth]{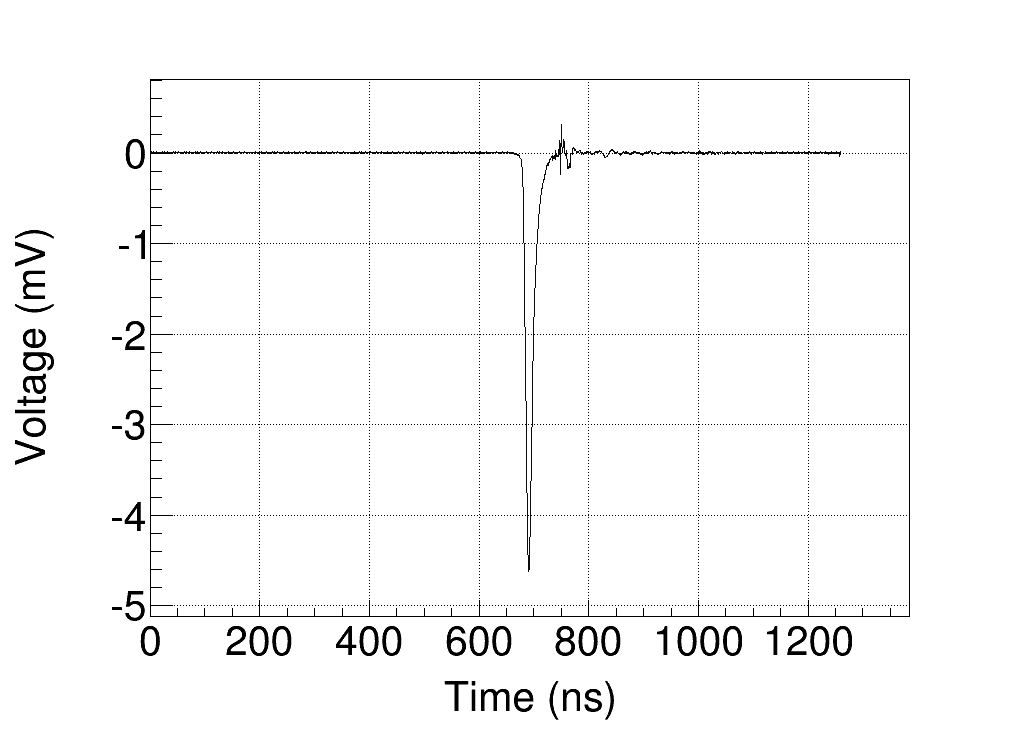}
			\put(15,28){\includegraphics[width=0.35\textwidth]{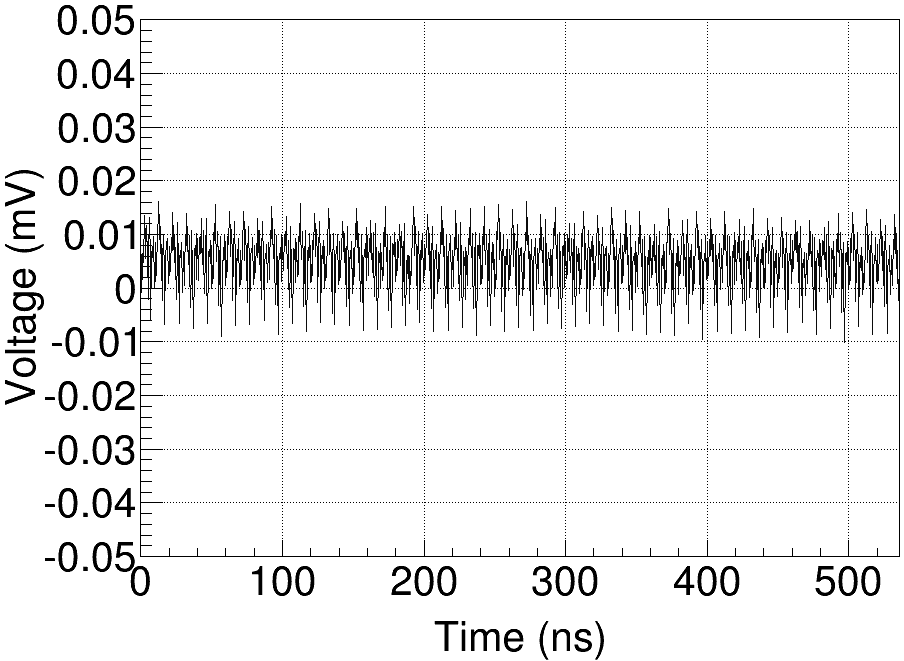}}
		\end{overpic}
        \end{subfigure}
        \caption[Mean pulses with external trigger vs. software trigger]{Mean pulses with zoomed baseline using software trigger (left) and external trigger with pretrigger computed pedestals (right). Note the different vertical scales in zoomed images.\label{fig:PedestalMeanETPretrigIRQ}}
\end{center}
\end{figure}

It is worth noting the circular nature of the MATACQ memory (as seen in Figure~\ref{fig:MatacqChip}) and the random position of the trigger in this memory. The effect seems to be caused by a subtle crosstalk between trigger and digitization systems. In all previous plots the pedestal calibration was done with auto (software) trigger because it is the faster way of getting all needed samples. Different methods from calibrating pedestal were compared. First using the ANAIS trigger (see Section~\ref{sec:TriggMod}) as external trigger but in OR mode giving a reduction from 0.15 mV to 0.05 mV observing a little positive bias that can be attributed to photoelectrons. For this reason, the pedestal computing was modified taking advantage of the known trigger strategy: only data from the first points of the pretrigger zone is used to obtain the mean value. A final improvement was achieved by doing the acquisition with IRQ notification instead of active poll loop. This behavior seems to be related to the VME polling induced noise covered in Section~\ref{sec:Baseline}. The effect in the final conditions is below 25 $\mu$V peak-to-peak that given the 0.4 mV of initial effect is an improvement of more than one order of magnitude.
\paragraph{}
The RMS baseline value was also tested in the above conditions but no improvement could be reported in Figure~\ref{fig:RMSPedestals}. The total mean pulse was computed with external trigger and computing only pretrigger points with the IRQ acquisition method and the result is compared in Figure~\ref{fig:PedestalMeanETPretrigIRQ} with the previous mean pulse obtained with software trigger: a much thiner baseline and an almost negligible matrix effect can be observed. A subtle remaining effect is still observed: a 0.2 mV peak-to peak spike at trigger time. It is negligible in normal operation conditions due to the presence of signal with much more amplitude. For a complete description of Single Electron Response see Section~\ref{sec:PMTSignal}.
\paragraph{}
The previous results are fundamental to establish criteria to routinely calibrate pedestals. The software developed to calibrate it and the protocol to do it are covered in Section~\ref{sec:SWPedestals}. 

\subsection{Trigger modules}\label{sec:TestTriggerStrat}
Various alternative modules to trigger the acquisition are available and they were checked in this section. First, the comparison of constant fraction discriminator with low threshold discriminator was performed (see Section~\ref{sec:TriggMod} for an introduction to these discriminators). Next, the stability of the constant fraction discriminator with temperature variations was tested. 
\subsubsection{Constant Fraction Discriminator (CFD) delay settings}
\begin{figure}[h!]
     \begin{center}
	\begin{subfigure}[b]{0.5\textwidth}
                \centering
                \includegraphics[width=\textwidth]{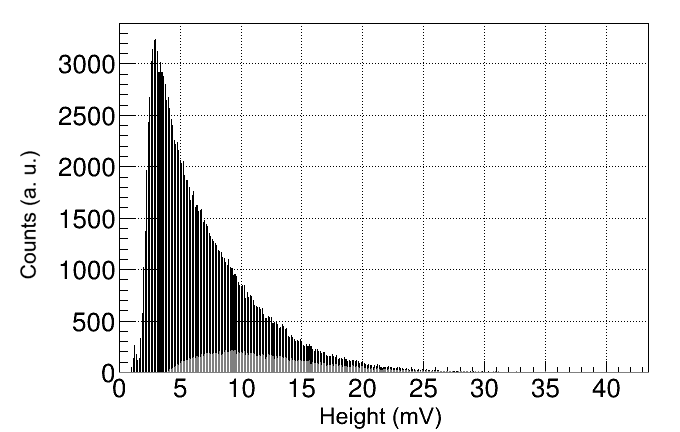}
		\caption{20 ns.}
                \label{fig:CFDTrigger20ns}
        \end{subfigure}%
        ~ 
        \begin{subfigure}[b]{0.5\textwidth}
                \centering
                \includegraphics[width=\textwidth]{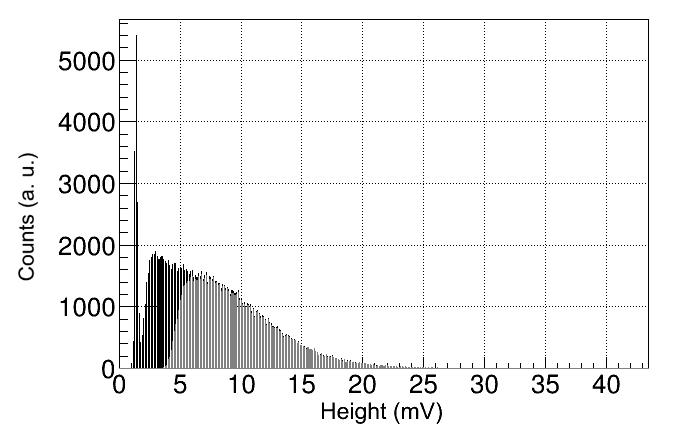}
		\caption{8 ns.}
                \label{fig:CFDTrigger8ns}
        \end{subfigure}
        \caption[CFD trigger efficiency with different delays]{CFD trigger efficiency with different delays. Pulse minimum distribution (mV) for all (black) and CFD triggered (gray) pulses.}
     \end{center}
\end{figure}

A basic set-up with a CFD discriminator, a MATACQ board and a PMT was mounted and the MATACQ internal trigger was used to store both PMT signal and discriminator output. The MATACQ internal trigger threshold was set to trigger near the baseline noise and to see as much photoelectron distribution as possible and discriminator behavior with those signals.
\paragraph{}
The CFD test showed that its default 20 ns delay configuration was not suitable for triggering with photoelectrons as can be seen in Figure~\ref{fig:CFDTrigger20ns} with 1 mV trigger level. The minimum value of the PMT signal is displayed for all events (black) and for those that CFD was triggered (gray). This plot gives a poor trigger efficiency with a marked non-threshold behavior. A very different trigger efficiency with a delay of 8 ns can be seen in Figure~\ref{fig:CFDTrigger8ns}, featuring a threshold at 5 mV. It is worth noting that settings for CFD threshold is the constant fraction of the real threshold (20\% fixed for this module), so 1 mV corresponds to 5 mV as it can be seen in Figure~\ref{fig:CFDTrigger8ns}.
\subsubsection{Constant Fraction Discriminator (CFD) vs. Low Threshold Discriminator (LTD)}
\begin{figure}[ht!]
     \begin{center}
	\begin{subfigure}[b]{0.5\textwidth}
                \centering
                \includegraphics[width=\textwidth]{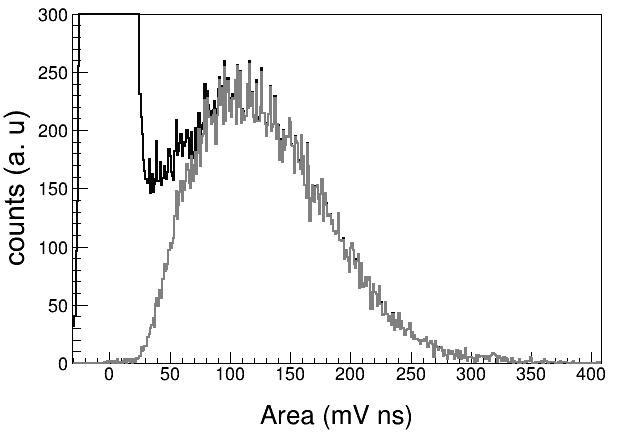}
                \label{fig:SERCFD}
        \end{subfigure}%
        ~ 
        \begin{subfigure}[b]{0.5\textwidth}
                \centering
                \includegraphics[width=\textwidth]{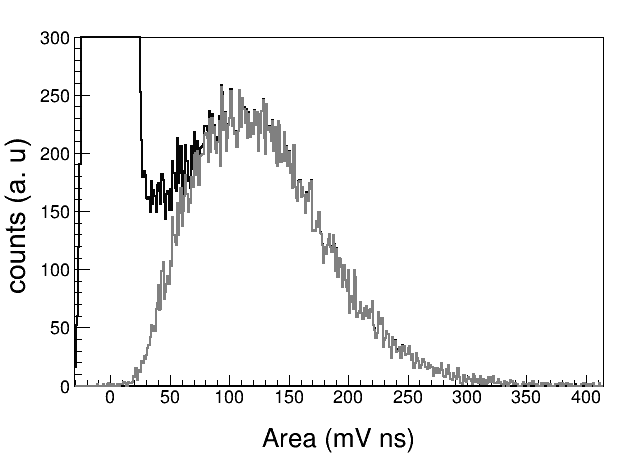}
                \label{fig:SERLTD}
        \end{subfigure}
	\caption[CFD and LTD trigger efficiency]{CFD (left) and LTD (right) trigger efficiency with single electron response signals. Triggered event area distribution is shown in gray and the distribution of all events in black.}\label{fig:SERDiscr}
     \end{center}
\end{figure}

The SER characterization set-up (see Section~\ref{sec:PMTTesting}) was also used to test different triggering strategies at photoelectron level. This set-up allowed to compare the trigger for the two different modules. The discriminator output was digitized for both models in order to determine the trigger behavior with a photoelectron as input signal. An example of this test can be seen in Figure~\ref{fig:SERDiscr} showing a very similar behavior once the CFD was configured with the shorter delay.

\paragraph{}
The similar SER triggering seen here, the better behavior of the CFD with DC level variations (see next section) and the more convenient CFD output window width were the reasons to choose the CFD modules for the ANAIS experiment.
\subsubsection{Trigger versus temperature}
\begin{figure}[h!]
  \begin{center}
    \includegraphics[width=0.65\textwidth]{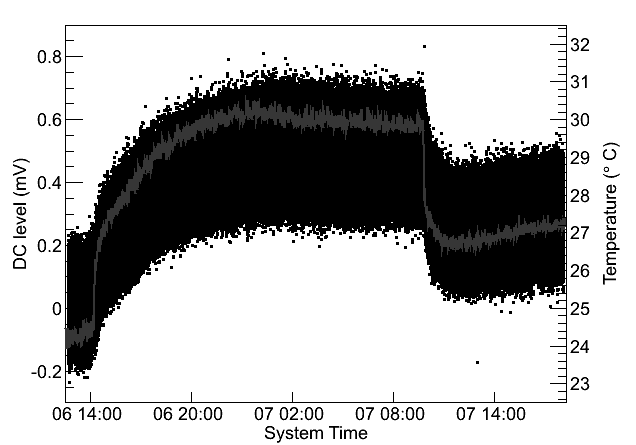}
    \caption[DC versus temperature]{\label{fig:DCvsTemp}DC level (black) and temperature (gray) over time switching off the air conditioning system.}
  \end{center}
\end{figure}

\begin{figure}[h!]
     \begin{center}
	\begin{subfigure}[b]{0.5\textwidth}
                \centering
                \includegraphics[width=\textwidth]{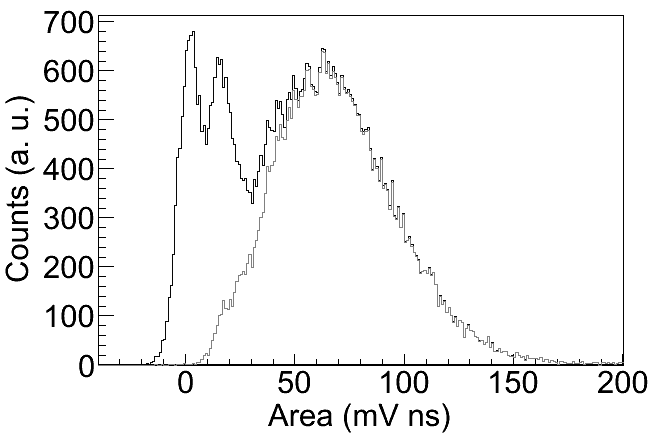}
		\caption{\label{fig:SERTrigg25D}}
        \end{subfigure}%
        ~ 
        \begin{subfigure}[b]{0.5\textwidth}
                \centering
                \includegraphics[width=\textwidth]{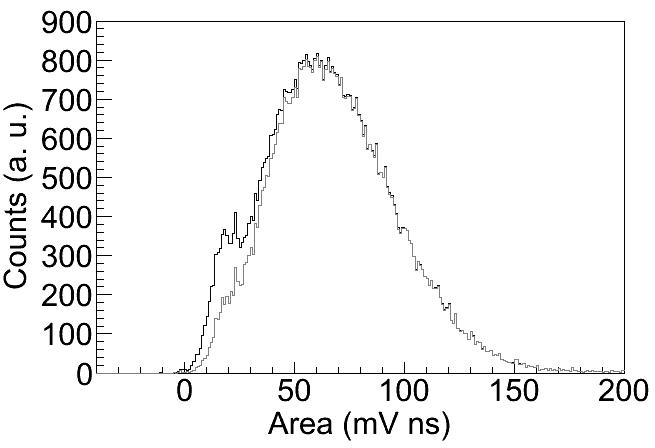}
		\caption{\label{fig:SERTrigg30D}}
        \end{subfigure}
	\caption[CFD behavior at different temperatures]{CFD behavior at different temperatures: 25 ºC (a) and 30 ºC (b) (triggered in gray, all events in black).\label{fig:SERTrigg}}
\end{center}
\end{figure}

The trigger behavior with temperature variations has also been studied. A set-up consisting of a PMT in dark environment, an ANAIS preamplifier, a CFD and a MATACQ digitizer was mounted at the laboratory of the University of Zaragoza. The data acquisition was configured to be triggered by the internal MATACQ level discriminator. The MATACQ board was configured to acquire the waveform of the PMT and the output of the CFD. The MATACQ trigger was set to trigger as low as possible above the baseline in order to study the CFD behavior. The temperature were controlled using the air conditioning system and it can be seen in gray in Figure~\ref{fig:DCvsTemp} showing an increase of more than four degrees Celsius. A DC drift can be reported correlated with the temperature variations. Such a DC drift on the digitized signal, of the order of 0.4 mV, could have impact in trigger and that was the reason to carefully test their behavior.
\paragraph{ }
The charge distribution at different temperatures was constructed to see the effect in the CFD trigger and it can be seen in Figure~\ref{fig:SERTrigg}. The signals triggering the MATACQ internal trigger can be seen in black and the CFD triggered signals can be seen in gray. Figure~\ref{fig:SERTrigg25D} shows the data at 25º~C and Figure~\ref{fig:SERTrigg30D} at 30~ºC. A clear reduction of the low area events can be noted due to the low threshold nature of the MATACQ discriminator. The CFD trigger exhibits a more robust behavior with the temperature triggering an almost full SER in both cases.
\paragraph{ }
The very similar behavior of the two kind of modules in terms of SER efficiency, the very convenient width of the CFD trigger output window and the good behavior of the CFD even with baseline shift with temperature were the reasons to choose CAEN N843 as ANAIS discriminator.
\subsection{Baseline characterization and improvements}\label{sec:Baseline}
A low threshold experiment has to be extremely careful about the electric noises which may affect both trigger and resolution. The present section is a catalog of detected noises in the ANAIS DAQ system. It covers all detected noises, describes the origin and the potential harm and explains the measures taken to prevent the generation and/or mitigate the effect.

\subsubsection{Preamplifier baseline effect }\label{sec:PreampBaselineTest}

\begin{figure}[h!]
     \begin{center}
	\begin{subfigure}[b]{0.5\textwidth}
                \centering
                \includegraphics[width=\textwidth]{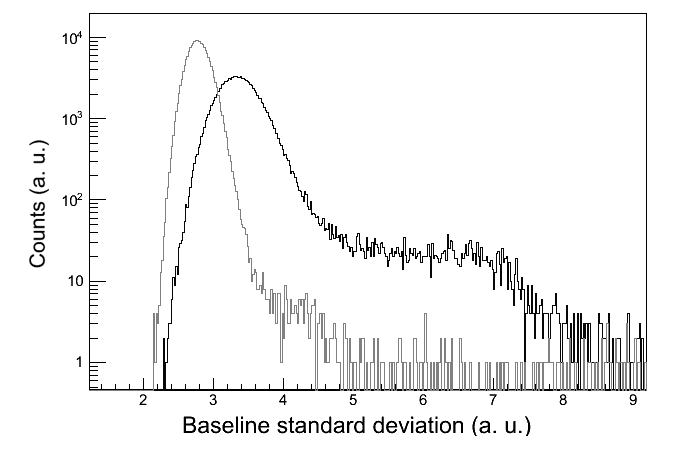}
                \label{fig:PreampRMS0}
        \end{subfigure}%
        ~ 
        \begin{subfigure}[b]{0.5\textwidth}
                \centering
                \includegraphics[width=\textwidth]{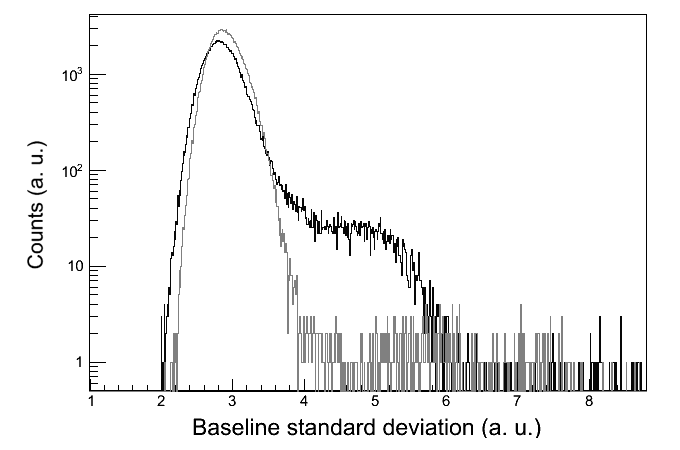}
                \label{fig:PreampRMS1}
        \end{subfigure}
        \caption[Baseline RMS with AD8009 based preamplifier]{Baseline RMS with CAEN NIM amplifier (black) and with AD8009 based preamplifier (gray).}\label{fig:PreampRMS}
\end{center}
\end{figure}

The home-made preamplifier replaced a CAEN $\times$10 amplifier located in a NIM crate outside the ANAIS hut. Two preamplifiers were installed outside the shielding but inside the hut, near the shielding. The cable length was shortened from 5 m to 1 m. The RMS from the baseline, calculated as described in Section~\ref{sec:PulseChar}, was plotted for both CAEN and AD8009 amplifiers as it can be seen in Figure~\ref{fig:PreampRMS}. This plot takes into account the different gain from both amplifiers ($\times$10 vs. $\times$5). The events to plot were pure baseline ones selecting events triggered by the other crystal in the ANAIS-25 set-up. The figure shows the RMS of the baseline of two PMTs. It can be seen a mean reduction of the RMS and it also can be observed a distribution shape change. A population near but separated from the main baseline peak was almost suppressed.

\begin{figure}[h!]
     \begin{center}
	\begin{subfigure}[b]{0.5\textwidth}
                \centering
                \includegraphics[width=\textwidth]{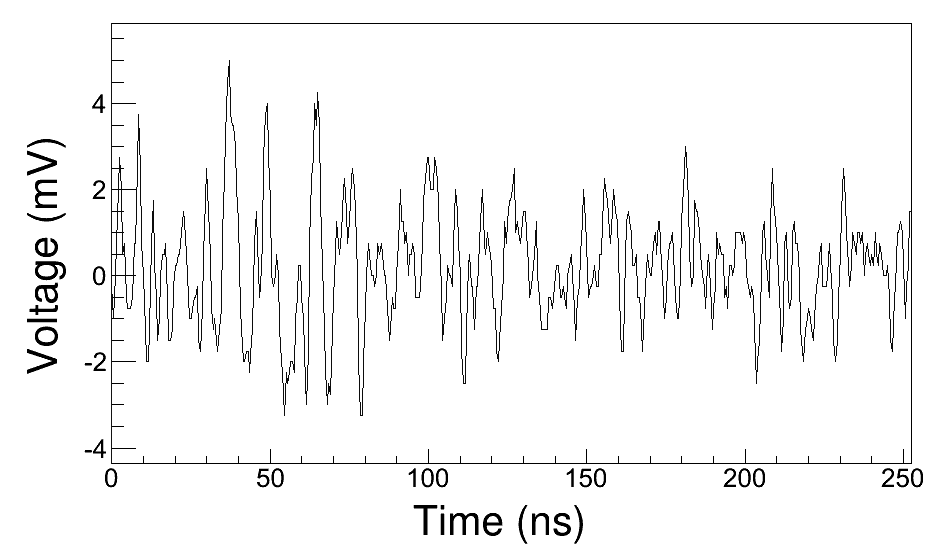}
        \end{subfigure}%
        ~ 
        \begin{subfigure}[b]{0.5\textwidth}
                \centering
                \includegraphics[width=\textwidth]{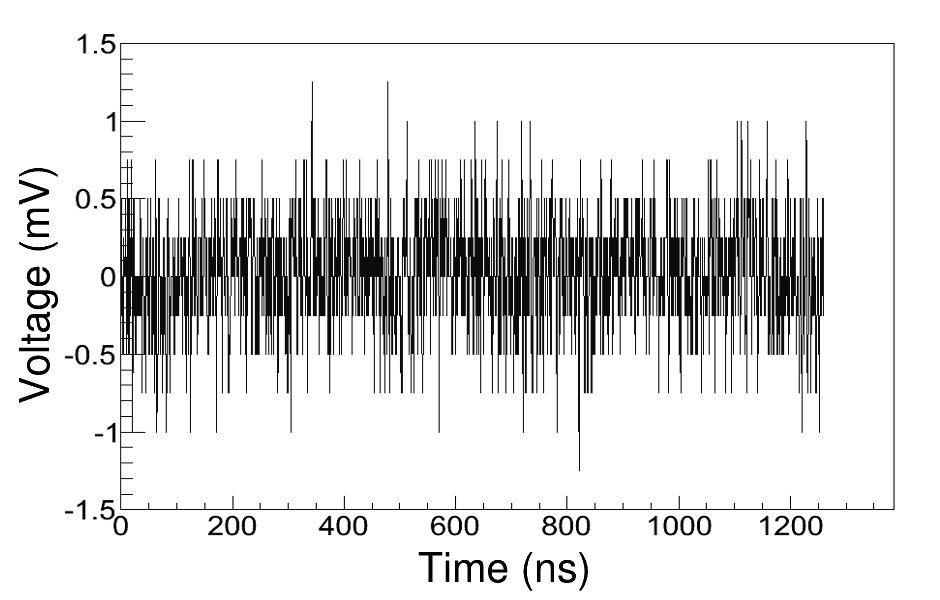}
        \end{subfigure}
\end{center}
    \caption[Noise without preamplifier]{\label{fig:NoisePreamp} Example of noise suppressed with AD8009 based preamplifier (left) and comparison with normal baseline with the use of preamplifier (right) (vertical mV, horizontal ns).}
\end{figure}

An example of the suppressed population can be seen in Figure~\ref{fig:NoisePreamp}. The left Figure shows a baseline from this population and it is compared with a normal baseline using the preamplifier (right Figure). The amplitude of this noise was not so relevant but DC calculation and energy estimators (QDC, area) can be affected at low energies.

\paragraph{}
\subsubsection{High voltage power supply noises}\label{sec:HVNoises}
\begin{figure}[h!]
     \begin{center}
	\begin{subfigure}[b]{0.5\textwidth}
                \centering
                \includegraphics[width=\textwidth]{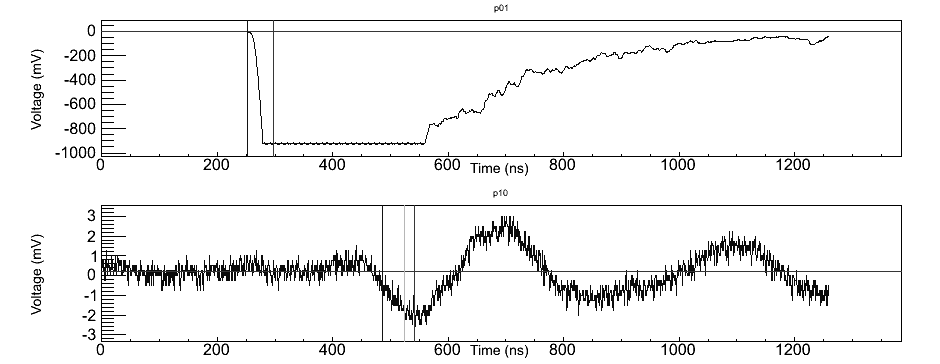}
                \caption{Crosstalk among channels using LC filter.}
                \label{fig:CrosstalkwoPI}
        \end{subfigure}%
        ~ 
        \begin{subfigure}[b]{0.5\textwidth}
                \includegraphics[width=\textwidth]{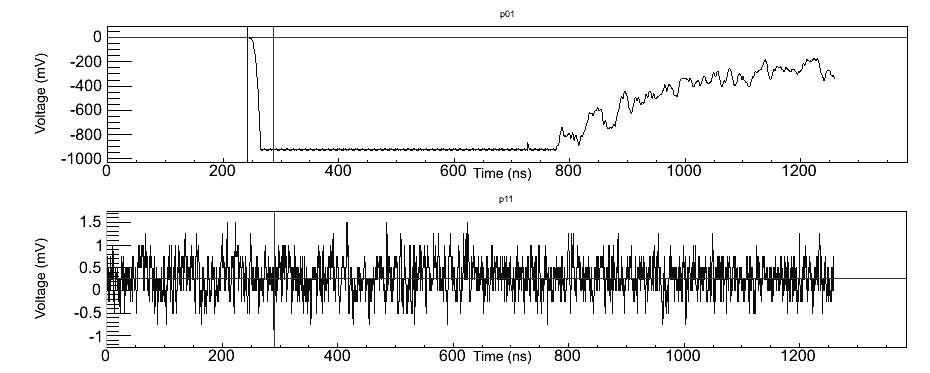}
                \caption{A similar event using $\pi$ filter.}
                \label{fig:CrosstalkPI}
        \end{subfigure}%
	
	\begin{subfigure}[b]{0.5\textwidth}
                \centering
                \includegraphics[width=\textwidth]{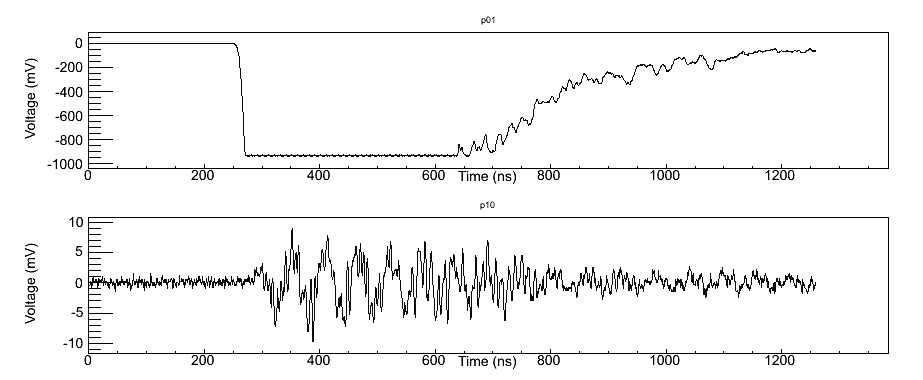}
                \caption{Crosstalk among channels without ferrite beads.}
                \label{fig:CrossTalkLemo}
        \end{subfigure}%
        ~ 
        \begin{subfigure}[b]{0.5\textwidth}
                \includegraphics[width=\textwidth]{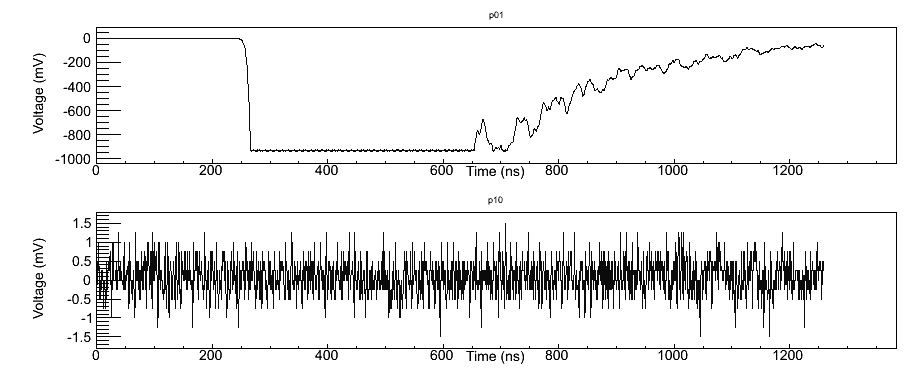}
                \caption{A similar event with ferrite beads removing crosstalk.}
                \label{fig:CrossTalkLemoNo}
        \end{subfigure}%


        \caption{High voltage crosstalks.}
\end{center}
\end{figure}

First ANAIS runs using Hamamatsu R6956MOD PMTs with recommended voltage divider (see Section~\ref{sec:VoltageDivider}) were unexpectedly noisy. The origin of this noise was identified at CAEN SY 2527 HV Power Supply level. The noisy nature of the Power Supply and the lack of any filtering in voltage divider was the cause of a poor baseline and episodes of triggering noise. A LC filter was installed in order to cut higher frequencies coming from HV Power Supply. 
A reduction of the noise population and a much more stable triggering were achieved.
\paragraph{}
A significant crosstalk among signals was also observed. High energy events produced oscillations in all other PMT signals as can be seen in Figure~\ref{fig:CrosstalkwoPI}. The amplitude of this effect is similar to photoelectrons so it can even trigger the event detection and give false coincidence events although its contribution to energy estimators (Area, QDC) was almost negligible. The source of this effect was identified at the level of A1833BP CAEN HV module hosted in CAEN SY 2527 HV mainframe by monitoring the signal from HV outputs through an HV capacitor. The effect was not filtered by the aforementioned LC filter.
\paragraph{}
A new $\pi$ filter has been designed in order to avoid such a crosstalk (470 pF, 1nF, 1 mH) and replace the LC filter in between HV Power Supply and every voltage divider. This filter removed the effect of the crosstalk as it can be seen in Figure~\ref{fig:CrosstalkPI}.
\paragraph{ }
Another crosstalk was identified at HV filter level. The box that housed all filters was modified to allow easy connection, but the cable from filter the PCB to the connector and later to the PMT was sensitive to big signals in other channels as it can be seen in Figure~\ref{fig:CrossTalkLemo}. The effect was totally removed by using ferrite beads in such cables as it can be seen in Figure~\ref{fig:CrossTalkLemoNo}.
\paragraph{}

\begin{figure}[h!]
     \begin{center}
                \includegraphics[width=0.6\textwidth]{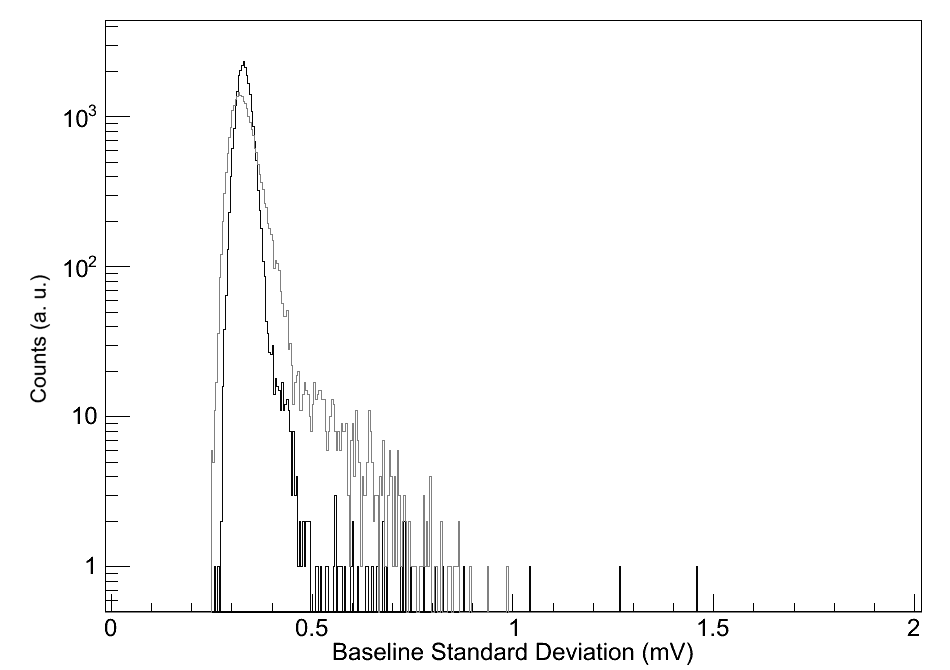}
                \caption{Baseline with A1833BP (black) and A1535 (gray).}
                \label{fig:RMSA1833vsA1535}
\end{center}
\end{figure}

The last HV Power Supply related issue was to study the two different available Power Supply modules, A1833BP and A1535. The results can be seen in Figure~\ref{fig:RMSA1833vsA1535} evidencing the A1535 a noisier nature. The A1833BP module has enough channels for a nine detector experiment, so this module is preferred. If the number of detectors increases, a new filter design like a double-$\pi$ filter could be needed.

\subsubsection{VME polling}\label{sec:VMEPoll}
\begin{figure}[h!]
     \begin{center}
	\begin{subfigure}[b]{0.4\textwidth}
                \centering
                \includegraphics[width=\textwidth]{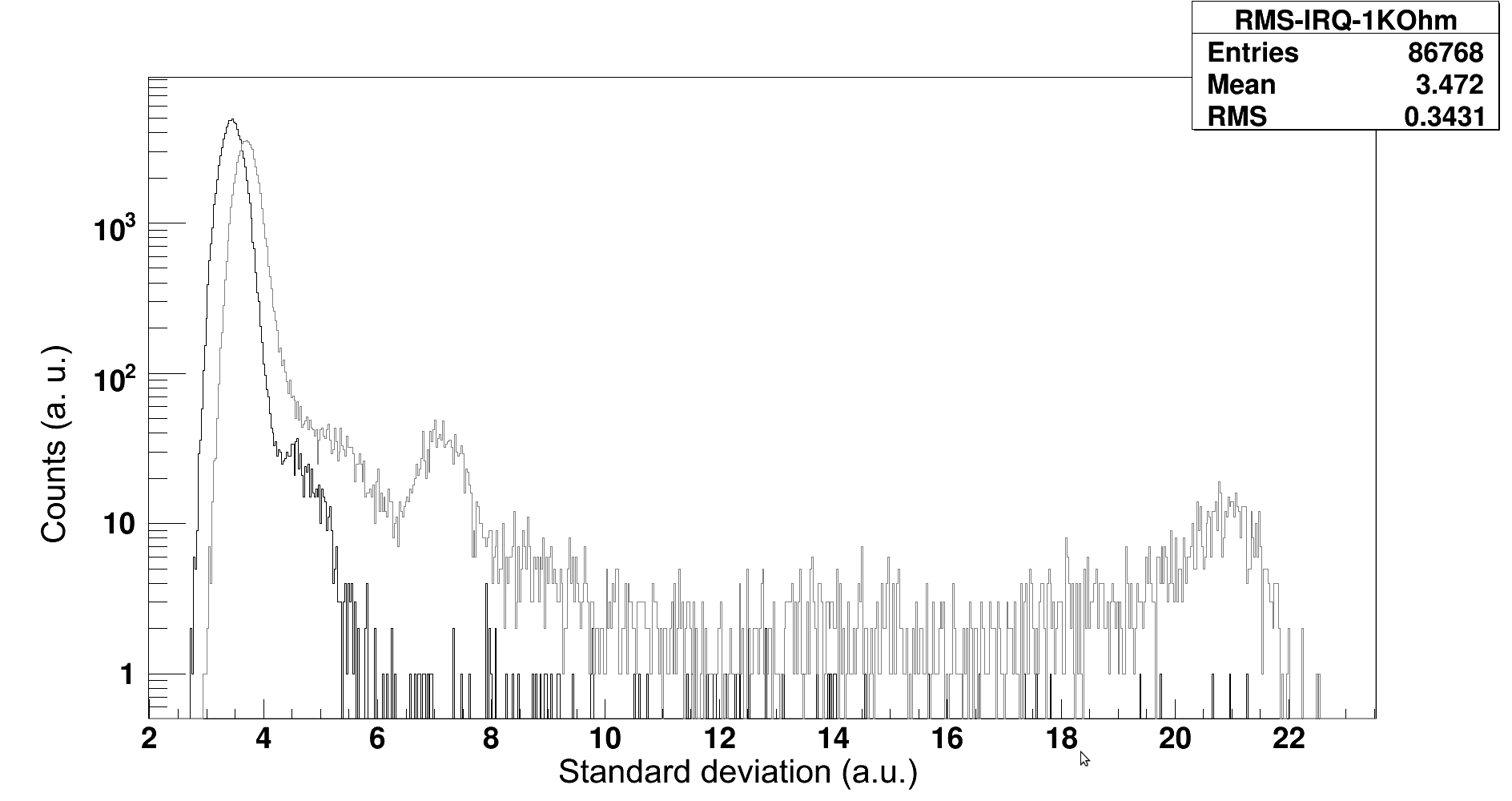}
                \caption{RMS baseline distribution for IRQ (black) and poll (gray).}
                \label{fig:RMSIRQvspoll}
        \end{subfigure}%
        ~ 
       	\begin{subfigure}[b]{0.4\textwidth}
                \centering
                \includegraphics[width=\textwidth]{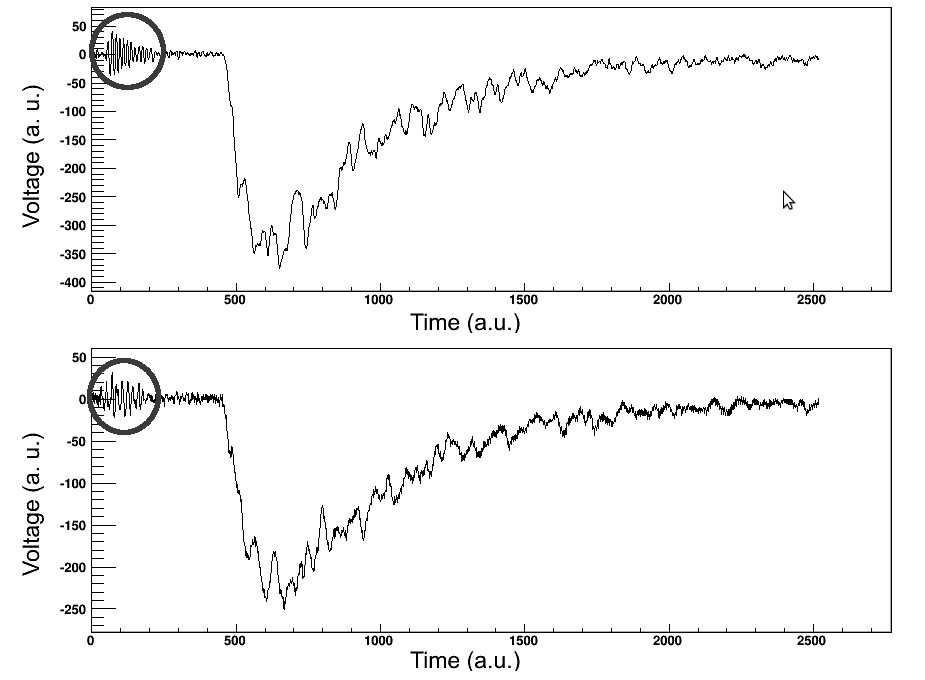}
                \caption{Noise in baseline due to VME polling.}
                \label{fig:PollNoiseWF}
        \end{subfigure}%

	\begin{subfigure}[b]{0.4\textwidth}
                \centering
                \includegraphics[width=\textwidth]{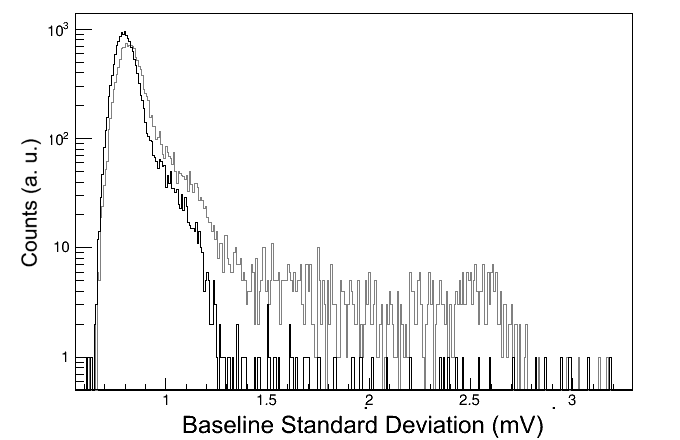}
                \caption{RMS baseline distribution (14 bit MATACQ).}
                \label{fig:RMSIRQvspoll14}
        \end{subfigure}%
        ~ 
       	\begin{subfigure}[b]{0.4\textwidth}
                \centering
                \includegraphics[width=\textwidth]{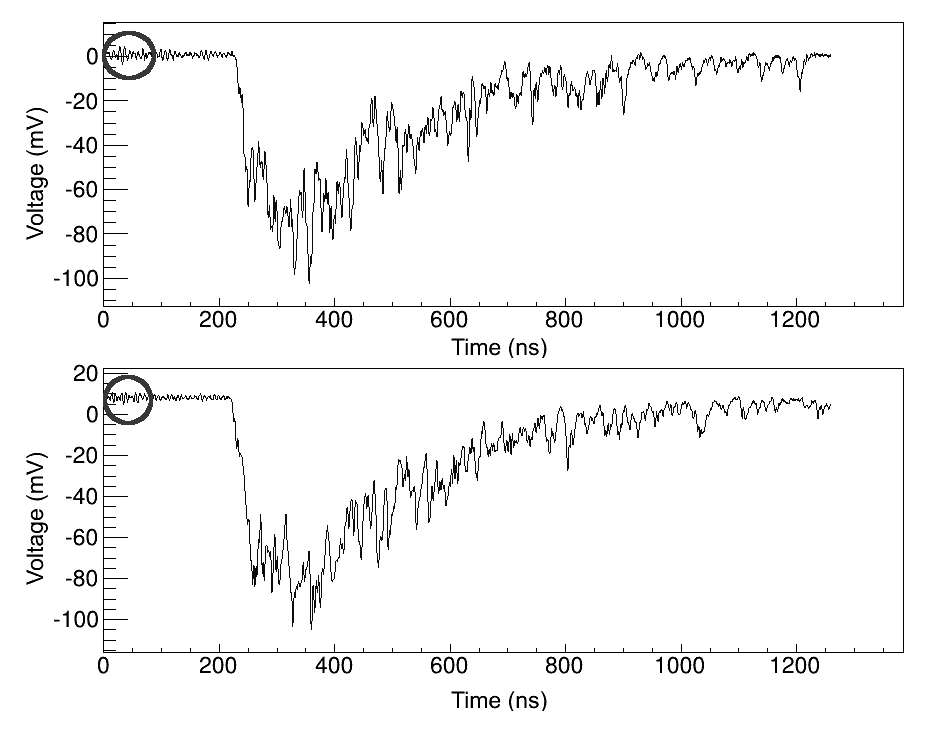}
                \caption{Noise in baseline due to VME polling (14 bit MATACQ).}
                \label{fig:PollNoiseWF14}
        \end{subfigure}%

	\caption[Poll generated noise]{Poll generated noise.}\label{fig:PollNoise}
\end{center}
\end{figure}

An unexpected effect was also discovered when using poll acquisition strategy (see Section~\ref{sec:IRQvspoll}). A noise associated to VME bus activity was identified in different RMS baseline distribution for IRQ and Poll strategies as it can be seen in Figure~\ref{fig:PollNoise}. This figure shows the RMS baseline sum distributions displaying relevant populations of higher RMS sum with poll strategy (Figure~\ref{fig:RMSIRQvspoll}) corresponding to signals like those seen in Figure~\ref{fig:PollNoiseWF}. This effect can be attributed to a MATACQ analog front-end poor isolation from the digital sections. This kind of noise can be particularly harmful because it can confuse the starting point detection (\texttt{t0}) of the signal described in the pulse characterization algorithm (see Section~\ref{sec:PulseChar}) and it can be detrimental to the rest of the algorithm. This effect can also affect to DC calculation, a key parameter in all the algorithm as well as \texttt{t0}.
\paragraph{}
The same test was performed with the 14-bit MATACQ. The RMS baseline sum distribution can be seen in Figure~\ref{fig:RMSIRQvspoll14} showing a better behavior in both amplitude and relative frequency with normal baseline for poll strategy. The effect of polling is visible anyway and it causes a poorer baseline.

\paragraph{}
The existence of this effect made impossible the use of polling strategy. Section~\ref{sec:DeadTimeMeas} describes the impact of not using poll in terms of dead time and how this was dealt in order to absorb this supplementary dead time.

\subsubsection{Fan-In/Fan-Out}
\begin{figure}[h!]
     \begin{center}
	\begin{subfigure}[b]{0.5\textwidth}
                \centering
                \includegraphics[width=\textwidth]{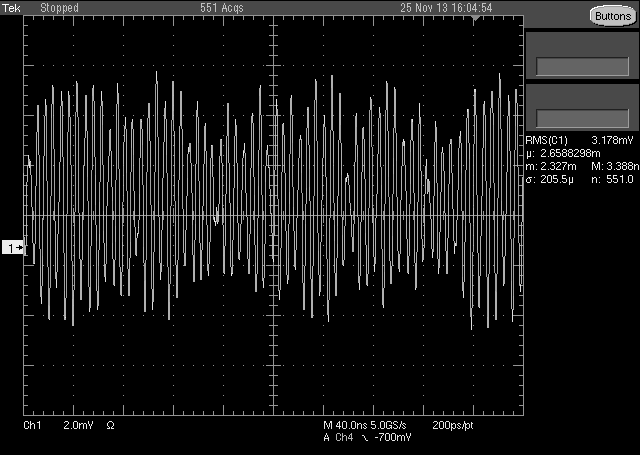}
                \caption{Baseline oscillation in Fan-In/Fan-Out module due to it discriminator.}
                \label{fig:NoiseFiFo}
        \end{subfigure}%
        ~ 
       	\begin{subfigure}[b]{0.5\textwidth}
                \centering
                \includegraphics[width=\textwidth]{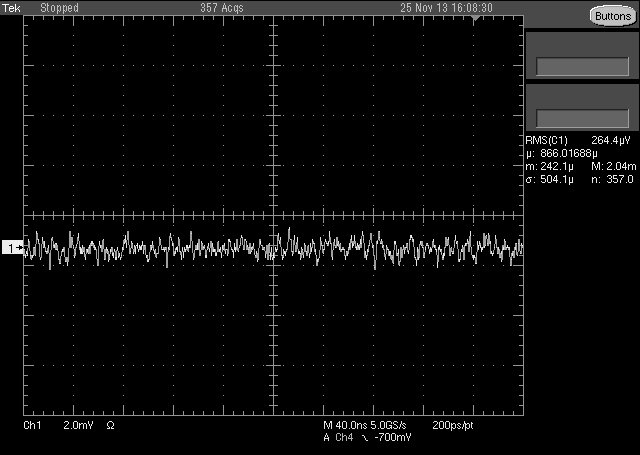}
                \caption{Fan-In/Fan-Out with higher lever discriminator.}
                \label{fig:NoNoiseFiFo}
        \end{subfigure}%
        \caption[Fan-In/Fan-Out discriminator noise]{}
\end{center}
\end{figure}

A very wide baseline was observed using Linear Fan-In/Fan-Out module as it can be seen in Figure~\ref{fig:NoiseFiFo}. This noisy baseline was caused by the (not used in ANAIS) internal discriminator of this module. A low default position of the threshold trimmer can cause constant triggering in the aforementioned discriminator and an undesired crosstalk to the baseline of all section. The solution to avoid this effect is to set the discriminator in a higher level. The baseline is notably thinner as can be seen in Figure~\ref{fig:NoNoiseFiFo}, 0.3 mV (RMS) versus 3 mV (RMS). 

\subsubsection{Preamplifier crosstalk}
\begin{figure}[h!]
     \begin{center}
	\begin{subfigure}[b]{0.5\textwidth}
                \centering
                \includegraphics[width=\textwidth]{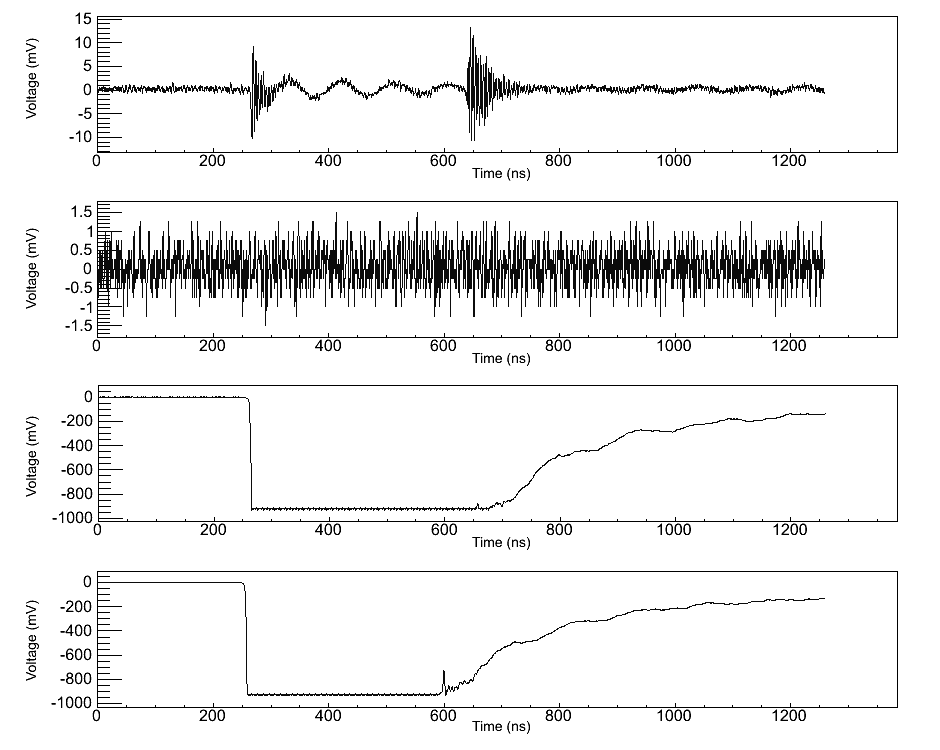}
        	\caption{\label{fig:PCBCrosstalkBefore}}
        \end{subfigure}%
        ~ 
       	\begin{subfigure}[b]{0.5\textwidth}
                \centering
		\includegraphics[width=\textwidth]{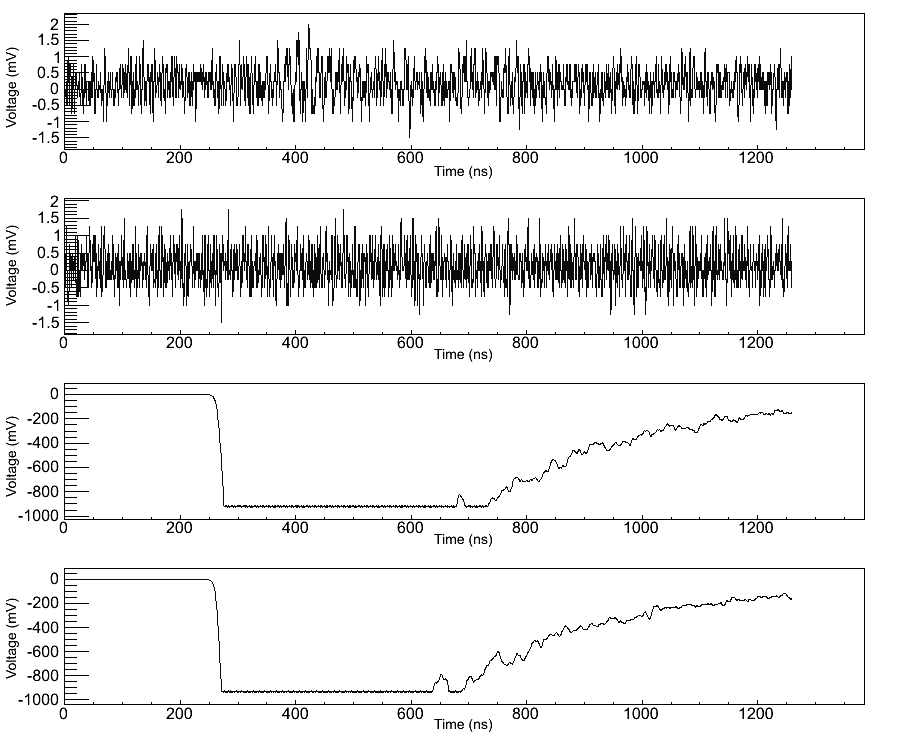}
        	\caption{\label{fig:PCBCrosstalkAfter}}
        \end{subfigure}%
\end{center}
        \caption[Preamplifier crosstalk]{Preamplifier crosstalk (a) and resolution with improved isolation (b).}
\end{figure}

A crosstalk was detected with very high energetic events between two preamplifiers in a subsequent design using the same PCB to two different AD8000. This crosstalk is shown in Figure~\ref{fig:PCBCrosstalkBefore} showing crosstalk between first and third signals that shared PCB.
\paragraph{ }
An effect like this one can be especially harmful because it can affect to coincidence studies like $^{40}K$ and $^{22}Na$ events used to determine the crystal contamination as seen earlier. The printed circuit was redesigned to improve the isolation between amplifiers. The result can be seen in Figure~\ref{fig:PCBCrosstalkAfter} showing no crosstalk with an event of very high energy.

\subsubsection{Improper EMC preamplifier power supply}

\begin{figure}[h!]
     \begin{center}
	\begin{subfigure}[b]{0.5\textwidth}
                \centering
                \includegraphics[width=\textwidth]{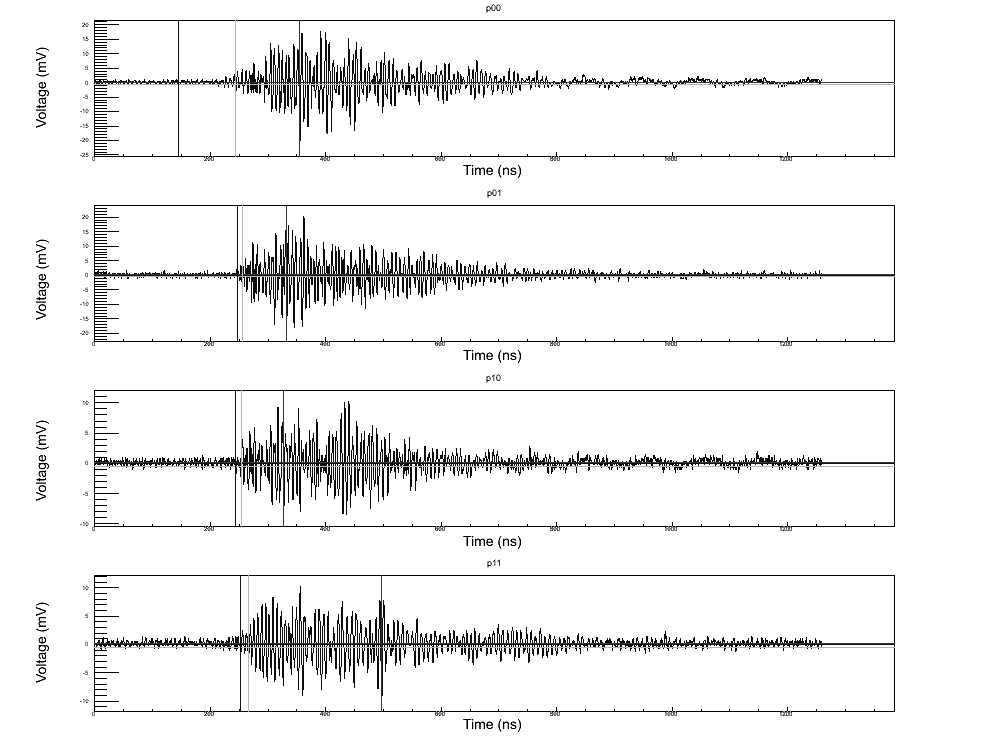}
        	\caption{\label{fig:NoiseEMCPreampSpike}}
        \end{subfigure}%
        ~ 
       	\begin{subfigure}[b]{0.5\textwidth}
                \centering
                \includegraphics[width=\textwidth]{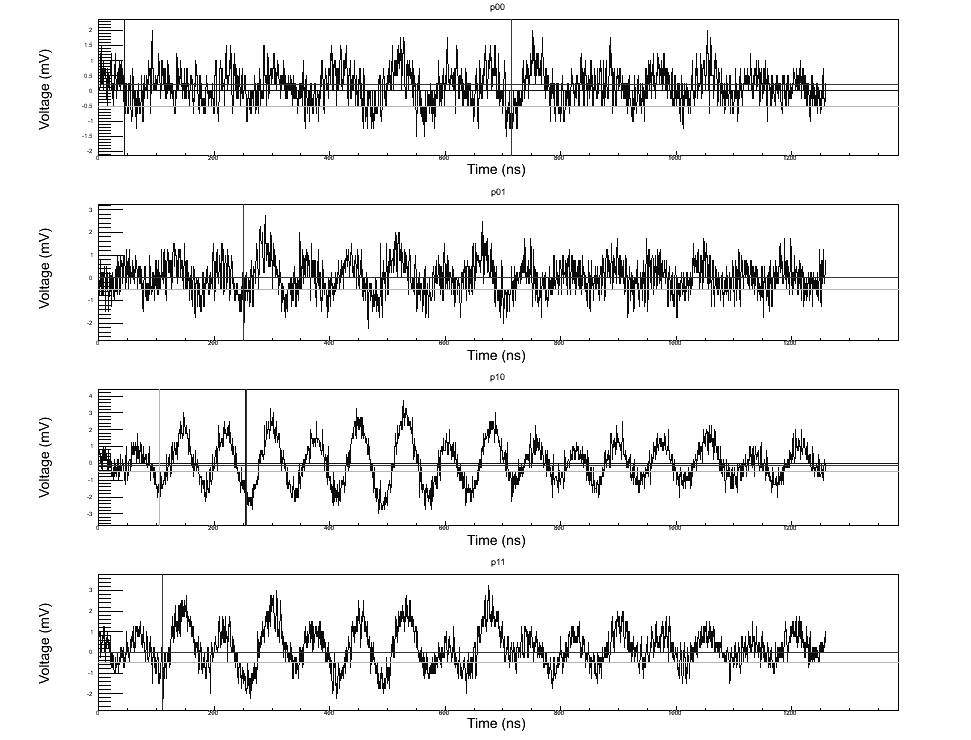}
        	\caption{\label{fig:NoiseEMCPreamp100Hz}}
        \end{subfigure}%
\end{center}
        \caption[EMC preamplifier noises]{EMC power supply preamplifier noises.}
\end{figure}

A series of noises were reported attributable to a poor EMC in the amplifier power supply in some test set-ups. The first type of noise consist of spikes with high frequency components and high amplitude correlated with flashes by broken fluorescent lamp. This spikes can be seen in Figure~\ref{fig:NoiseEMCPreampSpike}. The second one was detected by analyzing the triggering time pattern from the data, as it can be seen in Section~\ref{sec:TemporalParams}, and detecting a 100~Hz component. This noise can be seen in Figure~\ref{fig:NoiseEMCPreamp100Hz}.

These noises were not very frequent. They could trigger the acquisition and pass some quality cuts but their area pattern and the amplitude area ratio make simple to filter during the data analysis. The preamplifier power supply was redesigned in order to avoid such effects and the crosstalk described in the previous subsection.

\chapter{Muon detection system}\label{sec:Veto}
\section{Muon underground interaction}
The underground muon flux has a measured annual modulation~\cite{bellini2012cosmic} and muon-related events could mimic the dark matter signal~\cite{nygren2011testable,bernabei2012no,davis2014fitting,bernabei2014no, klinger2015muon}. The system proposed, implemented and described in this chapter allows to correlate muon events with the NaI(Tl) measured signals in order to study and control these effects.
\begin{figure}[h!]
     \begin{center}
                \includegraphics[width=.8\textwidth]{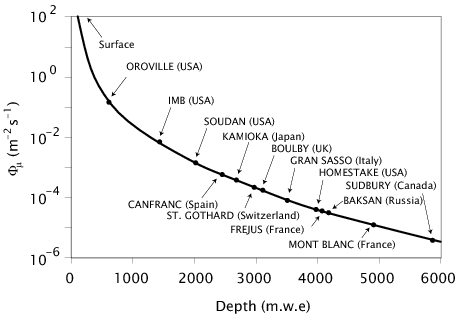}
		\caption[Underground laboratories muon flux]{Location of underground laboratories.\label{fig:UnderLabs}}
\end{center}
\end{figure}

\paragraph{ }
The residual muon flux at the Canfranc Underground Laboratory is approximately $5 \times 10^{-3} \mu$ s\textsuperscript{-1}m\textsuperscript{-2}~\cite{bettini2012talk} being more than four orders of magnitude lower than the surface flux, but enough to have some influence in the ANAIS detectors. The expected effects of muons in the NaI crystals are direct interaction in the crystal with very energetic depositions~\cite{knoll2010radiation} and other effects such as secondary neutrons~\cite{araujo2005muon}. Secondary neutrons can be very harmful for the direct search of dark matter because they can produce similar signals to the expected WIMP interaction.
\paragraph{ }

For this reason, a plastic veto system to tag muon-related events has been designed and tested in order to monitor possible systematic effects in the search of annual modulation in the data. The plastic scintillators characterization, the electronic front-end, the rate and the correlation of plastic events with NaI(Tl) events in both time and energy are covered in this chapter. 
\section{Muon detector set-up}
\begin{figure}[h!]
     \begin{center}
	     \includegraphics[width=0.7\textwidth]{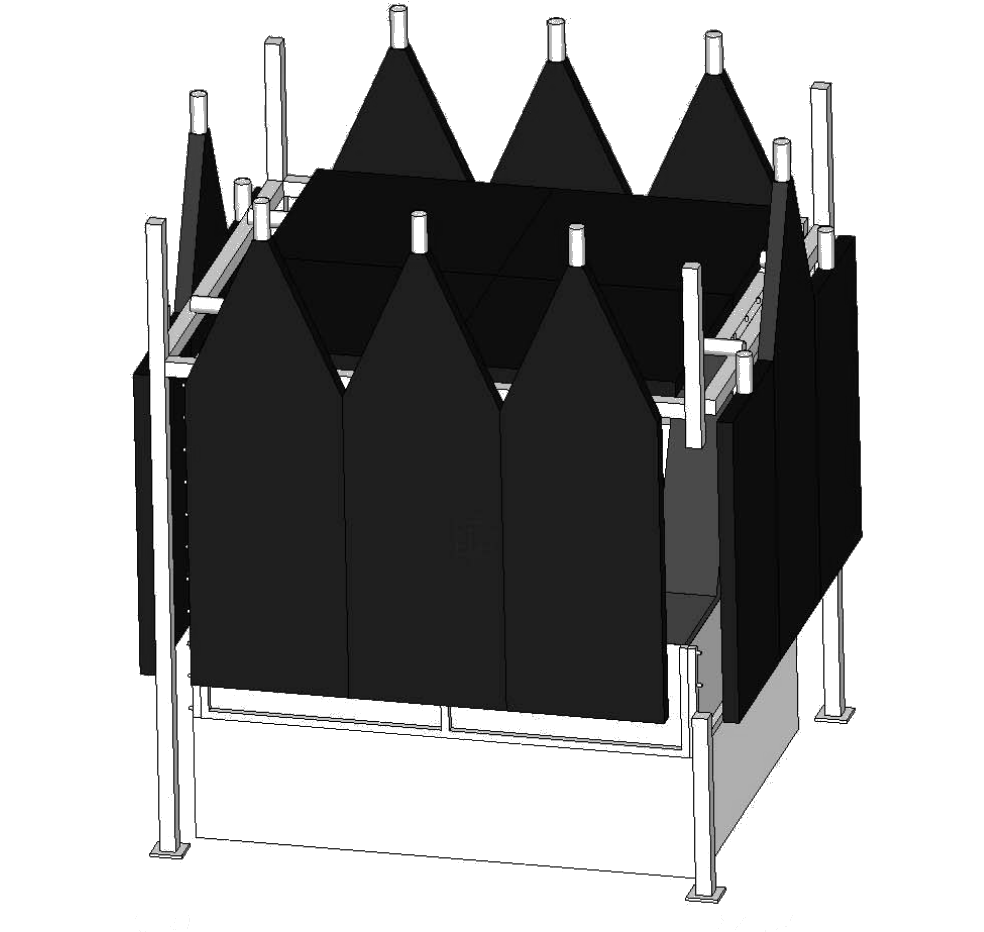}
		\caption[Muon detection system design]{Muon detection system design.\label{fig:VetoSetUp}}
\end{center}
\end{figure}

The muon detection system will consist of sixteen plastic scintillators (also called veto in this work) covering all the ANAIS set-up except the bottom face (see Figure~\ref{fig:VetoSetUp}). The role of these scintillators is to detect the residual muon flux arriving at the laboratory and to use their information to discard coincident events in the crystals and study the correlation between muon events and crystal events. For this study, a tagging strategy was carried out instead of a hardware vetoing strategy.

\paragraph{ }
The sixteen plastic scintillators have different dimensions and origins as it can be seen in Table~\ref{tab:VetoDim}. Eight units were constructed and used for the IGEX neutrinoless double-beta decay experiment~\cite{aalseth2002igex} and eight new Scionix ME2PX plastic scintillators were purchased to cover the ANAIS experiment.
\begin{table}[h!]
		\begin{center}
			\begin{tabular}{cccc}
				\toprule
				Plastic numbers&Vendor/Model&Dimensions (mm)\\
				\toprule
				1, 2, 7, 8&Scionix/R500*50B1000-2ME2PX&$1000\times500\times50$\\
				3 - 6&Scionix/R700*50B750-2ME2PX&$750\times700\times50$\\
				9 - 16&Homemade&$1000\times500\times50$\\
				\toprule
			\end{tabular}
			\caption[Plastic scintillators models and dimensions]{Plastic scintillators models and dimensions.} 
			\label{tab:VetoDim}
		\end{center}
\end{table}

\section{Plastic scintillator characterization}
Both plastic scintillators models were tested in order to characterize the muon signal. First, a simulation of the expected muon energy deposition in the plastics is presented. Next, the plastic scintillation test bench is described and finally the characterization of the homemade and Scionix plastics are shown.
\subsection{Plastic scintillator simulation}\label{sec:VetoSimulation}
\begin{figure}[h!]
     \begin{center}
                \includegraphics[width=0.9\textwidth]{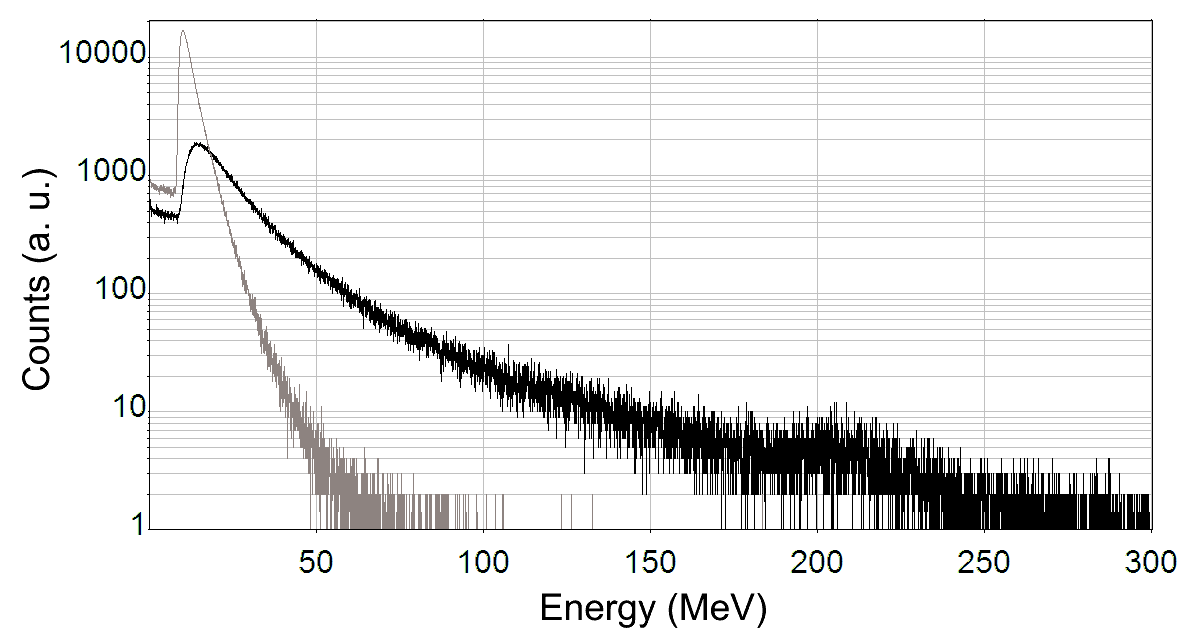}
		\caption[Simulated plastic spectra]{Simulated plastic spectra: horizontal (gray) and vertical (black) positions.\label{fig:VetoSimulation}}
\end{center}
\end{figure}

A simulation using Geant4 toolkit~\cite{agostinelli2003geant4} was performed as a guide to understand the expected signal in the plastic scintillators. Horizontal and vertical orientations were simulated given their positioning in the final set-up (see Figure~\ref{fig:VetoSetUp}). The interacting mechanisms considered were multiple scattering, ionization, bremsstrahlung and pair production. The considered underground angular distribution was $\propto cos^{3.6} \theta$ corresponding to 800 m of standard rock~\cite{stockel1969study} and the energy spectrum was taken from Lipari et al.~\cite{lipari1991propagation}. The results can be seen in Figure~\ref{fig:VetoSimulation}.
\paragraph{}
Figure shows the spectra of the horizontal (gray) and vertical (black) plastic orientations. Due to the plastic characteristics, both distributions of energy deposition show a \emph{muon peak} near 10 MeV. The spectrum of the vertical one features a wider peak and a longer tail. The peak corresponds to an energy deposition of $\sim$2 MeV per cm, in agreement with the expected interaction of a muon in this kind of plastic~\cite{knoll2010radiation}.
\paragraph{ }
This simulation can also be used to estimate the effect of the threshold location. A threshold level below the peak will tag most of the muons still rejecting the $\gamma$ background. Nevertheless, the continuum below the muon peak accounts the 6.6\% of the events in the horizontal position and an 8.9\% in the vertical position.

\subsection{Plastic scintillator test bench}
\begin{figure}[h!]
     \begin{center}
                \includegraphics[width=\textwidth]{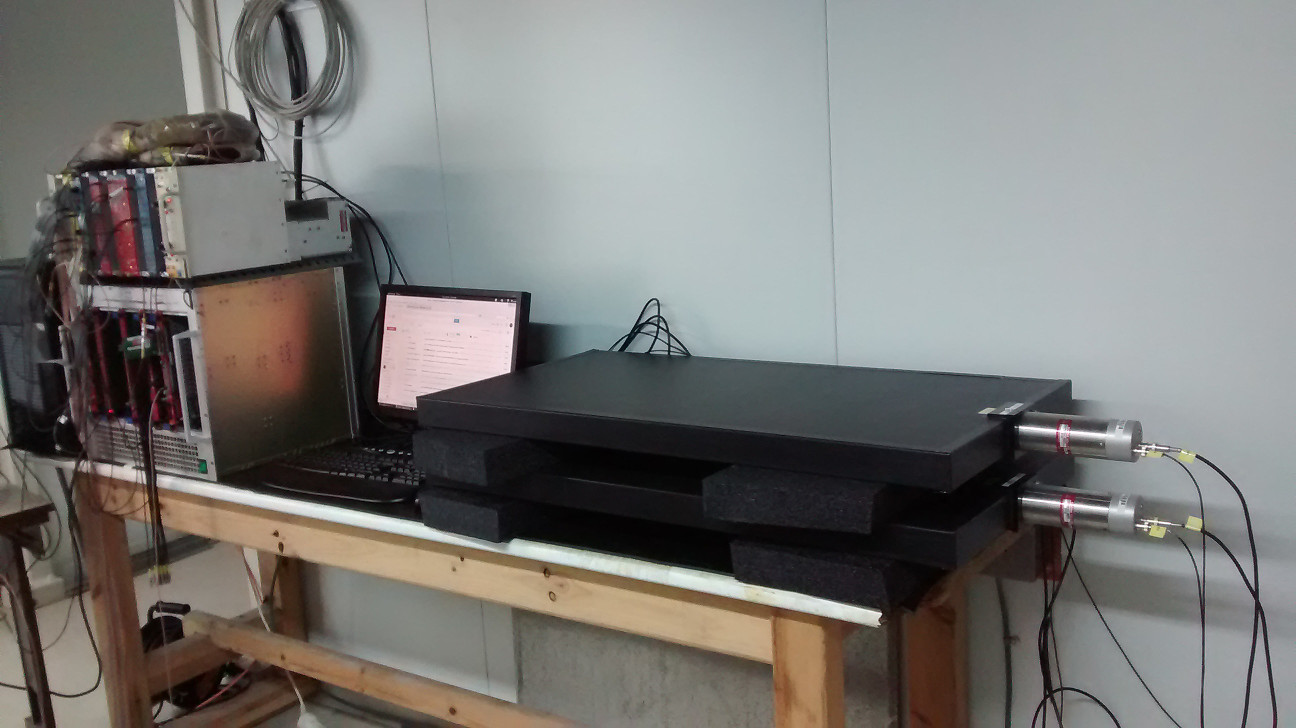}
		\caption[Plastic test bench]{Plastic scintillators test bench.\label{fig:PlasticTestBench}}
\end{center}
\end{figure}

A test bench (see Figure~\ref{fig:PlasticTestBench}) with two stacked plastics was mounted to digitize signals with an OR trigger. This test allowed to study the PMT electric signal of coincident and anti-coincident events.
\paragraph{}

The non-coincident events consist of natural radioactive background and PMT noise. The coincident events are mainly muons ionizing both detectors giving a spectrum predicted by the previous simulations. Anyway, the characterization of these populations was needed because of the geometry of the final set-up (see Figure~\ref{fig:VetoSetUp}) which does not allow to use coincidence to select muons because most of the expected trajectories pass through only one detector. This assumption is reviewed later, in Section~\ref{sec:MuMulti}. Once characterized the different signals for different populations, the features needed by the electronic front-end for the correct muon identification and tagging can be defined.

\subsection{Homemade plastic scintillators}
\begin{figure}[h!]

\begin{subfigure}[b]{0.5\textwidth}
  \begin{center}
  \includegraphics[width=\textwidth]{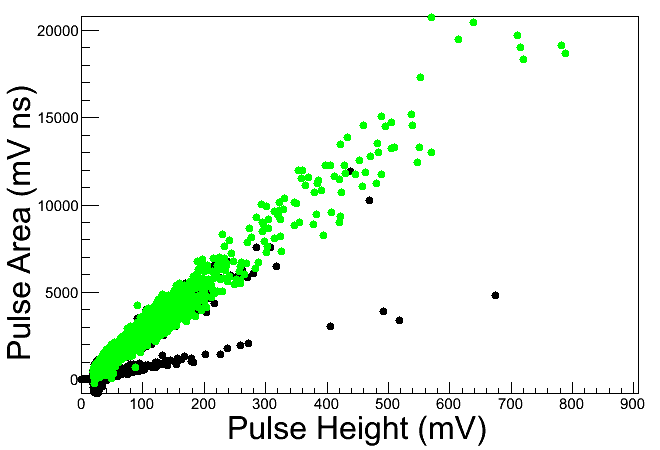}
  \caption{\label{fig:AreavsHMuHM}}
  \end{center}
  \end{subfigure}
        ~ 
\begin{subfigure}[b]{0.5\textwidth}
  \begin{center}
  \includegraphics[width=\textwidth]{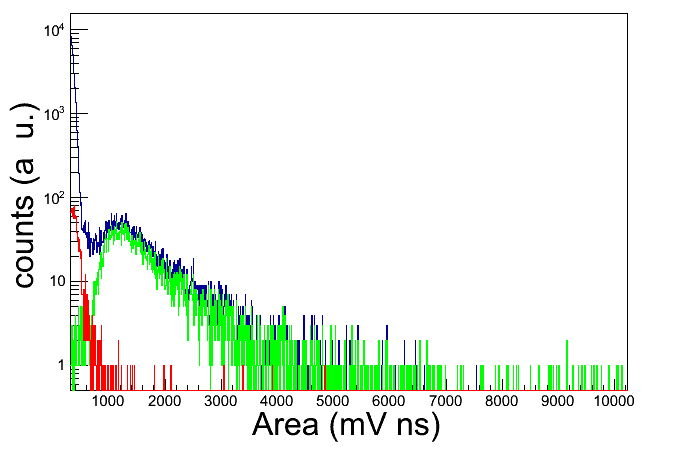}
  \caption{\label{fig:AreaMuHM}}
  \end{center}
  \end{subfigure}

\begin{subfigure}[b]{0.5\textwidth}
  \begin{center}
  \includegraphics[width=\textwidth]{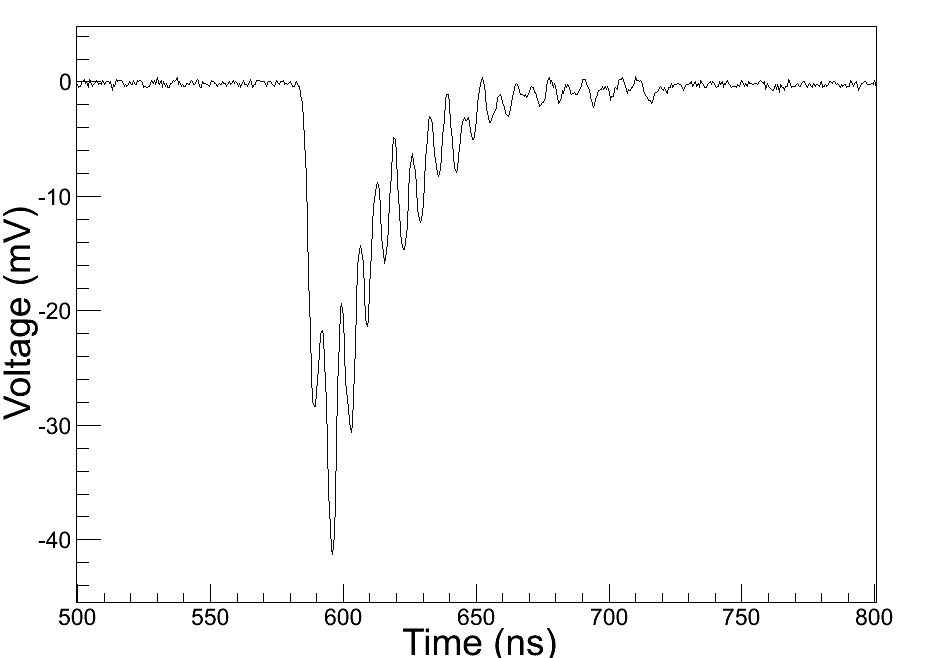}
  \caption{\label{fig:VetoHMSlowEvent}}
  \end{center}
  \end{subfigure}
        ~ 
\begin{subfigure}[b]{0.5\textwidth}
  \begin{center}
  \includegraphics[width=\textwidth]{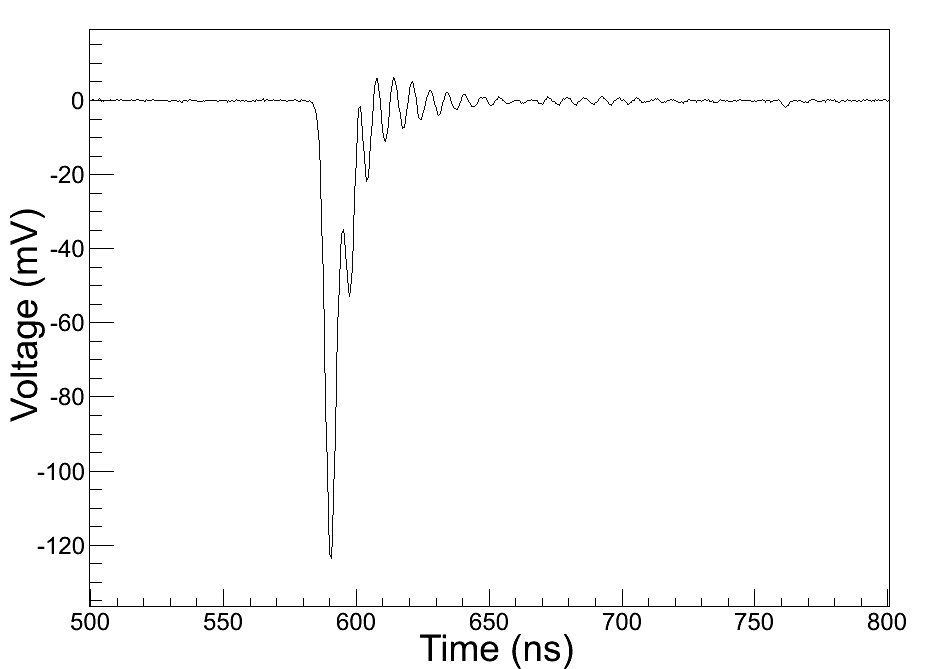}
  \caption{\label{fig:VetoHMFastEvent}}
  \end{center}
  \end{subfigure}

  \caption[Homemade plastic tests]{Scatter plot (pulse area vs. pulse height) (a) with coincident events in green and spectrum (b) with fast events in red and coincidences in green. Underground data.\label{fig:MuonsAndFast} Example of slow (c) and fast (d) events with similar collected charge.}
\end{figure}

The homemade plastic scintillators have been working in ANAIS since ANAIS-0 prototype~\cite{CCUESTA}. Two of these plastic scintillators were stacked and tested underground as described earlier in order to characterize their behavior.
\paragraph{ }
Figure~\ref{fig:AreavsHMuHM} shows the scatter of signal area (proportional to the event charge collected) versus the signal amplitude, with coincident events in green and non-coincident population (in black). It can be noted the faster nature of this later population (note its lower area/height ratio) compared with the coincident muon population. Figure~\ref{fig:AreaMuHM} shows the charge spectrum of almost one month of underground data in black, the coincident event in green and the fast population in red. Setting a hardware trigger in charge (area) discards most of the fast population and the other non-coincident background events. This trigger can be achieved by using an integrating shaping amplifier. The trigger level must be set below the muon peak to maximize the tagged muons but above the end of the natural radioactive background. This trigger strategy greatly reduces the triggering of the fast population. For example, setting the trigger at 700 $mV\cdot ns$ gives 60 events above the area threshold in almost one month, $2.5 \times 10^{-5} Hz$, two orders of magnitude below the muon counts. Examples of muon and fast events with similar collected charge are shown in Figures~\ref{fig:VetoHMSlowEvent} and \ref{fig:VetoHMFastEvent} respectively (note the different vertical range).

\subsection{Scionix plastic scintillators}
\begin{figure}[h!]
\begin{subfigure}[b]{0.5\textwidth}
  \begin{center}
  \includegraphics[width=.9\textwidth]{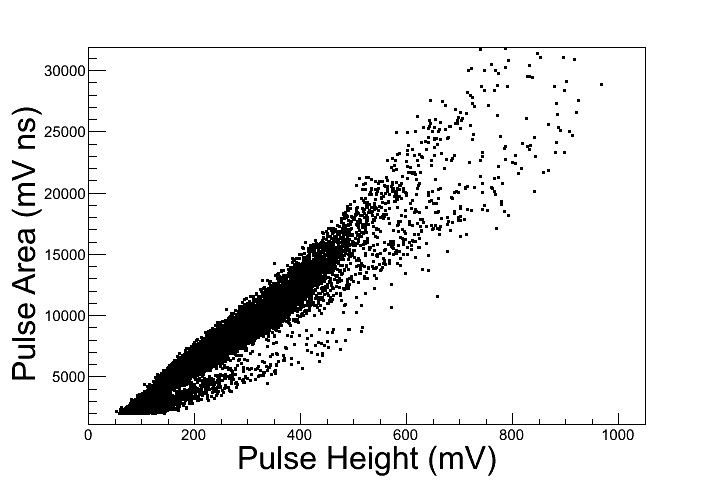}
  \caption{\label{fig:AreaMu0115}}
  \end{center}
  \end{subfigure}
        ~ 
\begin{subfigure}[b]{0.45\textwidth}
  \begin{center}
  \includegraphics[width=\textwidth]{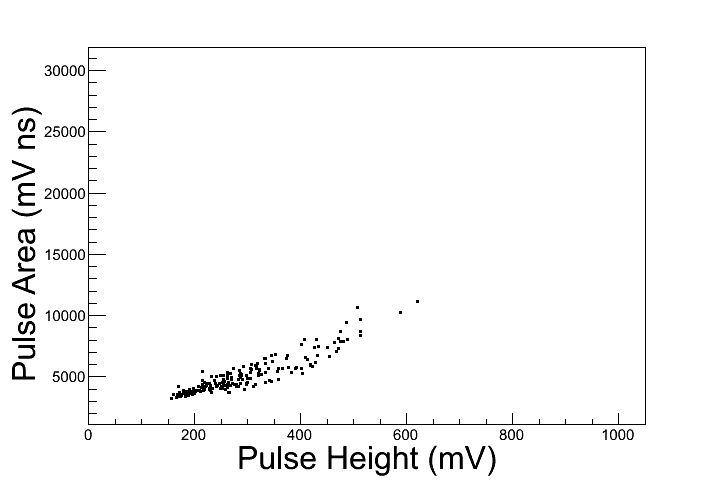}
  \caption{\label{fig:AreaMu0122}}
  \end{center}
  \end{subfigure}

\begin{subfigure}[b]{0.5\textwidth}
  \begin{center}
  \includegraphics[width=\textwidth]{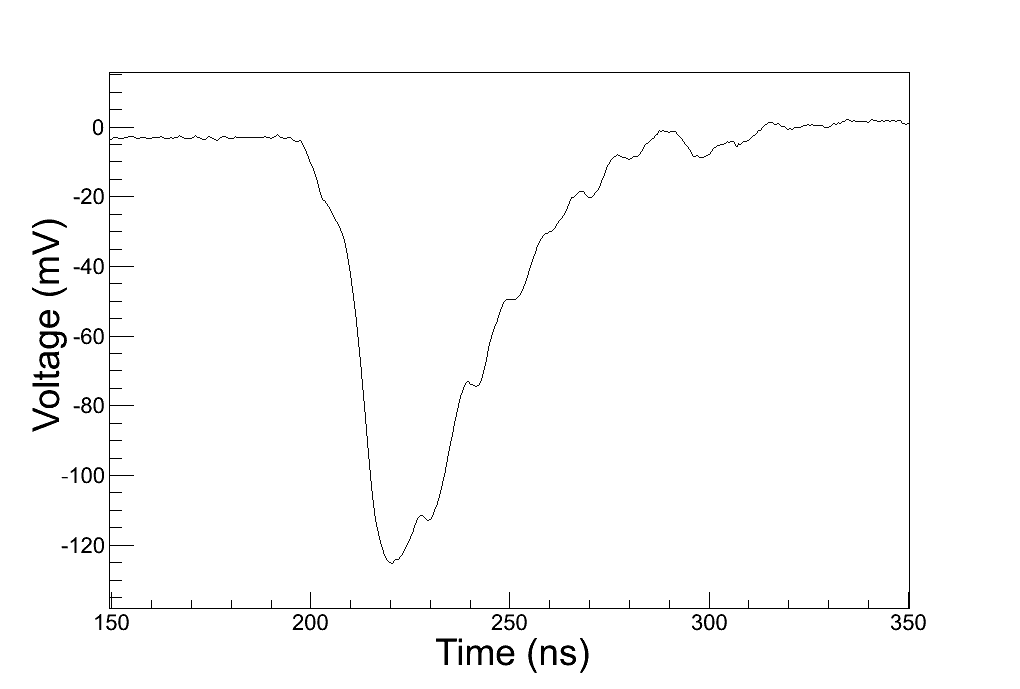}
  \caption{\label{fig:VetoSlowEvent}}
  \end{center}
  \end{subfigure}
        ~ 
\begin{subfigure}[b]{0.5\textwidth}
  \begin{center}
  \includegraphics[width=\textwidth]{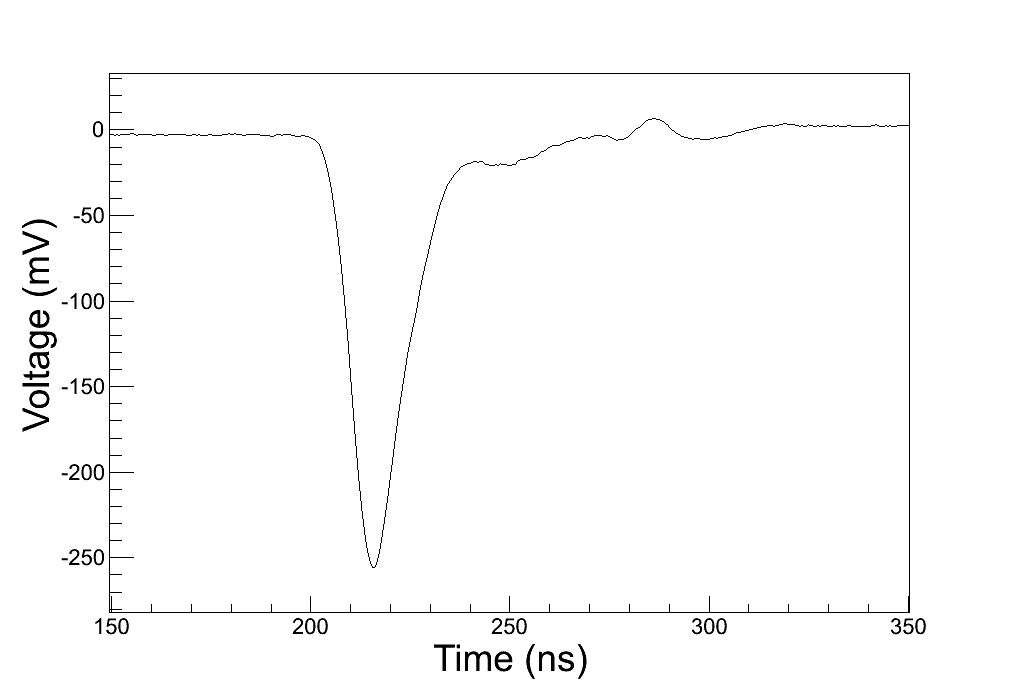}
  \caption{\label{fig:VetoFastEvent}}
  \end{center}
  \end{subfigure}

  \caption[Scionix plastics tests]{Scatter plot (pulse area vs. pulse height taken) at surface (a) and underground (b) with fast events dominating underground rate\label{fig:MuonsAndFastHM}. Example of slow (c) and fast (d) events with similar collected charge.}
\end{figure}

The underground data of the Scionix plastic scintillators also revealed a population faster than muon events but with a rate substantially higher than expected, five order of magnitude compared with the homemade plastic scintillators. This fast population, which is also present at surface, is clearly dominating the underground rate as can be observed in Figure~\ref{fig:MuonsAndFastHM}. Figure~\ref{fig:AreaMu0115} shows the area/height scatter at surface in a short run of five minutes featuring muons and fast population. Figure~\ref{fig:AreaMu0122} shows an underground run of same duration, showing a very high rate of the fast population and a complete suppression of the muon rate. Figures~\ref{fig:VetoSlowEvent} and \ref{fig:VetoFastEvent} show events corresponding to a muon and a fast event with similar collected charge (note the different vertical range).
\paragraph{}
These figures illustrate the impossibility of the previous triggering strategy: the fast population would trigger much more frequently than real muons with a rate \hbox{$\sim$ 1 Hz} per plastic scintillator. The use of such a trigger would imply the tagging of all NaI events as muon-related events reducing the effective exposure time (live time) beyond acceptable levels (see ANAIS experimental requirements in Section~\ref{sec:ExperimentalRequirements}). Therefore, a new trigger strategy was designed to tag muons based on pulse shape analysis, as it can be seen in Section~\ref{sec:VetoFrontEnd}.
\section{Electronic front-end design and DAQ}\label{sec:VetoFrontEnd}
\begin{figure}[h!]
     \begin{center}
                \includegraphics[width=\textwidth]{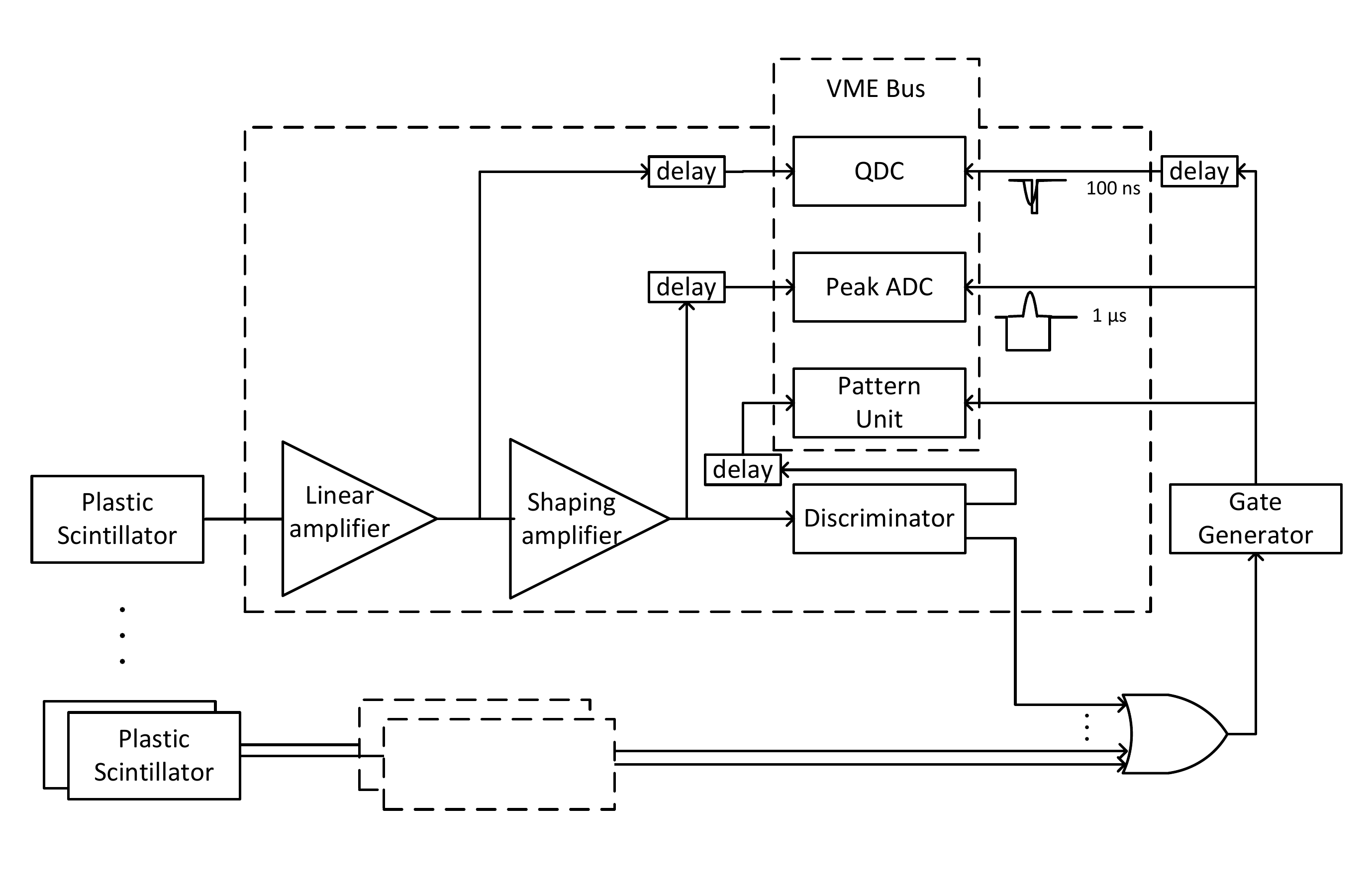}
		\caption[Muon tagging electronic set-up]{Muon tagging electronic set-up.}\label{fig:SetupVeto}
\end{center}
\end{figure}

The muon detection front-end was designed to characterize and offline discard the fast population due to its high rate in the Scionix plastics. This discrimination is based on the shape of the fast population. The design of the front-end electronics can be seen in Figure~\ref{fig:SetupVeto} showing the detailed scheme of one plastic channel. The key concept is to have two different parameters for each event: one proportional to the whole area and the other accounting for the last part of the pulse. This allows to reject the fast population based on the ratio of these two parameters.
\paragraph{ }
The original plastic signal is linearly amplified and duplicated. One signal goes through a 16 channel shaper, timing filter amplifier with constant fraction discriminator (Mesytec MSCF-16) providing an output with the amplitude proportional to the collected charge and a logical trigger signal for every channel. The module also provides an OR trigger that is used as global trigger. The output of this module is delayed and used as input to a Peak Sensing ADC (CAEN V785) that digitizes the maximum amplitude during the trigger window.
\paragraph{ }
The other linearly amplified signal is delayed and used as input of a QDC (see Section~\ref{sec:VMEModules} for a description of this module and other VME modules). The QDC digitizes the charge collected during the trigger window. This trigger window is generated by the global trigger with a convenient delay, collecting only the tail of the signal in such a way that the fast events collect much less charge than the slow events. The delay in QDC trigger is 320 ns from the global trigger given the shape of both populations (see Figures~\ref{fig:VetoSlowEvent} and \ref{fig:VetoFastEvent}) and the delay of the non-shaped signal (see Figure~\ref{fig:SetupVeto}).
\paragraph{}
Storing peak sensing data is not only necessary for fast population discrimination but it is also valuable to monitor effects such as gain or threshold instabilities that could affect the data analysis. The discrimination using both QDC and Peak Sensing values can be seen in the next subsection.
\paragraph{}
In addition to the analog to digital conversion, some other information is recorded in order to properly analyze and correlate the plastic data with the NaI scintillators data. Every event has also information of its multiplicity pattern and the real time clock of the event. The multiplicity pattern is obtained from the digital outputs of the MSCF-16 and stored by a CAEN 16 Bit Strobed Multihit Pattern Unit V259. The real time clock is shared with the clock of the NaI DAQ (see Section~\ref{sec:TimeScalers}), counted and stored by a CAEN 32 Channel Latching Scaler V830. The time distance between events is calculated in the time correlation analysis (see Section~\ref{sec:TemporalParams}). A test of such a correlation and its use to study the muon-related events can be seen in Section~\ref{sec:MuNaIEvt}.
\paragraph{}
The DAQ software is exactly the same developed for the NaI acquisition (see Chapter~\ref{sec:DAQSW}) with an appropriate configuration showing its versatility and general purpose nature. It is launched remotely by the NaI DAQ scripts (see Section~\ref{sec:DAQScripts}) in order to guarantee the same count in both latched scalers. The time correlation between both DAQs is studied in Section~\ref{sec:DAQsSynch}.
\subsection{Pulse shape analysis characterization}
\begin{figure}[h!]
  \begin{center}
  \begin{subfigure}[b]{0.5\textwidth}
  \includegraphics[width=\textwidth]{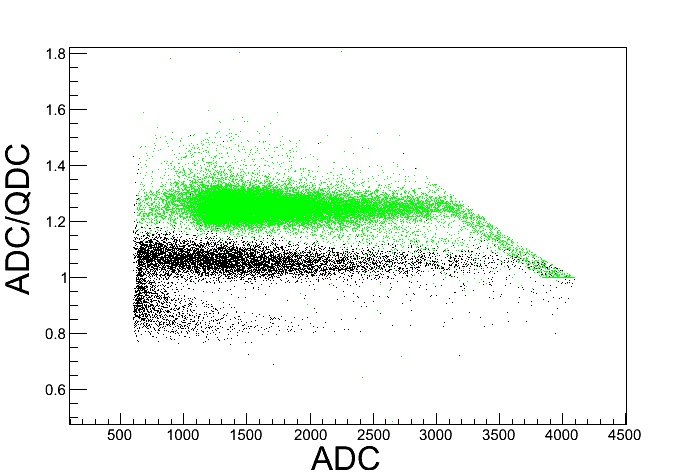}
  \end{subfigure}
  \end{center}
  \caption[Coincident events at surface]{Coincident events at surface.\label{fig:VetoADCQDCCoinc}}
\end{figure}

\begin{figure}[h!]
\begin{subfigure}[b]{0.5\textwidth}
  \includegraphics[width=\textwidth]{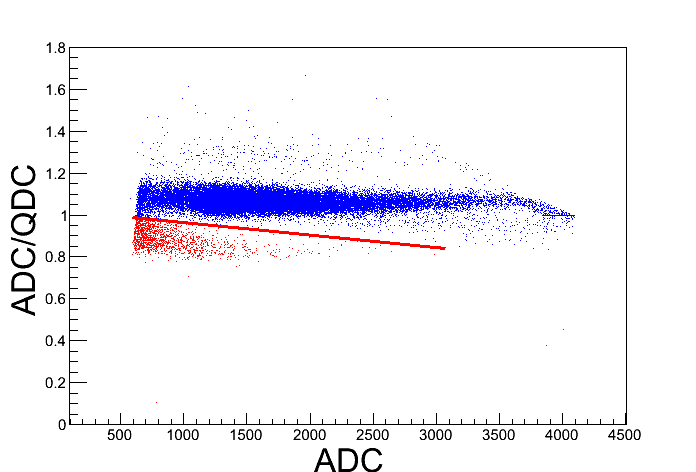}
  \caption{\label{fig:VetoADCQDCDiscr}}
  \end{subfigure}
  ~
 \begin{subfigure}[b]{0.5\textwidth}
  \includegraphics[width=\textwidth]{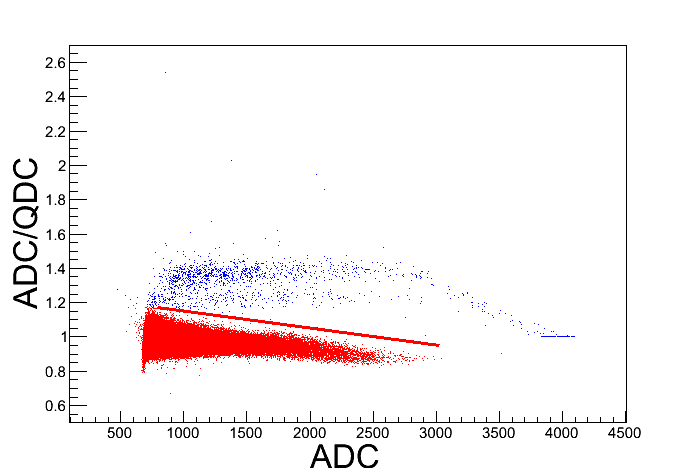}
  \caption{\label{fig:VetoADCQDCDiscrUnd}}
  \end{subfigure}
  \caption{Muons (blue) and fast events (red) discrimination at surface (a) and underground (b).\label{fig:VetoADCQDC}}
\end{figure}

The muon response of the plastic scintillators and the discrimination capabilities of the aforementioned front-end were studied by testing two of them at surface, at the LSC external building. Again, a simple set-up with two stacked plastic scintillators (as in Figure~\ref{fig:PlasticTestBench}) and the front-end was mounted. The muons were identified by selecting events with signal in both detectors. This population can be seen in green in Figure~\ref{fig:VetoADCQDCCoinc} showing the ratio between ADC and QDC values versus the ADC.
\paragraph{ }
The non-coincident events (in black in this picture) are the fast population (at the bottom of the figure) and non-coincident muons (intermediate population). This effect is due to a peculiarity of the Mesytec MSCF-16 trigger: it triggers 10 ns later in case of coincident event resulting in a lower QDC value and the consequent higher ADC/QDC ratio. This effect was confirmed by taking data with different overlapping horizontal plastic surface and checking the growing of the intermediate population and the decrease of the top population and by digitizing the digital trigger signal and calculating the distance in both cases. The impact of this effect in a non-overlapping set-up like the final ANAIS muon detector system is low due to the low probability of a random coincidence of two fast events. In any case, once located the muon population, the rest of the test is done in non-overlapping (non-stacked) position. The scatter plot in such a non-stacked position can be observed in Figure~\ref{fig:VetoADCQDCDiscr} showing a possible discrimination strategy.
\paragraph{ }
The same configuration of two stacked plastic scintillators was mounted underground. The results can be seen in Figure~\ref{fig:VetoADCQDCDiscrUnd} showing a very suppressed muon population (blue) as opposed to the dominant fast population (red). These plots show the possibility of a discrimination strategy using these values. 
\paragraph{}
Table~\ref{tab:FastRate} shows the similar rate of the fast population in surface and underground conditions, pointing to an internal contamination in the scintillator, the photomultiplier or the optical coupling as origin of such a population. The small differences in the population rate can be attributed to the slight differences in PMT operating voltage due the use of an analog power supply.
\begin{table}[h!]
\centering
\begin{tabular}{ccc}
\toprule
Plastic number & Surface (Hz) & Underground (Hz)  \\
\hline
\#3 &  1.40$\pm$0.05 & 1.31 $\pm$ 0.01 \\
\#4 &  1.55$\pm$0.05 & 1.26 $\pm$ 0.01 \\
\toprule
\end{tabular}
\caption{Fast population rate.\label{tab:FastRate}}
\end{table}
\paragraph{ }
Two different criteria can be applied for muon selection depending on their use for muon tagging or for muon flux determination. The muon selection for flux estimation strategy should minimize the false positive acceptance. This can be achieved using a conservative cut and estimating their efficiency. In addition to this cut, the efficiency of the plastic scintillator along their surface must be studied and quantified. This measurement will be performed in the near future and for this reason the total muon flux is not estimated in this work. Anyway, the results without applying any efficiency correction are similar to previous measurements~\cite{bettini2012talk} as we will see in Table~\ref{tab:A37MuonFlux}.
\paragraph{}
The muon tagging event selection must ensure the identification of all possible events with an accurate timing information. This tagging is used to identify and remove NaI events from the dark matter analysis. Having this fact in mind, some false positive events can be accepted in order to minimize the rejection of real muons. This scheme was used to select muons in the rest of this chapter.
\section{ANAIS-37 plastic scintillators set-up}
\begin{figure}[h!]
	\begin{center}
                \includegraphics[width=.75\textwidth]{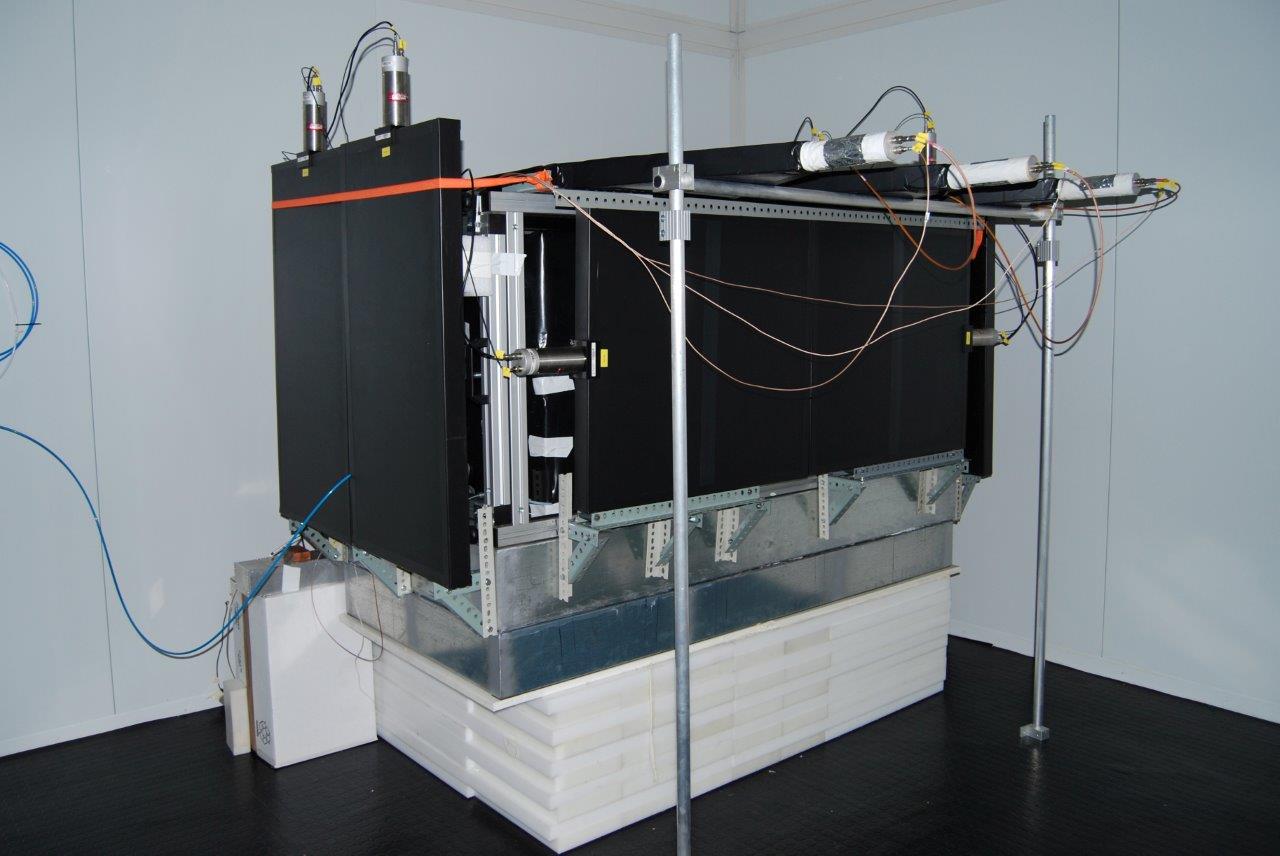}
		\caption[ANAIS-37 plastic scintillators set-up]{ANAIS-37 plastic scintillators set-up.\label{fig:VetosA37}}
	\end{center}
\end{figure}

\begin{figure}[h!]
	\begin{center}
        	\includegraphics[width=.9\textwidth]{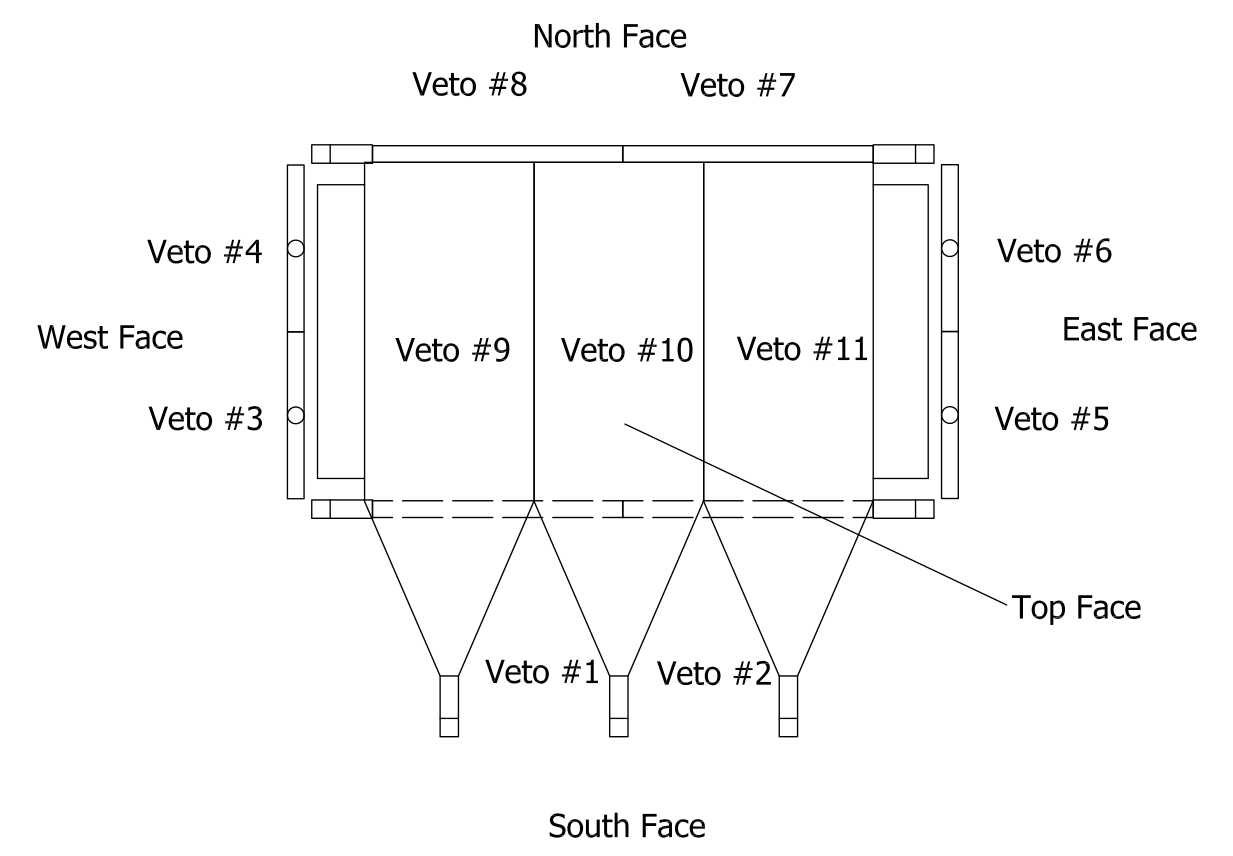}
		\caption[ANAIS-37 plastic scintillators faces]{ANAIS-37 plastic scintillators numbering and face grouping\label{fig:VetosA37Design} (top view).}
\end{center}
\end{figure}

A preliminary set-up was mounted with eleven plastic scintillators covering the shielding of ANAIS37 set-up (see Section~\ref{sec:ANAIS37}). Eight Scionix scintillators were placed at the lateral faces, two of them in each face and three homemade scintillators were placed at the top of the shielding as it can be seen in Figures~\ref{fig:VetosA37}. The face grouping and the veto numbering can be seen in Figure~\ref{fig:VetosA37Design}.

\begin{figure}[h!]
  \begin{subfigure}[b]{1\textwidth}
  \includegraphics[width=\textwidth]{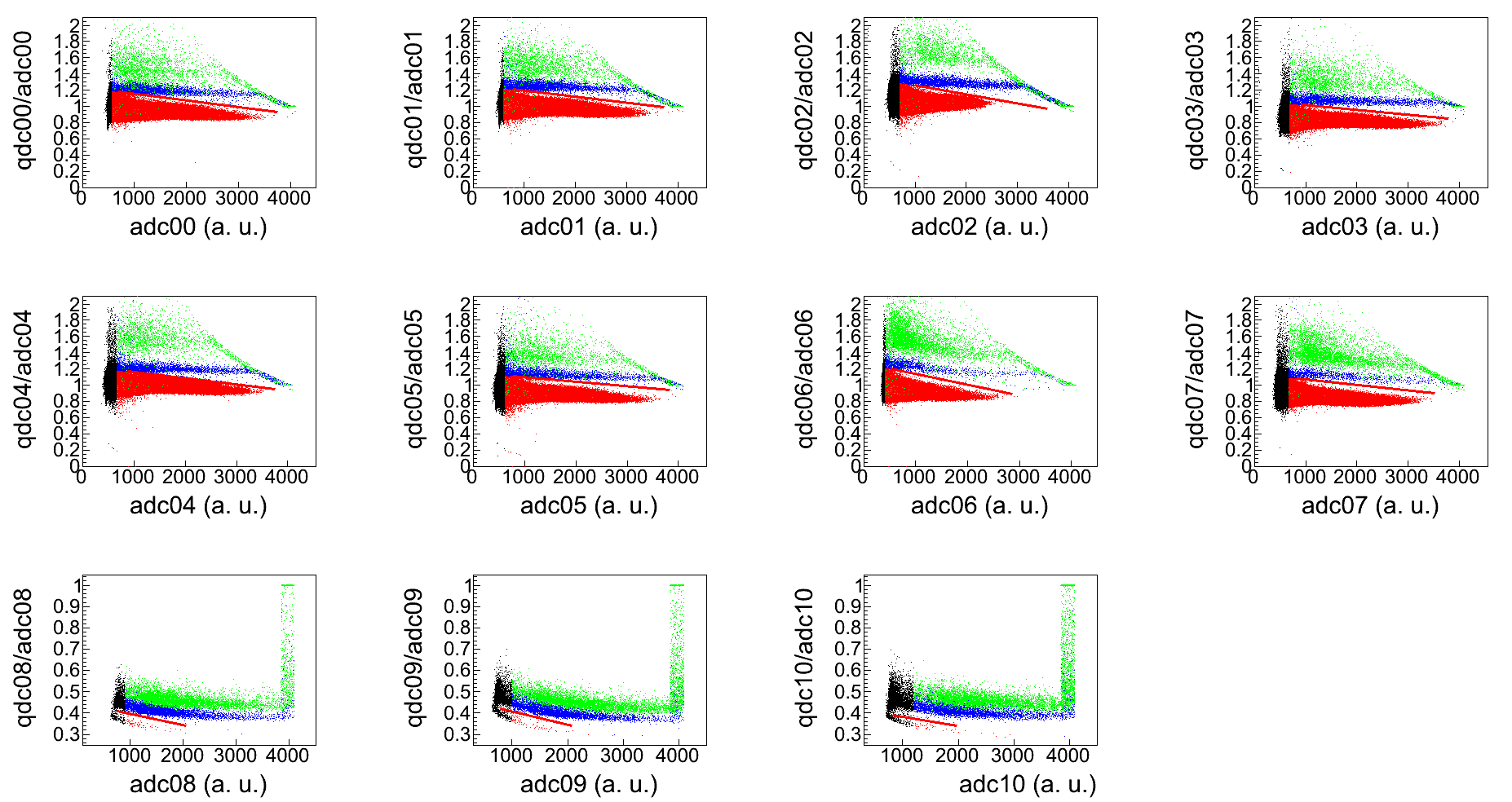}
  \caption{\label{fig:MuonDiscr11}}
  \end{subfigure}
 
  \begin{subfigure}[b]{1\textwidth}
  \includegraphics[width=\textwidth]{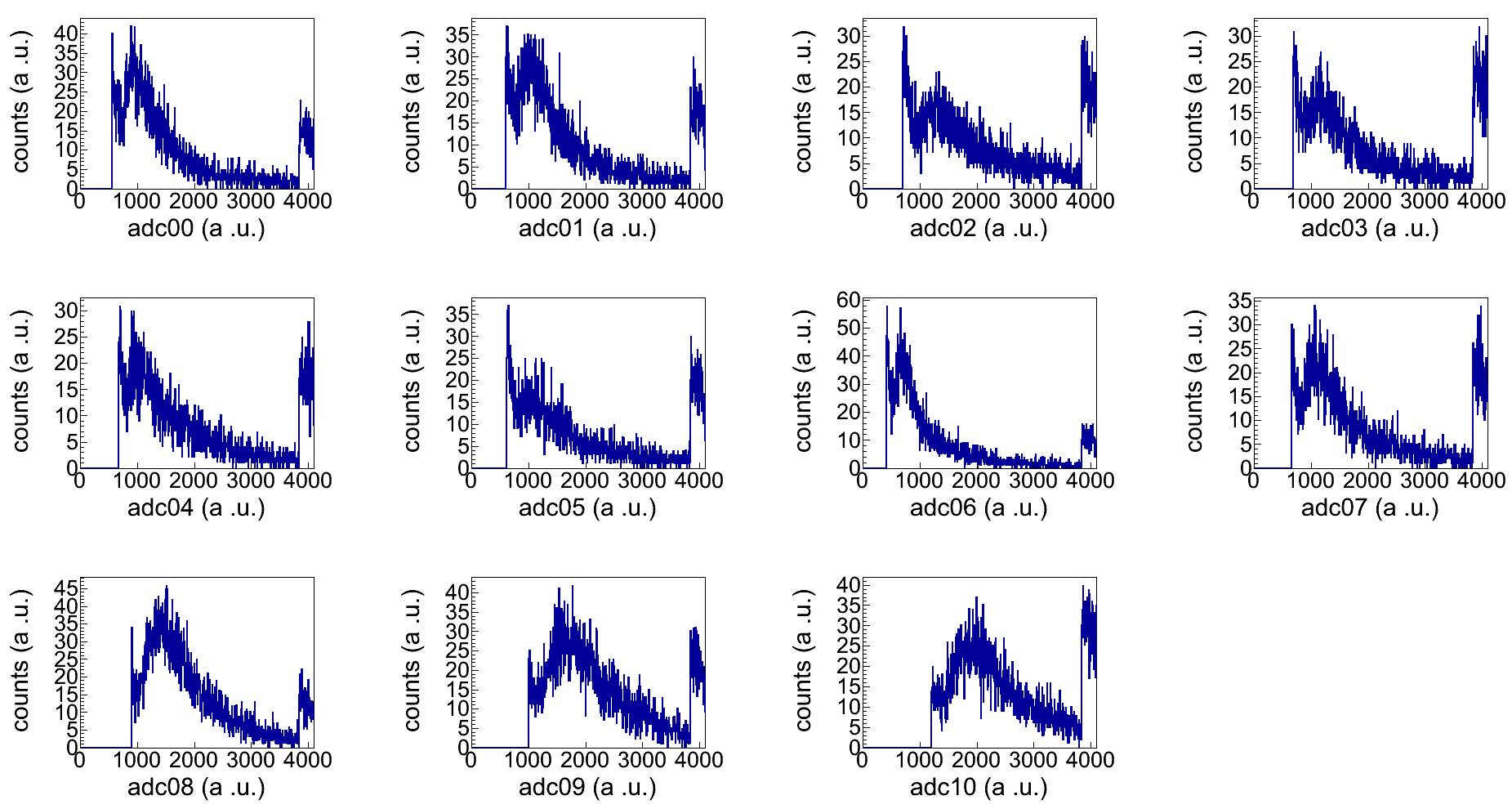}
  \caption{\label{fig:MuonSpectra11}}
  \end{subfigure}
  \caption[Muon discrimination and spectra]{\label{fig:MuonDiscrAndSpectra}Muon discrimination (a) and spectra (b).}
\end{figure}

\paragraph{ }
A preliminary result of a month of data taking can be seen in Figure~\ref{fig:MuonDiscrAndSpectra}. The discrimination and muon selection can be seen in Figure~\ref{fig:MuonDiscr11} and the resulting muon spectra is shown in Figure~\ref{fig:MuonSpectra11}. It can be observed the expected muon peak in every detector.

\paragraph{ }
Table~\ref{tab:A37MuonFlux} shows the preliminary results of the detection rate in every detector. It can be noted that the top detectors have more flux than vertical ones, in agreement with the expected angular distribution~\cite{stockel1969study}. Additionally, the south and north faces have more counts than east and west, suggesting a non-homogeneous flux with a preferred direction south-north, compatible with the Tobazo mountain profile. This preferred direction could be deduced with the multiplicity information (see Section~\ref{sec:MuMulti}). The total muon flux could be deduced using the previous rates, estimating the muons below the threshold and correcting with the efficiency of the plastic scintillators. This efficiency has not been yet determined as mentioned earlier and for this reason the data are given in measured muon per square meter and second.
\begin{table}[h!]
		\begin{center}
			\begin{tabular}{cccc}
				\toprule
				Face& Plastic&\begin{tabular}{@{}c@{}}Measured rate\\($\mu \times m^{-2}\times s^{-1}$)$\times 10^{-3}$\end{tabular}&\begin{tabular}{@{}c@{}}Measured rate\\($\mu \times m^{-2} \times s^{-1}$)$\times 10^{-3}$\end{tabular}\\
				\toprule
				\multirow{2}{*}{South} &\#1&$5.27\pm 0.10$&\multirow{2}{*}{$5.43\pm 0.07$}\\
				&\#2&$5.53\pm0.10$&\\
				\hline	
				\multirow{2}{*}{West} &\#3&$4.88\pm 0.10$&\multirow{2}{*}{$4.75\pm 0.07$}\\
				&\#4&$4.61\pm 0.10$&\\
				\hline	
				\multirow{2}{*}{East} &\#5&$4.53\pm 0.10$&\multirow{2}{*}{$4.54\pm 0.07$}\\
				&\#6&$4.54\pm 0.10$&\\
				\hline	
				\multirow{2}{*}{North} &\#7&$5.05\pm 0.10$&\multirow{2}{*}{$4.91\pm 0.07$}\\
				&\#8&$4.77\pm 0.10$&\\
				\hline	
				\multirow{3}{*}{Top} &\#9&$7.23\pm0.12$&\multirow{3}{*}{$7.36\pm 0.07$}\\
				&\#10&$7.54\pm 0.12$&\\
				&\#11&$7.32\pm 0.12$&\\
				\toprule

			\end{tabular}
			\caption[Muon detection rate]{Muon detection rate.} 
			\label{tab:A37MuonFlux}
		\end{center}
\end{table}
\subsection{Temporal behavior}\label{sec:MuMod}
\begin{figure}[h!]
     \begin{center}
                \includegraphics[width=0.6\textwidth]{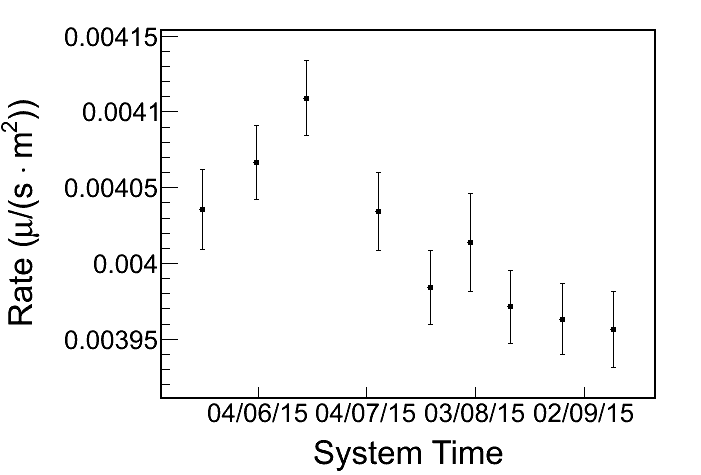}
	\caption[Rate vs. time]{Muon detection rate over time.}\label{fig:MuA37RatevsTime}
\end{center}
\end{figure}

The seasonal modulation of the muon flux at underground is a well measured effect~\cite{bellini2012cosmic,ambrosio1997seasonal,agafonova2011analysis,desiati2011seasonal, adamson2010observation} due to the seasonal variation of the temperature of the upper atmosphere. This temperature variation affects to the atmospheric density, which modifies the fraction of mesons decaying to muons energetic enough to reach underground detectors~\cite{barrett1952interpretation}.
\paragraph{ }
The temporal behavior of the muon detection rate can be seen in Figure~\ref{fig:MuA37RatevsTime}. It shows the counts per surface and time along several runs suggesting a modulation with a maximum in summer, at the end of June, compatible with those observed at the LNGS~\cite{bellini2012cosmic,ambrosio1997seasonal,agafonova2011analysis}. However, a complete seasonal cycle is needed to confirm and fit such a modulation.
\subsection{Multiplicity and face coincidences}\label{sec:MuMulti}
\begin{figure}[h!]
     \begin{center}
                \includegraphics[width=0.6\textwidth]{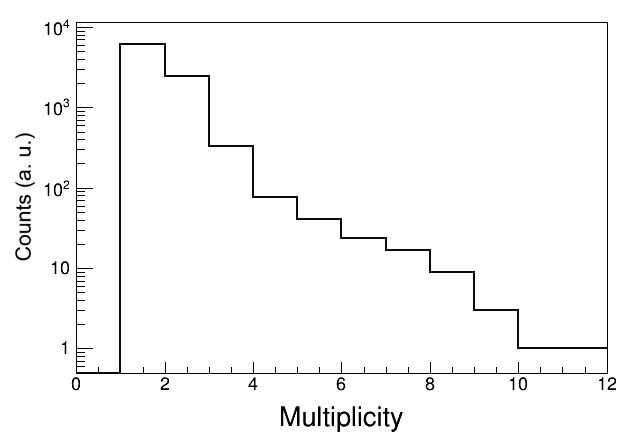}
	\caption[Plastic trigger multiplicity]{Plastic trigger multiplicity.}\label{fig:VetoMultiplicity}
\end{center}
\end{figure}

The study of the coincident events among plastic scintillators can give more information about the muon flux. Figure~\ref{fig:VetoMultiplicity} shows the total multiplicity of the events in a run, featuring a quick decrease of events with multiplicity but it is worth to note that there are events hitting almost all plastic scintillators which can be attributed to muon showers~\cite{davis1971phenomenology}. In any case, most of the muons leave signal in one plastic only, making impossible the use of the coincidence as the only criterion to identify muons and reject the fast population, in agreement with the earlier assumption.
\paragraph{ }

Additionally to the trigger multiplicity, the coincidence among different faces gives hints about the muon flux coming to the ANAIS experiment. A count of events hitting two faces was carried out and it can be seen in Table~\ref{tab:MuFaces}. The table is grouped in faces with equal geometry in order to compare the rates properly. The most prominent feature is the asymmetry North-South in coincidence with top face. It shows the muon flux being non-homogeneous in azimuthal angle, with a preferred direction South-North, in agreement with the Tobazo mountain profile and with the individual plastic rate seen at the beginning of this section. The North-South and East-West coincidences cannot be directly compared because of the rectangular geometry of the shielding.
\begin{table}[h!]
\centering
\begin{tabular}{ccc}
\toprule
Faces & Coincident rate (mHz)  \\
\hline
 Top-North&   $3.26\pm0.07$\\
 Top-South&   $1.05\pm0.04$\\
\hline
 Top-East&   $0.87\pm0.04$\\
 Top-West&   $0.95\pm0.04$\\
\hline
 North-South&   $0.37\pm0.02$\\
\hline
 East-West&   $0.14\pm0.01$\\
\hline
 North-East&   $0.48\pm0.03$\\
 North-West&   $0.34\pm0.02$\\
 South-East&   $0.45\pm0.03$\\
 South-West&   $0.47\pm0.03$\\
\toprule
\end{tabular}
\caption{Faces coincident rate.\label{tab:MuFaces}}
\end{table}

\subsection{Muon related NaI(Tl) events}\label{sec:MuNaIEvt}\label{sec:DAQsSynch}
\begin{figure}[h!]
  \begin{center}
  \includegraphics[width=0.6\textwidth]{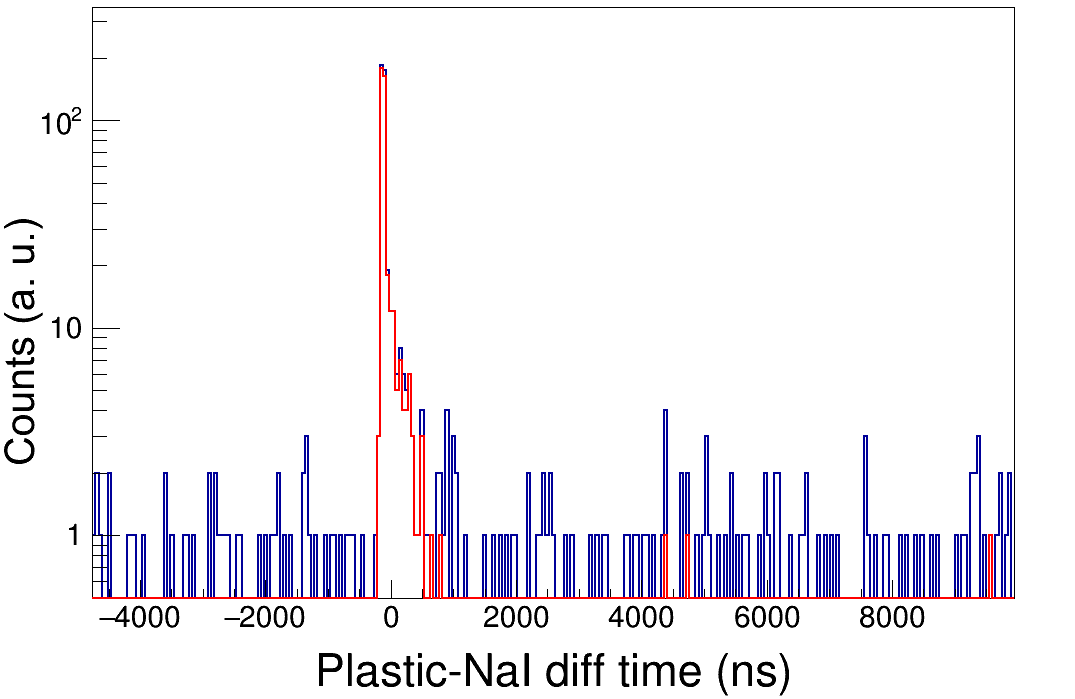}
  \end{center}
  \caption[Real time DAQ Correlation]{\label{fig:MuonPlasticNaI}Time difference between NaI and plastic events.}
\end{figure}

The synchronization between the NaI DAQ (see Chapter~\ref{sec:FrontEnd}) and the plastic DAQ (see Section~\ref{sec:VetoFrontEnd}) was checked by finding events with a real time difference less than 10~$\mu s$. This test should reveal the muon interaction in the plastic scintillators previous to a second interaction in the NaI crystals. This effect can be seen in Figure~\ref{fig:MuonPlasticNaI}. It shows all events in the considered time window (in blue) and the muon events (in red). It can be observed a high accumulation of events in two tick bins (almost simultaneous events) and a tail of delayed events in the crystal. The time correlation described here is used to examine the effects of the muon interactions in the NaI(Tl) data as it can be seen hereafter.
\paragraph{ }
The muon-related events in the NaI crystal are direct interactions with very energetic depositions and secondary neutrons among others. Such secondary neutrons can be very harmful because it can be confused with expected WIMP interaction as seen at the beginning of this chapter.
\paragraph{ }
The muon interaction tagged in both plastic scintillators and NaI crystals were studied. Approximately a 7\% of the plastic events have a subsequent crystal event within 1 $\mu$s interval, half of them being very high energy events with simultaneous events in both data acquisition systems corresponding to direct muon interaction in the crystal. Table~\ref{tab:MuVetPlast} shows the coincident events in 100~ns time window with higher energy than main alpha contribution (energy greater than 5.3 MeV due to $^{210}$Po). This condition was established in order to discard low energy coincidences. 
\begin{table}[h!]
		\begin{center}
			\begin{tabular}{cc}
				\toprule
				Detector& Coincident rate (events/day)\\
				\toprule
				D0 & $26.7 \pm 0.7$\\
				D1 & $26.3 \pm 0.7$\\
				D2 & $26.2 \pm 0.7$\\
				\toprule
			\end{tabular}
			\caption[]{Plastic and NaI coincident events at high energy.} 
			\label{tab:MuVetPlast}
		\end{center}
\end{table}
\paragraph{ }
A crosscheck for these numbers can be done by calculating the rate expected by the measured vertical flux, the main component of the total flux, in the NaI surface. The measured muon rate in the top face is $7\times 10^{-3}~\mu/(m^2\cdot s)$ and the detector section is 0.036~$m^2$ giving a contribution due to the vertical flux of the order of 22.9 $\mu/day$. The difference can be attributed to the non-vertical component of the total flux, which is relevant as seen with the coincident rate in the previous sections.
\paragraph{ }
\begin{figure}[h!]
  \begin{center}
  \includegraphics[width=0.6\textwidth]{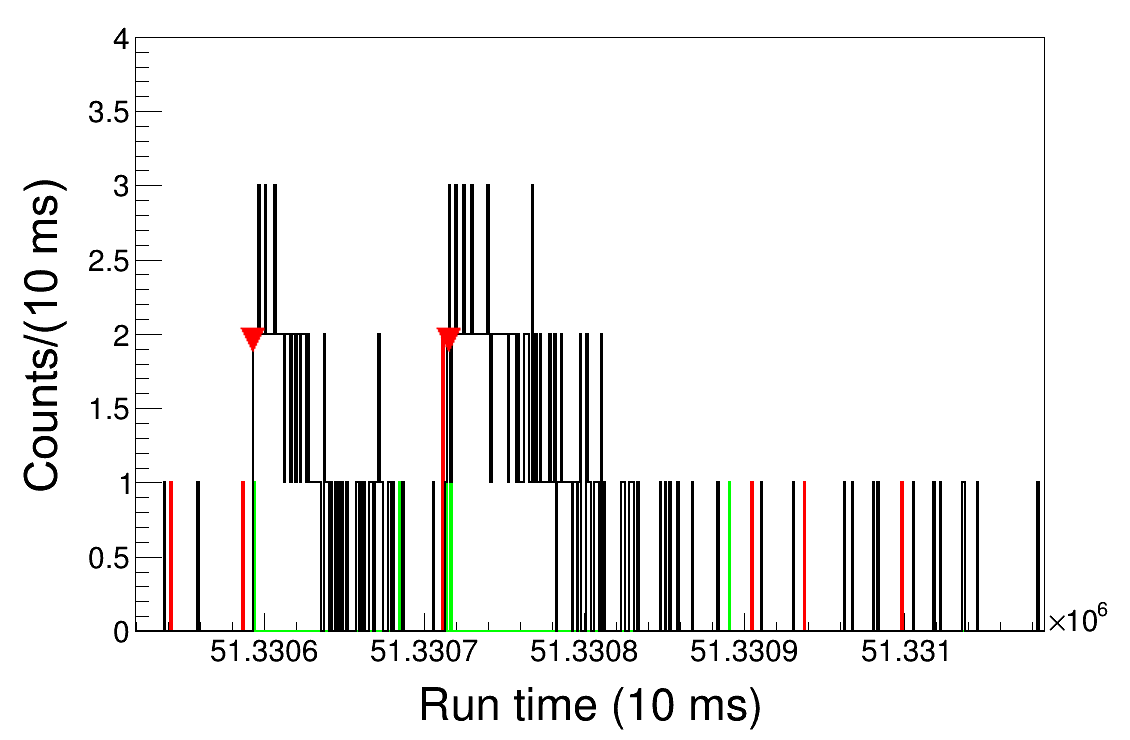}
  \caption[Muon related events]{\label{fig:MuonNaiRate}NaI event rate correlated with plastic events (red triangles).}
  \end{center}
\end{figure}

\begin{figure}[h!]
  \includegraphics[width=\textwidth]{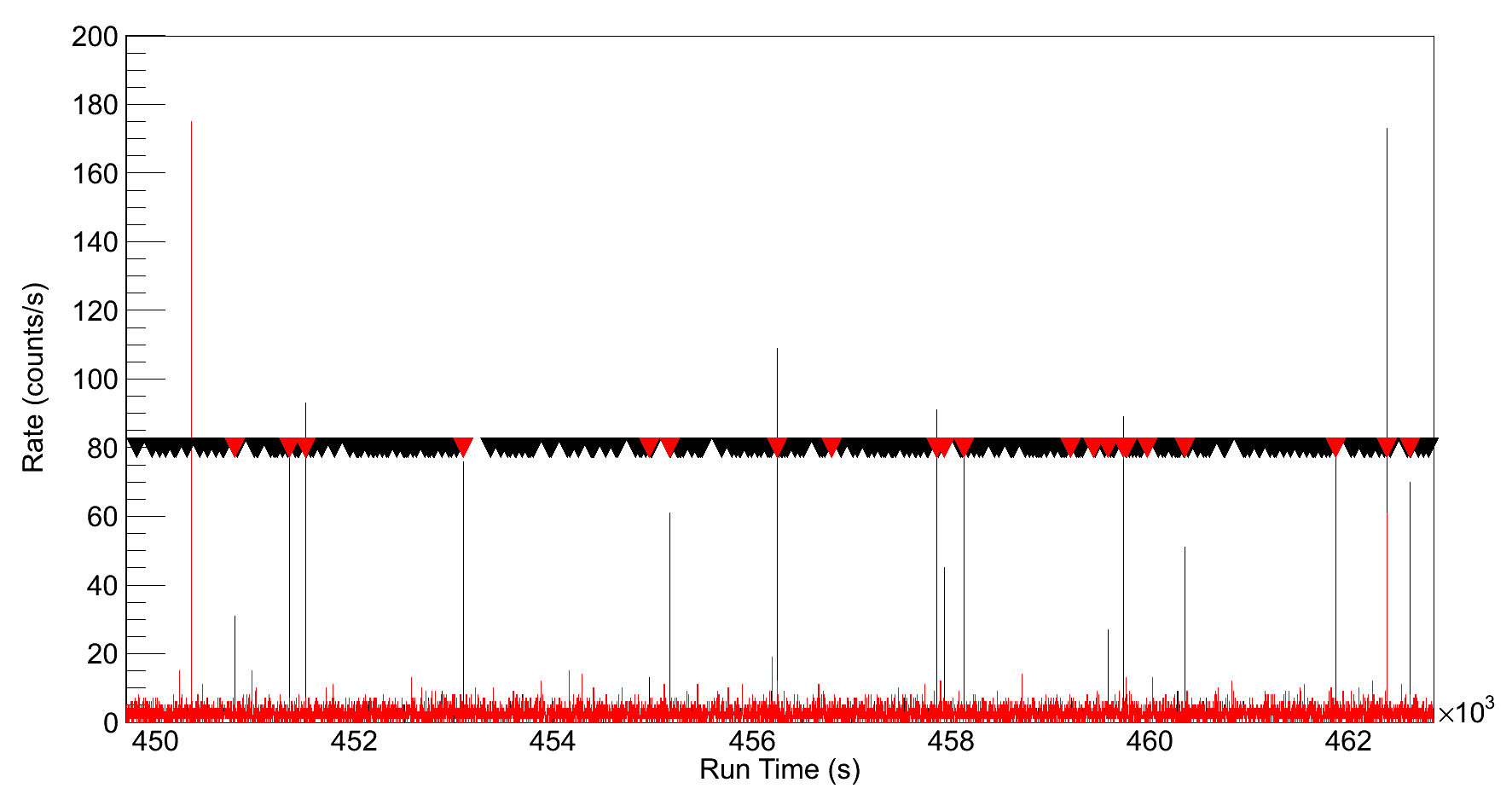}
  \caption[High rate and muon correlation]{\label{fig:RateMuAll} Event rate of all (black) and excluding plastic related events.}
\end{figure}

These events produce a crystal activation or afterglow as seen in Figure~\ref{fig:MuonNaiRate}. This figure shows a very infrequent succession of two muon events in less than two seconds. The NaI events are grouped in bins of 10~ms with three detector events (D0 red, D1 green, D2 black). The two events hit in the same detector (D2) with a high energy deposition and an afterglow of the order of tenths of seconds. The plastic tagged events can be seen as red triangles in the picture, at the beginning of the high rate interval.
\paragraph{ }
All these facts can be seen in Figure~\ref{fig:RateMuAll}. It shows the tagged muons with triangles over the events binned in one second bins. The red triangles are muons tagged with less than one $\mu$s interval with one NaI event. The black triangles are the remaining plastic events. The black line marks all NaI events and the red line discards all events in a 1 second window after tagged muon.
\paragraph{ }

\subsubsection{High rate episodes and very high energy events}
\begin{figure}[h!]
  \begin{center}
  \includegraphics[width=0.9\textwidth]{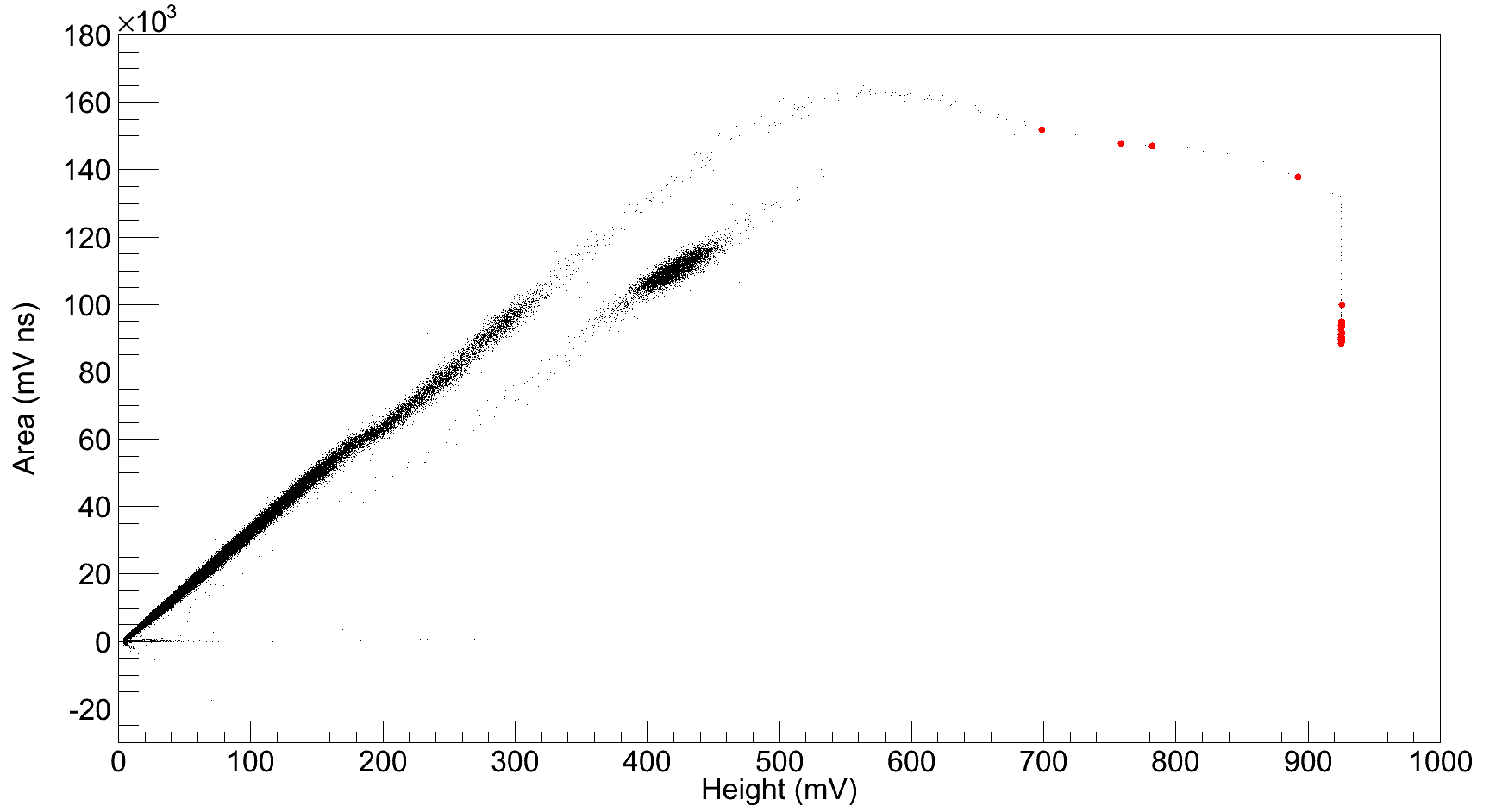}
  \caption[High rate without plastic event]{\label{fig:RateMuNoPlasticIdent} Area (charge) versus height for a background run (black) and highest energy events in one second window from a high rate episode without plastic scintillators tag (red).}
  \end{center}
\end{figure}

The existence of a very clear correlation between high rate episodes and plastic events was the reason to study all the high rate episodes. Most of the high rate episodes have a related plastic scintillator event ($\sim$ 86\%) but some are not tagged as it can be seen in Figure~\ref{fig:RateMuAll}. These episodes were identified and studied in order to establish their origin. Once identified, a search for high energy events up to one second before was performed. These events can be seen in red in Figure~\ref{fig:RateMuNoPlasticIdent}. This figure plots their area versus the height in the high energy range digitization. This scatter is used to discriminate $\alpha$ particles (the population in the middle) from $\gamma/\mu$ particles. The plot also shows the PMT saturation effect with very high energetic events. This effect can give less collected charge (or area) for a muon event than for an $\alpha$ interaction. The pulse shape analysis allowed to identify these high energy events as muons. 

\begin{figure}[h!]
  \includegraphics[width=\textwidth]{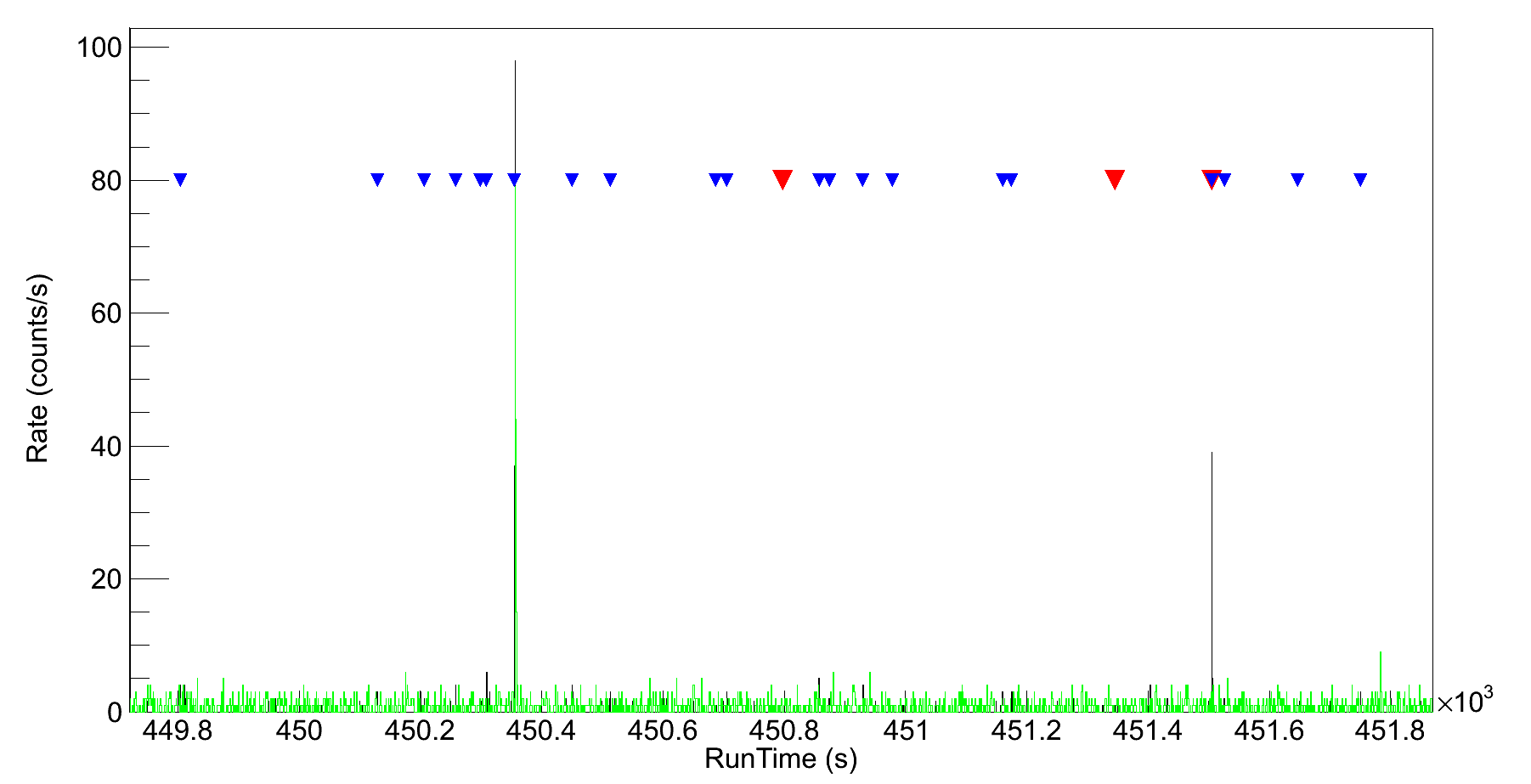}
  \caption[High rate and high energy events correlation]{\label{fig:RateHEMu}Event rate of all events (black) and rate excluding events after high energy or muon events (green)(blue: high energy events, red: plastic tagged event).}
\end{figure}

\paragraph{ }
A filtering strategy cutting events near a muon tagged in plastic vetoes and very high energy events in the NaI crystals were used with one second time windows as it can be seen in Figure~\ref{fig:RateHEMu} removing the high rate intervals (or afterglows) illustrating the effect of the slow NaI(Tl) constants already studied by the ANAIS team~\cite{cuesta2013slow}. The figure shows the rate of detector D2 and the identification of very high energy events (blue triangles) and plastic events (red triangles). It is worth to note that all high rate episodes can be explained with very high energy interactions, being discarded some other possible origins such as electric noise or other instabilities.

\subsubsection{Muons without subsequent high rate episode}
\begin{figure}[h!]
  \begin{subfigure}[b]{\textwidth}
  \begin{center}
  \includegraphics[width=.8\textwidth]{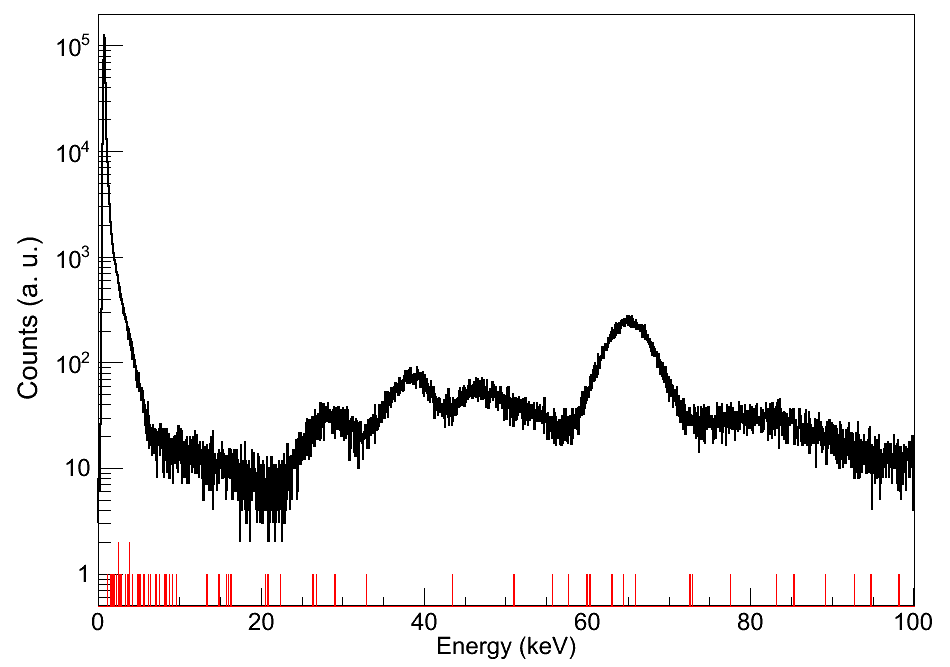}
  \caption{\label{fig:MuonTaggedEventsArea}}
  \end{center}
  \end{subfigure}
  
  \begin{subfigure}[b]{\textwidth}
  \begin{center}
  \includegraphics[width=0.8\textwidth]{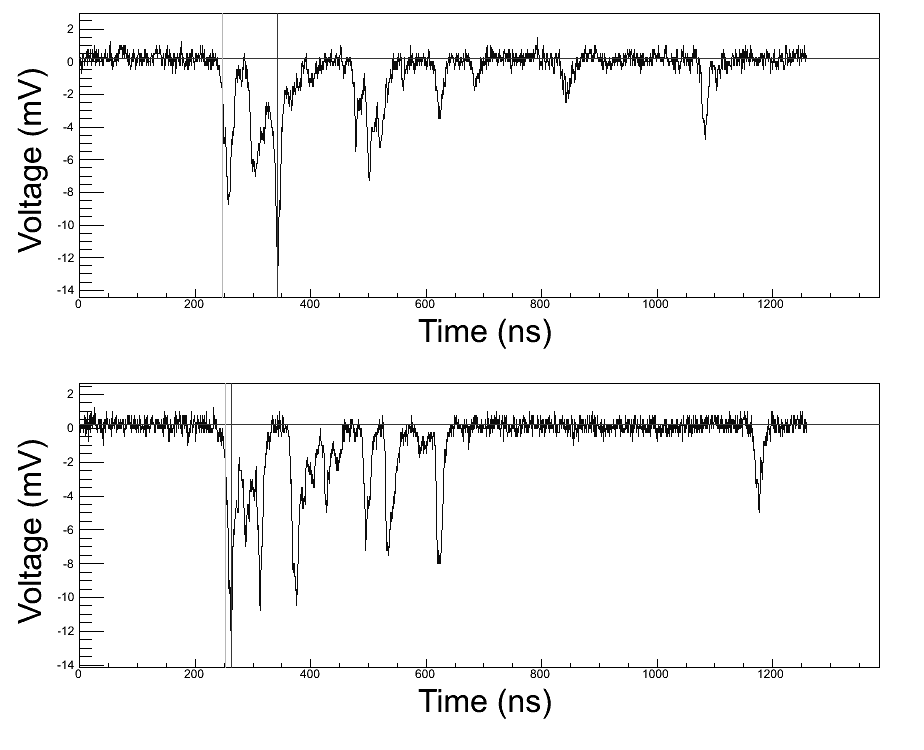}
  \caption{\label{fig:MuNoRateEvent}}
  \end{center}
  \end{subfigure}

  \caption[Muon related low energy events]{Muon related low energy events: \label{fig:MuonNorate} energy distribution in red with all background events in black (a) and a typical muon-related event (b).}
\end{figure}

In addition to very high energy events, low energy event population is expected as a result of muon interactions such as secondary neutrons described earlier.
\paragraph{}
Once the high rate episodes were identified, the information can also be used to select NaI(Tl) events related to plastic scintillator events that does not produce afterglow. The spectra of the events with one millisecond time windows between plastic and crystal time event can be observed in Figure~\ref{fig:MuonTaggedEventsArea} in red compared with the normal raw background in black.
\paragraph{ }
The more notable feature is a very low energy event population, lower than 20 keV. An event of such a population can be seen in Figure~\ref{fig:MuNoRateEvent}, showing a scintillation signal that would pass all quality cuts. All these events should be rejected for a dark matter analysis because of their possible cosmic radiation related origin. In addition to their rejection, the tagging and offline removing strategy allows to study the behavior of the muon-related events.
\paragraph{}
A preliminary study was done in order to quantify the influence of such a population. The events in the 1-6 keV energy window coincident with a tagged muon within a 1 ms window are $0.55\pm0.05$ counts/(kg day). This population decreases to $0.07\pm0.02$ counts/(kg day) with quality cuts (only one detector triggering, number of peaks, NaI(Tl) scintillation shape and symmetry between PMT signals, see Section~\ref{sec:EventSelection} for a description of ANAIS low energy event section). The random coincident population of events passing all quality cuts in 1 ms windows with a tagged muon is of the order of $8\times10^{-4}$ counts/(kg day). This value is an upper bound calculated by using the frequency of all tagged muons ($\sim 2.5\times 10^{-2}Hz$). For this reason, a clear correlation between these low energy events and muon can be established, suggesting a contribution of muon induced neutrons detected via nuclear recoil in the crystal. 
\paragraph{}
The background events in that region passing the same cuts are $\sim18$ count/(kg day) being the muon-related events a factor 0.004, so the influence in the total rate at low energy is negligible. Anyway, the low energy cuts are being improved and this factor should be recalculated with a better noise rejection. In any case, the study of these events will give a very valuable information of nuclear recoil events such as shape and possible modulation. It is worth to mention again that the most considered scenario for WIMP interaction is a nuclear recoil, and having a population of this kind of interaction allows to characterize the response of the detectors. For this purpose, a statistically significant amount of events is needed.





\chapter{Data acquisition software}\label{sec:DAQSW}
The requirements, implementation and tests of the acquisition software for the ANAIS experiment are covered in this chapter. First of all the requirements of the software are listed and discussed. Next, all the software used in the acquisition system is described including relevant design and implementation details.
\paragraph{}
The acquisition software has been developed expressly for this work except the classes that defines the interaction with the VME boards which have been completed with new modules (14-bit MATACQ, V830) and updated for new functionality and performance reasons.
\section{DAQ software requirements}\label{sec:DAQSWRequirements}
The most important software requirements are given by the characteristics of the experiment (see Section~\ref{sec:ExperimentalRequirements}) and it imposes some specific restrictions. The software requirements for the ANAIS experiment include the data acquisition and storing, a detector calibration mode, the dead time minimization, scalability in the number of signals, stability, flexibility, monitoring and robust storage. There are some other requirements given by the ease of use or for help the diagnose and debugging auch as easy configuration, debuggable output format and task automation in data synchronization and analysis. All these features are described and discussed in this section.
\subsubsection{Data acquisition and storing}The most important requirement is to acquire and store data. In addition, this software must detect the trigger and unlock the hardware in order to be available for a new trigger once the data have been acquired as described in Section~\ref{sec:DigitalFrontEnd}. This is needed because of the lack of more than one buffer in the digitizer. The software must also ensure that all data are correct, checking the data availability and performing the correct negotiation with every module. An error in this critical step can give subtle effects such as incorrect data or uncorrelated data in an event.
\subsubsection{Detector calibration mode}The software must allow a fast calibration of the detectors. In this mode the system must provide the energy spectra from the detectors. The easiest way to achieve this feature is to use only the QDC (see Section~\ref{sec:VMEModules}). In this way the acquisition system is fast and the generated data are minimized. As a plus, the system could be configured to do the spectra and to digitize the signals in order to test Pulse Shape Analysis algorithms with calibration source.
\subsubsection{Dead time minimization}The software can introduce dead time because of the lack of double buffers seen earlier. This time has to be minimized in order to maximize the live time of the system, a requirement of the global system (see Section~\ref{sec:ExperimentalRequirements}).
\subsubsection{Scalability}The whole system must be scalable because of the expected growing of the number of crystals in the future. The software must consider this growing as a key design guideline: performance (live time, DAQ frequency, output data flow) of the whole system must be appropriate.
\subsubsection{Stability}The access to the experiment is difficult because of the underground placement. The DAQ design must be done assuming that the resultant system is going to be unattended. This assumption adds a relevant requirement to the software: it must be very reliable and stable. This is a very critical feature taking into account the need of maximization of the live time of the complete system (see Section~\ref{sec:ExperimentalRequirements}).
\subsubsection{Flexibility}The DAQ system of the ANAIS experiment must allow adding and removing information from the written data. It is a very common need to add some counters of signals in the data on a temporary basis for debugging hardware or for physical reasons such as exploring energy regions outside the usual one. So the ideal system must allow to add data and must allow mixing data with and without this temporal data. Additionally, the possibility of conditionally download data from the boards and/or store events based on the first acquired values would be useful in order to limit useless transmission dead time and disk space and allow less stringent hardware trigger.
\subsubsection{Monitoring}The DAQ itself must be monitored to find and notify any hardware or software incidence. The monitoring system must ensure that all the detectors provide signals and the acquisition rate is normal. The monitoring system must notify those incidences via Internet as soon as possible. It is also desirable that the DAQ system sends reports for an early incident detection.
\subsubsection{Robust storage}The storage system must be robust and redundant in order to avoid data loss and allow fast transfer to the University of Zaragoza servers for the analysis and backup processes.
\subsubsection{Debuggable output format}The debugging process of the DAQ system is easier if the output format is easy to view, query and analyze.
\subsubsection{Task automation}There are tasks that are needed in a periodic and repetitive way: data synchronization with the University of Zaragoza servers and execution of data analysis software. These processes can be automated and the results can generate reports which can detect any anomaly or malfunction.
\section{DAQ software design}
The design of the DAQ software is described in the present section. First, all the software components are presented. Next the design of the DAQ software is discussed. 
\subsection{Software Components}\label{sec:SWComponents}
\begin{figure}[h]
  \begin{center}
    \includegraphics[width=.85\textwidth]{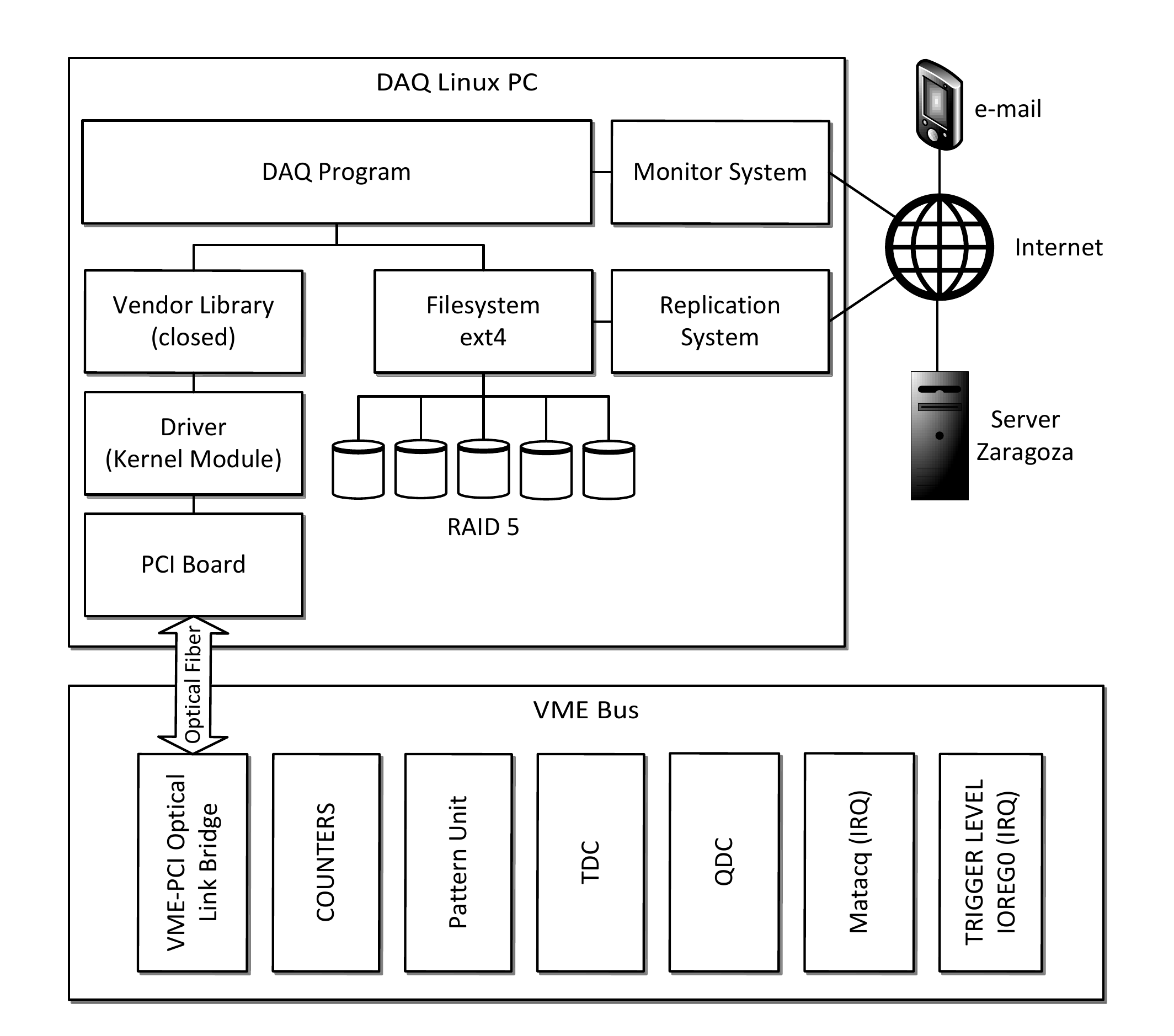}
    \caption[DAQ Software components]{\label{fig:SWStack}DAQ Software components and their interactions.}
  \end{center}
\end{figure}

In this section the components of the whole DAQ software system are covered. In Figure~\ref{fig:SWStack} these components and its interactions are graphically described. 

\subsubsection{DAQ program} 
\begin{figure}[t!]
  \begin{center}
    \includegraphics[width=1\textwidth]{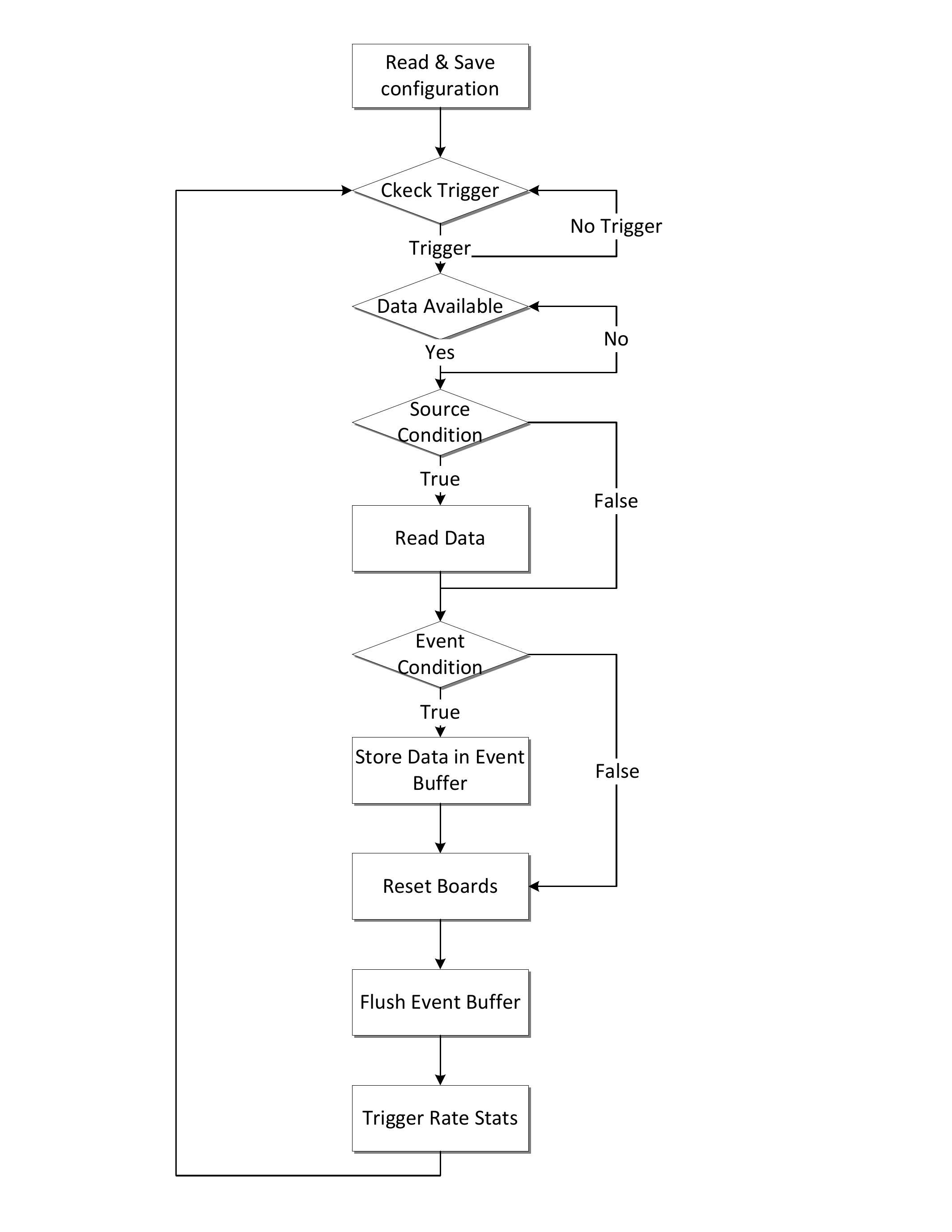}
    \caption[DAQ Program flow chart]{\label{fig:DAQFlowChart}DAQ Program flow chart.}
  \end{center}
\end{figure}

This program acquires and stores the data from the VME bus. The most important activities are:
\begin{itemize}
\item Read and interpret the configuration file. This file describes the boards used by the DAQ system and other relevant parameters. In this step all parameters are saved in the data file so the data acquired are saved together with the actual hardware configuration.
\item Check for a new event. The system is able to do it with polling or waiting for an IRQ. In Section~\ref{sec:DeadTimeMeas}, both options are reviewed, discussed and compared.
\item Read timestamps from real time, live time and dead time counters (see Section~\ref{sec:TimeScalers}) as well as the system time timestamp in order to properly tag the events and allow time with other DAQ system (such as NaI and Plastic scintillators correlation seen in Section~\ref{sec:DAQsSynch}).
\item Check for the validity and availability of the data in every electronic board.
\item Check, if configured, the logical conditions to store the values.
\item If the associated condition to a value is true, read its value.
\item Prepare boards for a new acquisition.
\item Allow the arrival of new triggers by resetting the I/O Register (see Section~\ref{sec:DigitalFrontEnd}).
\item Store data.
\item Generate statistics about rate with Real Time counters .
\end{itemize}
This activities and their sequence are summarized in Figure~\ref{fig:DAQFlowChart}.

\paragraph{}
The code was written in C++ in order to find the right level of abstraction without suffering a big performance penalty. The design of this piece of software is described in next section. It was written for being executed in a Linux PC.
\subsubsection{Kernel module and vendor library} The vendor of the VME/PCI Bridge (see Section~\ref{sec:VMEModules}) provides two different components to manage the VME bus from a PC. The first component is a kernel module with source code. This is needed by the nature of the drivers in Linux. It has some advantages because it can be easily debugged and instrumented as can be seen in Section~\ref{sec:IRQvspoll}. The second one is a closed library that implements the communication functionality. This library is called by the DAQ program and the library \emph{talks} directly with the driver via \texttt{ioctl} calls.
\subsubsection{Network Time Protocol (NTP)} Personal computers have unreliable clocks. The system has NTP installed in order to mitigate this known irregularities:  jitter and frequency wander. A comparison between hardware real time clock (see Section~\ref{sec:TimeScalers}) and PC clock with NTP can be seen in Section~\ref{sec:RealTimeTest}.

\subsubsection{ext4 and software RAID5}The chosen filesystem is \texttt{ext4}~\cite{EXT4} because it is the default in modern Linux distributions, it is well tested and it provides \emph{journaling}. This feature prevents data loss by ensuring consistence of filesystem metadata: the resulting filesystem is harder to corrupt. Additionally the storage system has a software RAID5 system in order to fulfill the redundancy requirement. RAID5 only allows one broken hard disk, if there is another broken disk the RAID cannot be reconstructed and the data are lost. To prevent this eventuality the system is configured to monitor the disks and RAID health. If there is any incidence the monitor system (\texttt{checkarray} from \texttt{mdadm} package~\cite{website:mdadm}) sends an email in order to solve the incidence as soon as possible.
\subsubsection{Data synchronization}\label{sec:DataSynch}The synchronization system was designed to be invoked on demand or in an automatic way. It uses \texttt{rsync}~\cite{Tridgell96rsync} via SSH~\cite{SSH}.

\subsection{DAQ Program Design}\label{sec:DAQConfig}\label{sec:DAQDesign}
The DAQ program design follows the scalability and performance guidelines seen in the previous section.
\subsubsection{Classes design}
\begin{figure}[h]
  \begin{center}
    \includegraphics[width=1\textwidth]{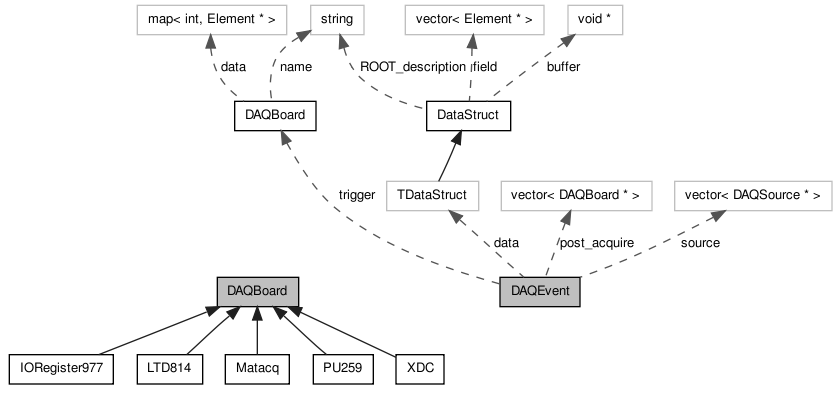}
    \caption[DAQ program classes design ]{\label{fig:classes}DAQ program classes design.}
  \end{center}
\end{figure}

The design of the program is guided by the need of easily configure a system that has to grow in the number of crystals. The easiest way to achieve that is to encapsulate functionality in concepts such as Data Source, Data, Board or Event. This concepts are implemented as C++ classes and Figure~\ref{fig:classes} shows its relationships. 

\paragraph{}
\texttt{DAQEvent} is the nuclear base of the DAQ system. It acquires the data coming for a source (\texttt{DAQSource}), waits for a trigger coming for a board (\texttt{DAQBoard}) and prepares boards (\texttt{DAQBoard} again) for a new acquisition. \texttt{DAQBoard} can implement several features depending on the features of the real board: it can wait for a trigger, it can be prepared (if needed) for a new acquisition and it acquires the data that can be accessed with a \texttt{DAQSource} instance.

\subsubsection{Configuration}\label{sec:DAQConfig} It is easy to instantiate a new board or complete a new event with new data once defined and implemented the concepts above described. But instantiating manually all classes require program modification and it can be error prone and a tedious task. For that reason a configuration system was designed in order to make an easy and legible configuration file. It is based in YAML~\cite{YAML} (YAML Ain’t Another Markup Language) in order to be easy for humans and programs. 

\paragraph{}
The data acquired and stored when an event has triggered can be configured in a straightforward way. An example of event with data, trigger and post acquire actions can be seen in Listing~\ref{lst:yaml}. In this example the trigger is generated by an IRQ generated in the VME bus \texttt{vme1}, the data to store come from some channels of a QDC, TDC and MATACQ boards and for some counters. Additionally the actions to perform before a new acquisition are configured in the \texttt{post\_acquire} clause.

\begin{lstlisting}[label=lst:yaml, caption=Event configuration example.]
- event: Event
  trigger: vme1
  write_mode: asynch
  source: [RealTime, clock, LiveTime, DeadTime]
  source: [qdc1_0, qdc1_1, qdc1_2, qdc1_3, qdc1_4, qdc1_5]
  source: [matacq1_0, matacq1_1]
  source: [tdc1_0, tdc1_1]
  source: [pattern_unit]
  post_acquire: [qdc1, matacq1, tdc1, scalerl, io_reg_0]
\end{lstlisting}

This configurability makes trivial to add and remove data from an event and insert debugging information on it. Additionally, this system is versatile enough to allow to configure two distinct acquisition set-ups such as ANAIS main set-up (see Chapter~\ref{sec:FrontEnd}) and the plastic scintillators muon detector (see Chapter~\ref{sec:Veto}).

\subsubsection{Output Format} The ROOT~\cite{ROOT} format was chosen as output format because of the ease of querying data and the compression feature. ROOT is a data analysis library developed at CERN in the context of particle physics. It allows to plot and analyze data in a very fast way. The DAQ program, in the \texttt{DAQEvent} class, generate dynamically a data structure (\texttt{TDataStruct}) that allows the ROOT library to store the data in its format. So the output data is determined by the configuration and there is no need of additional recompilation or reconfiguration. The compression feature allows to decrease the output data size but it needs more CPU time. This increase in CPU consumption can be excluded from the dead time using asynchronous data storing described in this section.
\paragraph{}
The actual output format and the naming conventions used for the ANAIS experiment are described in Section~\ref{sec:DAQNamingConventions}.
\subsubsection{Conditional data acquisition} This feature was designed taking advantage of ROOT features and with data and dead time minimization in mind. Every data (every \texttt{DAQSource}) can be configured to evaluate a condition. If this condition is true the data is acquired and stored, if it is false a default value is stored without trying to acquire the real data. This has two advantages: it saves acquisition time and it can save disk space. \paragraph{}
This feature can be used to discard data with a condition based on other data. For example the DAQ system can acquire and store only digitization from the signals that has triggered. In Listing~\ref{lst:store_trigger} it can be seen how to configure a source (MATACQ in this case) in order to avoid acquisition if the crystal has not triggered. The condition is that the data of the pattern unit (see Section~\ref{sec:VMEModules}) has the first bit on, so the first crystal has triggered.

\begin{lstlisting}[label=lst:store_trigger, caption=Conditional data acquisition.]
- source: matacq_0
  board: matacq
  condition: "(pattern_unit&1)!=0"
  channel: 0 
\end{lstlisting}
\paragraph{}
The key point of the design of this feature is the use of \texttt{TFormula} (or the more versatile \texttt{RooFormulaVar}). This ROOT class implements formula evaluation at runtime in a very efficient way. So this feature can be used for many purposes. For example it can be used to digitize only a narrow range of energy values. This can be useful to accumulate digitized signals from interesting but scarce ranges in a calibration without the need of huge amount of disk space. With this technique the statistical significance of pulse shape analysis can be increased for some energy ranges. In Listing~\ref{lst:store_qdc} it can be seen an example of such a configuration, in this case the condition is low energy events that triggers.

\begin{lstlisting}[label=lst:store_qdc, caption=Conditional data acquisition with energy condition.]
- source: matacq_0
  board: matacq
  condition: "qdc1_0<400 && ((pattern_unit&1)!=0)"
  channel: 0 
\end{lstlisting}
\subsubsection{Conditional event storing} This feature is implemented in a very similar way than the previous one. It allows to skip valueless data. Every \texttt{DAQEvent} can be configured to evaluate a condition based on the related source values. The event is saved if the condition is true and it is discarded otherwise.
\paragraph{ }
The Listing~\ref{lst:time_store} shows a rather unconventional but simple example. This configuration stores events the fist second of every hour very useful to characterize stability parameters in the long run, such as the temperature measurements seen in Section~\ref{sec:TestTriggerStrat}.
 \begin{lstlisting}[label=lst:time_store, caption=Conditional event storing.]
- event: Event1
  trigger: vme1
  condition: "fmod(clock,3600)<1"
  source: [clock, matacq1_0]
  post_acquire: [matacq1]
 
\end{lstlisting}

\subsubsection{Asynchronous data storing}\label{sec:asynch_store} 
\begin{figure}[h]
  \begin{center}
    \includegraphics[width=1\textwidth]{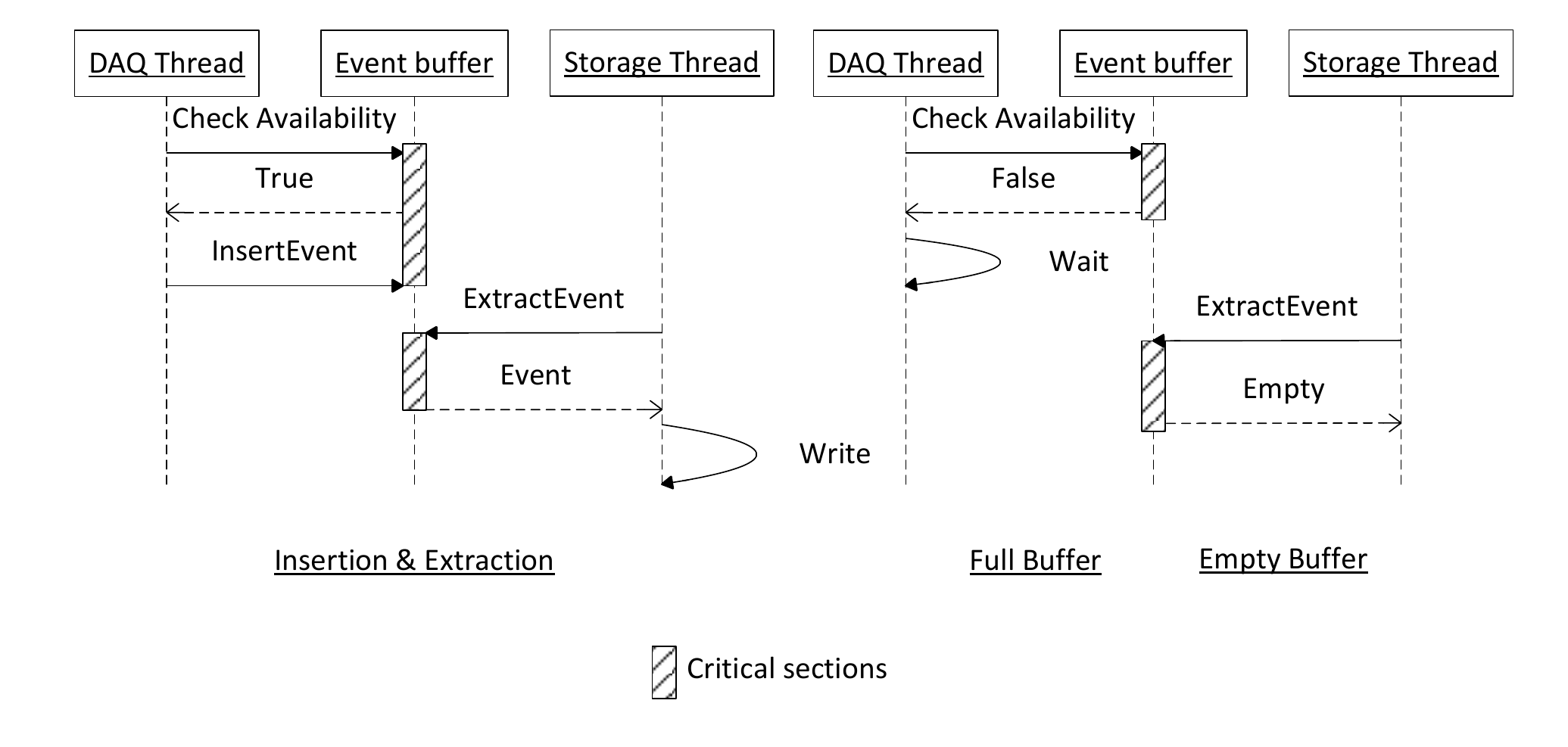}
    \caption[Asynchronous data storing diagram]{\label{fig:AsynchSeqDiagram}Asynchronous data storing diagram. Left: normal insertion and extraction. Right: empty buffer and full buffer conditions.}
  \end{center}
\end{figure}

The process of data storing in hard disk may incur in high non-deterministic latencies. An asynchronous data storing process can avoid such latencies and allow new triggering as soon as possible. The design of this mechanism consist of a circular data buffer and a storage thread. The acquisition saves data in the buffer and the storage thread writes the new data. This feature was implemented as optional and it allows to compare different storing strategies (see Section~\ref{sec:DeadTimeMeas}). The asynchronous storage sequence diagram can be seen in Figure~\ref{fig:AsynchSeqDiagram}. The normal process is shown in the left side and the buffer empty and buffer full conditions are shown in the right side. The critical section is needed in every thread in order to avoid race conditions accessing to the Event Buffer can also be seen in the figure.

\subsubsection{DAQ Monitoring} 
\begin{figure}[h!]
     \begin{center}
	\begin{subfigure}[b]{0.6\textwidth}
                \centering
                \includegraphics[width=\textwidth]{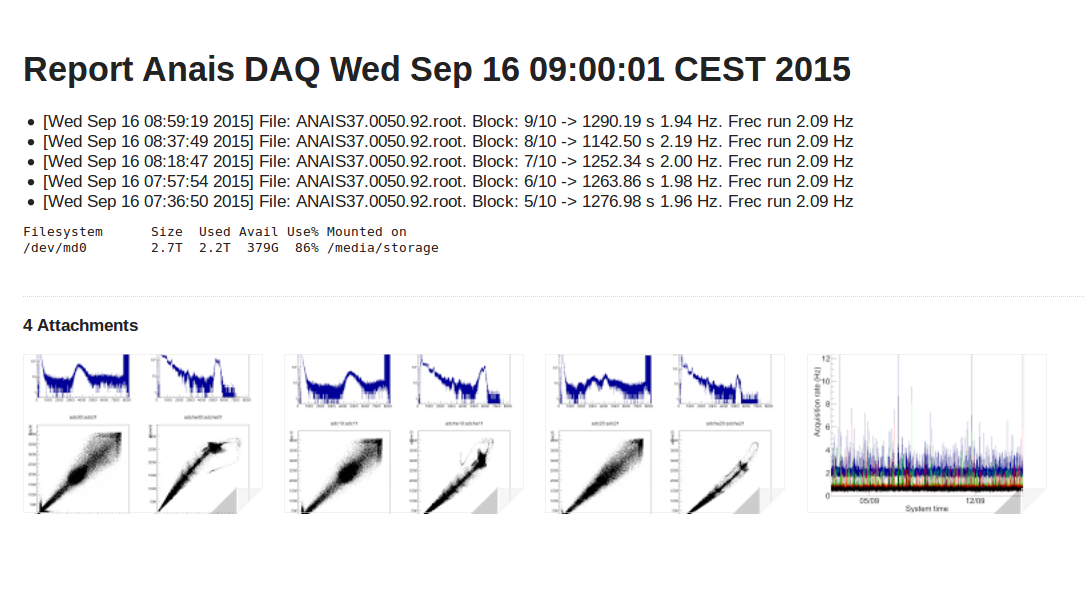}
                \caption{}
                \label{fig:DAQemailbody}
        \end{subfigure}%
        ~ 
        \begin{subfigure}[b]{0.4\textwidth}
                \centering
                \includegraphics[width=\textwidth]{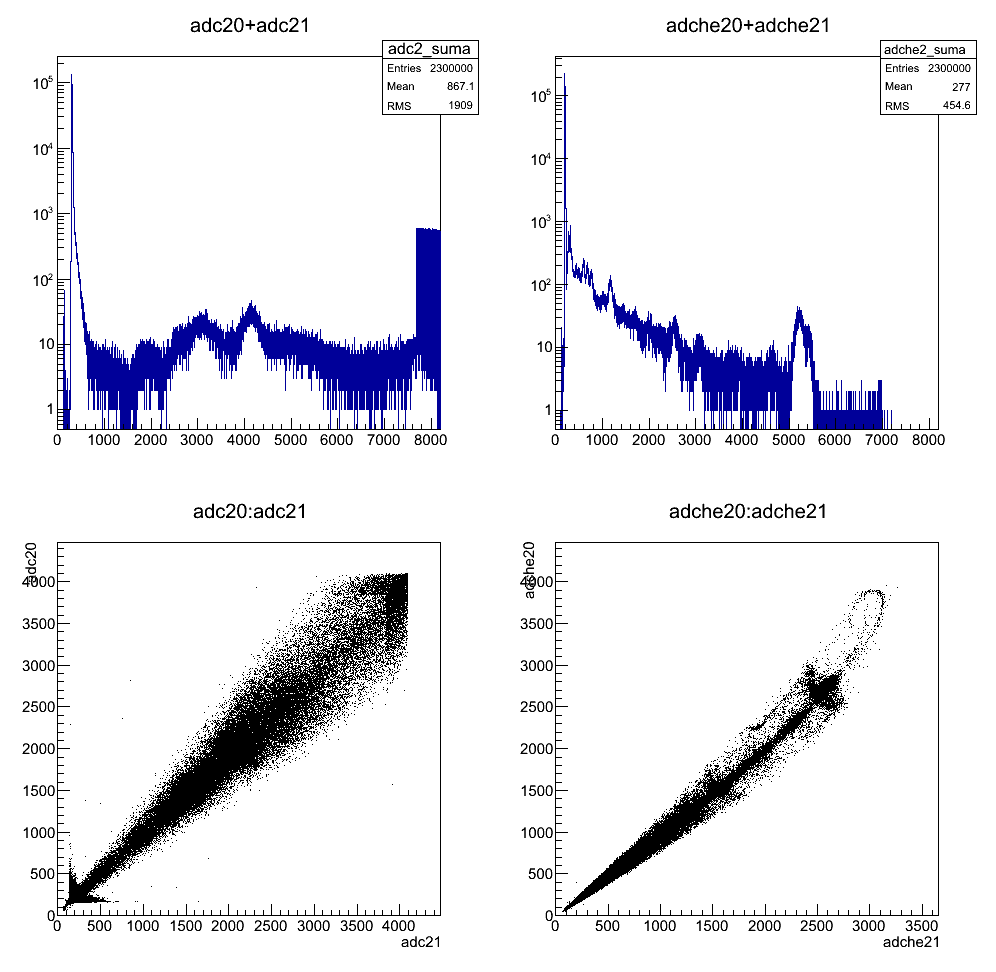}
                \caption{}
                \label{fig:DAQemail0}
        \end{subfigure}

\begin{subfigure}[b]{0.6\textwidth}
                \centering
                \includegraphics[width=\textwidth]{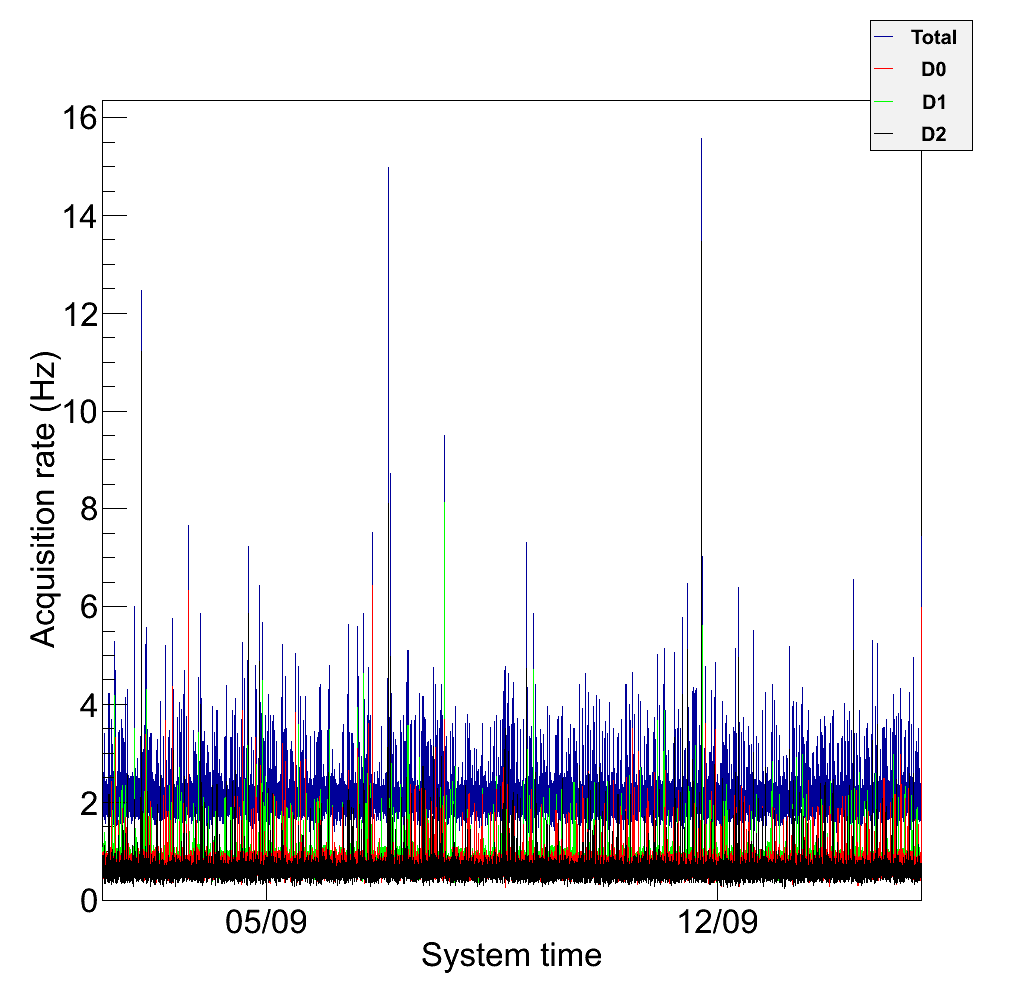}
                \caption{}
                \label{fig:DAQemailrate}
        \end{subfigure}
%

        \caption{DAQ report email. Body (a) and attatched reports: spectra and asymmetry plots from a detector (b) and rate report (c).}\label{fig:DAQemail}
\end{center}
\end{figure}

A subsystem of DAQ monitoring is needed to be able to know the acquisition rate in real time. This subsystem is designed to write statistics in shared memory and read this statistics via filesystem. The use of shared memory is fast and it does not incur in performance penalty. In addition this statistics are stored in write time so it does not add dead time directly in asynchronous data storing configuration.
\paragraph{}
An external \texttt{cron} task monitors the execution of the DAQ process and their statistics via filesystem. It sends an email in case of DAQ process failure and in case of abnormal acquisition rate. Additionally it sends a daily easily customizable email report with rate statistics and energy, asymmetry and rate plots. The \texttt{cron} process is spawned at minimum priority in order to minimize the interference with the acquisition thread. A typical report content for a system of two detectors can be seen in Figure~\ref{fig:DAQemail}. The email body with global acquisition rate information can be observed in Figure~\ref{fig:DAQemailbody}. The report script generates a plot for every configured detector with QDC spectra in all the configured energy ranges (see Figure~\ref{fig:DAQemail0}). Additionally, scatter plots with QDC signals are sent in order to detect abnormal asymmetries caused by signal malfunction. Finally a plot with the rate of all detectors and total rate is sent (Figure~\ref{fig:DAQemailrate}). This plot is generated by a script that calculates the frequency with a minute temporal binning. The script is configurable and it allows to compute rates in other timebases than minutes and it is ready to compute the rate of new detectors added to the system.
\section{DAQ software implementation}
The program implementation may seem straightforward once the main loop is designed (see Figure~\ref{fig:DAQFlowChart}). But the specific requirements about stability impose a very careful implementation of every hardware negotiation. For example, all operations on every board must have a proper timeout mechanism in order to avoid unwanted DAQ stops. This mechanism must be implemented alongside a proper error checking and a logging system in order to be able to detect possible infrequent errors that can cause malfunction or even stall the whole system. 
\subsection{Output file structure and naming conventions}\label{sec:DAQNamingConventions}
Due to the flexible DAQ system the output file format is determined by the event configuration. This is a very convenient feature because it allows testing configurations without development. But this flexibility can result in a disorganized output without some discipline. In this section the naming convention used to configure DAQ is covered.
\paragraph{}
The aim of the naming convention is to establish useful criteria to the rest of the software chain in order to produce output easy to analyze for users and programs. The guidelines used for the naming convention are:
\begin{easylist}
& naming detector or signal parameters in a compact way
& making easy and predictable the parameter names
& avoiding capital letters and subscript operators to aid typing
\end{easylist}
\paragraph{}
For these reasons the criterion of adding numbers with crystal and signal information at the end of the parameter was used: \texttt{name\_parameter+ncrystal+nsignal} (e.g. \texttt{tdc21} is detector 2 signal 1 TDC value). Similarly, ROOT aliases were configured to get compact information about detectors: for example replacing the final signal digit with an \texttt{s} to express sum (e.g \texttt{adc0s} meaning \texttt{adc00+adc01}). All signal parameters for the default configuration are listed in Table~\ref{tab:NamingConventionSignals} showing name compositions. These naming criteria are also used in signal parameter extraction as it can be seen in Section~\ref{sec:AnalysisNames}.
\begin{table}[h!]
	\begin{center}
		\begin{tabular}{ l p{2cm} p{6cm} p{4cm} }
			\toprule
			Origin & Name & Description & Examples \\
			\toprule
			Digitizer & \texttt{p} & The signal digitized pulse & \texttt{p31}: Pulse, detector 3, signal 1 \\
			\hline
			Digitizer & \texttt{np} & The digitized pulse size &\texttt{np20}: Pulse size, detector 2, signal 0\\
			\hline
			QDC & \texttt{adc} & Charge-to-digital data at low energy&\texttt{adc1s}: alias for \texttt{adc10+adc11}, the QDC sum for detector 1\\
			\hline
			QDC & \texttt{adche}, \texttt{adcvhe} & Charge-to-digital data at high and very high energy &\texttt{adche0s}: QDC sum for detector 0 at high energy \\
			\hline
			TDC & \texttt{tdc} & Time-to-digital data for the PMT trigger&\texttt{tdc30-tdc31}: time span between two signal triggers for detector 3\\
			\toprule
		\end{tabular}
		\caption[Naming convention for signals parameters]{Naming convention for signals parameters.} 
		\label{tab:NamingConventionSignals} 
	\end{center}
\end{table}
\paragraph{}
The global parameters for the default configuration are listed in Table~\ref{tab:NamingConventionGlobal}.

\begin{table}[h!]
	\begin{center}
		\begin{tabular}{ l l p{9.2cm} }
			\toprule
			Origin & Name & Description \\
			\toprule
			Scaler& \texttt{RT} & Real time clock (50 ns clock tick)\\
			\hline
			Scaler& \texttt{LT} & Live time clock (50 ns clock tick)\\
			\hline
			PC clock& \texttt{clock} (alias, \texttt{utc}) &  PC time in seconds (UNIX time, seconds elapsed since 1970)\\
			\hline
			PC clock& \texttt{clock\_ns} & PC time in nanoseconds (nanoseconds elapsed since 01/01/2013) \\
			\toprule
		\end{tabular}
		\caption[Naming for global parameters]{Naming for global parameters.} 
		\label{tab:NamingConventionGlobal} 
	\end{center}
\end{table}
\subsection{Run routine and run script}\label{sec:DAQScripts}
The needed stability in the data taking forces to routinely calibrate the whole system. This calibration routine also marks the length of a data taking run and it is used to do some others routine tasks such as data synchronization (see Section~\ref{sec:DataSynch}) and data processing (see Chapter~\ref{sec:AnalysisSW}). This section describes the steps needed by the aforementioned routine:

\begin{easylist}[enumerate]
& Stop data taking. It stops both NaI crystal DAQ and plastic scintillators DAQ.
& Synchronize data.
& Energy calibration with X-ray and $\gamma$ sources.
& Pedestal calibration of the MATACQ digitizer boards.
& Execute data taking again. It remotely (asynchronously via \texttt{ssh}) run the plastic scintillators DAQ and then starts the NaI DAQ preserving this execution order in order to correctly maintain the synchronization of both DAQs (see Section~\ref{sec:VetoFrontEnd} for details).
\end{easylist}
\paragraph{}
The user is prompted by the system in order to follow the above steps. Step 3 is performed by the \texttt{CAL} script and steps 4 to 2 are guided by the \texttt{DAQ} script.
\subsubsection{Detectors energy calibration: \texttt{CAL} and \texttt{CALPULS} scripts}
The energy calibration is performed with X-ray and $\gamma$ sources introduced inside the shielding for this purpose. Once the sources are correctly located the calibration script can be invoked. There are two different scripts: \texttt{CAL} and \texttt{CALPULS}. The difference is that \texttt{CALPULS} uses a configuration that includes pulse digitization and it is very similar to background configuration whereas the \texttt{CAL} configuration excludes all pulse digitization. For this reason \texttt{CALPULS} has more dead time and it is used to calibrate pulse analysis algorithms described in Chapter~\ref{sec:AnalysisSW}. \texttt{CAL} is used to routinely calibrate the system.
\paragraph{}
Both scripts take the same optional algorithms: \texttt{--source} and \texttt{--comment}. The \texttt{source} parameter specifies the $\gamma$ source(s) used and the \texttt{comment} parameter saves any remarkable fact related to the particular calibration process such as source position or non-standard hardware configuration. Both parameters are saved in run log file as described later.
\subsubsection{\texttt{DAQ} script: MATACQ pedestal calibration, background acquisition and data synchronization}\label{sec:SWPedestals}
This script launches the background data acquisition and all related tasks: pedestal calibration, background acquisition and data synchronization.
\paragraph{}
First of all it starts the MATACQ pedestal calibration using the same configurations as the DAQ program and calibrating only the MATACQ boards configured in it. Following the characterization described in Section~\ref{sec:MatacqCharac} the pedestal calibration process is done with the normal acquisition conditions: signal plugged in the board, the preamplifiers switched on and the PMT high voltage in the normal working values. This calibration uses only the pretrigger points as described in the aforementioned section. The pedestal calibration is stored in plain-text file with the MATACQ configuration name as filename. The DAQ later reads this pedestal calibrations in order to subtract them and obtain the actual signal.
\paragraph{}
Next, the \texttt{DAQ} script spawns the DAQ program and waits until it finishes. It removes the shared memory file (described in Section~\ref{sec:DAQConfig}) when the user stops the execution. It is used to detect abnormal process termination if this file exists and the process is not running triggering a warning email notification.
\paragraph{}
The last step is to launch the data synchronization process described in Section~\ref{sec:DataSynch} that triggers automatic data analysis (see Section~\ref{sec:AnalysisAuto}).
\subsection{Run log} The DAQ program can take several parameters in order to store information about the data taking run as described in the previous section. This information can be the calibration source used or any other relevant comment. This information is stored with other run statistics in the run log file. This file is plain text file with tab separated values. The output fields are described in Table~\ref{tab:LogFileFields}.
\begin{table}[h!]
	\begin{center}
		\begin{tabular}{ l p{10 cm} }
			\toprule
			Field name & Description\\
			\toprule
			Filename & The run base filename. It includes the nun number and the set-up name. \\
			\hline
			Number of files & Number of files written in the run\\
			\hline
			Source & This field stores the source used in calibration process or \texttt{Background} otherwise\\
			\hline
			Start & Run start time and date in the format Year-month-day hour:minute:second in local time\\
			\hline
			Stop & Run stop time and date in the same format\\
			\hline
			Events & The number of events stored in the run\\
			\hline
			Real Time & The real time measured with scaler\\
			\hline
			Live Time & The live time measured with scaler\\
			\hline
			Rate & The number of events divided by the Live Time\\
			\hline
			Comment & The comment provided at execution time. It is useful to document some special event related to the run\\
			\toprule
		\end{tabular}
		\caption[Log file fields]{Log file fields.} 
		\label{tab:LogFileFields} 
	\end{center}
\end{table}



\chapter{Data analysis software}\label{sec:AnalysisSW}
The ANAIS data analysis software extracts parameters from the primary acquired data seen in Section~\ref{sec:DAQNamingConventions}. The current algorithms and its implementation are the result of the evolution of the analysis from previous prototypes~\cite{CPOBES,TesisMaria,CCUESTA}. The original parts developed in this work are the adaptation of the algorithms to the new DAQ data types and structures, the reimplementation of the peak detection and the peak based algorithms (Sections~\ref{sec:PeakDetection} and \ref{sec:PeakParams} respectively), the improvement of the baseline characterization algorithms (see Section~\ref{sec:DCCalc}) and the configuration and automation systems (see Section~\ref{sec:AnalysisConfigAut}).
\paragraph{ }
The analysis software reads the file generated by the DAQ software extracting the information from the analysis of the waveform and the combination of different acquired parameters. It uses the DAQ configuration parameters as input of the analysis algorithms and it writes information in ROOT format in order to aid data analysis and scripting. 
\paragraph{ }
The output of these algorithms can be classified into three groups: signal, detector and event parameters. The signal parameters are extracted from the digitized waveform of the individual PMT (such as the area or the pulse onset), the detector parameters are extracted from the two module signals and the event parameter are not specifically related to one signal or detector (such as the trigger time or the event dead time).
\paragraph{ }
The presented algorithms are still improving in order to get better knowledge of the low energy region. In particular, the construction of new energy estimators at very low energy is ongoing and the filtering protocols (see Section~\ref{sec:EventSelection}) are expected to be improved with the blank module information (see Section~\ref{sec:ANAIS37}).
\section{Algorithms}\label{sec:Algorithms}
The algorithms developed for the event, detector and signal analysis are described in this section. First, the basic pulse parameter extraction explaining details about their design and implementation are explained. Next a description of peak detection and peak related parameters is given. Finally the temporal event parameters are calculated. Further details about the implementation and parametrization of this analysis can be found in Section~\ref{sec:AnalysisConfig}.
\subsection{Basic signal characterization}\label{sec:PulseChar}
\begin{figure}[h!]
  \begin{center}
    \includegraphics[width=.7\textwidth]{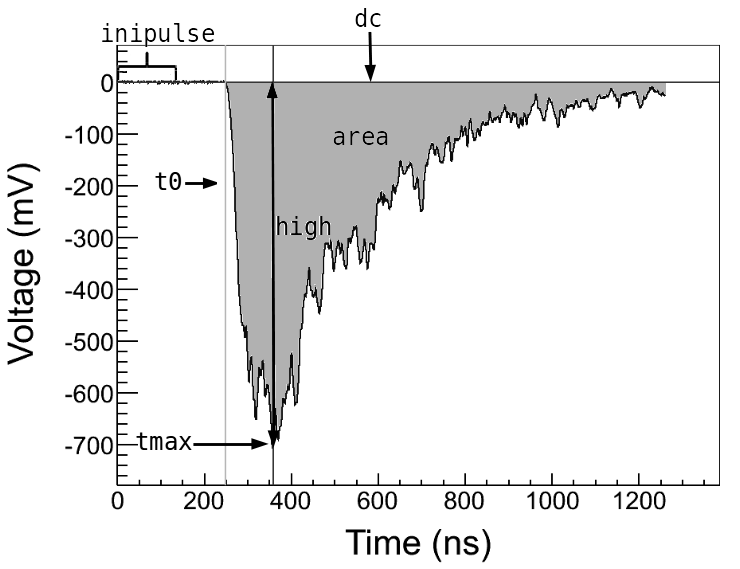}
    \caption[Bogus DC calculation]{\label{fig:PulseParamsN}Basic pulse parameter extraction.}
  \end{center}
\end{figure}

The signal for every PMT is digitized to extract important parameters for energy estimation or noise rejection. The first process in pulse analysis consists of finding the minimum, maximum or area and other basic parameters of the pulse. This section covers the introduction to the pulse parameter calculation explaining algorithms and dependencies among parameters. Figure~\ref{fig:PulseParamsN} illustrates the most important parameters covered in this section.
\paragraph{}
All the parameters described below are calculated in physical units to be able to mix data from varied hardware or distinct hardware configurations such as \hbox{12-bit/14-bit} MATACQs and 1 or 2 GS/s sampling rates. The analysis software reads the hardware configuration and converts arbitrary units to physical ones. The basic units are \emph{mV} for voltage and \emph{ns} for time. 
\paragraph{}
The signal is a fixed length array of voltage values, sometimes called \texttt{pulse} in this work. This length is 2520 points as seen in Section~\ref{sec:VMEModules}, named \texttt{pulse\_depth} in the following. The parameters extracted from the signal are reviewed below:
\paragraph{Baseline level (\texttt{dc}):} This parameter measures the baseline level before the pulse. It takes advantage of digitization pretrigger:  the value is computed taking the first half of the pretrigger points. These pretrigger points are referred in this work as \texttt{inipulse}. This variable needs a correct \texttt{inipulse} value in order to get a valid result. \texttt{dc} is a very critical parameter because many others depend on it. Therefore, several algorithms were compared. The selected one takes information from skewness to refine \texttt{dc}, \texttt{rms} and \texttt{skw} itself as described in Section~\ref{sec:DCCalc}.
\paragraph{Baseline mode (\texttt{mode}):}This parameter is the statistical mode of the first half of pretrigger points.
\paragraph{Baseline weighted mode (\texttt{wmode}):}This parameter is the weighted mean of the five more frequent values of the first half of pretrigger points. It is calculated to get a center of baseline distribution points better than the mode in multimodal distributions. A detailed use of this parameter is described in Section~\ref{sec:DCCalc}.
\paragraph{Baseline standard deviation (\texttt{rms}):} This parameter is calculated as the standard deviation of the same points used to calculate \texttt{dc}. The distribution of this parameter is a good estimator of the baseline quality as it can be seen in Section~\ref{sec:Baseline}.
\paragraph{Baseline Skewness (\texttt{skw}):}It is the skewness of the same points used to calculate \texttt{dc} as it can be seen in Section~\ref{sec:DCCalc}.
\paragraph{Maximum (\texttt{max}):} It is the maximum value of all points of the digitized signal.
\paragraph{Height (\texttt{high}) and its time (\texttt{tmax}):} The parameter \texttt{high} is the minimum point from the digitized signal. The parameter \texttt{tmax} is the time when this minimum is reached, giving a relation between them:
\begin{equation} pulse_{tmax}  = high\end{equation}
\texttt{high} is calculated without any other dependence. The amplitude of a pulse can be evaluated as \texttt{dc-high} (with the alias \texttt{amp}), introducing \texttt{dc} dependence in that way. It is worth to note that \texttt{high} is calculated without any filtering and therefore amplitude may be biased by high frequency components. Anyway, amplitude is only used in PSA algorithms to discriminate the incident particle (with very good results as it can be seen in Section~\ref{sec:MuNaIEvt}) and it is not used as energy estimator.
\paragraph{Pulse onset (\texttt{t0}):} The \texttt{t0} parameter is the time where the signal overtakes a threshold marking the start of a triggered signal. The simplest and default way to calculate it is using an external parameter as software threshold: \texttt{threshold\_sw}. The value \texttt{t0} is extracted from the condition:
\begin{equation} pulse_{t0}<dc-threshold\_sw\end{equation}
If the threshold is not reached, \texttt{t0} value is set to \texttt{pulse\_depth} marking a lack of \emph{software trigger}. This calculation uses \texttt{dc} parameter as seen above. It is critical to this and subsequent steps to set carefully the threshold value to the actual hardware trigger in order to obtain accurate \texttt{t0} values.
\paragraph{}
A varied set of algorithms has been implemented to calculate \texttt{t0} due to the different trigger strategies used in this work. A complete list of such algorithms can be seen in Section~\ref{sec:AnalysisConfig}.
\paragraph{Pulse area (\texttt{area}):} Pulse area is one of the most important parameters extracted in the analysis software because of its expected linear behavior with the energy yield. It is calculated as shown in the following equation:
\begin{equation} 
area = dc \cdot ( pulse\_depth - t0 ) - \sum_{i=t0}^{pulse\_depth} pulse_i  
\end{equation}
This calculation subtracts the baseline using \texttt{dc} and it begins to sum values in \texttt{t0} so its value rely on both parameters.
\paragraph{Low energy area (\texttt{abe}):} This parameter is an attempt to avoid baseline noise in area computation for very low energy events. It is computed similarly to \texttt{area} but it only sums the values that pass the threshold (an \texttt{rms\_threshold} described in Section~\ref{sec:AnalysisConfig}).
\paragraph{First momentum (\texttt{fm}), second momentum (\texttt{sm}) and third momentum (\texttt{tm}):}
These three parameters are the first, second and third momenta calculated from \texttt{tmax} to the end of the pulse, defined as follows:

\begin{equation*} 
	fm = \sum_{i=tmax}^{pulse\_depth} (dc - pulse_i)\cdot(i - tmax) 
\end{equation*}
\begin{equation} 
sm = \sum_{i=tmax}^{pulse\_depth} (dc - pulse_i)\cdot(i - tmax) ^2 
\end{equation}
\begin{equation*} 
tm = \sum_{i=tmax}^{pulse\_depth} (dc - pulse_i)\cdot(i - tmax) ^3
\end{equation*}
These parameters have useful temporal information that can be used for pulse shape analysis.
\paragraph{\texttt{p1} and \texttt{p2}:} The \texttt{p} parameters are area ratios in different time-bases. Their definition is:
\begin{equation*} 
p1 = \frac{area(t0+100ns,t0+600ns)}{area(t0,t0+600ns)}
\end{equation*}
\begin{equation} 
p2 = \frac{area(t0,t0+50ns)}{area(t0,t0+100ns)}
\label{eq:PDefinition}
\end{equation}
\paragraph{}
The integration time intervals are selected having the NaI scintillation constant in mind.
\subsection{Peak detection}\label{sec:PeakDetection}
This algorithm identifies negative peaks in the pulse. This is very useful given the ability of resolving single photoelectrons in low energy events~\cite{CPOBES} (see Section~\ref{sec:LESignal} for a description of low energy signals). The peak information is used to discriminate noises from real NaI scintillation~\cite{CCUESTA}. Additionally, the number of peaks are also strongly correlated with the energy at low energies and its temporal distribution can be very useful to discriminate bulk scintillation from other sources of light or noise.
\paragraph{}
For this work, two algorithms were compared: an algorithm based on \texttt{TSpectum} ROOT class and its \texttt{Search} method, usually used to find peaks in $\gamma$-ray spectra and a simple algorithm based on finding minima in the pulse.
\subsubsection{Spectrum based algorithm}
The method \texttt{TSpectrum::Search} uses a well know algorithm to identify peaks in a spectrum~\cite{morhavc2000identification}, but it is used in an unconventional way. The pulse is inverted in order to allow the mentioned algorithm to find maximum. Next, sigma of the expected peaks and peak minimal threshold is supplied to the algorithm via \texttt{nrms} value (see Section~\ref{sec:AnalysisConfig}). The algorithm calculates smoothed waveform from source waveform based on Markov chain method~\cite{silagadze1996new}. Finally, a list with peaks with position and height is produced as a result of the algorithm. This algorithm is very costly from a computational point of view.
\subsubsection{Minimum algorithm}
\begin{figure}[h!]
  \begin{center}
    \includegraphics[width=.7\textwidth]{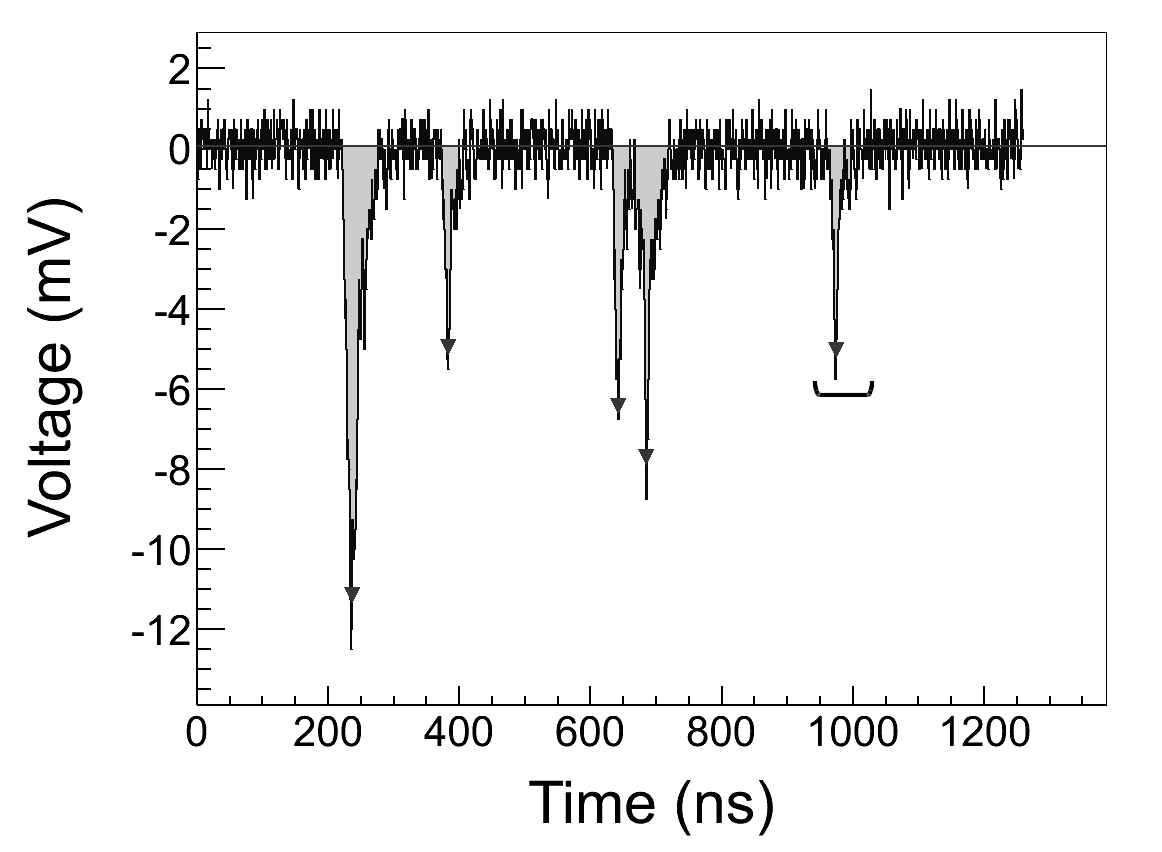}
    \caption[Bogus DC calculation]{\label{fig:PulseParamsPeaks}Peak detection and peak based parameters (see Section~\ref{sec:PeakParams}).}
  \end{center}
\end{figure}

The aim of this algorithm is to find photoelectrons by locating minima. The first step of this algorithm is to filter the higher frequencies of the signal with a low pass filter and eliminate fluctuations. The cut frequency is set to 100 MHz. The second step is to calculate the derivative pulse and find the minima with the derivative sign change. The implementations of this algorithm also changes the postprocessing of the detected peaks. This algorithm requires a height to the peak below a software threshold (see Section~\ref{sec:AnalysisConfig} for details).
\paragraph{ }
The last algorithm is the currently used for its fast implementation and its very similar performance. Figure~\ref{fig:PulseParamsPeaks} shows the peak detection of a very low energy event.
\subsection{Peak based parameters}\label{sec:PeakParams}
The last step in pulse characterization is to extract parameters taking advantage of peak extraction information. This information can be useful to characterize parameters in very low energy events and it is also used to record information about the last peak. This last peak information was extracted to get additional information about SER without the bias of trigger cut. This bias was detected trying to get SER information with one peak pulses seeing a photoelectron distribution abruptly cut in trigger level~\cite{CCUESTA}.
\paragraph{}
The peak based parameters are:
\paragraph{Number of detected peaks (\texttt{n}):} This parameter is the number of peaks detected by the previous algorithm.
\paragraph{Area around peaks (\texttt{areap}):} The algorithm that calculates \texttt{areap} uses peaks in order to sum area values only near them in a fixed time window in order to reduce the effect of the baseline in the integrated charge (area). This area can be seen in gray in Figure~\ref{fig:PulseParamsPeaks}.
\paragraph{Last peak area (\texttt{arealp}):} This parameter sum area from the last peak in a fixed time window. This parameter is very useful to characterize the single electron response with the signal of real events as it can be seen in Section~\ref{sec:LY}. This will account the gray zone under the last peak seen in Figure~\ref{fig:PulseParamsPeaks}. The integration interval is marked with a brace.
\paragraph{Last peak height (\texttt{hlp}):} The $y$ (voltage) value of last peak. It does not take into account the \texttt{dc} value. It is useful to characterize the single electron response amplitude distribution.
\paragraph{Last peak position (\texttt{poslp}):} The $x$ (temporal) position of last peak. Used to be able to select non-triggered signals in the SER, avoiding the trigger bias.
\subsection{Time parameters}\label{sec:TemporalParams}
This group of algorithms constructs some derived temporal parameters relative to an event from raw temporal data. The more important temporal parameters are distance to previous event (\texttt{drt}), event dead time (\texttt{ddt0}) and temporal distance to the last plastic tagged event (veto) (\texttt{vts}).
\paragraph{}
The temporal distance to previous event can be used to detect temporal triggering patterns attributed to noise as described in Section~\ref{sec:DeltaRT}. The event dead time gives information about the dead time distribution among events (see Section~\ref{sec:DeadTimeMeas}). The distance to veto signal is used to subtract these muon related events which was already explained in Section~\ref{sec:DAQsSynch}.
\paragraph{}
The distance to a plastic tagged event must be done by correlating the two acquisition data. This is why the veto data are loaded together with the NaI data in order to identify the nearest precedent veto event and calculate the time distance to it, \texttt{vts} in seconds and \texttt{vt} in (50 ns) ticks. Another useful variable is the distance to a very energetic event in the same crystal. This variable can be used to remove the afterglow events due to a very energetic event. The variable is named \texttt{heticks} in seconds (and \texttt{hetick} in ticks) for every crystal (\texttt{hetick2} for the detector 2). The use of the distance to a muon tagged event and to a very energetic event can be seen in Section~\ref{sec:MuNaIEvt}.
\section{Baseline algorithms}\label{sec:DCCalc}
\begin{figure}[h!]
  \begin{center}
    \includegraphics[width=.8\textwidth]{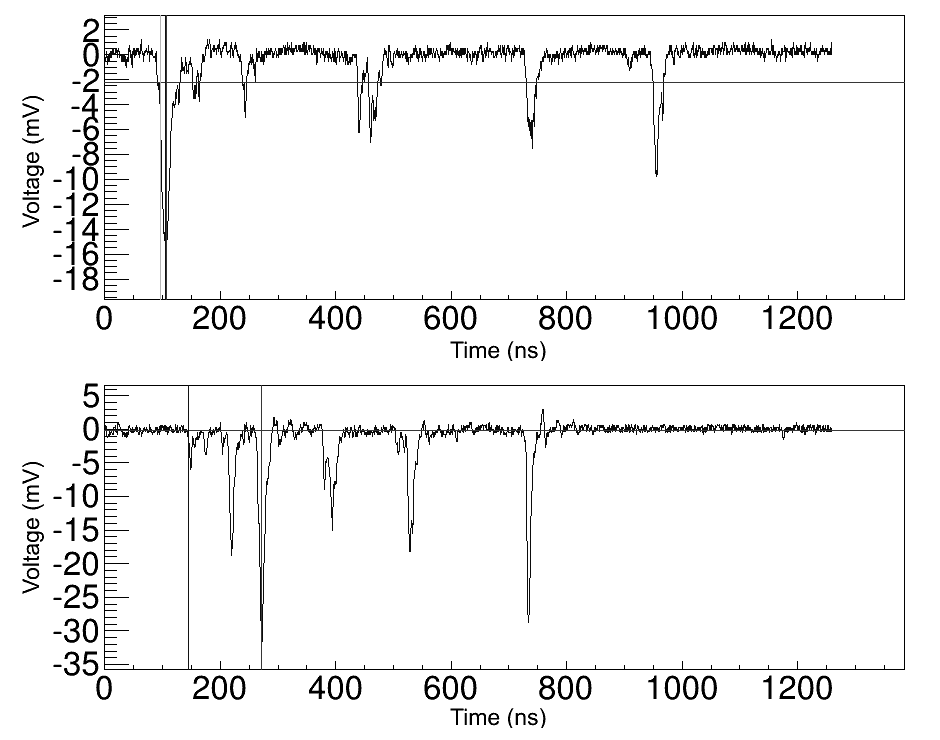}
    \caption[Bogus DC calculation]{\label{fig:PulseInDcPoints}Pulse with bogus calculated \texttt{dc} (horizontal red line in top pulse) (Horizontal scale in ns, vertical in mV).}
  \end{center}
\end{figure}

\begin{figure}[h!]
 \begin{center}
\begin{subfigure}[b]{0.5\textwidth}
                \centering
                \includegraphics[width=1\textwidth]{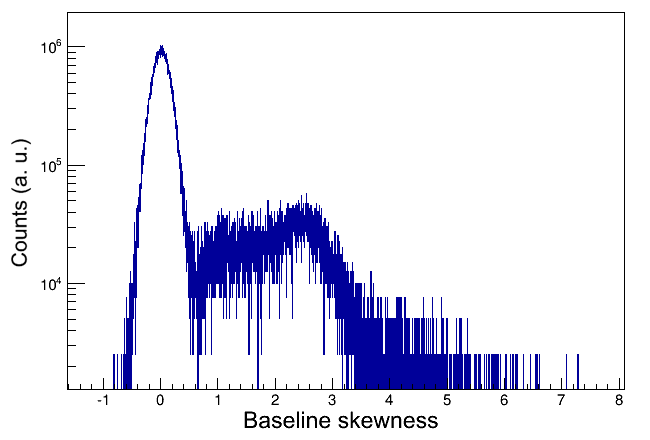}
		\caption{\label{fig:skw}Baseline skewness. The right population corresponds to photoelectrons in baseline.}             
        \end{subfigure}
        ~ 
        \begin{subfigure}[b]{0.5\textwidth}
                \centering
                \includegraphics[width=1\textwidth]{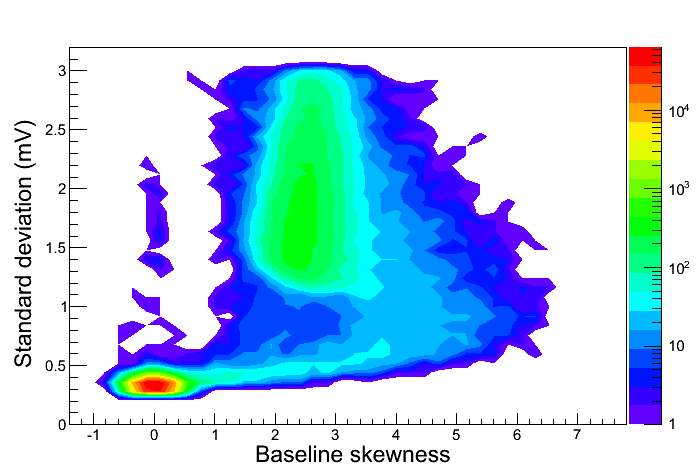}
		\caption{\label{fig:skw_rms_scatter}Skewness vs. standard deviation. Photoelectrons are in the top right population.} 
        \end{subfigure}

        \caption[Baseline skewness and RMS distribution]{Baseline skewness and RMS distribution.}
\end{center}
\end{figure}

\begin{figure}[ht!]
 \begin{center}
\begin{subfigure}[b]{1\textwidth}
\centering
    \includegraphics[width=.8\textwidth]{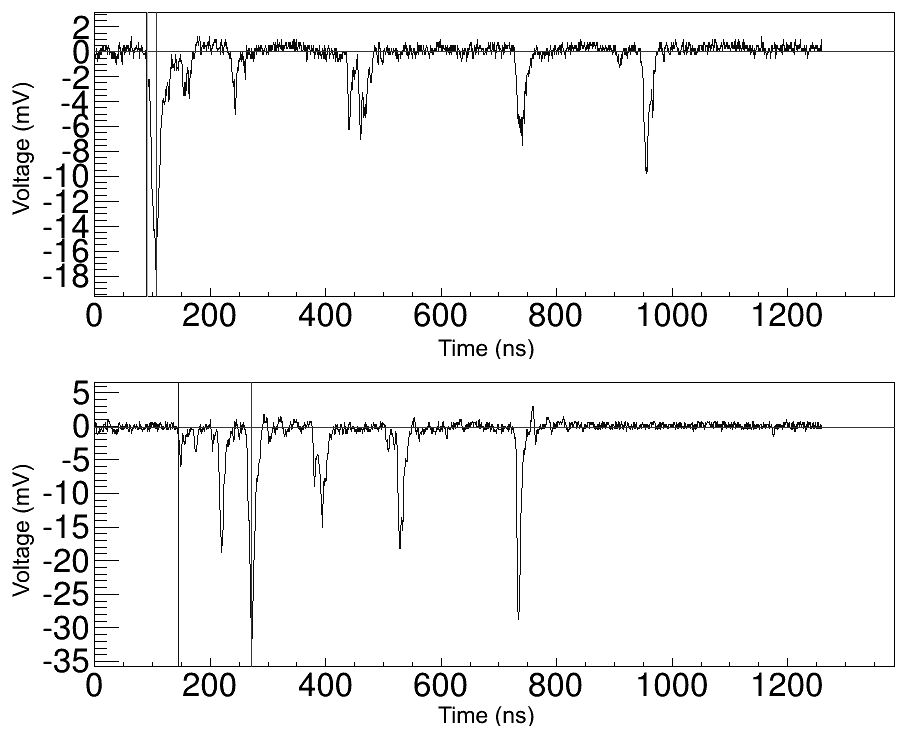}
    \caption[Refined DC calculation]{\label{fig:PulseInDcPointsRef}\texttt{dc} calculation with photoelectron exclusion.}

\end{subfigure}

\begin{subfigure}[b]{1\textwidth}
\centering
                \includegraphics[width=.75\textwidth]{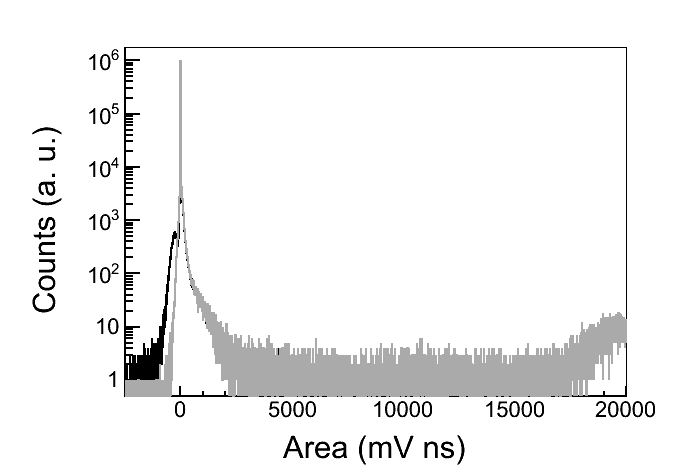}
		\caption{\label{fig:skw_area_comp}Area without (black) and with photoelectron exclusion (gray).}      
\end{subfigure}
%
        \caption[Photoelectron exclusion results]{Photoelectron exclusion results.}
\end{center}
\end{figure}

Baseline parameter extraction is a key phase in the pulse characterization. These parameters (\texttt{dc}, \texttt{rms}, \texttt{skw}, etc.) are used to characterize the pretrigger zone and they are used (especially \texttt{dc}) in all steps of pulse related algorithms. A wrong calculation is propagated to all other parameters. Hence, a careful study of the parameter extraction algorithms was performed and some improvements were implemented and tested.
\paragraph{}
A naive implementation of \texttt{dc} and \texttt{rms} calculations revealed some populations of bogus \texttt{dc} with negative area mostly consisting in pulses with photoelectrons in the points used to calculate baseline parameters as it can be seen in Figure~\ref{fig:PulseInDcPoints}.
\paragraph{}
Skewness (\texttt{skw}) was used for the study of such a population. It quantifies the asymmetry of a distribution and it can be used to identify the presence of peaks in baseline points. In Figure~\ref{fig:skw} it can be seen a typical baseline skewness distribution: right population corresponds to events with photoelectrons in the baseline. An algorithm to refine baseline parameters can be carried out given such a distribution. In addition, the algorithm uses the baseline root mean square value in order to identify photoelectrons in baseline more robustly. This is possible given the scatter distribution of these two parameters as it can be seen in Figure~\ref{fig:skw_rms_scatter}. 
\paragraph{}
The refined algorithm consists of recalculating \texttt{dc}, \texttt{rms} and \texttt{skw} in the case of big skewness and root mean square. First, all the three parameters are calculated (being \texttt{dc\_points} the first half of pretrigger \texttt{inipulse}). \texttt{mode} and \texttt{wmode} parameters are also calculated being the most frequent value and the weighted average for the five most frequent values in \texttt{dc\_points}. Next, \texttt{skw} and \texttt{rms} are checked with thresholds configured given by its distribution (see Section~\ref{sec:AnalysisConfig}). The refinement cuts off the extreme points using \texttt{wmode} and \texttt{trigger\_level} as reference and establishing windows of 5 ns when the asymmetric zone is detected. Once the points are removed, the calculations are recomputed with the remaining population.
\paragraph{}
The results of a better \texttt{dc} identification can be observed in Figure~\ref{fig:PulseInDcPointsRef} and it can be compared with the previous one as seen in Figure~\ref{fig:PulseInDcPoints}. 
Additionally, the area spectrum was compared with and without refinement showing difference at random coincidence population only as it can be seen in Figure~\ref{fig:skw_area_comp}. This spectrum shows an energy interval from 50 keV to noise peak and it remains identical with the exception of the aforementioned noise peak. 
\section{Implementation, configuration and automation}\label{sec:AnalysisConfigAut}
A subtle feature of data analysis is algorithm parametrization. It has to scale-up along the experiment life cycle and must allow flexibility and ease of configuration. This parametrization can become a tedious and error-prone process having in mind the increase of the number of signals. Hence, some mechanisms were implemented in order to reduce manual parameter introduction, avoid data replication and automate processes where possible.
\paragraph{}
In this section the analysis software invocation is covered and all features such as configuration or automation are described.
\subsection{Algorithm parametrization}\label{sec:AnalysisConfig}
All algorithms used in analysis software need parametrization as seen above. Some of this parameters can be extracted from the DAQ configuration (see Section~\ref{sec:DAQConfig}), others from SER characterization (\ref{sec:PMTSignal}) and the remaining ones from characterization algorithm process (\ref{sec:Algorithms}).
\subsubsection{DAQ parameters}
The parameters read from DAQ configuration are sampling rate and least significant bit value (the voltage value of digitization units). Both of them are used to get the result of the algorithms in physical units. Sampling rate is read from the digitizer configuration and the vertical resolution is deduced from the configured board: 250~$\mu$V/point for 12-bit MATACQ and 125~$\mu$V/point for 14-bit MATACQ.
\paragraph{}
Critical parameters from DAQ configuration are hardware threshold and coincidence window both of them configured in CAEN N843 constant fraction discriminator. This module does not allow either parameter reading or remote configuration and for this reason the values have to be changed carefully and in a documented way, modifying configuration parameters as described in the next section.
\subsubsection{Input parameters}
This section covers all input parameters required by the analysis software which can be configured via configuration file. It describes all possible valid values for the parameters and their meaning.
\paragraph{\texttt{inipulse}:} The number of points considered as pretrigger. It is used to calculate the baseline level. See Section~\ref{sec:PulseChar}.
\paragraph{\texttt{len\_of\_interest}:} This parameter is mostly used in unconventional acquisitions. It defines (in nanoseconds) the time windows to integrate the area and to obtain other parameters like \texttt{areap}, \texttt{fm}, \texttt{sm} or \texttt{tm}. It is used when the acquisition windows is wider than the width of the expected signal. A special case for the use of this parameter is the SER characterization given the width of a single photoelectron (see Section~\ref{sec:PMTSignal}). The use of \texttt{len\_of\_interest} prevents considering extra baseline when needed. The default value is all the pulse depth, in nanoseconds, using all points of the waveform.
\paragraph{\texttt{trigger}:} This parameter configures the \emph{trigger software} parameter. It must match the hardware trigger level value. It is expressed in millivolt.
\paragraph{\texttt{lp\_trigger}:} This parameter configures the \emph{trigger software} for the last peak algorithm. It is very useful to distinguish this value from the \texttt{trigger} value because using the same value will cause trigger bias in the SER extraction (see Section~\ref{sec:LY}).
\paragraph{\texttt{skw\_thr}:} This value is used as skewness threshold of refined \texttt{dc} calculation as described in Section~\ref{sec:DCCalc}. Setting this parameter to 0 disables this refinement (default value)
\paragraph{\texttt{rms\_thr}:} This parameter is used combined with the previous \texttt{skw\_thr}. See Section~\ref{sec:DCCalc}.
\paragraph{\texttt{adche\_thr}:} This parameter is used as threshold for counting the time since a very energetic event occurred. The output variable \texttt{hetick} is set to 0 when this threshold is exceeded, otherwise is set to clock distance to the last event above this threshold. See Section~\ref{sec:TemporalParams}.
\paragraph{\texttt{nrms}:} This value is used as threshold of peak identification described in Section~\ref{sec:PeakDetection}. This threshold is calculated as \texttt{dc-nrms}$\cdot$\texttt{rms} and all peaks below it are discarded. It uses 3 as default value.
\paragraph{\texttt{trigger\_mode}:}It allows to change the default algorithm to determine \texttt{t0} using (or not) the aforementioned \texttt{trigger} value. It is useful in case of using different hardware trigger methods and it allows software algorithm to behave in the same way. The valid algorithms are:
\begin{easylist}[itemize]
& \texttt{relative}: This is the default value of this parameter and the most frequently used. It calculates \texttt{t0} as described in Section\ref{sec:Algorithms}. The constant fraction discriminator has a relative trigger and this algorithm has to be used in this case. 
& \texttt{absolute}: This algorithm does not use extracted \texttt{dc} value as estimation of the baseline level. It uses \texttt{trigger} as an absolute level. This configuration is useful in the case of using a threshold discriminator.
& \texttt{rms}: The \texttt{rms} method uses \texttt{dc-nrms}$\cdot$\texttt{rms} as trigger level. It is worth to note that this algorithm uses different level for every pulse. This trigger level has the advantage of adapting the level to the noise of the baseline and it can be useful in noisy runs. It has the disadvantage of being different of the hardware one.
& \texttt{t0\_fixed}: This is a special method that assigns \texttt{t0} to \texttt{inipulse} value. It can be useful in the case of having the signal of interest in a fixed time relating to digitizer trigger. It can be used in special cases in conjunction with \texttt{len\_of\_interest} value.
\end{easylist}
\subsubsection{Configuration file}
All the above parameters can be configured via configuration file. It uses YAML format in the same way as DAQ program (see Section~\ref{sec:DAQConfig}). It has a section to override the default values in a global way, set values to a set of signals or in a per signal way. This scheme was designed in order to avoid to set all parameters of all signals but allowing individual configuration. This last feature is needed by subtle differences between signals given the PMT parameter dispersion (see Section~\ref{sec:PMTResults}).
\paragraph{}
The file consists of several sections. Every section can configure all parameters mentioned in the previous section. The name of the section determines the signals affected by the configuration. There are three different ways to configure the parameters:
\begin{easylist}[itemize]
& \texttt{default} section: In this section the override of the default parameters can be done and it applies to all signals unless it is override by a more specific section.
& regular expression section: This section overrides the previous section in the case that the name of the signal matches with a regular expression~\cite{thompson1968programming}. A regular expression section can be used to configure a set of signals that shares similar properties. It has been useful in the case of test signals and especially with attenuated signals as those used in ANAIS-25 and ANAIS-37 set-ups.
& signal name section: This section configures the parameter used to a particular signal overriding all other possible applying sections.
\end{easylist}
\paragraph{}
This mechanism can be observed in the Listing~\ref{lst:yaml_config}. The example uses default and regular expression sections only. It sets skewness threshold, trigger level and inipulse by default, and it overrides trigger level and skewness threshold for signals that matches \texttt{p[2-3].?} regular expression. The result of this matching can be \texttt{p20}, \texttt{p21}, \texttt{p30} and \texttt{p31} taking into account signal naming conventions as it can be seen in Section~\ref{sec:AnalysisNames}.

\begin{lstlisting}[label=lst:yaml_config, caption=Analysis configuration example.]
- default:
  trigger: -1.2
  skw_thr: .5
  inipulse: 580.
- p[2-3].?:
  trigger: -10
  skw_thr: 0
\end{lstlisting}
This parametrization is critical for obtaining accurate results. Knowing the settings used in this analysis step is very convenient and hence all configuration parameters are written in the output file with the obtained parameters. This file is described in the next section.
\subsection{Output file and Naming conventions}\label{sec:AnalysisNames}
All parameters extracted and calculated from the DAQ data are written in an output file. This file includes all DAQ data excluding the waveform data for performance reasons. The new file has all information related to an event: DAQ and analysis parameters and the path to pulse information. This path is used for the Pulse Inspection Tool (see Section~\ref{sec:PIT}). This file also contains the analysis configuration mentioned in the previous section. Table~\ref{tab:NamingAnalysis} lists all analysis parameters added to the DAQ parameters mentioned in Tables~\ref{tab:NamingConventionSignals} and \ref{tab:NamingConventionGlobal} as summary of all algorithms described in Section~\ref{sec:Algorithms}. 

\begin{table}[h!]
	\begin{center}
		\begin{tabular}{ l p{6.5cm} p{4.2cm}}
			\toprule
			Name & Description & Type \\
			\toprule
			\texttt{max} & Maximum pulse value & Signal parameter\\
			\hline
			\texttt{high} & Minimum pulse value & Signal parameter\\
			\hline
			\texttt{tmax} & Minimum pulse time & Signal parameter\\
			\hline
			\texttt{dc} & baseline level & Signal parameter\\
			\hline
			\texttt{mode} & baseline statistical mode & Signal parameter\\
			\hline
			\texttt{rms} & baseline standard deviation & Signal parameter\\
			\hline
			\texttt{skw} & baseline skewness & Signal parameter\\
			\hline
			\texttt{t0} & pulse onset time & Signal parameter\\
			\hline
			\texttt{area} & pulse area & Signal parameter\\
			\hline
			\texttt{abe} & low energy area & Signal parameter\\
			\hline
			\texttt{fm} & first momentum from \texttt{tmax} & Signal parameter\\
			\hline
			\texttt{sm} & second momentum from \texttt{tmax}& Signal parameter\\
			\hline
			\texttt{tm} & third momentum from \texttt{tmax}& Signal parameter\\
			\hline
			\texttt{p1},\texttt{p2},\texttt{p3},\texttt{p4}  & Signal area ratios in different time-bases& Signal parameters\\
			\hline
			\texttt{pnm\_peaks}  & Identified peaks of \texttt{n} detector \texttt{m} signal stored in a \texttt{TPolyMarker} & Peaks\\
			\hline
			\texttt{n}  & Number of peaks& Peak based parameter\\
			\hline
			\texttt{areap} & Area around peaks& Peak based parameter\\
			\hline
			\texttt{arealp} & Last peak area & Peak based parameter\\
			\hline
			\texttt{hlp} & Last peak height & Peak based parameter\\
			\hline
			\texttt{poslp} & Last peak time (position in pulse) & Peak based parameter\\
			\hline
			\texttt{drt} & Delta Real Time & Event Parameter\\
			\hline
			\texttt{ddt} & Delta Dead Time & Event Parameter\\
			\hline
			\texttt{vts} & Time in seconds since veto event & Event Parameter\\
			\hline
			\texttt{heticks} & Time in seconds since a very energetic event& Detector Parameter\\
			\hline

			\texttt{p1ns},\texttt{p2ns},\texttt{p3ns},\texttt{p4ns}  & Detector \texttt{n} area ratios in different time-bases& Detector Parameters\\
			\toprule

		\end{tabular}
		\caption[Naming for analysis parameters]{Naming convention for analysis parameters.} 
		\label{tab:NamingAnalysis} 
	\end{center}
\end{table}

Following the naming convention described in Section~\ref{sec:DAQNamingConventions} all pulse parameters are followed by its detector and signal number. For example, the area of the detector 2 signal 0 is identified as \texttt{area20}. All parameters also allows the \emph{sum notation} (e.g. \texttt{n3s} for total peaks in detector 3, an alias for \texttt{n30+n31}).
\paragraph{}
It is worth to note the special nature of \texttt{p} parameters \emph{sum notation}. In these parameters the sum ratio is defined as shown in expression~\ref{eq:PDefinition} with areas being the sum of the areas of the two signals from a detector. For this reason, these parameters are defined as \emph{Detector Parameters} in Table~\ref{tab:NamingAnalysis}.

\
\subsection{Automation and report notification}\label{sec:AnalysisAuto}
Launching the analysis software and checking the results can be a tedious and error prone task in a multisignal multidetector experiment. Hence, the process has been automated as much as possible.
\paragraph{}
This automation relies on the data synchronization system (see Section~\ref{sec:DataSynch}). The analysis is automatically triggered when a new run has been synchronized via inotify Linux feature~\cite{love2005kernel} and \texttt{incrond} daemon~\cite{website:incron}. The new run is analyzed, several tests are performed and reports are generated in order to detect possible malfunction and undesirable conditions such as unexpected noise. The implementation of all stability checks described in Section~\ref{sec:DataStab} in an automatic way is ongoing.
\paragraph{ }
Some other reports are currently performed for its visual inspection such as baseline parameter distribution (\texttt{dc}, \texttt{rms}, \texttt{skw}, allowing the discovery of baseline anomalies as those reported in Section~\ref{sec:Baseline}), QDC and area stability along one run and the pulse onset difference. Two report examples are described in this section: the delta trigger time distribution and the correct coincidence tagging crosschecked with the pulse onset difference.
\subsubsection{Delta trigger time distribution}\label{sec:DeltaRT}
\begin{figure}[h!]
    \begin{subfigure}[b]{0.5\textwidth}
    \includegraphics[width=\textwidth]{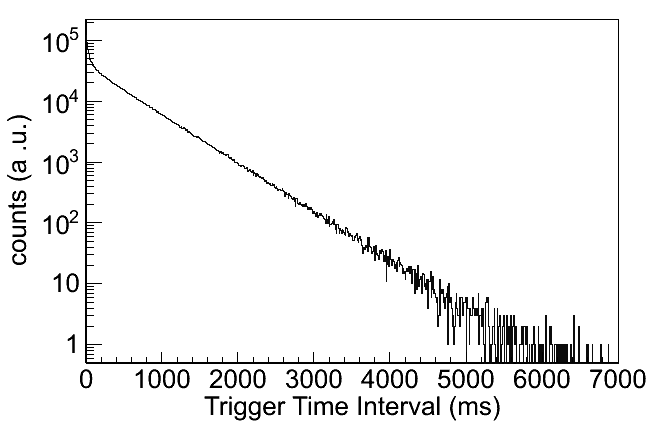}
    \caption[Delta trigger time in normal operation]{Delta trigger time in normal operation.\label{fig:DRTNorm}\\\hspace{\textwidth}}
    \end{subfigure}
    \begin{subfigure}[b]{0.5\textwidth}
    \includegraphics[width=.95\textwidth]{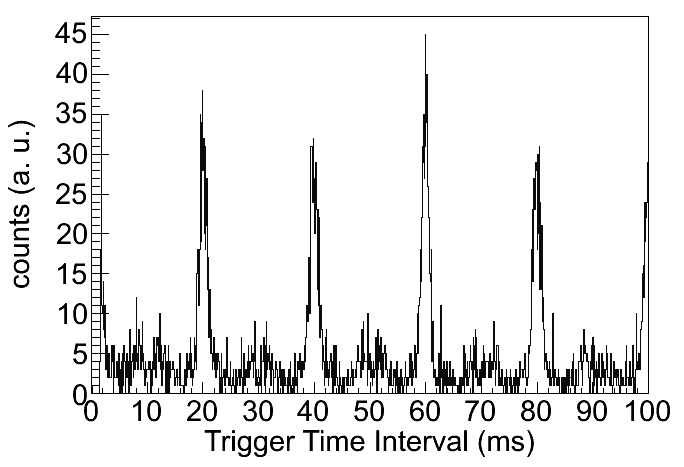}
    \caption[Delta trigger time distribution]{Delta trigger time distribution with a triggering 50 Hz noise.\label{fig:DRT50Hz}}
    \end{subfigure}
  \begin{center}
    \caption[Delta trigger time distribution]{Delta trigger time distributions.\label{fig:DeltaRT}}
  \end{center}
\end{figure}

Both radioactive background and possible dark matter signal are Poissonian processes. For such processes, the probability of measuring $x$ events having a mean rate $r$ in a time $t$ is:
\begin{equation}
	P(x,rt) = \frac{(rt)^{x}}{x!}e^{-rt}
\end{equation}
The probability of having two consecutive events in $dt$ is:
\begin{equation}
	P_c dt = P(0,rt)P(1,r dt) = r e^{-rt}dt
\end{equation}
which corresponds with an exponential decay with constant $r$. Therefore, all Poissonian processes will contribute with an exponential distribution and any departure of this behavior should be considered as effect of other kind of spurious events such as electric noise.
\paragraph{}
The typical delta time distribution can be seen in Figure~\ref{fig:DRTNorm} for a particular ANAIS-37 run. It exhibits a clear exponential decay with a mean rate of 1.8 Hz for this particular run. Additionally, it can be seen another constant at the beginning of the distribution caused by the long phosphorescence constants of the NaI(Tl) crystals~\cite{cuesta2013slow} explaining the 2.1 Hz trigger rate measured in the run.
\paragraph{ }
Typical noises have constant frequency distributions and can be easily detected in such a way. For example, the noise effect can be seen in Figure~\ref{fig:DRT50Hz}. A noise with base frequency of 50 Hz can be deduced from this figure and it corresponds to HV power supply noise, now filtered as described in Section~\ref{sec:HVNoises}.

\subsubsection{T0 vs. Pattern Unit}\label{sec:T0vsPU}
The information given by the \texttt{t0} parameter can be used to infer PMT coincidence trigger. For this reason, comparison between \texttt{t0} and the Pattern Unit can be a good crosscheck for both extraction algorithm and hardware module. The trigger from each detector is produced when the signals from both PMTs are above the threshold in the defined coincidence window. This condition can be translated in terms of \texttt{t0} and coincidence window. The crosscheck can be done in two ways: try to detect signals in both PMTs that do not give signal in the Pattern Unit (\texttt{pu}) or signals marked by the pattern unit but without a valid \texttt{t0}.
\paragraph{}
The search of events that fulfills these conditions gives information about the \texttt{t0} algorithm quality (including improper settings like \texttt{threshold\_sw}), Pattern unit hardware problems and differences between trigger and digitization lines, including noises in either line.
A high number of anomalous events can aid to identify acquisition problems: \texttt{pu} without \texttt{t0} in the waveform can reveal noises in trigger signal and \texttt{t0} coincidences without \texttt{pu} tagging could be caused by noises in digitization signal only.
\subsection{Analysis tools}
A selection of tools developed in order to aid analysis process are covered in this section. 
\subsubsection{Pulse inspection tool}\label{sec:PIT}
\begin{figure}[h!]
  \begin{center}
    \includegraphics[width=1\textwidth]{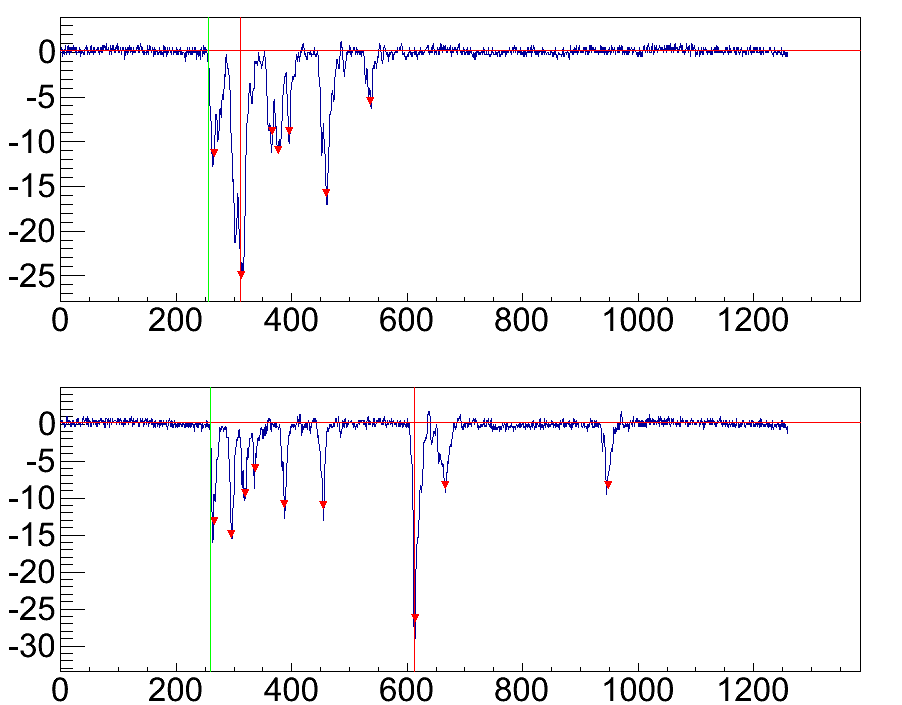}
    \caption[Pulse inspection tool]{Pulse inspection tool.\label{fig:PIT}}
  \end{center}
\end{figure}

A direct inspection of the waveform is not trivial due to the ROOT file format. For this reason a tool to visualize the signals was developed. In addition, a set of features was implemented for adding visual information and for parameter extraction check. An example of a low energy event displayed with the Pulse inspection tool can be seen in Figure~\ref{fig:PIT}. The parameters are drawn in the pulse with vertical and horizontal lines. The default values displayed are \texttt{t0} (green), \texttt{tmax} (vertical red) and \texttt{dc} (horizontal red). The identified peaks are displayed as an option.
\paragraph{}
This tool is based in a similar tool for the previous ANAIS DAQ. It was adapted to match the different data format and it evolved to cover other features and to scale-up with the increase of signals. The most important feature was the calculation of mean pulse of a previous selected list. The mean pulse can be calculated centering all pulses in the same parameter (\texttt{t0}, \texttt{tmax} or \texttt{poslp}) useful to obtain SER mean pulses (as it can be seen in Section~\ref{sec:LY}) or scintillation constants~\cite{cebrian2012background}. Another feature is an optional parameter to specify the detector to display, very useful in an increasing number of signals scenario. The tool also supports exporting the resultant pulses to allow external data processing.
\subsubsection{ScatInspector}
\begin{figure}[h!]
  \begin{center}
    \includegraphics[width=1\textwidth]{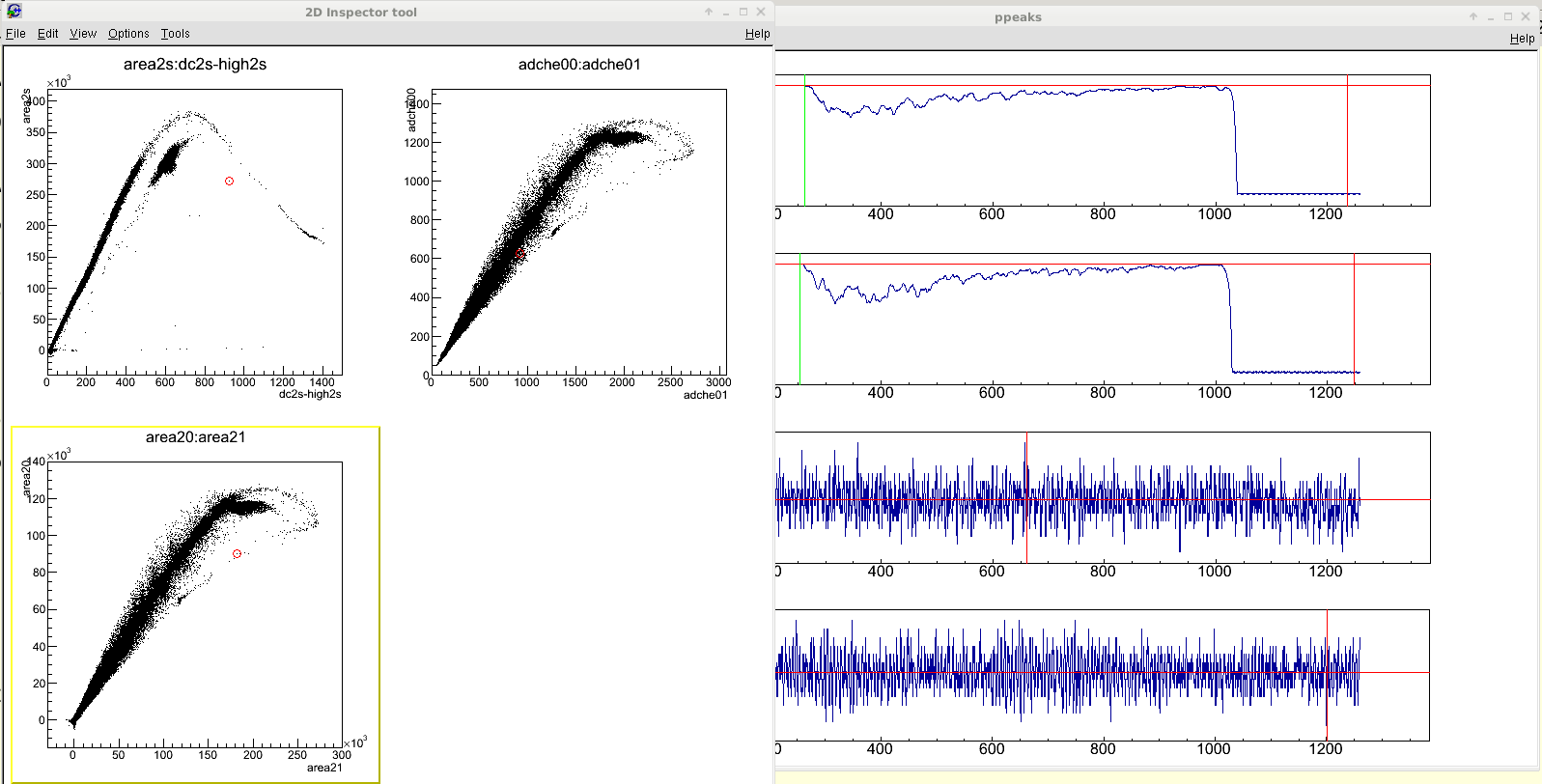}
    \caption[ScatInspector tool]{ScatInspector tool. Selecting an event in a scatter plot highlights it in other plots and shows the pulses.\label{fig:ScatInspector}}
  \end{center}
\end{figure}

Another closely related tool is ScatInspector. This tool is designed to inspect scatter plots in a very useful way. The user can select any point in a scatter plot and the waveform of each PMT is drawn using the aforementioned pulse inspector tool. In addition, ScatInspector highlights the same event in the other canvases as it can be seen in Figure~\ref{fig:ScatInspector}. This feature is very useful to explore unknown populations displaying the same event in several scatter plots. 
\subsubsection{ANAIS data browser}
\begin{figure}[h!]
 \begin{center}
\begin{subfigure}[b]{0.5\textwidth}
                \centering
                \includegraphics[width=1\textwidth]{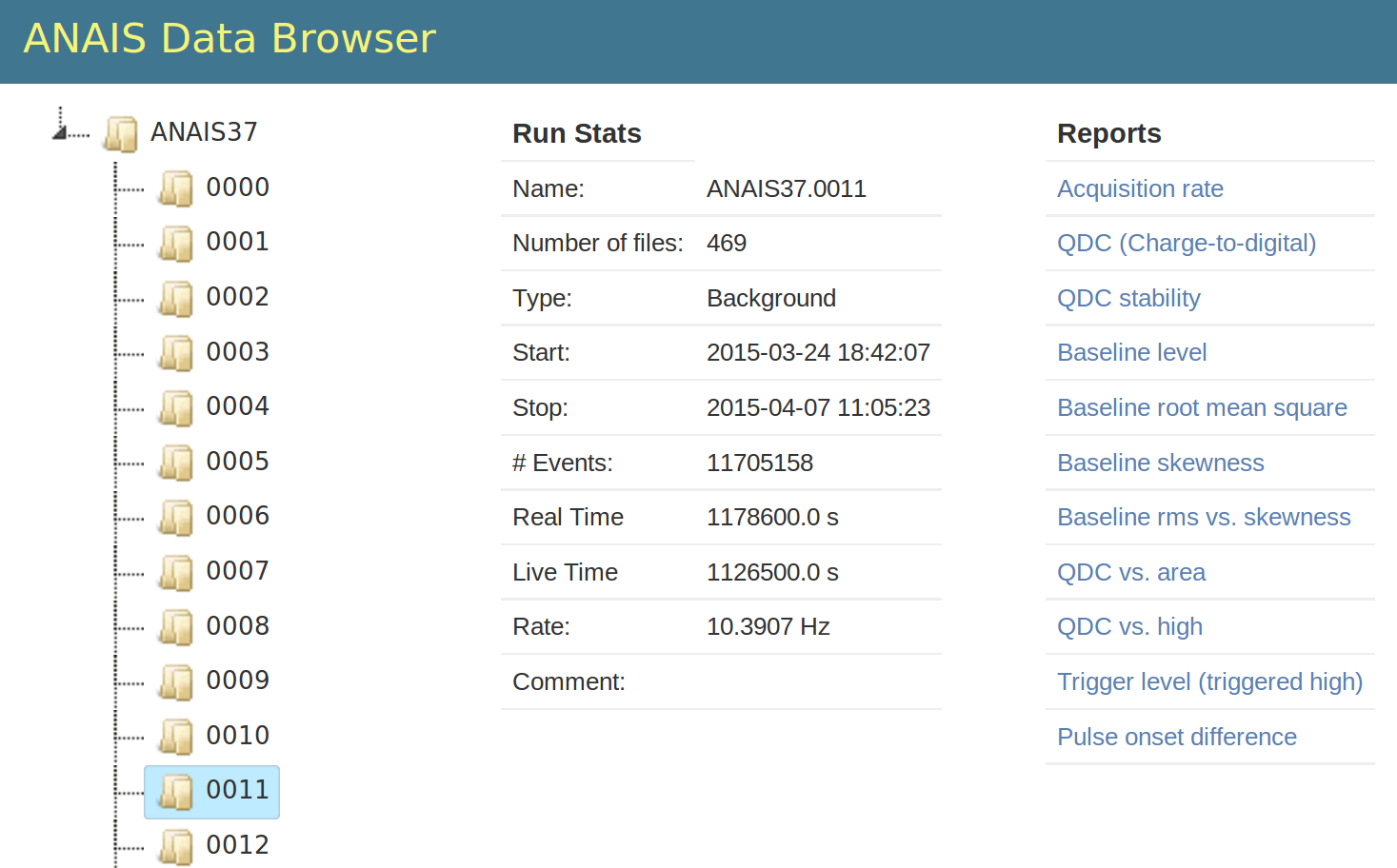}
        \end{subfigure}
        ~ 
        \begin{subfigure}[b]{0.465\textwidth}
                \centering
                \includegraphics[width=1\textwidth]{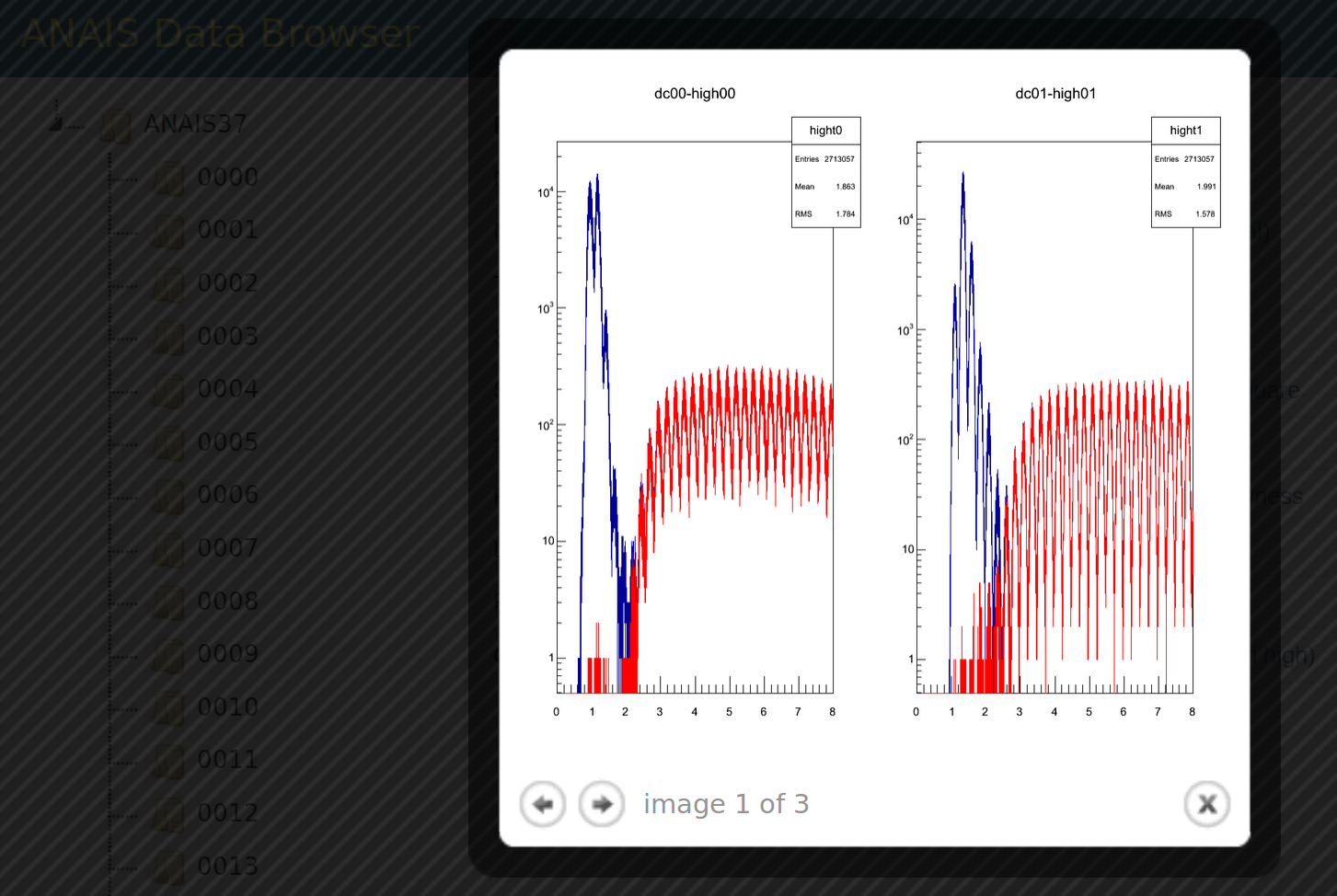}
        \end{subfigure}
        \caption[ANAISDataBrowser]{ANAIS Data Browser web application.\label{fig:ADB}}
\end{center}
\end{figure}

A web application was designed and implemented using the django framework~\cite{website:django} in order to aid the ANAIS data browsing. The graphic user interface uses a hierarchical tree with set-ups and runs in every set-up, showing all automatic information of every run in both text and graphical ways as it can be seen in Figure~\ref{fig:ADB} and the results of the automatic tests described in Section~\ref{sec:AnalysisAuto}. It also allows to follow the stability of the data along a set-up by plotting the parameters described in Section~\ref{sec:DataStab}. This interface could be expanded to give public access to the ANAIS data in the future. 

\paragraph{}
Other non-visual tools have been created such as a skim file creator in order to store filtered events together allowing to manage data cuts in a fast and easy way and making persistent lists of events, or the event counter per live time in regular time bins described in Section~\ref{sec:TimeBinning}. 
\section{Low energy NaI event selection}\label{sec:EventSelection}
The previously described analysis algorithms are essential for the energy estimation, the event selection and the noise removal. The most significant sources of noise at low energy masking the bulk NaI(Tl) scintillation are the scintillation of the surrounding materials such as quartz~\cite{amare2014study} or PMT glass and dark events in the photomultiplier. Additionally, electric noise can also be harmful despite its usual almost zero area (energy) because of their more symmetric nature (see Section~\ref{sec:Baseline}). The current selection cuts of low energy bulk scintillation events, based on previous works~\cite{cuesta2014bulk,CCUESTA} taking advantage of the experience of the operation of previous prototypes, are presented in this section.
\paragraph{ }
The different temporal behavior of these noises as compared with the bulk NaI scintillation is used as strategy to remove them. Very fast scintillation tends to pile-up several photoelectrons and the peak detection algorithm does not resolve them. This fact allows to impose the condition to number of peaks \texttt{n} in PMT signal (\texttt{n>2} in each signal). The cut also removes most of the dark random coincidences (see Section~\ref{sec:100nsNoise}) in addition to the fast scintillation.
\paragraph{ }
Next, cuts based on the specific temporal parameters \texttt{p1s} and \texttt{p2s} taking into account the NaI scintillation constant are applied. A third kind of cuts is sensitive to the symmetry of light collection between the PMT signals in a detector. The spurious events tend to be much more asymmetrical than the bulk scintillation. The applied cut is performed in the ratio between the area difference between signals versus total area (\texttt{area0-area1}\texttt{/}\texttt{area0+area1}).
\paragraph{ }
The distribution of these parameters and the efficiency of the cuts have been characterized with genuine $^{109}Cd$ events in very long calibrations storing the very low energy events. This process takes advantage of the conditional data storing described in the previous chapter. 
\paragraph{ }
In addition to the noise removal, other cuts are performed in the low energy region in the search for dark matter interaction. The events correlated with muons and other high energy events must be avoided because of their non-WIMP related nature (see Chapter~\ref{sec:Veto}, especially Section~\ref{sec:MuNaIEvt}, for further information about this kind of events). A time after any muon interaction will be rejected from the dark matter analysis because of this reason. The afterglow caused for very energetic events is also rejected. These two cuts are easily performed by using the \texttt{vts} and \texttt{heticks} described earlier. 
\paragraph{ }
These cuts will be reviewed with the understanding of the PMT events. This understanding will be highly improved with the ongoing data taking of a blank module (see Section~\ref{sec:ANAIS37}) allowing to characterize and reject as much PMT events as possible and helping to reach an energy threshold of 1 keVee before filtering. 

\chapter{Data acquisition characterization}\label{sec:FullCharac}
The whole data acquisition system was tested during its development at the University of Zaragoza and has been characterized with the ANAIS-25 and ANAIS-37 prototypes at the LSC. The characterization of the time parameters such as real time, live time and duty cycle are presented in Sections~\ref{sec:RealTimeTest}, \ref{sec:DeadTimeMeas} and \ref{sec:LiveTimeMeas} respectively.
\paragraph{ }
In addition, the hardware trigger efficiency has been studied. First, the energy calibration is presented in Section~\ref{sec:Calibration}. Next, the single electron response (SER) has been characterized onsite and compared with the characterized at the University of Zaragoza PMT test bench in Section~\ref{sec:SERLP}. Additionally the SER trigger efficiency has been estimated. The light collection has been measured for ANAIS-25 and ANAIS-37 combining the SER with the calibration. Finally, the hardware trigger efficiency has been simulated with different coincidence time windows taking into account the previously measured light collection and SER trigger efficiency. The simulation result has been crosschecked with measurements for very low energy events as it can be seen in Section~\ref{sec:HWTriggEff}. The effect in random coincidence triggers due to the coincidence window reduction can be seen in Section~\ref{sec:100nsNoise}.
\section{Real time clocks}\label{sec:RealTimeTest}
Real time clocks play a key role in the ANAIS experiment. The time binning needed to test the annual modulation and some other stability parameters (see Section~\ref{sec:TimeBinning}) rely on real time clocks. The DAQ system has two clocks: a hardware 50 ns tick clock (see Section~\ref{sec:TimeScalers}) and the software clock with 1 ns nominal resolution. The accuracy and stability of these clocks are characterized in this section. PC clocks are known to be irregular and unreliable. The Network Time Protocol~\cite{mills1991internet} (NTP) was used to mitigate the effect allowing clocks to be synchronized within a few milliseconds.
\subsection{32-bit latched vs. 64-bit non-latched counter}\label{sec:RT-RT0}
\begin{figure}[ht!]
  \begin{center}
    \includegraphics[width=.6\textwidth]{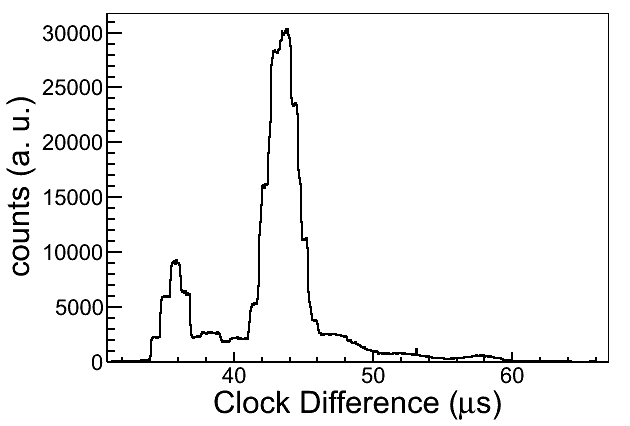}
    \caption[Non-latched/latched scalers time difference]{\label{fig:RT-RT0}Non-latched/latched scalers time difference.}
  \end{center}
\end{figure}

The real time clock is stored in two different scalers: a 32-bit latched scaler and a 64-bit non-latched scaler. The real time and live time measurement strategy has been presented in Section~\ref{sec:TimeScalers}. The improvement of the use of a latched scaler is reviewed in this section as well as the reasons to keep the old 64-bit scaler.
\paragraph{ }
The time difference of the two scalers can be observed in Figure~\ref{fig:RT-RT0}. The latched scaler stores via hardware the counter value at the trigger as opposed to the value of the non-latched scaler that keep increasing the value. The time difference accounts the time needed to wake-up the DAQ software and the time needed to download the scaler value. This time difference is always positive showing the non-latched nature of the 64-bit scaler. Additionally, the time difference distribution shows a mean of 43 $\mu$s and exhibits a binning pattern of 1 $\mu$s revealing a 1 MHz clock somewhere in the communication process with a minimum value of the order of 20 $\mu$s, fully compatible with the IRQ latency measured in Section~\ref{sec:IRQvspoll}.
\paragraph{ }
The more precise latched measurement of the trigger is chosen for the real time event tagging, and its value is referred as real time clock in the following. Nevertheless, the non-latched 64-bit scaled was maintained in operation because it gives information about the DAQ software response. Additionally, its higher bit depth is useful to hold the long temporal information in the 32 most significant bits.
\subsection{NTP system}\label{sec:NTP}
\begin{figure}[h]
     \begin{center}
\begin{subfigure}[b]{0.5\textwidth}
                \centering
                \includegraphics[width=\textwidth]{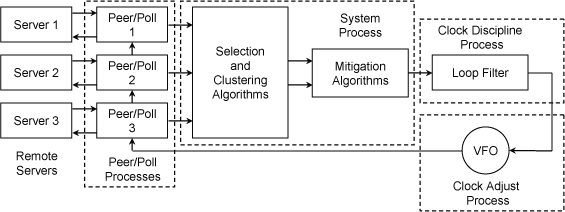}
                \caption{NTP global architecture.}
                \label{fig:NTPGlob}
        \end{subfigure}%
        ~ 
        \begin{subfigure}[b]{0.3\textwidth}
                \centering
                \includegraphics[width=\textwidth]{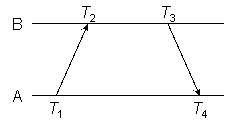}
                \caption{NTP Timestamps.}
                \label{fig:NTPtimestamp}
        \end{subfigure}
	\caption[NTP architecture]{NTP architecture. From~\cite{website:ntpdoc}.}\label{fig:NTPArchitecture}
\end{center}
\end{figure}

The NTP architecture is shown in Figure~\ref{fig:NTPArchitecture}. The PC behaves as client for NTP servers. Packets are exchanged between the client and server using NTP On-Wire Protocols. These protocols are resistant to lost, replayed or spoofed packets. The information obtained passes sanity tests\footnote{It checks packet, date and time format, packet duplication and distance threshold among other tests.}. This packet interchange obtains timestamps (see Figure~\ref{fig:NTPtimestamp}): the origin timestamp T1 upon departure of the client request, the receive timestamp T2 upon arrival at the server, the transmit timestamp T3 upon departure of the server reply, and the destination timestamp T4 upon arrival at the client. These timestamps are used to calculate the clock offset and roundtrip delay samples:
\refstepcounter{equation}\label{eq:NTPOffsetDelay}
\begin{align}
offset = [(T2 - T1) + (T3 - T4)] / 2 \tag{\theequation a}\\
delay = (T4 - T1) - (T3 - T2)\tag{\theequation b}
\end{align}
This offset variable is a good approximation to the clocks distance between client and server if roundtrip delay is symmetric, due to the cancellation given its definition. The delay variable is a good approximation to the roundtrip delay if the clock offset is small.
\paragraph{ }
Therefore, the offset calculation quality depends on the quality of the network between client and server because it assumes symmetric propagation delays. Asymmetric routes, network congestion or long route paths can cause a deterioration of synchronization accuracy as it can be seen in next section. 
\paragraph{ }
This calculations for one or more servers are processed by mitigation algorithms selecting the offset and delay samples most likely to produce accurate results and choosing one of them as the system peer and producing the final offset. This offset is used by clock discipline algorithm to adjust the PC clock time and frequency. The clock offset and frequency are recorded in the \texttt{loopstats} file and it is used in this work as quality reference.
\begin{figure}[h]
  \begin{center}
    \includegraphics[width=.6\textwidth]{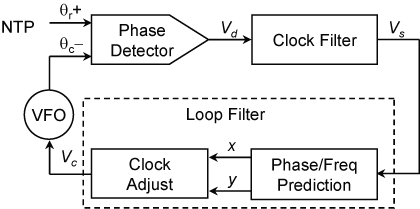}
    \caption[NTP Clock Discipline Algorithm]{\label{fig:NTPdiscipline}NTP Clock Discipline Algorithm. From~\cite{website:ntpdoc}.}
  \end{center}
\end{figure}
\paragraph{ }
It is important to note that NTP was conceived to minimize both offset and network use. For this reason, the clock adjustment is not performed in a naive way by subtracting the offset: it tries to regulate the local clock taking into account the past data and guessing its future behavior. A block diagram of this clock discipline algorithm is shown in Figure~\ref{fig:NTPdiscipline}. The timestamp of a reference clock of remote server is compared with the timestamp of the system clock, represented as a variable frequency oscillator (VFO), to produce a raw offset sample $V_d$. Offset samples are processed by the \emph{clock filter} to produce a filtered update $V_s$. The \emph{loop filter} implements a type-2 proportional-integrator controller (PIC). The PIC can minimize errors in both time and frequency using predictors $x$ and $y$, respectively. The \emph{clock adjust} process samples these predictors once each second to produce the system clock update $V_c$.

\subsection{Real Time clocks characterization}

\begin{figure}[h!]
  \begin{center}
	\begin{subfigure}[b]{0.5\textwidth}
        \centering
	\includegraphics[width=\textwidth]{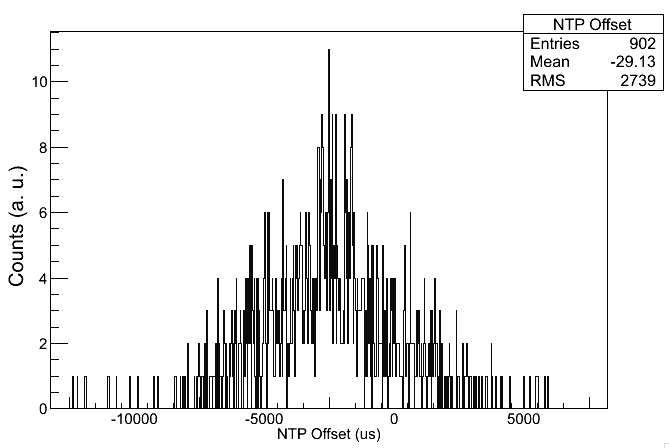}
    	\caption{\label{fig:NTPOffsetHist}NTP offset (Default servers).}
	\end{subfigure}%
	~ 
	\begin{subfigure}[b]{0.5\textwidth}
        \centering
	\includegraphics[width=\textwidth]{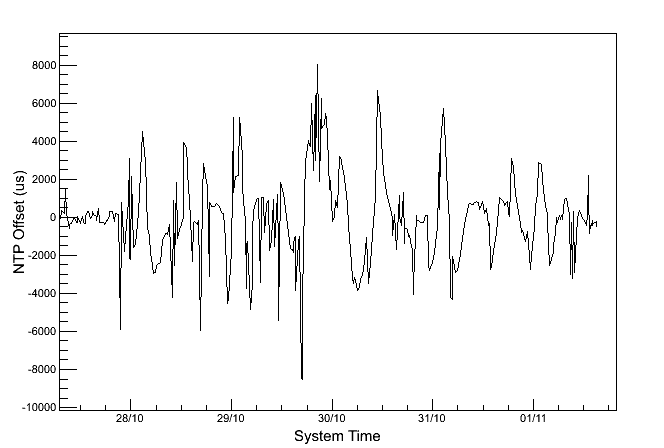}
    	\caption{\label{fig:NTPOffsetvsClock}NTP offset over time (Default servers).}
	\end{subfigure}

	\begin{subfigure}[b]{0.5\textwidth}
	\centering
	\includegraphics[width=\textwidth]{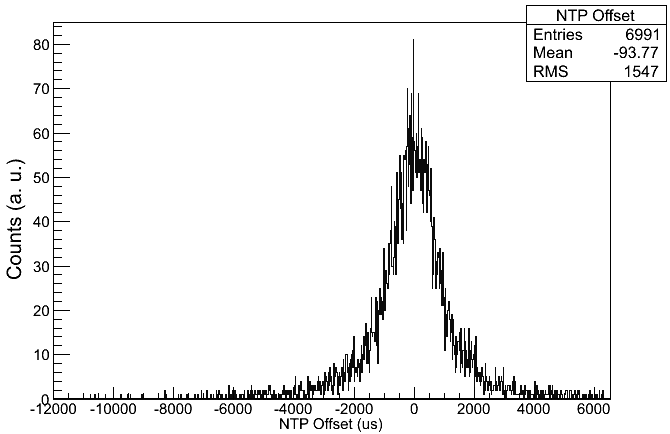}
    	\caption{\label{fig:NTPOffsetHistes}NTP offset (Spanish servers).}
	\end{subfigure}%
	~ 
	\begin{subfigure}[b]{0.5\textwidth}
        \centering
	\includegraphics[width=\textwidth]{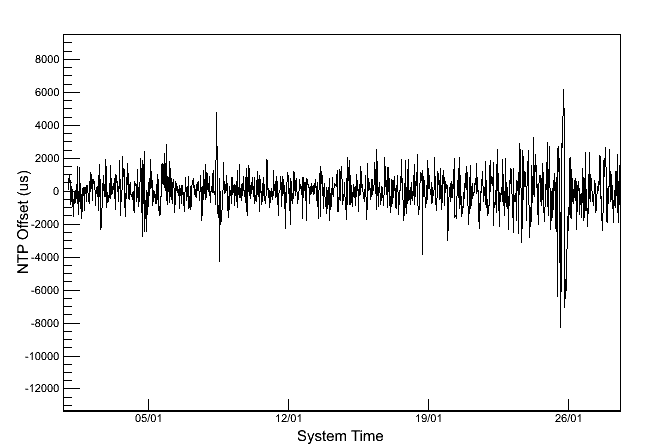}
    	\caption{\label{fig:NTPOffsetvsClockes}NTP offset over time (Spanish servers).}
	\end{subfigure}

        \caption[Default NTP servers vs. Spanish pool NTP servers]{NTP Offset statistics:  default NTP servers (top) vs. Spanish pool NTP servers (bottom).\label{fig:NTPOffset}}
  \end{center}
\end{figure}

The reliability of the PC System Clock can be estimated by collecting NTP statistics. The \texttt{loopstats} file contains date in Modified Julian Calendar (MJC), time in seconds (millisecond precision) from the beginning of the day and clock and frequency offsets used to compensate clock drifts and frequency wander (in seconds with nanosecond precision and parts per million respectively). The offset distribution and its behavior over time can be seen in Figure~\ref{fig:NTPOffset}. The top Figures show the offset distribution (\ref{fig:NTPOffsetHist}) and offset over time (\ref{fig:NTPOffsetvsClock}) with default Linux distribution NTP servers. It can be seen that the offset distribution gives a FWHM of 2.7 ms with many offsets in the order of ten milliseconds, worse than NTP claim of few milliseconds~\cite{mills1991internet}. As seen in previous section, the accuracy of NTP synchronization is dependent on the quality of network. A new NTP configuration was used with Spanish NTP pool (\texttt{es.pool.ntp.org}). A closer pool of servers could improve network quality connection. The result can be seen in the bottom figures: offset distribution (\ref{fig:NTPOffsetHistes}) and offset over time (\ref{fig:NTPOffsetvsClockes}). Offset distribution shows a better FWHM of 1.5~ms with much more statistical significance and confirming the effect of long route paths between client and server in NTP protocol. The frequency offset is almost constant giving a value of 30 ppm due to the slow nature of the PC system clock. It is worth to note that this offset is usually less than a very few milliseconds, so the PC clock can be used as reliable clock in the long term using NTP synchronization and the offset logs are stored to detect bigger differences due to network congestion or network failures.

\begin{figure}[h]
     \begin{center}
	%
        \begin{subfigure}[b]{1\textwidth}
       		\centering
	        \includegraphics[width=0.6\textwidth]{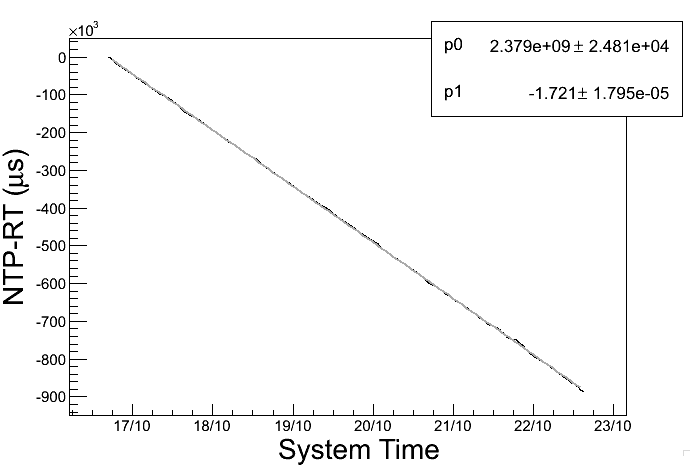}
		\caption[Hardware clock drift Fit]{\label{fig:RTwNTPFit}Hardware clock drift linear fit.}
        \end{subfigure}

       \begin{subfigure}[b]{1\textwidth}
                \centering
                \includegraphics[width=0.6\textwidth]{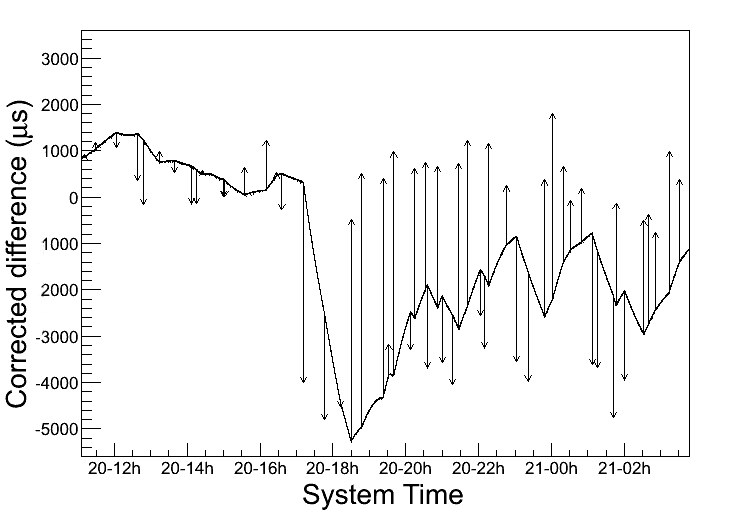}
                \caption[NTP effect in the clock difference]{NTP offset effect.}
                \label{fig:RTNTPOffsetArrow}
        \end{subfigure}
        \caption[NTP Clock vs. Hardware Clock]{NTP Clock vs. Hardware Clock: clock differences, drift estimate, difference subtracting drift and the effect of NTP offset.}\label{fig:RTNTPOffset}
\end{center}
\end{figure}

\begin{figure}[h!]
  \begin{center}
    \includegraphics[width=\textwidth]{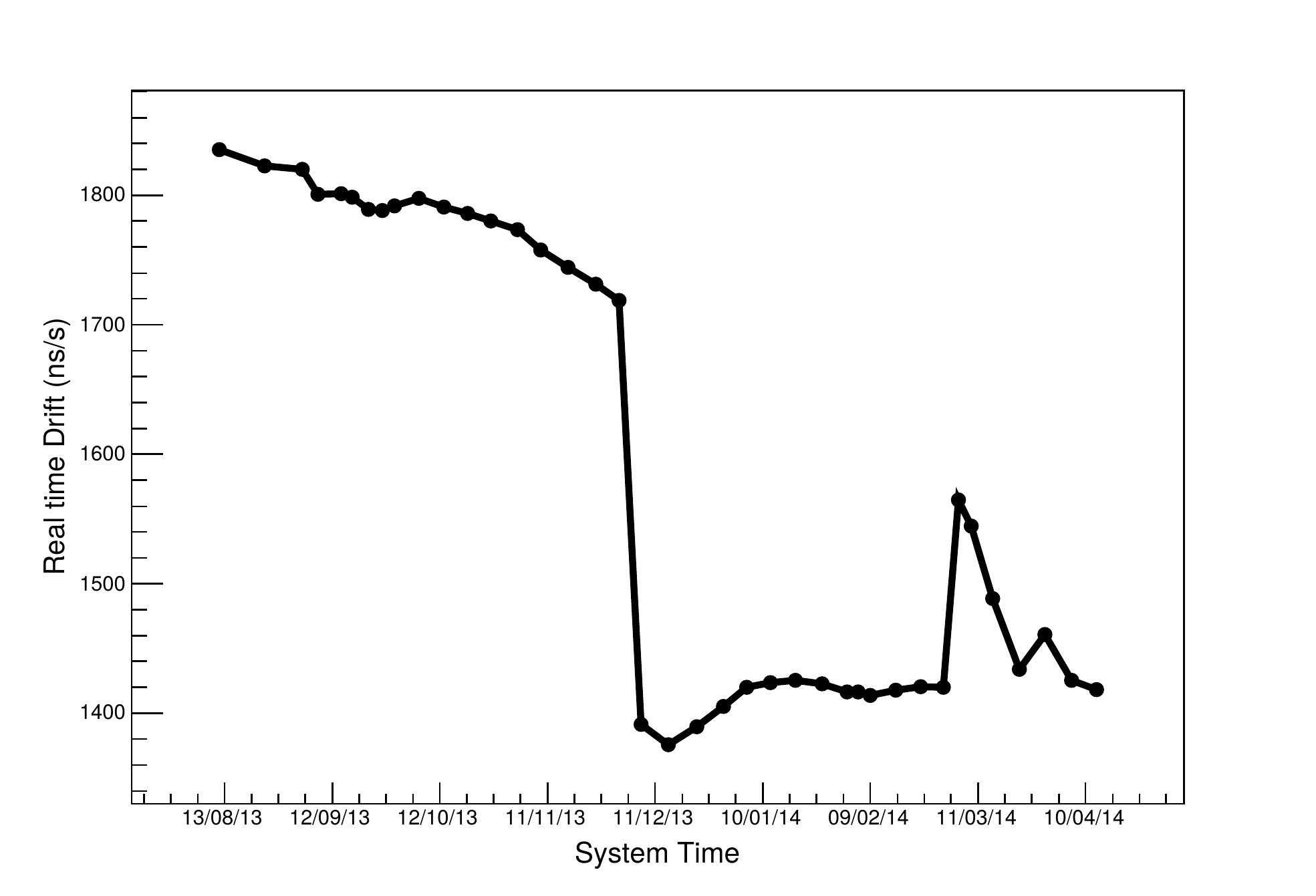}
    \caption[Drift over time]{\label{fig:MetaDrift} Real time drift over time.}
  \end{center}
\end{figure}

\paragraph{}
The Hardware clock behavior using NTP controlled PC clock as reference can be seen in Figure~\ref{fig:RTwNTPFit}. It shows a noticeable drift. This drift is very linear, in this case with a slope of 1.72 $\mu$s/s: a systematic drift of 1.7 ppm. Such a drift can be discounted in order to see the relative clock oscillations as can be seen zoomed in Figure~\ref{fig:RTNTPOffsetArrow}. The peak-to-peak oscillation is in the order of few milliseconds and it is compatible with the NTP corrections as it can be observed in NTP statistical log (Figure~\ref{fig:NTPOffsetHistes}). These data were combined with typical Real Time data from DAQ in order to explain such an oscillation. The result of this combination can be observed in Figure~\ref{fig:RTNTPOffsetArrow}. It can be seen the offset represented as an arrow. The length of this arrow is the NTP offset (in $\mu$s) and it points up or down if it is negative or positive respectively. The figure shows that the slope changes are correlated with the offsets in both time and amplitude. This offset driven behavior also confirms the PIC nature of the Clock Discipline Algorithm seen in Section~\ref{sec:NTP}. Therefore, the most important component of the oscillation of the difference between real time clocks can be attributed to NTP. It shows a good stability for the hardware clock and its instabilities must be below the millisecond range.  
\paragraph{}
This can be used to monitor the hardware real time clocks in the long term as it can be seen in Figure~\ref{fig:MetaDrift}. This figure shows run drifts over time using the data from eight months runs. It can be observed two big shifts in the measured drift. Both changes happened after maintenance tasks involving adding or removing VME bus modules.

\paragraph{}
This clock characterization gives as conclusions:
\begin{easylist}[itemize]
& Hardware clock has a systematic drift of very few ppm
& PC clock is synchronized via NTP with a precision of few milliseconds 
& Both of them are stable enough for their purpose:
&& Hardware clock can be used as timestamp of events and for computing Live Time within a run. The typical drift in a week run will be of the order of one second (1.02 s/week given 1.7 $\mu$s/s)
&& PC clock can be used as absolute clock with millisecond precision
& They can be monitoring each other in order to control long term stability
\end{easylist}

\section{Dead time characterization}\label{sec:DeadTimeMeas}
\begin{figure}[h]
  \begin{center}
    \includegraphics[width=.8\textwidth]{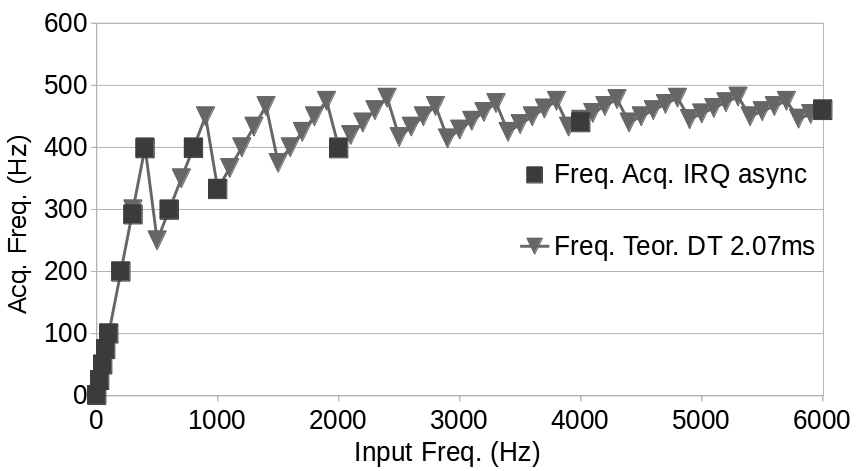}
    \caption[Dead Time frequency behavior]{\label{fig:DTfrec}Measured acquisition frequency behavior (squares) and fixed dead time model (triangles, see text) versus input trigger frequency.}
  \end{center}
\end{figure}

The system dead time has been characterized by using the live time and dead time counters previously described in Section~\ref{sec:TimeScalers}. The counters behavior was crosschecked using a pulser as trigger. The input frequency was varied in order to observe the acquisition frequency behavior. The acquisition frequency $F_a$ was modeled as:
	\begin{equation} 
		F_a = \frac{1}{\ceil*{DT/T_i}T_i}\label{eq:AcqFreq}
	\end{equation}
	taking into account the periodic nature of the input signal of period $T_i$. This expression considers the acquisition dead time as a fixed time, in which the acquisition does not accept any new event: $\ceil*{DT/T_i}$ are the number of periods in which the input trigger will be rejected because of the dead time $DT$, giving a total acquisition period of $\ceil*{DT/T_i}$. The comparison between the predicted and the measured acquisition frequency is observed in Figure~\ref{fig:DTfrec} using the mean of the measured dead time as $DT$. It shows a good agreement with a subtle difference explained by the non-constant nature of the dead time.

Once tested, the dead time measurements were used to compare several set-ups and configurations in order to minimize it.
\subsection{Synchronous vs. asynchronous data storing}\label{sec:AsynchTest}
\begin{figure}[h]
  \begin{center}
    \includegraphics[width=.6\textwidth]{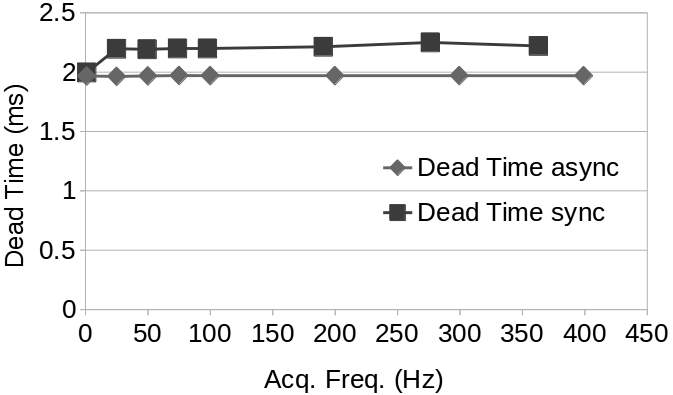}
    \caption[Dead time for synchronous and asynchronous data storing at different acquisition frequencies.]{\label{fig:DTAsync}Dead time for synchronous and asynchronous data storing at different acquisition frequencies.}
  \end{center}
\end{figure}

The two different strategies used to store the data seen in Section~\ref{sec:asynch_store} were tested with the aforementioned pulser trigger in order to measure the dead time differences. Figure~\ref{fig:DTAsync} shows these measurements displaying up to 10\% less dead time at high acquisition frequencies for the asynchronous data storing. The use of the asynchronous data storing isolates the data acquisition from the unwanted latencies of the filesystem writing.
\paragraph{}
The asynchronous strategy was selected to its use in ANAIS experiment given their better performance at high trigger frequency. In this way, the advantage will have a positive impact at calibration time and in the afterglow episodes after a very high energy events.

\subsection{IRQ waiting, board order reading and dead time}\label{sec:IRQvspoll}
The IRQ trigger strategy was preferred over polling because of the noise associated to the latter strategy (see Section~\ref{sec:VMEPoll}). However, the process of waiting to the interrupt could increase the dead time. An evaluation of the order of magnitude of the cost of this process was performed by instrumenting the kernel module from CAEN and calculating the time difference between the interrupt time of arrival to kernel space and the time when the kernel will be passed to the userspace. This time was always between 10-15 $\mu$s~\cite{MATFM}. Having in mind the MATACQ board conversion time ($\simeq 650 \mu s$), the IRQ waiting latency can be absorbed in this time.
\paragraph{ }
The order of the negotiation and data reading of the modules were rearranged for this optimization and IRQ and polling strategies gave similar dead time results. This rearrangement was also used to make use of the MATACQ conversion time to read the data of the other boards, giving as a result a lower dead time.
\subsection{Live time clock versus dead time measured with latched scaler}
\begin{figure}[!h]
	 \begin{subfigure}[b]{0.50\textwidth}
            \centering
	    \includegraphics[width=1\textwidth]{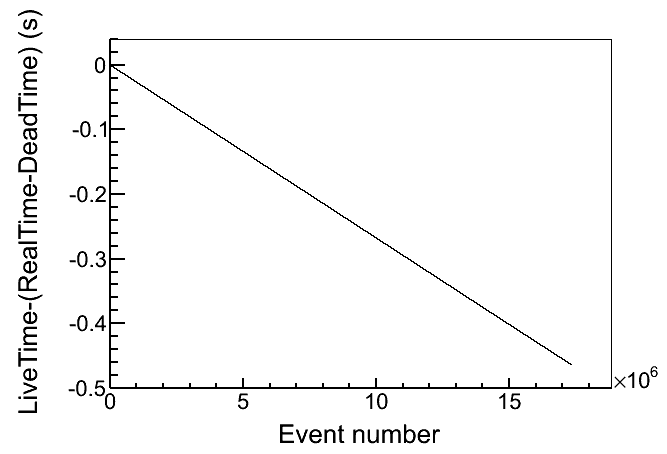}
	    \caption{\label{fig:LT-RT+DT} }
	 \end{subfigure}
	 ~
	 \begin{subfigure}[b]{0.50\textwidth}
            \centering
	    \includegraphics[width=1\textwidth]{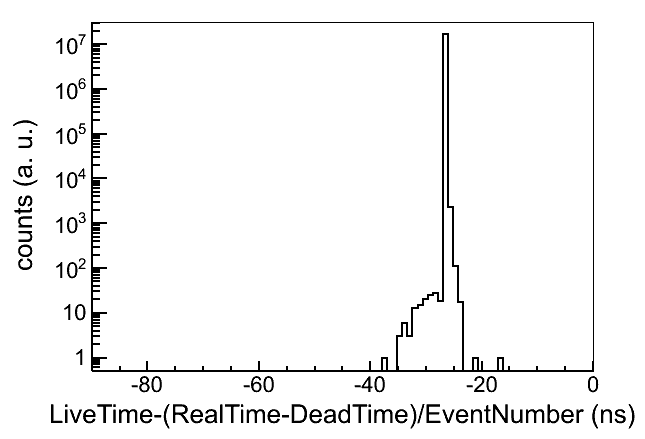}
	    \caption{ \label{fig:LT-RT+DTperEvt}}
	 \end{subfigure}
  \begin{center}
	  \caption[Latched vs. non-latched DT comparison.]{\label{fig:LTLatchvsNolatchs}Time difference between live time measured using live time counter.}
  \end{center}
\end{figure}

The previous results were obtained using the non-latched counter and the live time clock measuring the live time. The latched scaler was acquired and the comparison of the two methods was performed in order to choose the better. Figure~\ref{fig:LT-RT+DT} shows the difference between the live time measured with non-latched counter using the live time clock and the dead time measured with the latched scaler as described in Section~\ref{sec:TimeScalers}, and taking into account that $RealTime=LiveTime+DeadTime$. The systematic (very low, 0.5 parts per million) drift of the non-latched counter can be explained with a skipped clock tick starting the live time again. This can be observed in Figure~\ref{fig:LT-RT+DTperEvt}: it shows the live time difference between clocks divided by the number of the event. The mean of such a distribution tends to be 25 ns, a half of the clock tick (50 ns), probing the equal probability of lose or count the starting live time clock tick. This test evinces a good agreement between the two methods revealing a very subtle systematic effect using the live time separate counter clock. Therefore, the latched scaler measurement was considered as live time/dead time reference and it is used in the following.
\subsection{Dead time measurements and prospects }
\begin{figure}[!h]
  \begin{center}
	 \begin{subfigure}[b]{0.48\textwidth}
            \centering
	    \includegraphics[width=1\textwidth]{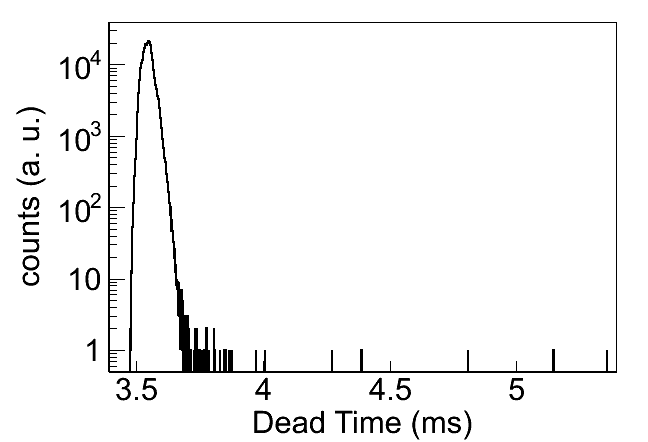}
	    \caption{\label{fig:DTdistA25} }
	 \end{subfigure}
	 ~
	 \begin{subfigure}[b]{0.48\textwidth}
            \centering
	    \includegraphics[width=1\textwidth]{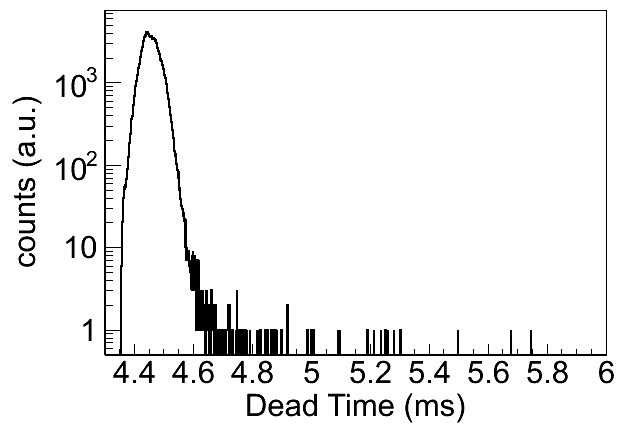}
	    \caption{ \label{fig:DTdistA37}}
	 \end{subfigure}
    \caption[Dead time event distribution.]{\label{fig:DTdist}Event dead time distribution for ANAIS-25 (a) and ANAIS-37 (b) set-ups.}
  \end{center}
\end{figure}

\begin{figure}[!h]
  \begin{center}
	    \includegraphics[width=.6\textwidth]{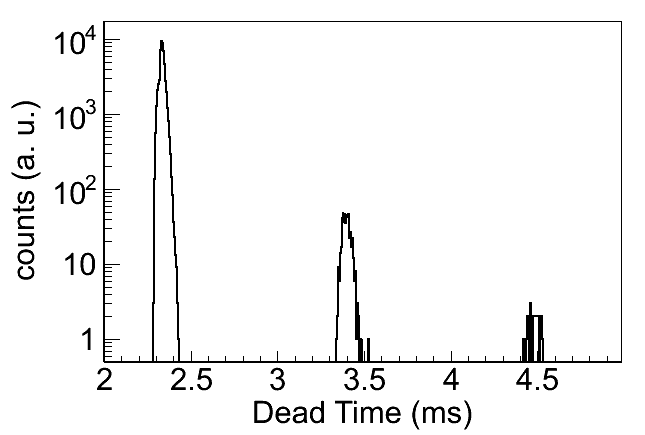}
    \caption[Dead time event distribution with trigger condition downloading.]{\label{fig:DTdistA37Blank}Event dead time distribution with trigger condition for transfer and data storing.}
  \end{center}
\end{figure}

The dead time has been measured thanks to the latched scaler for every individual event. The distribution of this dead time for an ANAIS-25 run can be seen in Figure~\ref{fig:DTdistA25} with a mean value of 3.5 ms. The dead time has overall low dispersion but exhibits some higher values due to the non-deterministic nature of a non-realtime kernel but with very little contribution to the total.
\paragraph{ }
The mean value of the dead time for ANAIS-37 is 4.5 ms. The increase of dead time is due to the additional MATACQ board and the subsequent penalty for its data reading. It is worth to note that these prototypes have twice digitizer channels per detector than the final set-up due to the high energy signals digitization. The dead time for the ANAIS full experiment will increase with two additional MATACQ boards and it will be of the order of 5 ms taking into account the use of 14-bit MATACQs which allow the use of 64-bit block mode transfer and the use of a VME/PCI bridge with CONET2 protocol reducing the data reading time.
\paragraph{ }
The 5 ms dead time per event would give a 5\% of dead time with a 10 Hz acquisition frequency. Two strategies will be used in the final experiment to reduce its impact: reduce the coincidence window and configure conditional acquisition.
\paragraph{ }
The change to the coincidence window from 200 ns to 100 ns would reduce the dark rate random coincidence by half. The rate reduction using such a coincidence window is studied in Section~\ref{sec:100nsNoise}. The study of the hardware trigger efficiency in Section~\ref{sec:HWTriggEff} also quantifies the expected effect of a 100 ns window in the hardware trigger efficiency of the NaI scintillation events.
\paragraph{ }
The conditional acquisition stores data depending on configurable runtime conditions as described in Section~\ref{sec:DAQDesign}. This technique can be applied to reduce the dead time by avoiding the transfer of irrelevant data such as baselines from non-triggered detectors. The selection can be done by checking the trigger pattern (via Pattern Unit module) and read the triggered data only. This strategy would be feasible due to the high hardware trigger efficiency (see Section~\ref{sec:HWTriggEff}). The conditional data acquisition has been already tested in long $^{109}Cd$ calibrations performed to have statistical significance at low energies with big improvements in both dead time and disk usage terms. It has been also tested in very high acquisition rate test run with radon inside the shielding (see Section~\ref{sec:N_2_Flux}). The condition for trigger was used in order to save disk space and have less dead time. The distribution of the dead time with such conditions is shown in Figure~\ref{fig:DTdistA37Blank}. It can be observed a multimodal distribution accounting for the events that transfer data for one, two or three MATACQ boards. The vast majority of the events are \emph{single-hit} events and for this reason the mean dead time is roughly the dead time for one MATACQ (and the rest of modules such as pattern unit, QDC and counters) giving a value of 2.4~ms. 

\paragraph{}
This dead time should be reduced by installing the new 14-bit MATACQ board which allows the faster A32/D64 VME transfer mode giving a mean below 2 ms even for the full ANAIS experiment taking into account the fact that most triggers are \emph{single-hit} events and hence only one MATACQ data board has to be transferred.  
\paragraph{}
Anyway, the final experiment dead time should be of the order of the 2\% taking a conservative 2.4 ms per event and a trigger rate of 1 Hz per detector (see Figure~\ref{fig:SetupsRate} to see the data supporting the last assumption).
\section{Live time, dead time and down time}\label{sec:LiveTimeMeas}
\begin{figure}[h!]
     \begin{center}
	\begin{subfigure}[b]{.7\textwidth}
		\centering
                \includegraphics[width=\textwidth]{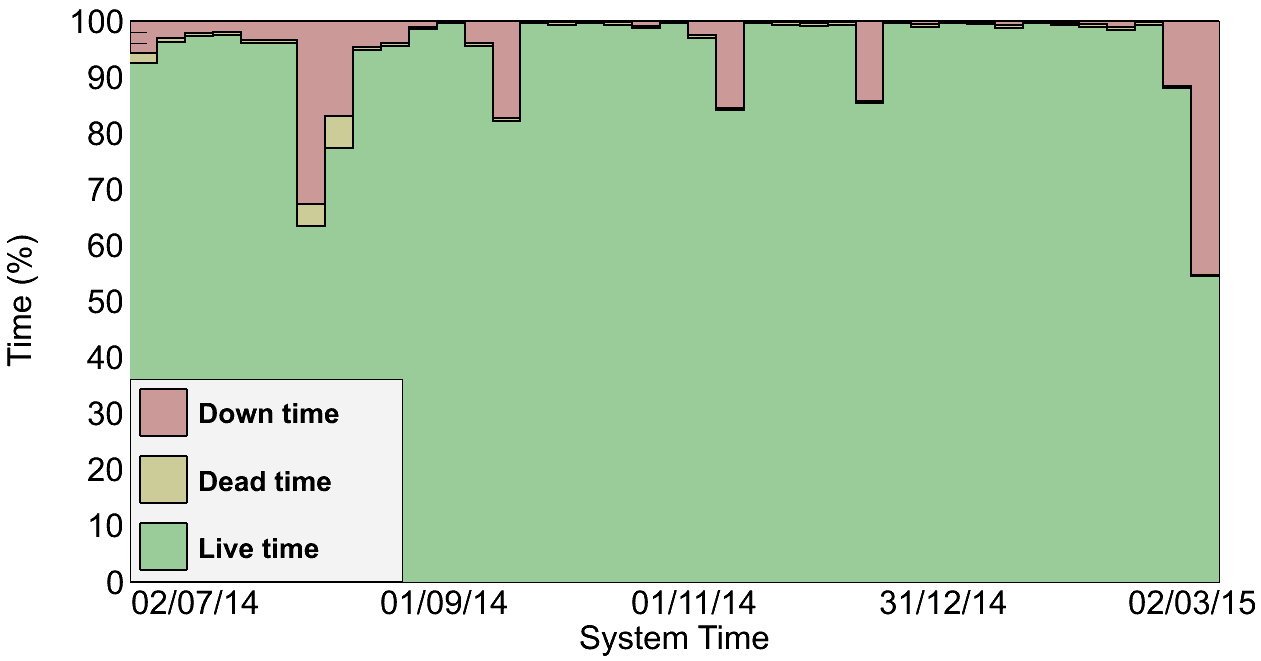}
		\caption{\label{fig:DTDTLTA25_III}}
	\end{subfigure}

	\begin{subfigure}[b]{.7\textwidth}
		\centering
                \includegraphics[width=\textwidth]{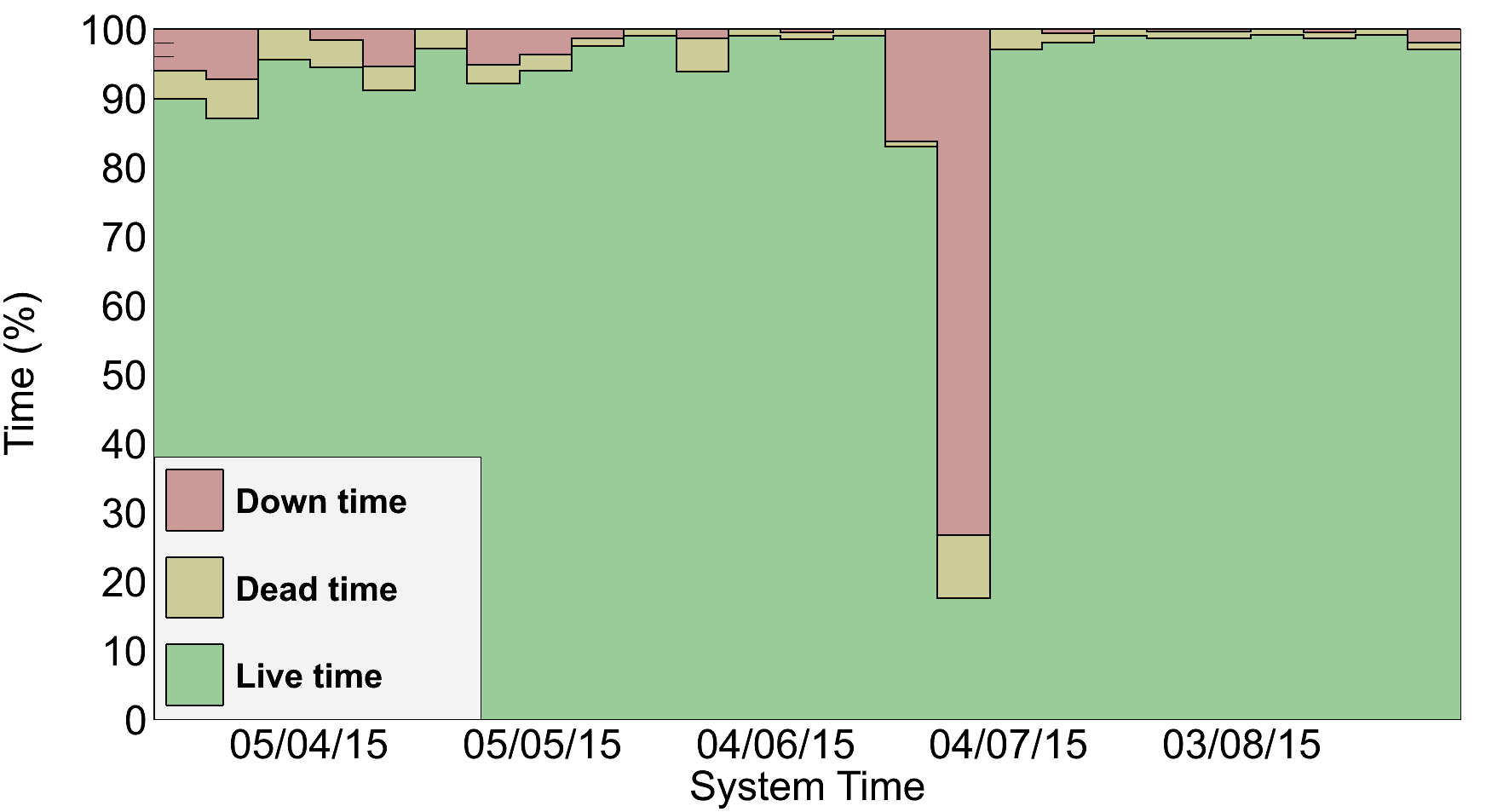}
		\caption{\label{fig:DTDTLTA37}}
	\end{subfigure}
                
	\caption{Down time, dead time and live time over system time in weekly binned basis for ANAIS-25 (a) and ANAIS-37 (b). \label{fig:DTDTLT}}
\end{center}
\end{figure}

The live time is a key parameter in the ANAIS experiment because of its impact in the exposure. In this section, the evolution of the live time along a typical series of runs is studied in order to test the validity of the whole system and routine, including the scheduled stops for system calibration.
\paragraph{}
The evolution of the live time of ANAIS-25  and ANAIS-37 set-ups can be seen in Figure~\ref{fig:DTDTLT} showing also the dead time and the down time contribution to total time. The most important down time in the ANAIS-25 set-up (see Figure~\ref{fig:DTDTLTA25_III}) was due to the maintenance labors in the LSC electrical facility. The other downtimes are due to the installation of the veto system and to long pulse calibrations. Anyway, the total live time percentage in this run was over 95\% of the time and the dead time a 0.8\% of the total time. 
\paragraph{}
The currently running ANAIS-37 set-up (see Figure~\ref{fig:DTDTLTA37}) shows only a big stop due to the change of D2 PMT covers. The total live time percentage was about 91\%, the down time a 5\% and dead time of a 3\%. This dead time increase is due primarily to the rate increase caused by the PMT light exposure and for a LSC $N_2$ nitrogen flux cut (see Section~\ref{sec:N_2_Flux}). It is worth to note that the reduction of live time for events related to muon and high energy events has not been accounted in this analysis, but this reduction can be estimated in the order of 1\% cutting one second after every such event.
\paragraph{ }
Anyway, these live time percentages are above the live time reported by DAMA~\cite{bernabei2013final} ranging from 58\% to 83\% depending on the year. The data also show a good homogeneous distribution of the live time along the set-up time, mandatory for a proper modulation analysis. 
\subsection{Live time measurements}\label{sec:TimeBinning}
It is essential to compute the live time in arbitrary regular time intervals to properly calculate the evolution of the rate of events of interests such as alpha rate or single hit low energy events. This live time binning algorithm uses the two software and hardware clocks to account the real and live time of a particular time bin. Additionally, it can take a list of events and count them in every bin.
\paragraph{ }
The NTP controlled PC clock is used at the beginning of every run in order to establish the absolute time and the time event tagging is used to accumulate the real and live times to each bin. The algorithm allows to establish the bin time width as parameter and can be used to test the temporal behavior of especial events as well as compute time statistics as seen in this section.

\section{ANAIS-25 and ANAIS-37 calibration}\label{sec:Calibration}
\begin{figure}[h!]
     \begin{center}
     \includegraphics[width=.5\textwidth]{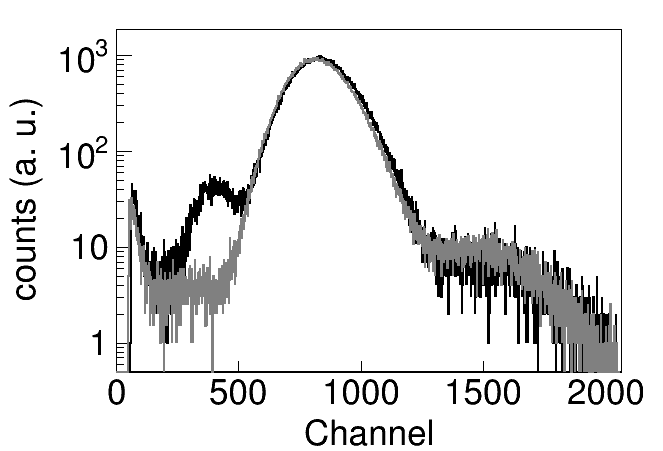}
     \caption{$^{109}Cd$ source without (gray) and with (black) heat-shrink tube showing the K-shell Br line.\label{fig:CalibBrLine}}
\end{center}
\end{figure}

The ANAIS-25 energy calibration consisted of $^{22}Na$ and $^{57}Co$ data taking with QDCs and occasional pulse calibration with radioactive sources in order to calibrate the area energy estimator~\cite{CCUESTA}.
\paragraph{ }
The calibration system changed in ANAIS-37 (see Section~\ref{sec:ANAIS37}) and the calibration routine was also changed in order to calibrate with $^{109}Cd$ only. This calibration allows to quickly calibrate at low energies due the intensity of 22.6 keV line. In addition to this line, the calibration spectrum features a peak due to the presence of a cover of heat-shrink tube with $Br$ content. The tube is used to ensure the junction of the sources to the wire. Figure~\ref{fig:CalibBrLine} shows the calibration results with and without the cover proving its origin. The line corresponds to 11.87 keV, the $Br$ K-shell X-rays mean. 

\begin{figure}[h!]
     \begin{center}
     \includegraphics[width=\textwidth]{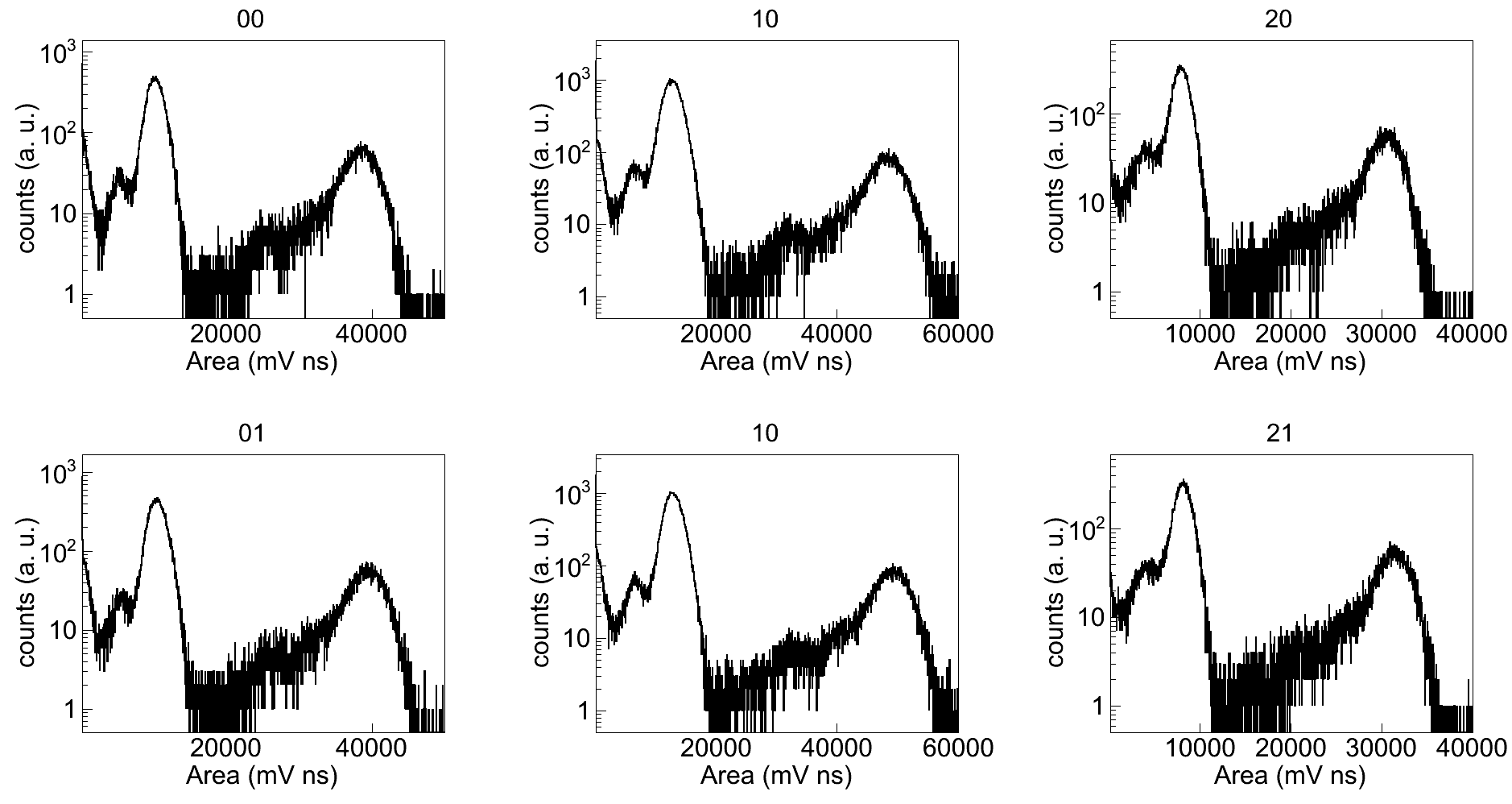}
     \caption{ANAIS-37 area calibration with $^{109}Cd$ source. It features the 22.6 keV and 88.0 keV peaks as well as the 11.9 keV of the $Br$ K-shell X-ray mean (see text).\label{fig:A37CalibSpectra}}
\end{center}
\end{figure}

\begin{figure}[h!]
     \begin{center}
        \begin{subfigure}[b]{0.5\textwidth}
        \includegraphics[width=\textwidth]{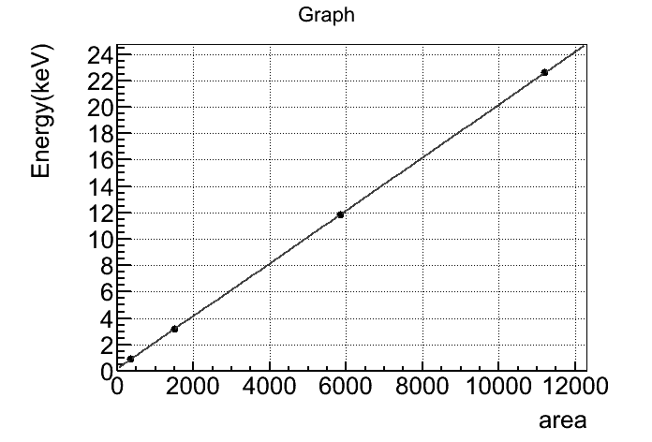}
	\caption{\label{fig:A25D1Calib}}
        \end{subfigure}%
	~
	\begin{subfigure}[b]{0.5\textwidth}
        \includegraphics[width=\textwidth]{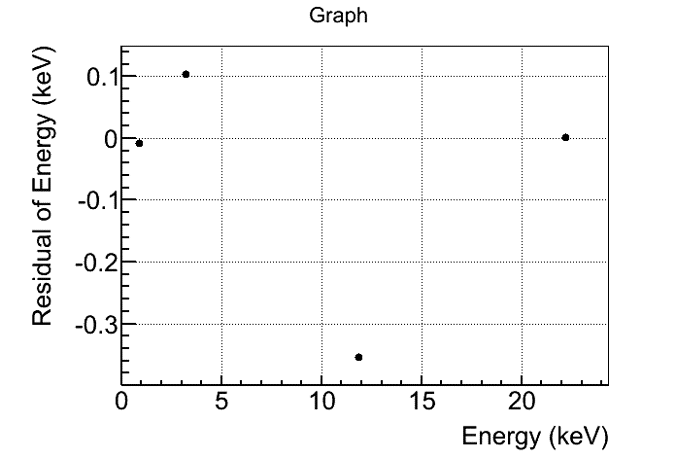}
	\caption{\label{fig:A25D1Res}}
        \end{subfigure}%
	\caption{Calibration fit and residuals using $^{109}Cd$ external source and $^{40}K$ and $^{22}Na$ internal lines.\label{fig:CalibA25}}
\end{center}
\end{figure}

\paragraph{}
Additionally, all calibrations are performed with pulse acquisition for area calibration in order to have routinely calibrated the two more used energy estimators, area and QDC. The area spectra for $^{109}Cd$ is shown in Figure~\ref{fig:A37CalibSpectra} for all ANAIS-37 PMT signals (signal ij corresponds to PMT j of detector Di).
\paragraph{ }
In this set-up the operating voltages of the D0 and D1 detectors were increased in order to better study the low energy region and hence the SER values are higher. The voltage of the new detector (D2) was selected to have a $10^6$ gain value in both PMTs in order to explore the high energy region being able to study the $\alpha$ rate contribution to the background (as seen in Sections~\ref{sec:ANAIS25} and \ref{sec:ANAIS37}). These gain differences can be observed in Figure~\ref{fig:A37CalibSpectra}. The global low energy calibration is performed by adding two internal lines, $^{40}K$ and $^{22}Na$ seen in Section~\ref{sec:ANAIS25} as shown in Figure~\ref{fig:CalibA25}. The accurate energy calibrations of ANAIS-25 and ANAIS-37 prototypes supports the idea of the scale-up of the same system and sources for the ANAIS full experiment.
\paragraph{ }
The very good resolution of the Alpha Spectra modules was measured in the first ANAIS-25 set-up~\cite{CCUESTA} and it can be reported again by measuring the $^{109}Cd$ 22.6 keV width peak. The results for the ANAIS-25 set-up after the replacing of D1 R11065SEL PMTs for R12669SEL2 (Table~\ref{tab:A25CdRes}) and ANAIS-37 (Table~\ref{tab:A37CdRes}) were calculated for every signal checking possible asymmetries. The global detector resolution was calculated with the sum signal. 
\begin{table}[h!]		
	\parbox{.45\linewidth}{
		\begin{center}
			\begin{tabular}{ccc}
				\hline
				PMT& \begin{tabular}{@{}c@{}}PMT \\ $\sigma$/E (\%)\end{tabular}&\begin{tabular}{@{}c@{}}Detector\\$\sigma$/E (\%)\end{tabular}\\
				\hline
				00& $10.93\pm0.03$&\multirow{2}{*}{$8.51\pm0.03$}\\
				01&$11.24 \pm 0.04$& \\
				10& $11.33\pm0.05$ &\multirow{2}{*}{$8.59\pm0.04$}\\
				11&$10.90\pm0.05$&\multicolumn{-2}{c}{} \\
				\hline

			\end{tabular}
			\caption[ANAIS-25 resolution at 22.6 keV]{ANAIS-25 resolution at 22.6 keV.} 
			\label{tab:A25CdRes} 
		\end{center}
	}
	\hspace{.9em}
	\parbox{.45\linewidth}{
		\begin{center}
			\begin{tabular}{ccc}
				\hline
				PMT& \begin{tabular}{@{}c@{}}PMT\\ $\sigma$/E (\%)\end{tabular}&\begin{tabular}{@{}c@{}}Detector\\$\sigma$/E (\%)\end{tabular}\\
				\hline
				00& $10.97\pm0.04$&\multirow{2}{*}{$8.73\pm0.03$}\\
				01&$11.18 \pm 0.04$& \\
				10& $11.40\pm0.03$ &\multirow{2}{*}{$8.80\pm0.03$}\\
				11&$11.02\pm0.03$& \multicolumn{-2}{c}{}\\
				20& $11.46\pm0.07$ &\multirow{2}{*}{$8.99\pm0.05$}\\
				21&$11.40\pm0.08$ &\multicolumn{-2}{c}{}\\
				\hline

			\end{tabular}
			\caption[ANAIS-37 resolution at 22.6 keV]{ANAIS-37 resolution at 22.6 keV.} 
			\label{tab:A37CdRes}
		\end{center}
	}
	\end{table}
\section{Onsite SER extraction}\label{sec:SERLP}
\begin{figure}[h!]
     \begin{center}
        \begin{subfigure}[b]{0.5\textwidth}
        \includegraphics[width=\textwidth]{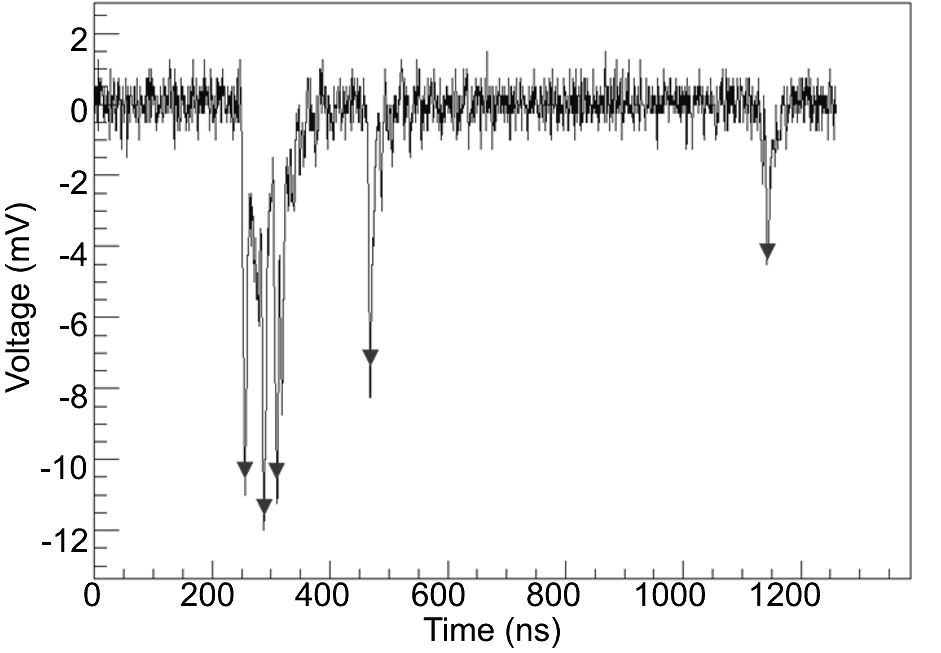}
	\caption{\label{fig:LP}}
        \end{subfigure}%
	~
	\begin{subfigure}[b]{0.5\textwidth}
        \includegraphics[width=\textwidth]{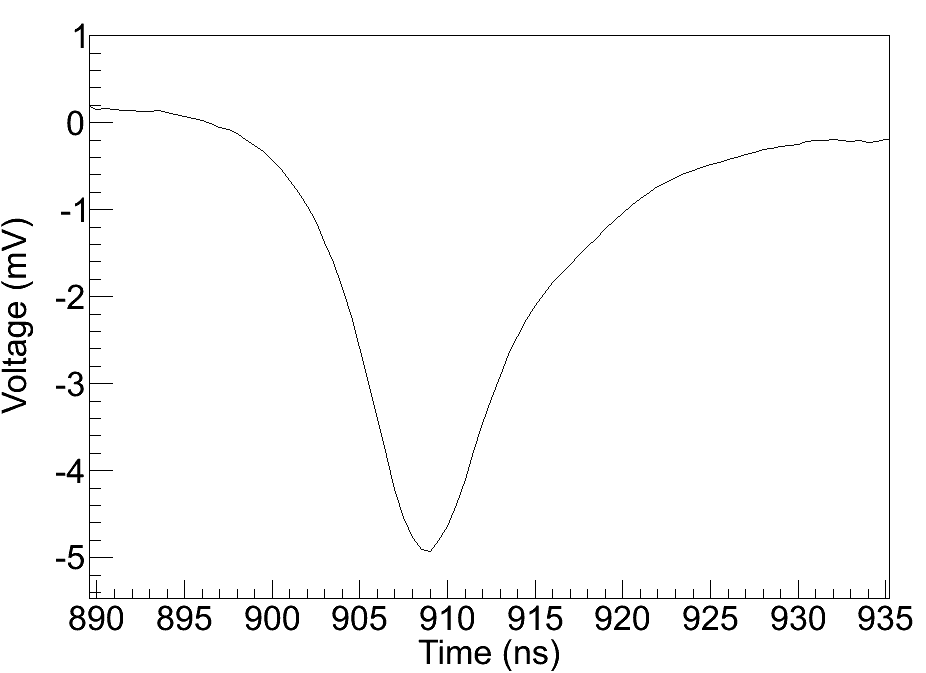}
	\caption{\label{fig:LPMean}}
        \end{subfigure}%
	\caption[]{Pulse with a low number of phe; peaks identified by the applied algorithm are shown with triangles (a). The last peaks of such kind of pulses reconstructs the Single Electron Response (SER). The SER mean pulse (b).\label{fig:LPPlots}}
\end{center}
\end{figure}

\begin{figure}[h!]
     \begin{center}
     \includegraphics[width=.6\textwidth]{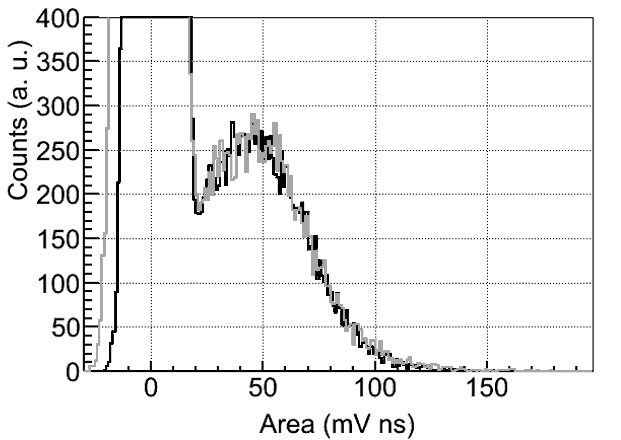}
	\caption{SER charge distributions derived at PMT test bench (gray) and along normal operation (black).\label{fig:SER_LP_LED}}
\end{center}
\end{figure}

The PMTs SER was measured at the University of Zaragoza test bench using an UV LED illumination of very low intensity, and triggering with the excitation LED signal as described in Section~\ref{sec:PMTTesting}. This characterization was very useful to compare and crosscheck to the measured SER of the same PMT at the LSC. This onsite SER extraction was performed using the peak identification algorithm (see Section~\ref{sec:PeakDetection}) with the data taken in the normal operation of the ANAIS detectors at the LSC.
\paragraph{}
The SER was built by selecting peaks at the end of the pulse of each PMT in order to avoid trigger bias in pulses having a low number of peaks to prevent the pile-up of several photoelectrons (phe). An example of a pulse fulfilling these conditions can be seen in Figure~\ref{fig:LP} and the mean pulse of a selection of this kind of events is shown in Figure~\ref{fig:LPMean}.  
\paragraph{}
The photoelectron area (proportional to charge) is integrated in a fixed time window around the peak maximum (see Section~\ref{sec:PeakParams})) in order to obtain the single electron response charge distribution. The SER charge distribution extracted for the same PMT by these two methods is compared in Figure~\ref{fig:SER_LP_LED} showing full agreement between both distributions.
\subsection{SER trigger efficiency}\label{sec:SERTrEff}
\begin{figure}[h!]
     \begin{center}
                \includegraphics[width=\textwidth]{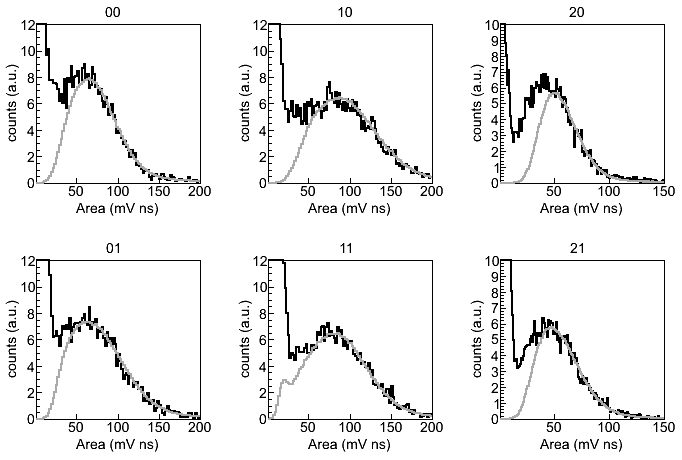}
	\caption{Single electron area distribution (black) compared with the area distribution of triggered events (gray).}\label{fig:TriggerAreaA37}
\end{center}
\end{figure}

The onsite single electron response extracted with the aforementioned last peak technique is compared with the distribution of triggered events in Figure~\ref{fig:TriggerAreaA37}. Both distributions are represented (conveniently scaled) in area for the ANAIS-37 set-up. The plots are constructed with the last peak algorithms for both populations. The SER is constructed by requiring non-triggering in order to avoid the trigger bias.
\paragraph{ }
The figures clearly show the previously mentioned higher voltage in D0 and D1 compared to the D2 lower value in ANAIS-37 set-up. It can be noted in Figure~\ref{fig:TriggerAreaA37} the almost full trigger efficiency of the photoelectron signals of the two first modules. The third module exhibits a high but not full photoelectron trigger efficiency. The method followed to quantify the trigger efficiency is to compare the area of the scaled distributions. The triggered distribution is integrated directly and the SER distribution is fitted to a Gaussian, the most important SER contribution as seen in Section~\ref{sec:PMTSignal}.
\begin{table}[h!]
\begin{center}
\begin{tabular}{ccc}

				\hline
				PMT&\% SER triggered\\
				\hline
				00& 80\%\\
				01& 89\%\\
				10& 88\%\\
				11& 97\%\\
				20& 71\%\\
				21& 75\%\\
				\hline
\end{tabular}
\caption[ANAIS-37 SER trigger efficiency]{ANAIS-37 SER trigger efficiency.} 
\label{tab:A37SEREff}
\end{center}
\end{table}
The SER trigger efficiency values seen in Table~\ref{tab:A37SEREff} are high considering that the hardware trigger level must be set above the baseline noise. It has to be noted the PMT 11 SER efficiency value seems to be overestimated because of the triggering of a dark count population as it can be observed in Figure~\ref{fig:TriggerAreaA37} (Signal 11). This effect is under study.
\paragraph{ }
Additionally, the statistical distribution of the scintillation photons (see Section~\ref{sec:LESignal}) and a good light collection (see next section) can give near 100\% trigger efficiency down to 1 keV as it can be seen in Section~\ref{sec:HWTriggEff}.

\section{Light collection}\label{sec:LY}
The light collection is another key parameter in a low threshold scintillation experiment because it has a direct influence on both energy threshold and resolution. For this reason, careful studies has been performed in ANAIS prototypes~\cite{CCUESTA}.
\paragraph{ }
First, the single electron response (SER) charge distribution was determined as described earlier. Next, the light collection (in photoelectrons per keV) can be inferred with the response to a known calibration line. In this section, SER extraction and light collection measurements for ANAIS-25 and ANAIS-37 set-ups are covered.

\paragraph{ }
The light collected by each of the PMTs coupled to the ANAIS-25 modules was calculated by dividing the charge distribution associated to a known energy deposition in the NaI crystal and the SER charge distribution derived as described in Section~\ref{sec:SERLP}. The 22.6 keV line from a $^{109}Cd$ calibration source was used for this study. The result of the SER charge spectrum and the $^{109}Cd$ line Gaussian fits can be seen in Table~\ref{tab:A25_Fits} (PMT ij corresponds to PMT j of detector Di). These results and the global light collection of the two ANAIS-25 detectors are summarized in Table~\ref{tab:A25LY}. The results confirm the prototypes outstanding light collection and its impact in resolution as seen earlier in Table~\ref{tab:A37CdRes}. The very good optical performance of the Alpha Spectra modules evidenced by these figures is very promising to reach a very low energy threshold of 1 keVee. 
\begin{table}[h!]
		\begin{center}
			\begin{tabular}{ c c c c c}
				\hline
				PMT& \begin{tabular}{@{}c@{}} SER mean\\mV·ns \end{tabular}& \begin{tabular}{@{}c@{}}SER $\sigma$\\mV·ns\end{tabular} & \begin{tabular}{@{}c@{}}22.6 keV mean\\mV·ns \end{tabular}& \begin{tabular}{@{}c@{}}22.6 keV\ $\sigma$\\mV·ns \end{tabular}\\
				\hline
				 00&$35.47\pm0.35$ & $21.73\pm0.25$ & $6122\pm2$ & $669\pm2$\\
				 01&$29.42\pm0.21$ & $18.81\pm0.24$ & $5057\pm2$ & $568\pm2$\\
				10&$41.20\pm0.25$ & $28.30\pm0.21$ & $7139\pm4$ & $809\pm4$\\
				11&$44.52\pm0.29$ & $24.36\pm0.24$ & $7570\pm4$ & $825\pm3$\\
				\hline
			\end{tabular}
			\caption[ANAIS-25 values for SER charge distribution and $^{109}Cd$ 22.6 keV line Gaussian fits]{ANAIS-25 values for SER charge distribution and $^{109}Cd$ 22.6 keV line Gaussian fits.} 
			\label{tab:A25_Fits} 
		\end{center}
	\end{table}
	\begin{table}[h!]
		\begin{center}
			\begin{tabular}{ccc}
				\hline
				PMT& \begin{tabular}{@{}c@{}}PMT\\ phe/keV\end{tabular}& \begin{tabular}{@{}c@{}}Detector\\ phe/keV\end{tabular}\\
				\hline
				00& $7.64\pm0.08$&\multirow{2}{*}{$15.24\pm0.09$}\\
				01&$7.61\pm0.05$& \\
				10& $7.67\pm0.05$&\multirow{2}{*}{$15.19\pm0.07$} \\
				11&$7.52\pm0.05$ & \\
				\hline

			\end{tabular}
			\caption[ANAIS-25 light collection]{ANAIS-25 light collection.\\\hspace{\textwidth}} 
			\label{tab:A25LY} 
		\end{center}
	\end{table}
\paragraph{ }
The same procedure was repeated with ANAIS-37 set-up data. The results of the SER charge distribution and the $^{109}Cd$ 22.6 keV line Gaussian fits can be seen in Table~\ref{tab:A37_Fits}. The light collection for every PMT and detector can be observed in Table~\ref{tab:A37LY}. The newly extracted values for D0 and D1 are compatible with those obtained for ANAIS-25. Good values for the new D2 (more than 15 phe/keV) have also been measured having again a good impact in terms of energy threshold and resolution, crucial for the sensitivity to WIMPs annual modulation.
\begin{table}[h!]
		\begin{center}
			\begin{tabular}{c c c c c }
				\hline
				PMT& \begin{tabular}{@{}c@{}} SER mean\\mV·ns \end{tabular}& \begin{tabular}{@{}c@{}}SER $\sigma$\\mV·ns\end{tabular} & \begin{tabular}{@{}c@{}}22.6 keV mean\\mV·ns \end{tabular}& \begin{tabular}{@{}c@{}}22.6 keV $\sigma$\\mV·ns \end{tabular}\\
				\hline
				 00&$61.47\pm0.36$ & $35.02\pm0.32$ & $10257\pm5$ & $1126\pm4$\\
				 01&$58.40\pm0.71$ & $43.06\pm0.51$ & $10425\pm5$ & $1166\pm4$\\
				10&$83.24\pm0.55$ & $46.52\pm0.51$ & $12820\pm5$ & $1463\pm4$\\
				11&$73.91\pm0.74$ & $42.04\pm0.52$ & $12740\pm5$ & $1404\pm4$\\
				20&$42.70\pm2.10$ & $25.42\pm1.79$ & $7928\pm5$ & $909\pm6$\\
				21&$44.57\pm2.10$ & $26.67\pm1.95$ & $8155\pm6$ & $930\pm6$\\
				\hline
			\end{tabular}
			\caption[ANAIS-37 value from SER charge distribution and $^{109}Cd$ 22.6 keV line Gaussian fits]{ANAIS-37 values from SER charge distribution and $^{109}Cd$ 22.6 keV line Gaussian fits.} 
			\label{tab:A37_Fits} 
		\end{center}
	\end{table}
	\begin{table}[h!]
		\parbox{.45\linewidth}{
		\begin{center}
			\begin{tabular}{ccc}
				\hline
				PMT& \begin{tabular}{@{}c@{}}PMT\\ phe/keV\end{tabular}& \begin{tabular}{@{}c@{}}Detector\\ phe/keV\end{tabular}\\
				\hline
				00& $7.38\pm0.04$&\multirow{2}{*}{$15.26\pm0.10$}\\
				01&$7.88\pm0.09$& \\
				10& $6.81\pm0.05$&\multirow{2}{*}{$14.44\pm0.09$} \\
				11&$7.62\pm0.08$ & \\
				20& $8.21\pm0.40$&\multirow{2}{*}{$16.31\pm0.56$} \\
				21&$8.09\pm0.38$ & \\
				\hline
			\end{tabular}
			\caption[ANAIS-37 light collection]{ANAIS-37 light collection.\\\hspace{\textwidth}} 
			\label{tab:A37LY} 
		\end{center}
	}
	\hspace{.9em}
	\parbox{.45\linewidth}{
		\begin{center}
			\begin{tabular}{cccc}
				\hline
				PMT& \begin{tabular}{@{}c@{}}QE\\\%\end{tabular}& \begin{tabular}{@{}c@{}}Incident\\ phe/keV\end{tabular}& \begin{tabular}{@{}c@{}}Incident\\ phe/keV\end{tabular}\\
				\hline
				ZK5902& 34.9&21.2&\multirow{2}{*}{44.1}\\
				ZK5908&34.4& 22.9&\\
				FA0018& 33.86&20.1&\multirow{2}{*}{41.1} \\
				FA0060&36.42&21.0&\\
				FA0034& 35.91&22.9&\multirow{2}{*}{43.7} \\
				FA0090&38.97& 20.8&\\

				\hline

			\end{tabular}
			\caption[ANAIS-37 Incident light]{ANAIS-37 Incident light.\\\hspace{\textwidth}} 
			\label{tab:A37IL} 
		\end{center}

	}
	\end{table}
	\paragraph{ }
	In addition to the light collected, the effect of the PMT quantum efficiency have been discounted in Table~\ref{tab:A37IL} in order to estimate the incident light to the PMT and test the optical homogeneity of the modules. The results show good similar values among PMTs and among modules.
	\paragraph{ }
	In summary, the PMTs single electron response was characterized onsite and compared with the previous PMTs measurements showing a full compatibility among them. Using this extraction, an excellent light collection for the three ANAIS detectors, of the order of $\sim$15 phe/keV, has been measured. Thanks to this, an energy threshold for the ANAIS experiment at 1 keVee is at reach, if populations at this level are understood and filtered. This would significantly improve the sensitivity of the ANAIS Project in the search for the annual modulation effect in the WIMPs signal (see Section~\ref{sec:SensitProsp}).
\section{Hardware trigger efficiency}\label{sec:HWTriggEff}
\begin{figure}[h!]
     \begin{center}
        \begin{subfigure}[b]{0.5\textwidth}
        \includegraphics[width=\textwidth]{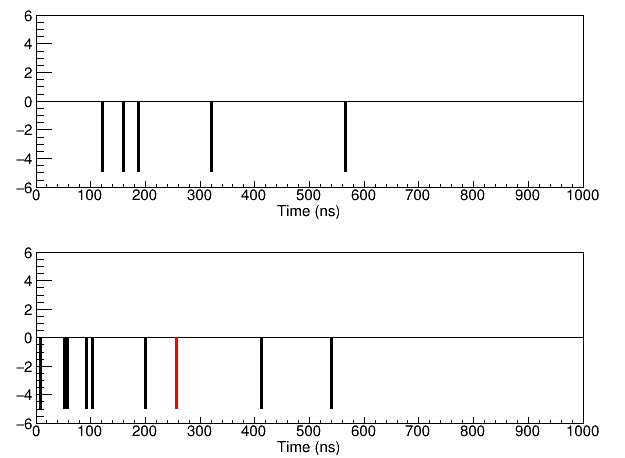}
	\caption{\label{fig:CoincTrigSimulT}}
        \end{subfigure}%
	~
	\begin{subfigure}[b]{0.5\textwidth}
        \includegraphics[width=\textwidth]{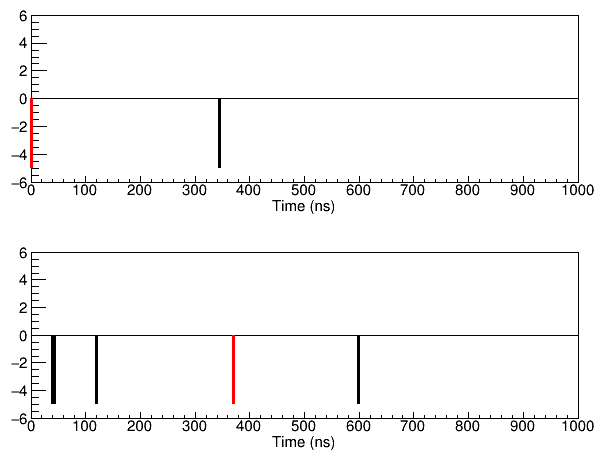}
	\caption{\label{fig:CoincTrigSimulNo}}
        \end{subfigure}%
	\caption[]{Simulation of the photoelectron time of arrival\label{fig:CoincTrigSimul} for 0.9 keV events. The event (a) would trigger with a 200 ns coincidence window, the event (b) does not. }
\end{center}
\end{figure}

\begin{figure}[h!]
     \begin{center}
        \begin{subfigure}[b]{0.5\textwidth}
        \includegraphics[width=\textwidth]{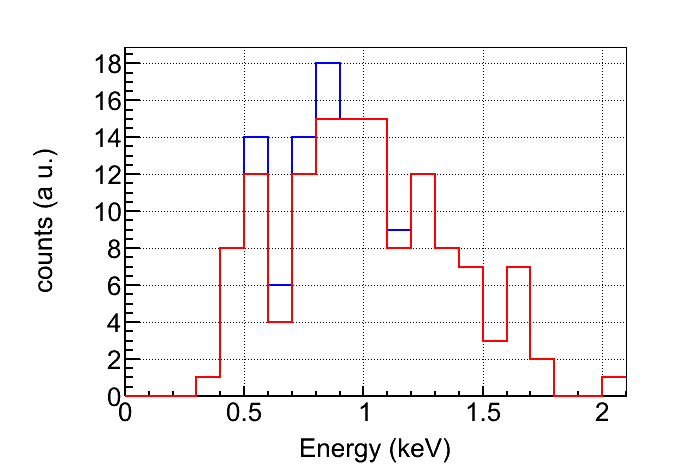}
	\caption{\label{fig:A37D0Na22TEff}}
        \end{subfigure}%
	~
	\begin{subfigure}[b]{0.5\textwidth}
        \includegraphics[width=\textwidth]{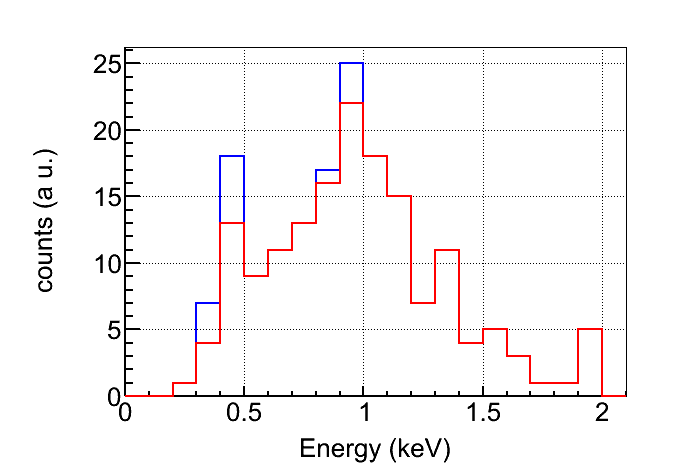}
	\caption{\label{fig:A37D1Na22TEff}}
        \end{subfigure}%
	
	\begin{subfigure}[b]{0.5\textwidth}
        \includegraphics[width=\textwidth]{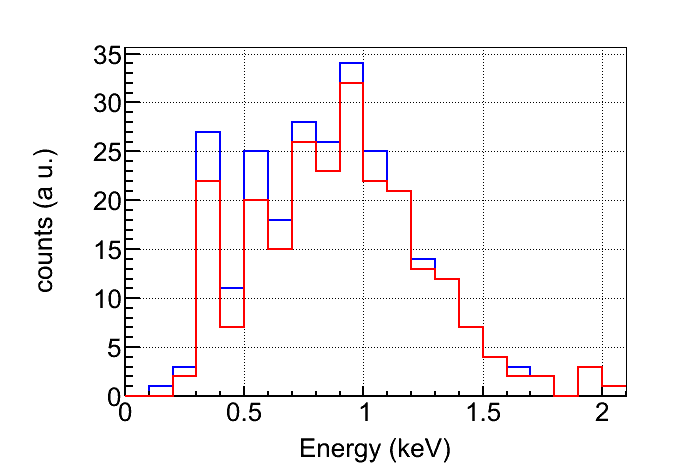}
	\caption{\label{fig:A37D2Na22TEff}}
        \end{subfigure}%
	\caption[]{Events with peaks in both signals recorded in coincidence $^{22}Na$ (black) and triggered (gray) in both detectors for D0 (a), D1 (b) and D2 (c)\label{fig:A37Na22TEff}. }
\end{center}
\end{figure}

The SER trigger efficiency seen in Section~\ref{sec:SERLP} and the computed light collection presented in Section~\ref{sec:LY} can be combined to simulate the hardware trigger efficiency behavior and can be compared later with the experimental data.
\paragraph{ }
The scintillation emission has been modeled as a Poissonian process of 40 photons per keV as seen in Section~\ref{sec:NaIScint}. The light collection was simulated by considering the probability of every photon to reach the PMT with the measured photoelectrons per keV. Once the phe number is obtained, the time of arrival of photoelectrons was simulated to follow the scintillation exponential distribution in every PMT. Next, the SER trigger efficiency was applied marking triggered photoelectrons using the measured values for $D0$, $D1$ and $D2$. Simulated signals of 0.9 keV can be seen in Figure~\ref{fig:CoincTrigSimul} with non-triggered photoelectrons marked as red.
\paragraph{}
Finally, a time window was applied to every event in order to test the trigger condition taking into account the dead time of the CAEN N843 constant fraction discriminator. The CFD does not retrigger for at least a minimum (configurable) dead time of 150 ns in which it does not generate another trigger window. The simulated event shown in Figure~\ref{fig:CoincTrigSimulT} triggers whereas the event in Figure~\ref{fig:CoincTrigSimulNo} do not fulfill the trigger condition with a 200 ns window. Different coincidence windows were tested as seen in Table~\ref{tab:CoincTrigSimul} for 0.9 keV events again.
\begin{table}[h!]
		\begin{center}
			\begin{tabular}{ c c c}
				\hline
				Detector& \begin{tabular}{@{}c@{}} Trigger efficiency\\(200 ns) \end{tabular}& \begin{tabular}{@{}c@{}} Trigger efficiency\\(100 ns) \end{tabular}\\
				\hline
				$D0$& $92.8\%$ & $83.8\%$ \\
				$D1$& $94.4\%$ &$86.4\%$ \\
				$D2$& $89.8\%$ &$79.4\%$ \\
				$D_{\textrm{perf}}$&$95.8\%$ & $89.3\%$ \\
				\hline
			\end{tabular}
			\caption[]{Simulated hardware trigger efficiency with different coincidence windows for 0.9 keV events.} 
			\label{tab:CoincTrigSimul}
		\end{center}
	\end{table}
\paragraph{ }
The resulting values show a very high hardware trigger efficiency for events of such low energy. Additionally, the efficiency of $D_{\textrm{perf}}$ use a perfect (100\%) SER trigger efficiency and it shows the irreducible fraction of non-triggering events due to the 0-photon Poissonian probability combined with the events with non-matched photons within the coincidence window.
\paragraph{}
These simulated results were crosschecked with real $^{22}Na$ events using the coincidence with 1274~keV events in adjacent modules (see Section~\ref{sec:ANAIS25}). The comparison was done considering the event rate of $^{22}Na$ with and without triggering and requiring peaks in both signals in order to avoid random coincidences with PMT noises. On one hand we consider the triggered events in coincidence by using the Pattern Unit information. On the other hand, we select all events acquired with a 1274~keV event in an adjacent detector with one peak or more in both PMT signals. The results of such a selection can be seen in Figure~\ref{fig:A37Na22TEff}. The triggered events are displayed in red and all events with peaks in blue.
\paragraph{}
The comparison of these triggering data with the simulation is shown in Table~\ref{tab:Na22Cross}. It is worth to note that these data are not the trigger efficiency because it does not take into account the events that do not have photoelectrons in one or both signals. The results are compatible given the statistical error. It can be observed a slight discrepancy in the D1 efficiency, caused by the SER trigger overestimation in signal 11 seen in Section~\ref{sec:SERTrEff}. Anyway, the overall compatibility supports the input parameters of the simulation: light collection and SER trigger efficiency.
\begin{table}[h!]
		\begin{center}
			\begin{tabular}{ c c c}
				\hline
				Detector& Measured efficiency& Simulated efficiency\\
				\hline
				$D0$& $93.5 \pm 2.0\%$ & $94.1\%$ \\
				$D1$& $92.3 \pm 1.6\%$ &$95.8\%$ \\
				$D2$& $89.4 \pm 1.7\%$ &$91.1\%$ \\
				\hline
			\end{tabular}
			\caption[]{Trigger efficiency in events with peaks.} 
			\label{tab:Na22Cross}
		\end{center}
\end{table}
\paragraph{ }
\section{Effect of noise reduction with a 100 ns coincidence window}\label{sec:100nsNoise}
\begin{figure}[h!]
\begin{center}
                \includegraphics[width=.6\textwidth]{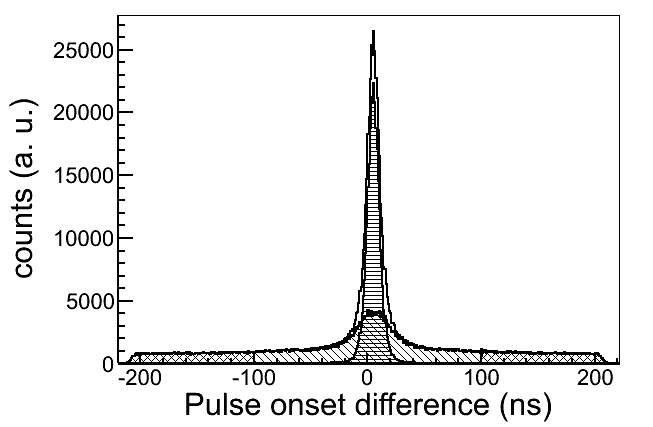}
		\caption{Onset time difference between signals 0 and 1 in one detector. Noise (diagonal) and genuine (horizontal) distributions.\label{fig:Coinc200vs100}}
\end{center}
\end{figure}

The previous section has concluded that ANAIS will have a very good hardware trigger efficiency, slightly reduced if the coincidence window is lowered from 200 ns to 100 ns. The noise reduction is studied in this section in order to see the full benefits of such a reduction.
\paragraph{ }
The temporal behavior of different kind of effects and the effect of the reduction can be observed in Figure~\ref{fig:Coinc200vs100}. It shows the distribution of the onset time difference between the signals of a module. Additionally, it plots those events having more than two peaks in both signals (filled with horizontal pattern) and those with less than or equal to two peaks in some of the signals (diagonal pattern). The selection of having two or more peaks in both signal is the first quality cut described in Section~\ref{sec:EventSelection} qualifying the diagonal region as noise. It can be seen that the noise accounts the vast majority of the plateau events. The effect of the coincidence window reduction can be quantified in such a way to computing the number of the events outside the (-100 ns,100 ns) interval and comparing with all noise events. This gives a noise reduction of 36\%, 35\% and 35\% for D0, D1 and D2 respectively, accounting around a 23~\% of all triggered events. This reduction can almost reach the 50\% in case of the activation due to $N_2$ flux cut (see Section~\ref{sec:N_2_Flux}) due to the random nature of such coincidences.
\paragraph{ }
It should be stressed that the reduction would be a 50\% if the noise was a totally random noise. It is clear that it is not the case because of the temporal distribution of such a population featuring a visible bump for lower pulse onset difference. The origin of such noises can be attributed to scintillation of some materials and other light emissions excluding dark noises. Anyway, a good reduction of random coincidences would be beneficial in terms of live time and disk space weighting the aforementioned low energy trigger efficiency loss.
\chapter{Environmental and data acquisition stability}\label{sec:Stability}
The system stability is a key feature for an annual modulation experiment as seen in Section~\ref{sec:ExperimentalRequirements}. Therefore, the temporal evolution of some relevant parameters is reviewed in this chapter. The slow control system, responsible for the monitoring and storage of environmental parameters, is presented in Section~\ref{sec:SlowControl} and its data evolution are show in Section~\ref{sec:SlowControlData}.
\paragraph{}
The evolution of some critical data acquisition parameters are also studied for ANAIS-25 and ANAIS-37 set-ups. The gain stability is checked in Section~\ref{sec:GainStability}. The evolution of the trigger level with the time and the temperature are shown in Section~\ref{sec:TriggerStability}. Finally, the window coincidence has been measured and a temperature dependency can be reported (see Section~\ref{sec:CoincWndStability}). The influence of such a dependency in trigger efficiency at very low energy has been computed.
\section{Slow Control}\label{sec:SlowControl}
\begin{figure}[h!]
  \begin{center}
    \includegraphics[width=.7\textwidth]{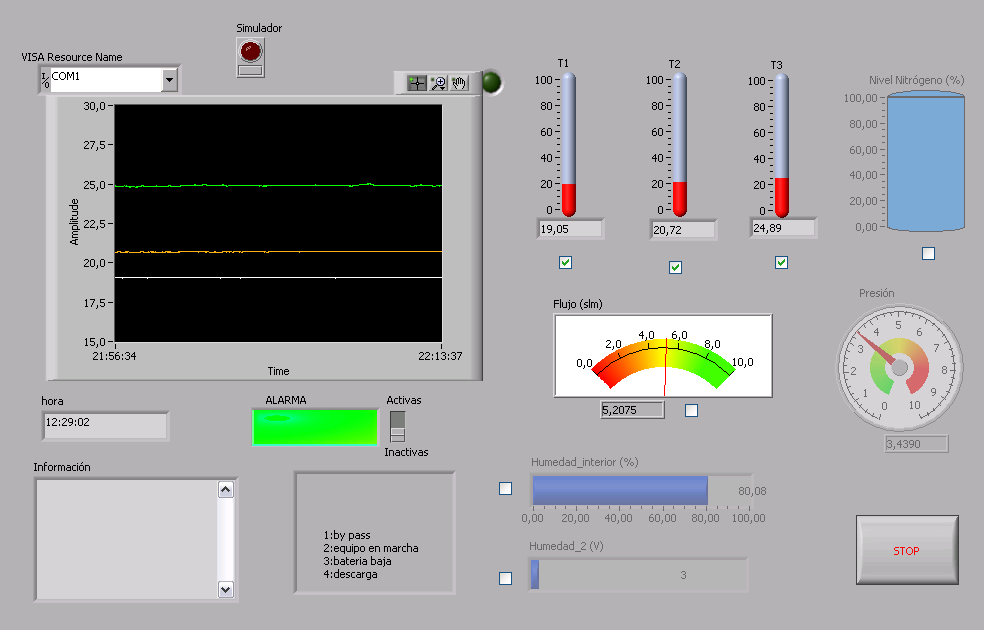}
    \caption[Slow control graphical interface]{\label{fig:SCGUI}Slow control graphical interface.}
  \end{center}
\end{figure}

The precise monitoring of environmental parameters is needed to check the stability of the system as described in Section~\ref{sec:ExperimentalRequirements}. This monitoring must save data from all relevant parameters such as temperature or nitrogen flux that can have influence in the data. The system must also give alarms in the case of abnormal parameter values in order to restore the standard conditions as soon as possible.
\paragraph{}
The list of parameters to monitor are:
\paragraph{Temperatures:} The temperature instability can cause a large number of effects in all the data acquisition stages. Therefore, the temperature must be monitored at different locations: inside the shielding, inside the hut and near the electronic chain.
\paragraph{Nitrogen flux:} The shielding is continuously flushed with boil-off $N_2$ flux in order to keep the experimental space as free as possible from radon and other possible atmospheric radioactive isotopes. This flux must be monitored registering any variation over the typical flux, including supply cut.
\paragraph{Radon concentration:} The radon activity is one of the most common sources of radioactive background as seen in Section~\ref{sec:Backgrounds}. The ambient concentration must be monitored in order to detect its variations despite the fact that above mentioned continuous boil-off $N_2$ flux reduces the internal activity at least a factor 100~\cite{cebrian2012background}. Therefore, measuring the radon activity inside the shielding would require a much more sensitive detector, far from the current sensitivity of commercial radon detectors.
\paragraph{PMT high voltage and intensity:} The PMT electrical signal is very dependent on the HV value as described in Section~\ref{sec:PMTTesting}. This value must be monitored in order to assure the gain stability of all PMTs. The current can also be monitored in order to detect any other problems in the PMT supply if overcurrent or current cut happens.
\paragraph{Crate health parameters:} The VME and NIM crates house the most of the electronic modules and hence monitoring these crates is crucial. The CAEN crates used in ANAIS allow to monitor several temperatures (power supply and module section temperature) as well as fan behavior and voltages and intensity of all crate power supply.
\paragraph{}
A system to monitor temperature and $N_2$ flux was designed and a LabVIEW program was developed~\cite{CPOBES}. Its graphic user interface can be seen in Figure~\ref{fig:SCGUI}. It monitors and stores the values from three PT100 temperature probes (inside the shielding, ambient temperature in Hall B and near the VME crate) and from an AWM3300 Honeywell flow sensor for $N_2$.
\paragraph{}
The radon concentration is measured by a Genitron (now Saphymo) AlphaGuard system that also records the pressure, humidity and temperature environmental parameters. Additionally, some utilities to monitor the intensity and voltage of the CAEN high voltage mainframe was later developed.
\section{Slow Control Data}\label{sec:SlowControlData}
\subsection{Temperatures}\label{sec:SCTemp}
\begin{figure}[h!]
 \begin{center}
	\begin{subfigure}[b]{0.4\textwidth}
                \centering
                \includegraphics[width=1\textwidth]{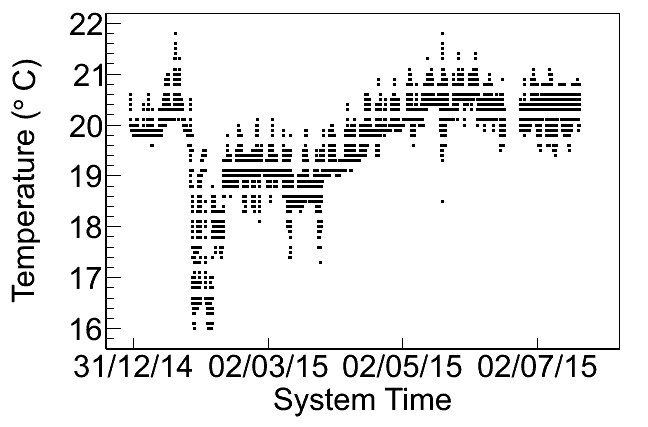}
		\caption{\label{fig:TempAG2015}2015 temperature at Hall measured by the AlphaGuard detector.} 
        \end{subfigure}
        ~ 
        \begin{subfigure}[b]{0.4\textwidth}
                \centering
                \includegraphics[width=1\textwidth]{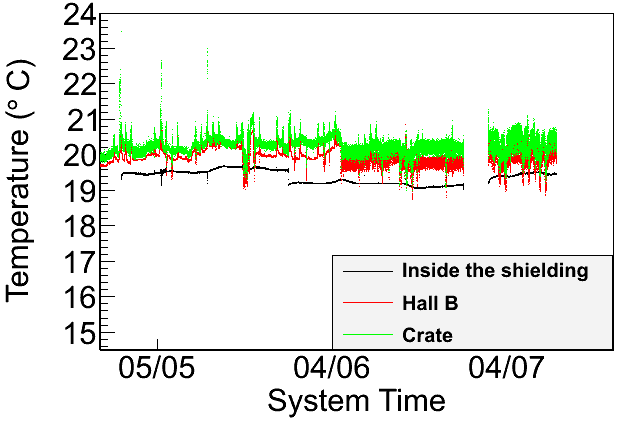}
		\caption{\label{fig:TempSCA37}Slow control probe temperatures in the ANAIS-37 set-up.}
        \end{subfigure}
	
	\begin{subfigure}[b]{0.4\textwidth}
                \centering
                \includegraphics[width=1\textwidth]{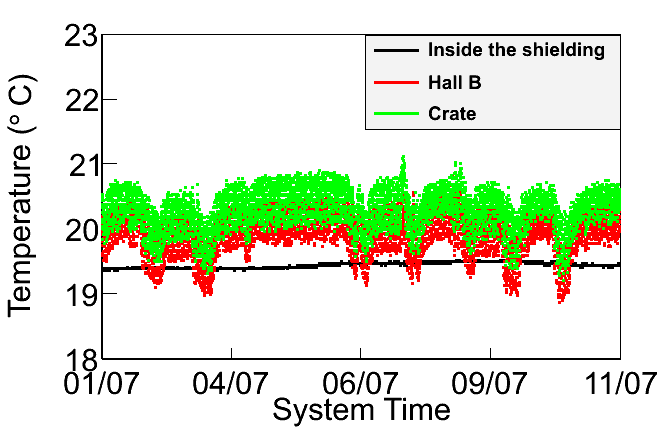}
		\caption{\label{fig:TempWeekA37}Temperature in a typical week.\\\hspace{\textwidth}}             
        \end{subfigure}
        ~ 
        \begin{subfigure}[b]{0.4\textwidth}
                \centering
                \includegraphics[width=1\textwidth]{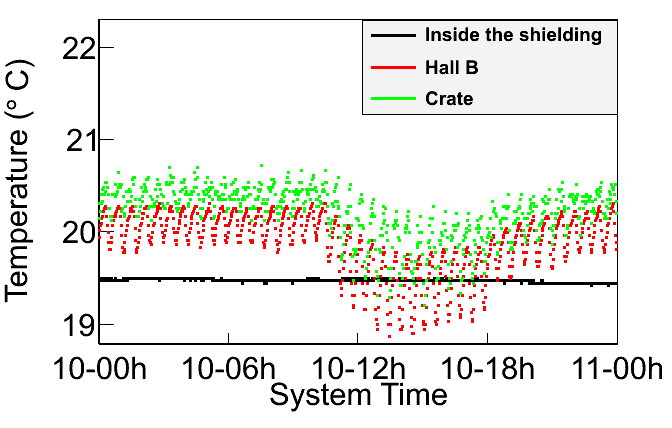}
		\caption{\label{fig:TempDayA37}Temperature during a typical working day.}
        \end{subfigure}

	\caption[Slow control temperature]{\label{fig:SCTemps}Slow control temperatures.}
\end{center}
\end{figure}

The overall temperature at the LSC Hall B is stable in two degree range as it can be seen in Figure~\ref{fig:TempAG2015}. The temperature taken by the AlphaGuard system shows the most notable variation at the beginning of the 2015 year due to LSC climate control system testing. The rest of the year, the Hall B room temperature ranges from 18 to 21 ºC. It can be noted a change in the thermostat behavior at the beginning of June, the time when the refrigeration was needed to keep the thermostat in its setpoint. Additionally, the weekly behavior can be observed with temperature variations (see Figure~\ref{fig:TempWeekA37}) associated to working activity (Monday to Friday) and more stable weekends. The room temperature variations are below 1 ºC in a typical day. These changes have very little impact inside the shielding due the thermal inertia, as shown in Figure~\ref{fig:TempDayA37}. The crate temperatures are also measured but their low precision make them suitable for health checking and as crosscheck purposes only.
\paragraph{ }
These data have been used to test temperature stability of the trigger level (Section~\ref{sec:TriggerStability}) and the stability of the coincidence window (Section~\ref{sec:CoincWndStability}).
\subsection{$Rn$ concentration, humidity and pressure}
\begin{figure}[!htb]
	\begin{subfigure}[b]{0.5\textwidth}
                \centering
                \includegraphics[width=1\textwidth]{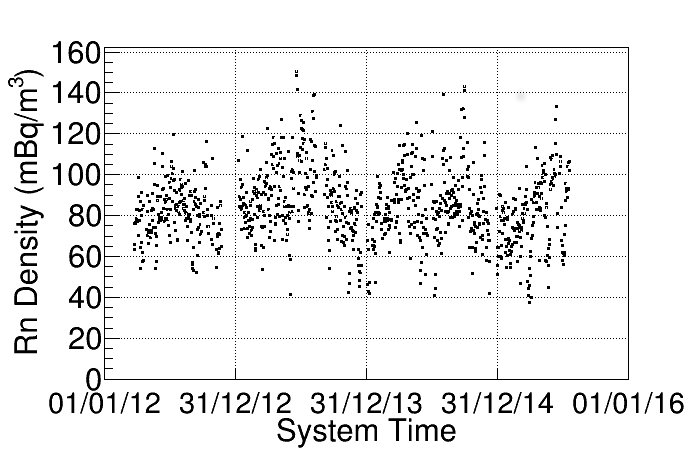}
		\caption{\label{fig:SCRn}.}
        \end{subfigure}
        ~ 
        \begin{subfigure}[b]{0.5\textwidth}
                \centering
                \includegraphics[width=1\textwidth]{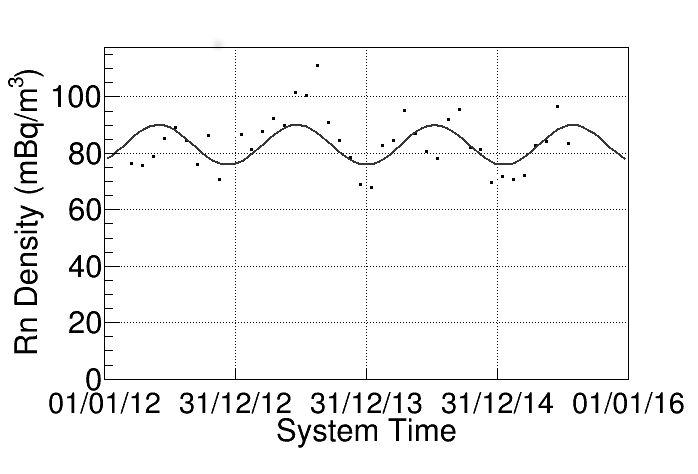}
		\caption{\label{fig:SCRnFit}.}
        \end{subfigure}

	\begin{subfigure}[b]{0.5\textwidth}
                \centering
                \includegraphics[width=1\textwidth]{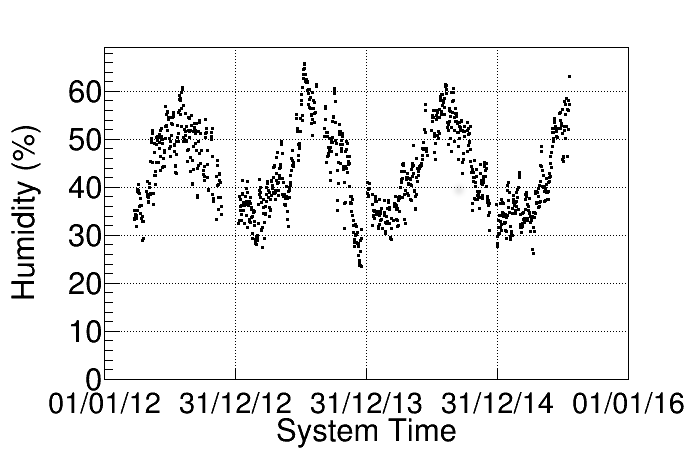}
		\caption{\label{fig:SCH}.} 
        \end{subfigure}
        ~ 
        \begin{subfigure}[b]{0.5\textwidth}
                \centering
                \includegraphics[width=1\textwidth]{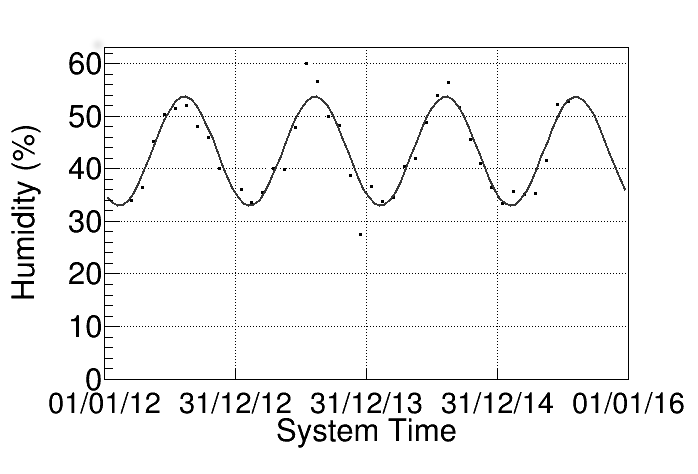}
		\caption{\label{fig:TempSCA37}}
        \end{subfigure}
	
	\begin{subfigure}[b]{0.5\textwidth}
                \centering
                \includegraphics[width=1\textwidth]{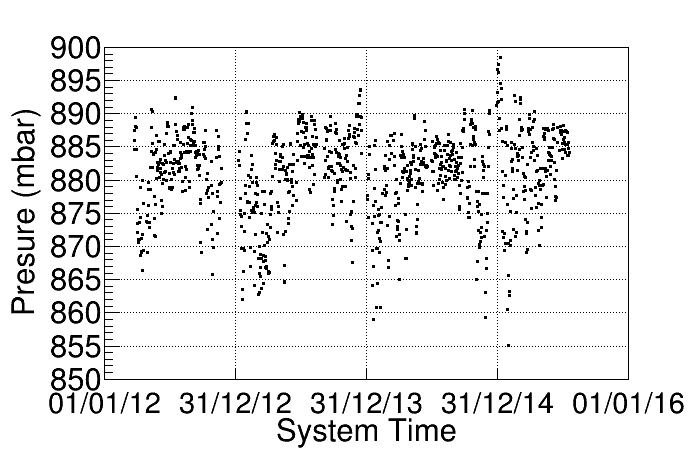}
		\caption{\label{fig:SCP}}             
        \end{subfigure}
        ~ 
        \begin{subfigure}[b]{0.5\textwidth}
                \centering
                \includegraphics[width=1\textwidth]{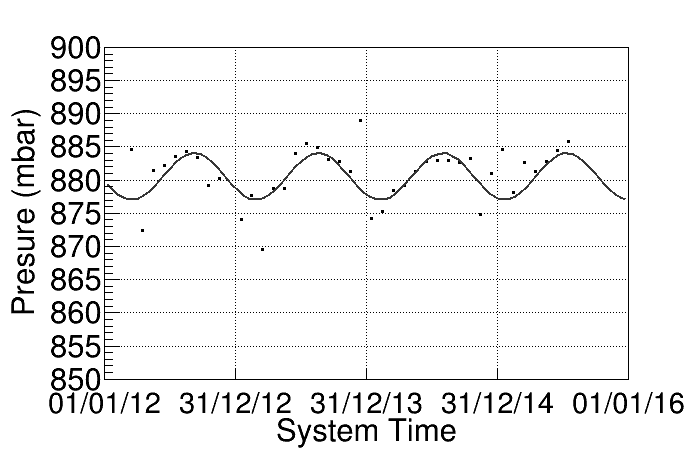}
		\caption{\label{fig:SCPFit}}
        \end{subfigure}

 \begin{center}
	\caption[Slow control temperature]{\label{fig:SCRnHP} Radon Density (a and b), humidity (c and d) and pressure (e and f) daily mean values (a, c and e) and monthly mean values with modulation fit (b, d and f).}
\end{center}
\end{figure}

Other environmental parameters measured by the AlphaGuard device are $Rn$ concentration, humidity and pressure. The temporal behavior of these parameters along the ANAIS-25 and ANAIS-37 set-ups can be seen in Figure~\ref{fig:SCRnHP}. The daily and monthly means are plotted over the time. It can be noted the seasonal variations of these parameters.  
A simple fit following a cosine function ($V_0+A\cdot cos(\frac{2\pi}{T}(t-t_M)$) has been performed using the monthly means and letting free the four parameters. Table~\ref{tab:SCFits} shows the extracted fit parameters.
\begin{table}[h!]
	\begin{center}
		\begin{tabular}{ c c c c c }
			\hline
			Parameter & Radon concentration& Humidity& Pressure \\
			\hline
			$V_0$& $82.98\pm 0.06$ $(mBq/m^3)$ & $43.35\pm0.01$ $(\%)$& $879.85\pm0.01$ $(mbar)$\\
			$A$ & $7.06\pm 0.09$  $(mBq/m^3)$& $10.35\pm0.01$ $(\%)$& $3.90\pm0.01$ $(mbar)$\\
			$T$  & $385 \pm 1$ days & $364\pm0$ days & $350\pm1$ days\\
			$t_M$ & \begin{tabular}{@{}c@{}}30\textsuperscript{th} May 2012\\($\pm$ 5 days)\end{tabular}& \begin{tabular}{@{}c@{}}10\textsuperscript{th} August 2015 \\($\pm$ 0.1 days) \end{tabular}& \begin{tabular}{@{}c@{}}25\textsuperscript{th} July 2014\\ ($\pm$ 0.2 days)\end{tabular}\\
			\hline
		\end{tabular}
		\caption[Environmental parameters fit]{Environmental parameters fit.} 
		\label{tab:SCFits} 
	\end{center}
\end{table}
The humidity data neatly modulates in a seasonal way, giving an almost exact year period. The other fits show less clear seasonal variations, with extracted periods near to one year. The same fit was applied to temperature data resulting compatible with no modulation due to the controlled nature of the parameter.
\paragraph{ }
Despite the seasonal external Rn variation, it does not imply radon inside the ANAIS shielding because of the flushing of boil-off $N_2$ system. There have been established upper limits to the radon inside in the ANAIS background prototypes~\cite{CCUESTA,cebrian2012background} and the new calibration system (see Section~\ref{sec:Calibration}) should have a good impact in the reduction of the residual radon content.   
\subsection{$N_2$ Flux}\label{sec:N_2_Flux}
\begin{figure}[h!]
        \begin{subfigure}[b]{0.5\textwidth}
                \centering
                \includegraphics[width=1\textwidth]{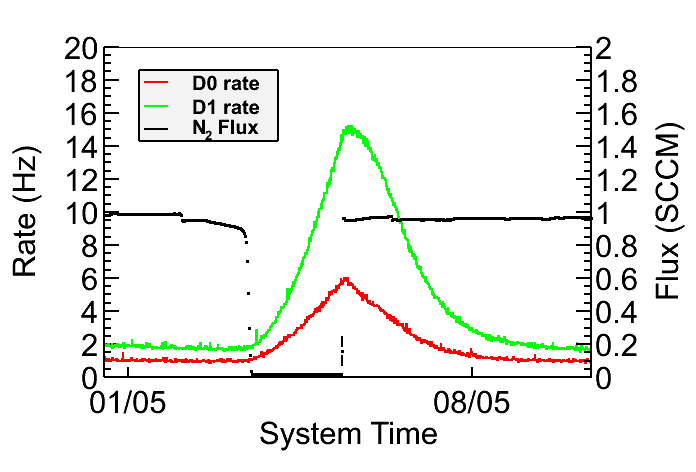}
		\caption{\label{fig:N2Cut_rate}Detector 0 (red), 1 (green) DAQ rate during $N_2$ flux cut (black).}
        \end{subfigure}
~
	\begin{subfigure}[b]{0.5\textwidth}
                \centering
                \includegraphics[width=1\textwidth]{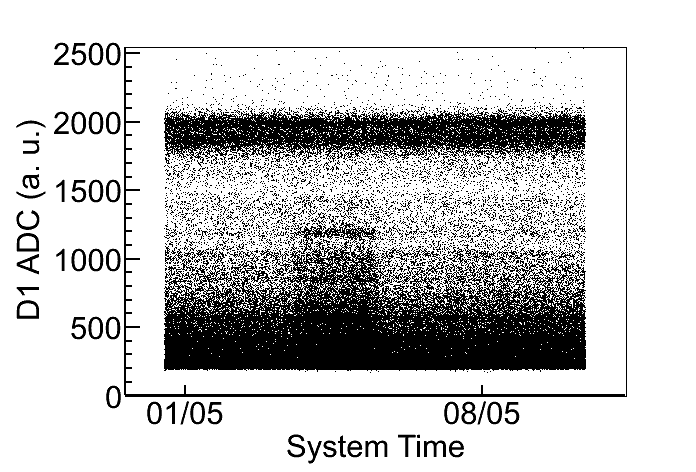}
		\caption{\label{fig:N2Cut_SpcvsTime}High energy over time showing appearing lines with flux cut.}             
        \end{subfigure}

	\begin{subfigure}[b]{0.49\textwidth}
                \centering
                \includegraphics[width=1\textwidth]{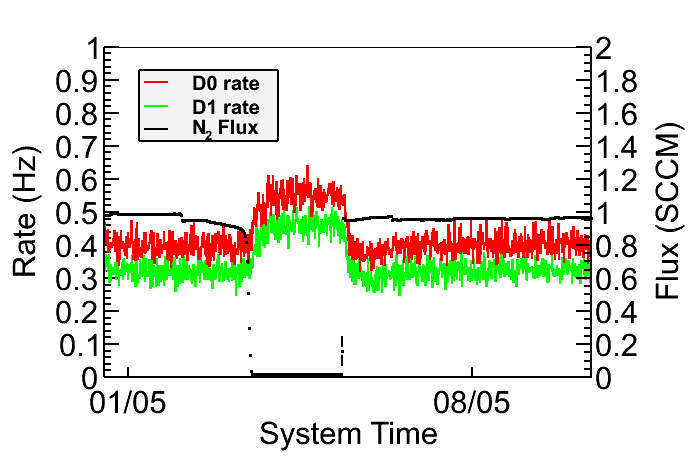}
		\caption{\label{fig:N2Cut_rate_gt3}Event rate with more than three peaks in both signals.\\\hspace{\textwidth}}
        \end{subfigure}
        ~ 
        \begin{subfigure}[b]{0.49\textwidth}
                \centering
                \includegraphics[width=1\textwidth]{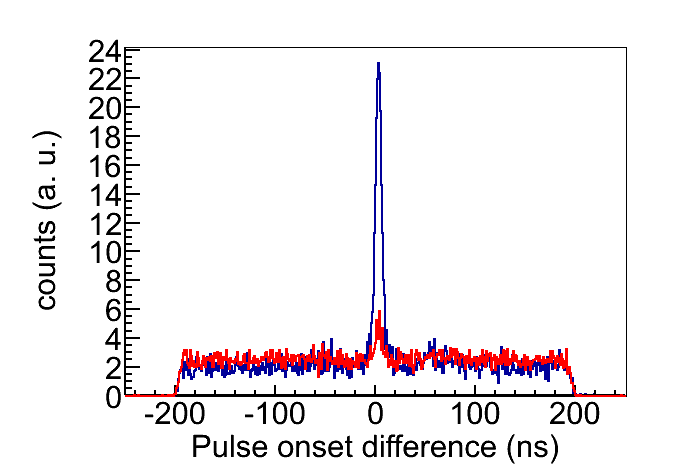}
		\caption{\label{fig:N2Cut_t0diff}Pulse onset difference in a normal period (blue) and with activation for $N_2$ flux cut (red).} 
        \end{subfigure}

	\caption[$N_2$ Flux and DAQ rate]{\label{fig:N_2_cut_effect}$N_2$ cut activation effect.}
\end{figure}

The ANAIS prototypes have seen changes in the acquisition rate associated to flux cuts. The effect of the lack of nitrogen flux can be observed in Figure~\ref{fig:N2Cut_rate}. A clear rate increase can be noted in both detectors correlated with the nitrogen cut. It is worth to note that the rate decrease is not immediate to flux restitution. Several lines appear in the spectrum over time almost simultaneously with the flux cut as can be observed in Figure~\ref{fig:N2Cut_SpcvsTime}. Therefore, the behavior of the whole rise and decay rate cannot be attributed directly to these lines, to $^{222}Rn$ or daughters, because the rate returns to the stable level at least three days after the flux restitution. The most important fraction of the rate increase is due to photoelectron random coincidence as it can be seen in the next two figures.
\paragraph{}
Figure~\ref{fig:N2Cut_rate_gt3} shows the rate of events having three or more peaks, the first quality cut for the low energy event selection (see Section~\ref{sec:EventSelection}). It shows an increase of those type of events that does not explain the full rate behavior, showing a fast decay after the flux restitution as opposed to the total rate (note the different vertical scale and shapes of Figures~\ref{fig:N2Cut_rate} and \ref{fig:N2Cut_rate_gt3}). Additionally, it can be seen the different temporal behavior of this kind of events. The onset pulse difference between the two PMT signals can be observed in Figure~\ref{fig:N2Cut_t0diff} in normal operation (blue) and with $N_2$ flux cut (red). The almost random temporal nature of the later suggests a de-activation mechanism of an activation caused by the radon. The origin of the photoelectron increase during nitrogen flux cut is under study but activation of the PMTs has been discarded because the blank module (a module with only PMTs, see Section ~\ref{sec:ANAIS37}) was not able to reproduce the effect.

\subsection{PMT power supply monitoring}\label{sec:HVPSData}
\begin{figure}[h]
  \begin{center}
	\begin{subfigure}[b]{0.5\textwidth}
        \centering
	\includegraphics[width=\textwidth]{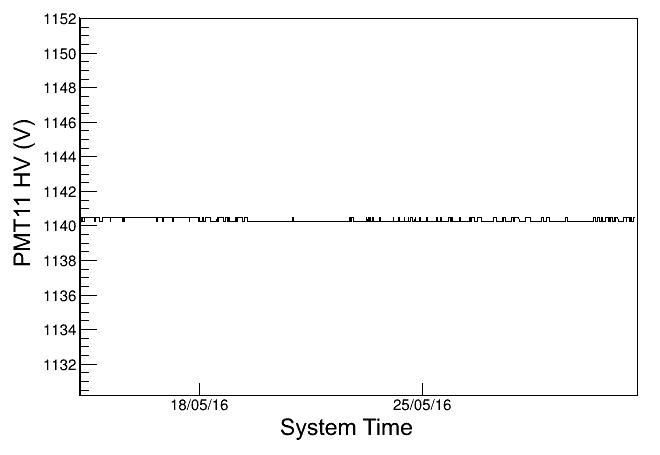}
	\end{subfigure}%
%
	~ 
	\begin{subfigure}[b]{0.5\textwidth}
        \centering
	\includegraphics[width=\textwidth]{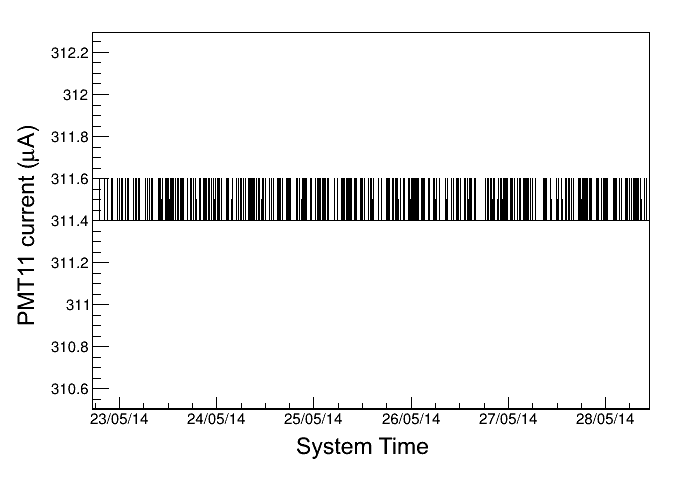}
	\end{subfigure}

	\caption[PMT HV Power Supply monitoring]{PMT Power supply monitoring: HV (left) and current (right).\label{fig:HVPSMonitor}}
  \end{center}
\end{figure}

The PMT power supply monitoring can detect malfunction and instabilities. The typical high voltage and current values for PMTs can be seen in Figure~\ref{fig:HVPSMonitor}. The current and voltage fluctuations in the normal operation are very small and they are in the limits of the least significant bit of the ADC. For this reason, the figures show a step behavior.

\section{Data taking stability}\label{sec:DataStab}
The evolution of critical acquisition parameters is studied in this section: the gain, the trigger level and the coincidence window. Variations in these parameters could mimic an annual modulation. For this reason, their temporal behavior must be monitored. Some other important parameters have been studied in the previous chapter, such as the single electron response and the light collection, which automatic stability control is ongoing.
\subsection{Gain stability}\label{sec:GainStability}\label{sec:GainStability}
\begin{figure}[h!]
     \begin{center}
	     
	\begin{subfigure}[b]{0.50\textwidth}
                \centering
                \includegraphics[width=\textwidth]{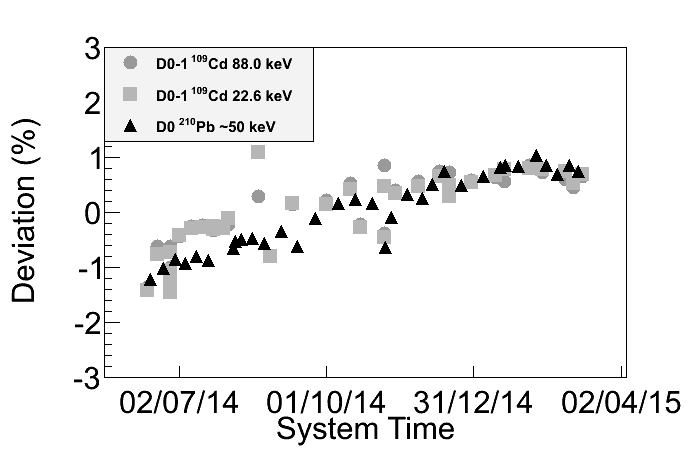}
                \caption{}
                \label{fig:A25D0CalibSum}
        \end{subfigure}%
        ~
	\begin{subfigure}[b]{0.50\textwidth}
                \centering
                \includegraphics[width=\textwidth]{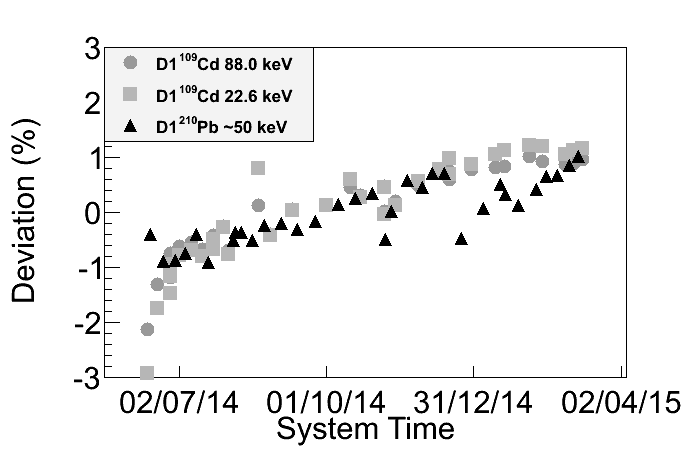}
		\caption{}
                \label{fig:A25D1CalibSum}
        \end{subfigure}
	\begin{subfigure}[b]{0.50\textwidth}
                \centering
                \includegraphics[width=\textwidth]{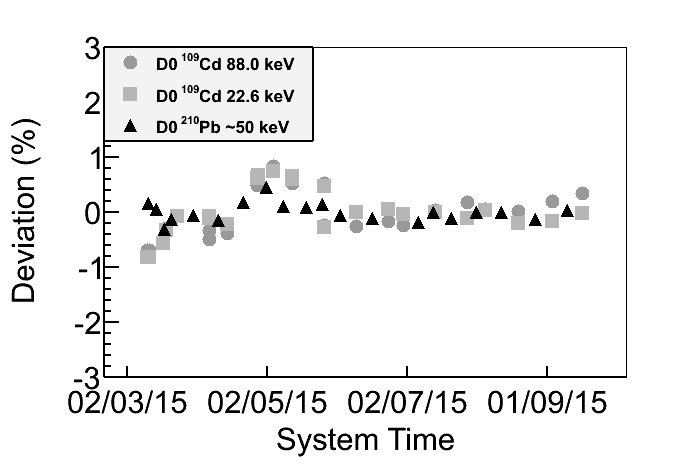}
                \caption{}
                \label{fig:A37D0CalibSum}
        \end{subfigure}%
        ~
	\begin{subfigure}[b]{0.50\textwidth}
                \centering
                \includegraphics[width=\textwidth]{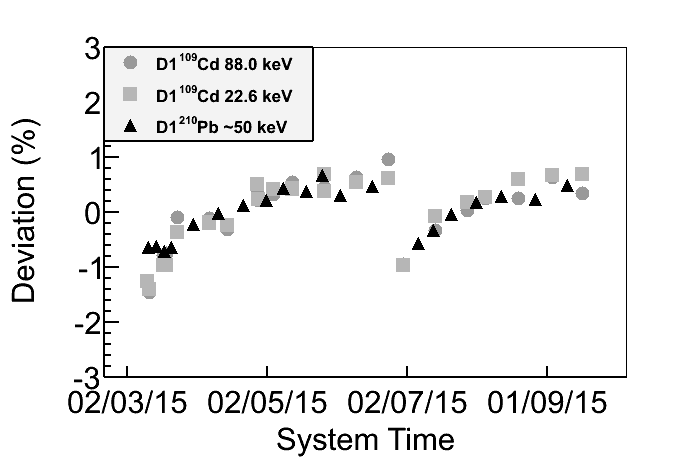}
		\caption{}
                \label{fig:A37D1CalibSum}
        \end{subfigure}
	
	\begin{subfigure}[b]{0.50\textwidth}
                \centering
                \includegraphics[width=\textwidth]{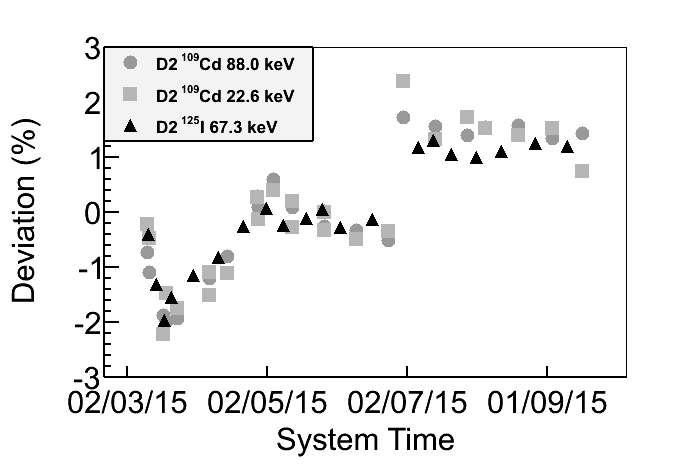}
		\caption{}
                \label{fig:A37D2CalibSum}
        \end{subfigure}

	\caption{Evolution of calibration and background lines (see text).\label{fig:A37CalibStabSum}}
\end{center}
\end{figure}

Monitoring the calibration of the energy estimators is a key feature needed for a long running experiment and even more critical in an annual modulation search. Therefore, the evolution of the calibration and background lines have been studied. Figure~\ref{fig:A37CalibStabSum} shows some calibration and background peaks over time. The evolution of the gain in the ANAIS-25 set-up is shown in Figure~\ref{fig:A25D0CalibSum} (D0) and Figure~\ref{fig:A25D1CalibSum} (D1). These figures show the percentage over their mean values of a series of parameters: two $^{109}Cd$ lines (22.6 keV and 88.0 keV) and the $^{210}Pb$ background peak (sum of a 46.5 keV gamma plus a $\beta^-$ accounting a mean value around 50 keV). 
\paragraph{ }
The calibration values have been calculated with QDC values and the background lines with the area energy estimator. The plotted values are extracted from a fixed window in the spectra around the maxima, avoiding in such a way the bias of a Gaussian fit with non-Gaussian peaks (some peaks are a weighted sum of several contributions). It is fair to mention the use of only one signal to the D0 calibration because of a temporal malfunction of the \texttt{qdc00} channel. Anyway, a very good stability can be reported, with a slight drift of the order of the 2\% of the mean value in both detectors and in both estimators, QDC and area. The ANAIS-25 routine was changed as seen in Section~\ref{sec:Calibration} digitizing also the calibration signals providing redundancy in energy estimators. 

%
\paragraph{}
The same procedure was applied for ANAIS-37 data with some little variations: all values are calculated with the area energy estimator and the background in D2 was calculated with the $^{125}I$ cosmogenic line, because of the reduced $^{210}Pb$ contamination of this module (see Section~\ref{sec:ANAIS37}). The $^{125}I$ decays with 100\% probability by electron capture to an excited state of the $Te$ daughter nucleus and the considered peak accounts this excited level (35.5 keV) and the binding energy of the K-shell of $Te$ (31.8 keV) resulting an energy deposition of 67.3 keV. The evolution of the calibration and background peaks can be seen in Figures~\ref{fig:A37D0CalibSum} (D0), \ref{fig:A37D1CalibSum} (D1) and \ref{fig:A37D2CalibSum} (D2) showing again a good stability with variations below the 2\% except a slightly higher variation in D2 due to the PMT cover replacement in June. This set-up modification also had impact in D1 gain to a lesser extent.
\paragraph{}
This study has shown a good stability of the gain and a good correspondence in the temporal behavior among calibration and background peaks. This will allow to control the gain evolution in the long run.


\subsection{Trigger level stability}\label{sec:TriggerStability}
\begin{figure}[h!]
\begin{center}
	\begin{subfigure}[b]{1\textwidth}
                \centering
                \includegraphics[width=.7\textwidth]{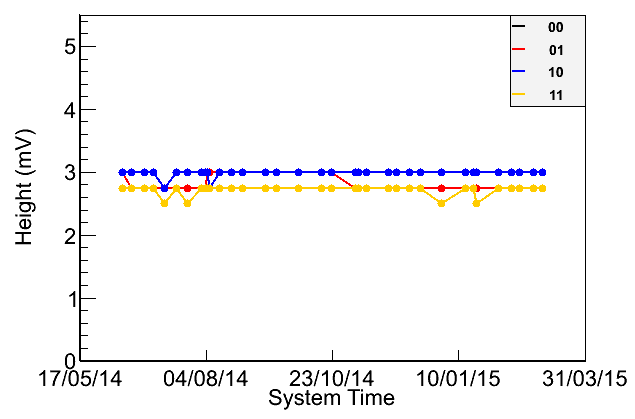}
                \caption{}
                \label{fig:TriggerLevelStabA25_III}
        \end{subfigure}%

	\begin{subfigure}[b]{1\textwidth}
                \centering
                \includegraphics[width=.7\textwidth]{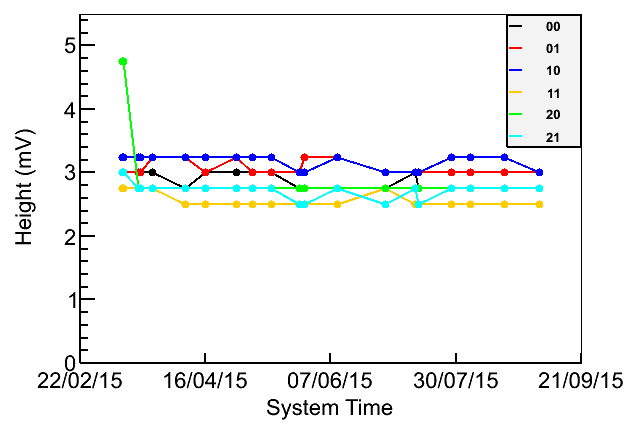}
                \caption{}
                \label{fig:TriggerLevelStabA37}
        \end{subfigure}%

		\caption{Trigger level evolution along the time for ANAIS-25 (a) and ANAIS-37 (see text for details).\label{fig:TriggerLevelStab}}
\end{center}
\end{figure}

The trigger stability is mandatory in a low threshold modulation experiment as well as a controlled gain. The single peak triggered population is first studied and compared with the onsite single electron response in order to verify the trigger level. Next, a study of the evolution of this trigger over the set-up taking data is presented.
\paragraph{}
\begin{figure}[h!]
     \begin{center}
                \includegraphics[width=\textwidth]{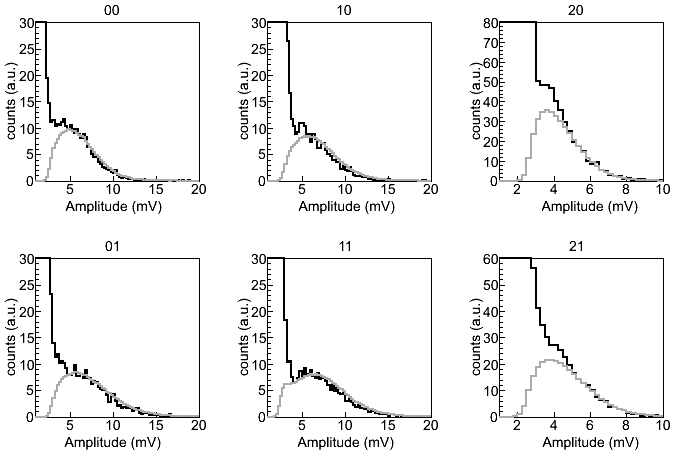}
	\caption{Single electron amplitude distribution (black) compared with the amplitude distribution of triggered events (gray).}\label{fig:TriggerHighA37}
\end{center}
\end{figure}
Figure~\ref{fig:TriggerHighA37} illustrates the effect of the trigger level in the amplitude discrimination. All PMT discriminators were configured with the same value giving a very similar discrimination curve. The stability of the trigger level has been studied by locating the transition where the triggered population clearly reduces. The criterion to locate the transition was chosen at one half of the maximum of the (1 mV, 5 mV) interval, the trigger transition zone seen in Figure~\ref{fig:TriggerHighA37}.
\paragraph{ }
The stability of the trigger transition can be observed in Figure~\ref{fig:TriggerLevelStab} using the previously described criterion. The results show a fairly constant transition point along the ANAIS-25 (Figure~\ref{fig:TriggerLevelStabA25_III}) and ANAIS-37 (Figure~\ref{fig:TriggerLevelStabA37}) set-ups moving only 0.25 mV, the 12bit MATACQ digitizer vertical resolution.
\paragraph{ }
There is a clear exception at the beginning of the ANAIS-37 set-up data taking corresponding to the detector D2. The D2 trigger was configured at higher level in the two first runs because of the exposure to light of their two PMTs vastly increasing the trigger rate at photoelectron level. Next, the trigger levels were set at the same values than the other detectors.
\subsubsection{Trigger level and temperature}
\begin{figure}[h!]
     \begin{center}
	\begin{subfigure}[b]{0.45\textwidth}
                \centering
                \includegraphics[width=\textwidth]{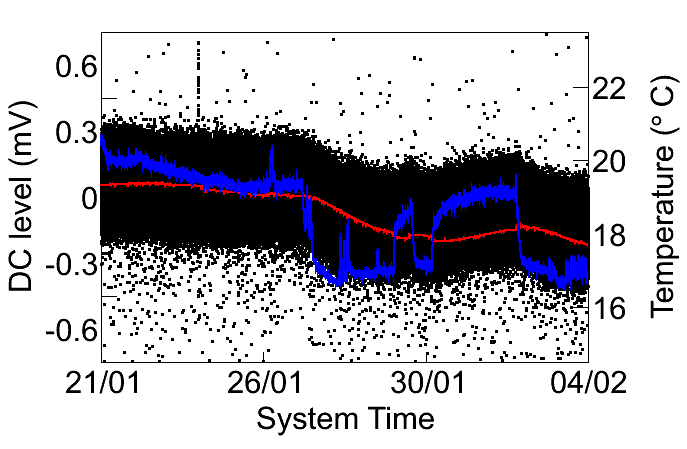}
                \caption{DC level (black) and temperatures (red shielding internal, blue VME crate) vs. time.}
                \label{fig:TriggerStabTempA25_III}
        \end{subfigure}%
        ~ 
        \begin{subfigure}[b]{0.45\textwidth}
                \centering
                \includegraphics[width=\textwidth]{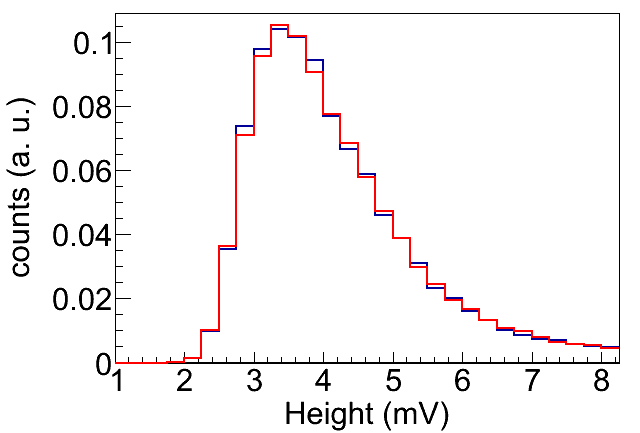}
		\caption{Triggered amplitude at high (\mbox{$\sim$20 ºC}, blue) and low temperature (\mbox{$\sim$17 ºC}, red).}
                \label{fig:TriggAmplTempA25_III}
        \end{subfigure}
        \caption{DC level and triggered height with temperature variations.}\label{fig:TriggStabTemp}
\end{center}
\end{figure}

The ANAIS temperature at the LSC is fairly constant in both inside the shielding and near the crates (see Section~\ref{sec:SCTemp}). However, some changes were made in the LSC air conditioning system and a short interval of ambient temperature instability happened. This interval was studied in order to describe its impact. It is worth to note that the effect of the temperature in the trigger at photoelectron level was studied in a somewhat reduced set-up in the Laboratory of the University of Zaragoza (see Section~\ref{sec:TestTriggerStrat}). A similar (subtle) DC drift in the pulse digitization can be seen during the temperature instability interval as it can be observed in Figure~\ref{fig:TriggerStabTempA25_III}. The study of the amplitude distributions of the triggered events gives no significant variations in the trigger level, far below 1\% in both mean and standard deviation (see Figure~\ref{fig:TriggAmplTempA25_III}). Additionally, no significant variation in the trigger level was measured.
\paragraph{ }
The very little measured trigger level variations over time or with the temperature have a very little impact in the SER trigger efficiency having in mind the SER amplitude distribution seen in Figure~\ref{fig:TriggerHighA37}. This impact is even lower for final conditions with higher gains (such as D0 and D1 in ANAIS-37) and taking into account the effect of the SER efficiency in the total hardware trigger efficiency with the excellent light collection of the ANAIS modules (see Section~\ref{sec:HWTriggEff}).
\subsubsection{Data acquisition rate stability}
\begin{figure}[h!]
     \begin{center}
	\begin{subfigure}[b]{.9\textwidth}
		\centering
                \includegraphics[width=\textwidth]{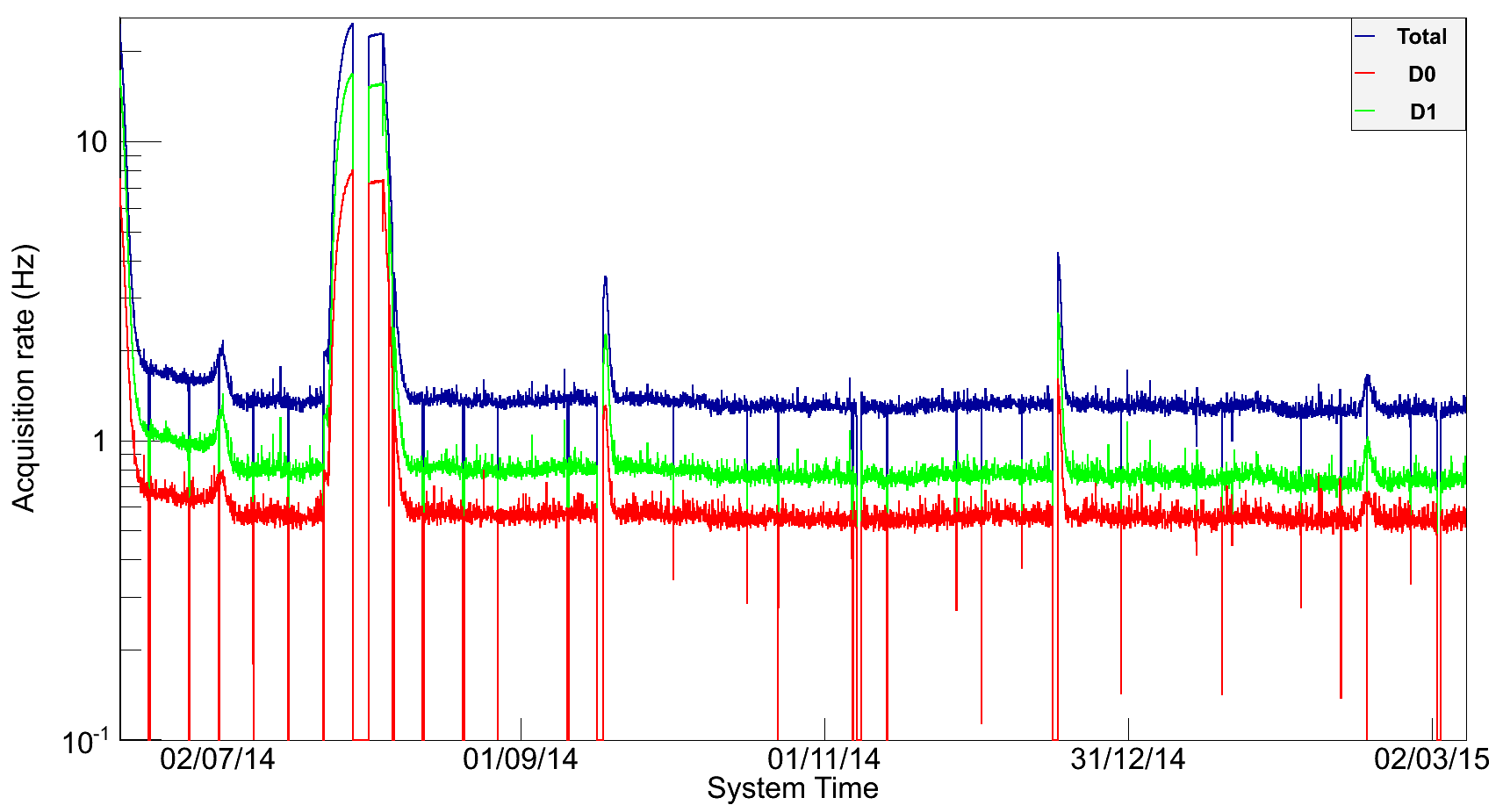}
	\end{subfigure}

                
	\caption{Trigger rate evolution for ANAIS-25.\label{fig:A25_IIIrate}\label{fig:SetupsRate}}
\end{center}
\end{figure}

The previously seen trigger level stability should mean a stable trigger rate in the long run since this rate is dominated by random photoelectron coincidences. As a general comment, the rate is unstable at the beginning of the set-ups due to the PMT light exposure and the radon entrance to the shielding, but once both effects decay, the other source of trigger instability is the $N_2$ flux cut as seen in Section~\ref{sec:N_2_Flux}. The ANAIS-25 set-up (see Figure~\ref{fig:A25_IIIrate}) exhibits all these features in its rate evolution with four $N_2$ flux cuts (the second and longest cut was deliberate in order to study the rate effect) and one anti-radon box aperture. The muon flux has influence in the data acquisition rate as seen in Section~\ref{sec:MuNaIEvt}. This effect is too fast to be observed in Figure~\ref{fig:SetupsRate} but it can be seen in Figure~\ref{fig:MuonNaiRate}. A seasonal modulation could have impact in this rate depending on its amplitude.

\paragraph{}
Despite the effect of the initial PMT activation and the Nitrogen flux cut, a good trigger stability has been achieved and an almost constant trigger rate is expected for the full experiment.
\subsection{Coincidence window stability}\label{sec:CoincWndStability}
\begin{figure}[h!]
\begin{center}
                \includegraphics[width=.6\textwidth]{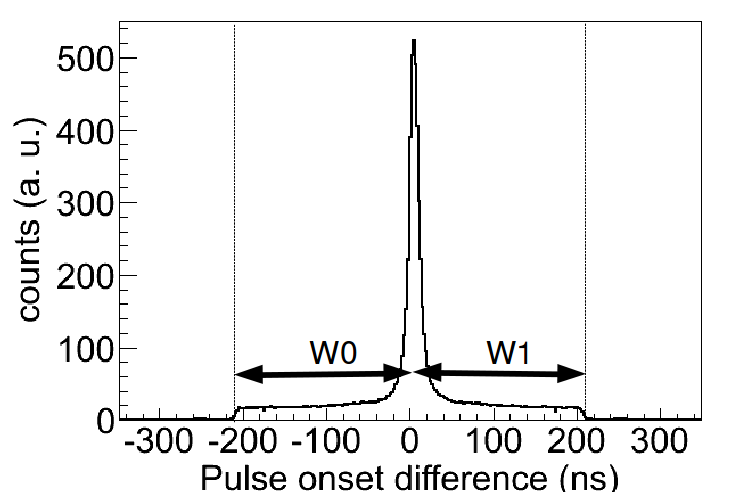}
		\caption{Onset time difference between signals 0 and 1 in one detector and coincidence windows extraction from discriminator 0 (W0) and discriminator 1 (W1).\label{fig:CoincWndT0}}
\end{center}
\end{figure}

\begin{figure}[h!]
     \begin{center}
	\begin{subfigure}[b]{0.5\textwidth}
                \centering
                \includegraphics[width=\textwidth]{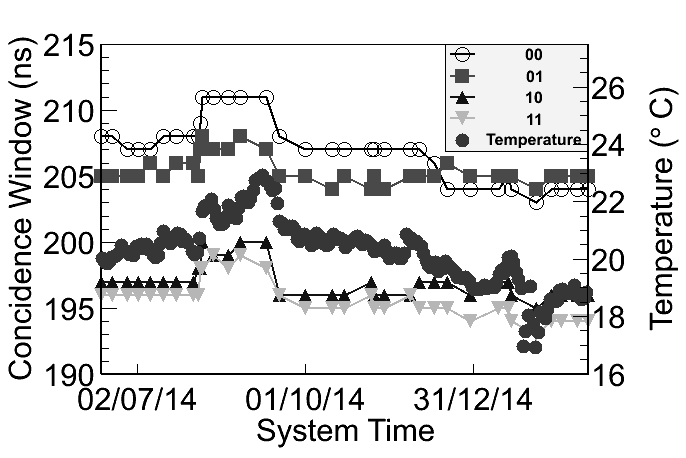}
                \caption{}
                \label{fig:CoincWndStab}
        \end{subfigure}%
        ~ 
        \begin{subfigure}[b]{0.5\textwidth}
                \centering
                \includegraphics[width=\textwidth]{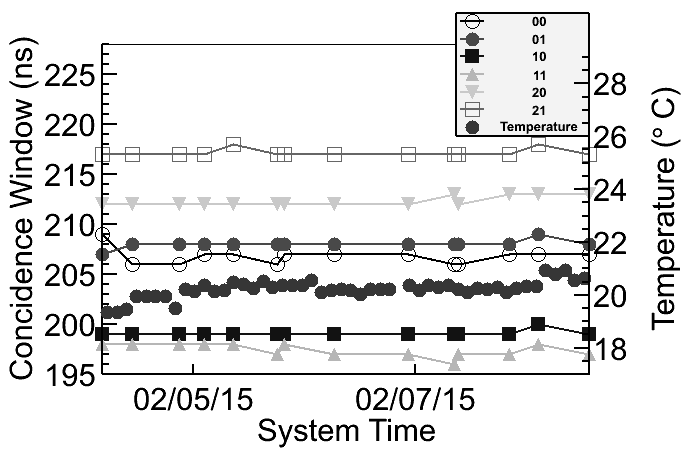}
		\caption{}
                \label{fig:CoincWndStabA37}
        \end{subfigure}
        \caption{Coincident windows and electronics temperature for ANAIS-25 (a) and ANAIS-37 (b) set-ups.}\label{fig:CointWndStabAll}
\end{center}
\end{figure}

Variations in the coincidence windows can affect the trigger efficiency at low energy as it can be seen in Section~\ref{sec:HWTriggEff}, the most harmful energy region for a low threshold experiment. For this reason, a measure of this coincidence window along the data taking would monitor the influence of this systematic effect.
\paragraph{}
The coincidence window can be measured from the ANAIS data in an indirect way by calculating the pulse onset difference between the two signals of a detector in triggering events. This measurement is possible thanks to the separate digitization of the two PMT waveforms. The distribution of the onset difference for one detector can be seen in Figure~\ref{fig:CoincWndT0}. 
\paragraph{ }
The CFD trigger window (W0 and W1 in Figure~\ref{fig:CoincWndT0}) can be deduced offline with the ANAIS data in such a way. The coincidence window evolution was measured considering the end of the time difference distribution where the counts are a half of the plateau. The data for all runs of ANAIS-25 can be seen in Figure~\ref{fig:CoincWndStab} showing some non-constant features with similar behavior in all windows. The evolution of the coincidence windows correlates with the temperature measured at the rack of the electronics as it can be seen in Figure~\ref{fig:CoincWndStab}. The correlation exhibits variations up to 6 ns in the temperature range 17~ºC - 23~ºC. The same procedure was applied for all background runs of ANAIS-37 set-up and the results can be seen in Figure~\ref{fig:CoincWndStabA37}. These results along this set-up were quite stable with differences not exceeding 3 ns with a more stable environmental temperature.
\paragraph{ }
The impact of such coincidence window variations was studied by simulating again the trigger efficiency as described in Section~\ref{sec:HWTriggEff} with offsets of 6ns. Table~\ref{tab:CoincWndShImp} shows the results for the simulated trigger efficiency for 0.9 keV events considering the D0 light collection and its SER trigger efficiency for each PMT signal showing an effect below the 1\% in trigger efficiency for 200 ns but in the order of the expected amplitude of the annual modulation for 100 ns.
\paragraph{}
\begin{table}[h!]
\begin{center}
\begin{tabular}{ccccc}

				\hline
				Coincidence window time (ns)& 200 & 206 & 100 & 106\\
				Trigger efficiency (\%) & 92.8&93.0&83.8&84.7\\
				\hline
\end{tabular}
\caption[ANAIS-37 SER trigger efficiency]{Simulated trigger efficiency for different coincidence windows at 0.9 keV.} 
\label{tab:CoincWndShImp}
\end{center}
\end{table}
It must be noted that the efficiency variations are highly dependent on the light collection and the coincidence window width and could mimic an annual modulation effect at low energies. Therefore, the monitoring of the coincidence windows in both direct and indirect ways is mandatory to guarantee the stability in the trigger efficiency. It is also worth to note that the trigger efficiency increases very quickly with the energy given the very high light collection of the ANAIS modules. In addition, a temperature stabilization system for the ANAIS front-end is under study.




%
\chapter*{Summary and conclusions}
\addcontentsline{toc}{chapter}{Summary and conclusions}
\markboth{Summary and conclusions}{Summary and conclusions}
There is an overwhelming number of hints pointing to the existence of dark matter. These evidences, ranging from galactic to cosmological scales, restrict the nature of this dark matter. It has to be very weakly interacting with other matter and it has to be non-relativistic (\emph{cold dark matter}). These facts exclude all known particles from the Standard Model of Particle Physics.
\paragraph{}
The aforementioned properties suggest methods to detect dark matter in both indirect and direct ways as well as the possibility of its creation in accelerators. The indirect detection of dark matter would imply to detect the byproducts of their co-annihilation or decay. The direct detection would be through the measurement of the direct interaction with detectors. A hypothetical WIMP (Weakly Interacting Massive Particle) is expected to interact producing a recoil with the subatomic particles composing the detector that can be measured with different techniques. 
\paragraph{}
Several detectors from the nineties have been devoted to the dark matter direct detection with different targets and techniques. The detector targets have ranged from cryogenic semi-conductors to noble gases TPCs. DAMA/LIBRA, an experiment carried out in the Laboratori Nazionali del Gran Sasso (LNGS), has claimed to have detected a signal of annual modulation in \emph{single-hit} events in the 2-6 keVee energy window. The annual modulation is a predicted effect considering a WIMP halo surrounding the Milky Way and the different relative velocity of the Earth with respect to this halo along the year. The DAMA/LIBRA result has not been confirmed by other experiments with nominal better sensitivity, although the comparison among different targets is controversial.
\paragraph{}
The ANAIS project has been a long time effort devoted to carry out an experiment to detect dark matter annual modulation with very low background NaI(Tl) detectors. This experiment could confirm the DAMA/LIBRA positive signal with the same target and technique. Such an experiment has a very stringent requirements in order to have enough sensitivity to an annual modulation at very low energy. These requirements are: very low energy threshold, a background as low as possible in the region of interest and a high enough target mass. In addition to these fundamental requirements, very good stability and control of environmental parameters have to be accomplished in order to avoid systematic effects to mimic the effect of the annual modulation. An experiment of more than one hundred kilograms of ultrapure NaI(Tl) has been conceived and it is being commissioned at the Canfranc Underground Laboratory (LSC).
\paragraph{}
This work was devoted to the design, implementation and characterization of a data acquisition system suitable for the ANAIS experiment, having in mind the previously mentioned requirements. It has described the Photomultiplier Tubes (PMTs) used by the ANAIS modules and the algorithms and protocol developed in order to pass quality tests to all units. It has presented the design and implementation of the electronic front-end for the ANAIS modules and muon tagging system along with the data acquisition software. The analysis software was adapted from the software of the previous prototypes to allow easy scale-up to the full experiment. Finally, the test of the optical performance of the Alpha Spectra modules and the tests of data acquisition and monitoring of environmental parameters were performed.
\paragraph{}
The Hamamatsu R12669SEL2 PMTs, the selected model for the ANAIS full experiment, were characterized. First, the features needed for ANAIS were reviewed: background as low as possible, high quantum efficiency and few dark counts. Theory and methods to characterize the single electron response (SER) have been reviewed. Additionally, algorithms to detect the incident light of a pulsed ultra violet LED and deconvolute the SER were developed, tested and used to characterize the gain and the relative quantum efficiency. The limitations of this method given the instabilities of the electronic circuit were pointed out and further improvements have been suggested. The dependency of the dark counts with the temperature and the de-excitation after ambient light exposure of the PMTs were measured. A suitable protocol to test all PMT units has been proposed and the results of gain (and \hbox{gain/voltage} curve), peak-to-valley, dark counts and relative quantum efficiency have been presented. All units tested have fulfilled the requirements in these parameters and in low background measurements performed in a low background HPGe detector at the LSC. All units are stored at the LSC and the quality control of the remaining ones is ongoing for their use in the full experiment.
\paragraph{}
The electronic front-end requirements have been reviewed following their related experimental goals. The conceptual design and the VME and NIM modules to be used were presented. A low noise linear preamplifier was also developed. The critical elements of the electronic chain such as the preamplifier, the constant fraction discriminator (CFD) and the MATACQ digitizer were thoroughly tested with special care in crucial aspects such as the quality of the digitized signal and the low trigger level. In addition, the baseline characterization and improvements were reviewed, showing specific problems to the data acquisition coming from the scale-up process. In particular, several crosstalk effects among signals coming across the HV power supply and the preamplifier were detected, addressed and solved.
\paragraph{}
The design and implementation of the muon detection system were also reviewed from its conception to the first preliminary data analysis. The rate of the detector noise was important at underground conditions and a specific offline pulse shape analysis was designed and tested. Total and coincidence counts were computed giving hints of the asymmetry of the muon flux due to the mountain profile. The muon effect in the NaI crystal data was studied observing both long afterglows with direct muon interaction and low energy events generated by the muon indirect interaction. This system will allow to tag the muon related events in order to study its influence and discard them from an annual modulation analysis.
\paragraph{}
A data acquisition program was designed and implemented for the ANAIS experiment. It was conceived with scalability, stability and performance requirements as guidelines. It was implemented in C++ in Linux and the output data is written in ROOT format making easier the debug and the analysis phases. The more important features of the program are the ease to add new detector and signals without having to rewrite/recompile neither acquisition nor analysis programs, the data acquisition configuration storage together with the acquired data, the asynchronous data storing allowing to absorb the disk latency and the runtime conditions evaluated to avoid reading and/or storing some values or events with significant disk and dead time savings. It is worth to note that both NaI and plastic scintillator data acquisition systems use the same program showing its versatility. Additionally, a monitor system was implemented to check the health of the acquisition system by testing the acquisition rate and sending daily reports. The DAQ software is currently running and ready for the full experiment.
\paragraph{}
The data analysis software has also been adapted for the growing number of signals and also tested with three detectors. All algorithms developed along the operation of the previous modules were reviewed and modified for the new data format and prepared for the final experiment. A new configuration system was developed in order to easily set the input parameters of the analysis and store this parametrization with the analyzed data. In addition to the data acquisition and the data analysis software, a system to synchronize, automatically analyze and pass quality test to the data and notify warnings in case of malfunction or unusual behavior was also implemented in order to quickly react to incidences and keep the duty cycle and the consequent exposure as high as possible.
\paragraph{}
The whole data acquisition system was characterized in order to know its performance in very relevant parameters for the ANAIS experiment such as dead time or hardware trigger efficiency. The characterization of the real time clock used in ANAIS was performed with a crosscheck with the PC system time synchronized via the Network Time Protocol (NTP) over Internet. This crosscheck showed a slight drift in the real time clock, but good stability and accuracy for the ANAIS purposes can be reported. Additionally, the NTP system was configured with Spanish servers with a much better time synchronization for the PC clock. The dead time was studied for several configurations. The acquisition process was configured for optimal operation with asynchronous data storing due to its lower impact in dead time. Additionally, the IRQ trigger strategy was preferred over polling due to the noise produced by the latter. The additional latency of the IRQ waiting was absorbed by reordering the acquisition and using the digitizer (MATACQ board) dead time to wait and download the rest of the modules. The dead time was also measured in the ANAIS-25 and ANAIS-37 prototypes noticing the increase of the dead time with the added digitizers. A strategy to keep the dead time controlled was designed by configuring the acquisition in order to store the triggered waveforms only, with a significant saving in both dead time and storage terms, giving as a result an almost fixed dead time of 2.4 ms with a foreseen improvement due to hardware improvements (the use of CONET2 VME/PCI bridge and 14bit-MATACQs). The accumulation of dead time was computed and considering the down time (which includes routine calibration and maintenance time), the live time was calculated for the ANAIS-25 and ANAIS-37 prototypes obtaining a notable duty cycle over the 95\% of the total time. These numbers should be achievable for the ANAIS full experiment given the foreseen dead time of the order of the 2\% with similar calibration and maintenance protocols.
\paragraph{}
The hardware trigger efficiency at low energy was also characterized. The single electron response (SER) was extracted onsite and the light collection was measured by using calibration lines, giving as a result an excellent light collection of the order of 14 photoelectron per keV. The trigger efficiency of the SER was computed for all PMTs. Using these results and the 200 ns coincidence window, a simulation was performed to obtain the trigger efficiency at very low energy. The result of the simulation was crosschecked with real 0.9 keV from $^{22}Na$ contamination in the bulk showing good agreement, giving an efficiency above 90\% at this energy. Additionally, a study of the coincidence window reduction to 100 ns and its impact in terms of noise reduction (35\% less noise) and trigger efficiency (of the order of 10\% lower) was performed.
\paragraph{}
The environmental and data acquisition stability has also been studied for the ANAIS-25 and ANAIS-37 prototypes. The slow control system was presented and the data for almost four years in the LSC Hall B were analyzed. Some parameters exhibit seasonal variations and they fit to a sinusoidal function with a near yearly period. Additionally, an increase in the rate correlated with the lack of Nitrogen flux was noted, pointing to an excitation of the NaI crystal as responsible of the emission of light with very slow constants triggered as random coincident photons. All these effects will be carefully monitored and mitigated where possible during the full experiment.
\paragraph{}
In addition to the environmental monitoring, the stability of data acquisition parameters that could mimic an annual modulation was studied along the ANAIS-25 and ANAIS-37 set-ups. First, the gain was checked along the time showing a good stability of the order of the 2\% in both calibration and background peaks. Next, the stability of the trigger level with time and temperature was studied finding no significant shift in it. The coincidence window was measured taking advantage of the separate signal digitization of both PMTs in every module. A dependency with the temperature was found in ANAIS-25 set-up (6 ns with a temperature shift of 6~ºC) that would be significant in the trigger efficiency at low energy. The ANAIS-37 set-up did exhibit a very little shift due to a more stable temperature conditions. A simulation confirms that 6 ns variations would be dangerous for 100 ns coincidence windows at very low energy, where annual modulation is expected. For this reason, the coincidence window will be monitored in detail and a temperature stabilization system for the ANAIS front-end is under study.
\paragraph {}
As a result of this work, a data acquisition system for the ANAIS experiment has been constructed and characterized. The system has been tested up to three detectors showing its versatility, scalability, reliability and good performance and makes it ready for the installation of six more detectors during the next months. Additionally, the stability of the environmental and data acquisition parameters has been studied showing good behavior. The protocols to monitor the full experiment have been developed. These good properties together with the very good performance of the modules in terms low background and light collection imply a very good potential to verify a possible dark matter annual modulation signal.
\chapter*{Resumen y conclusiones}
\addcontentsline{toc}{chapter}{Resumen y conclusiones}
\markboth{Resumen y conclusiones}{Resumen y conclusiones}
Existen numerosos indicios, a muy distintas escalas, desde la galáctica a la cosmológica, de la existencia de una gran cantidad de materia oscura en el Universo. Dichas evidencias además acotan las características de las partículas que la componen: deben interaccionar muy débilmente con el resto de la materia o de lo contrario el resultado de esas interacciones podría ser observado. Se trata de materia oscura no relativista (\emph{materia oscura fría}) que no encuentra candidatos a su composición en el modelo estándar de la física de partículas. Por lo tanto, la detección de materia oscura tendría implicaciones a todas las escalas y es un tema muy candente en la física tanto experimental como teórica.
\paragraph{}
Las propiedades de la materia oscura sugieren métodos para su detección tanto directa como indirectamente así como la posibilidad de su creación en aceleradores. La detección indirecta de materia oscura se realizaría mediante la observación de los productos de su aniquilación o desintegración. La detección directa se realizaría mediante la medida de su interacción con detectores. Un hipotético WIMP (\emph{weakly interacting massive particle}, partícula masiva que interacciona débilmente) interaccionaría produciendo un retroceso de alguna partícula subatómica del detector medible mediante diferentes técnicas.
\paragraph{}
Estos efectos se han tratado de observar desde los noventa en experimentos cada vez más sensibles con distintas técnicas y materiales blanco, desde detectores criogénicos basados en semiconductores hasta grandes TPCs de gases nobles. \hbox{DAMA/LIBRA}, un experimento llevado a cabo en el Laboratorio Nacional de Gran Sasso (LNGS), ha obtenido un resultado positivo de modulación anual en el rango energético de \hbox{2-6~keV} en eventos en anticoincidencia atribuido a materia oscura. La modulación anual es un efecto predicho dado el cambio estacional de la velocidad relativa de la tierra respecto al flujo de materia oscura. El resultado de \hbox{DAMA/LIBRA} no ha sido confirmado por ningún otro experimento, algunos de ellos con mejor sensibilidad nominal, aunque la comparación entre distintos materiales blanco es objeto de controversia.
\paragraph{}
El proyecto ANAIS, iniciado en los noventa, se ha dedicado a desarrollar un experimento de materia oscura con ioduro de sodio en el Laboratorio Subterráneo de Canfranc (LSC) y podría confirmar el resultado positivo de DAMA/LIBRA usando la misma técnica y el mismo material blanco. Un experimento de estas características posee unos requisitos muy estrictos para tener sensibilidad suficiente a la modulación anual: tener el menor fondo radioactivo posible en la zona de interés, poseer un umbral energético muy bajo y tener suficiente masa. Además de estos requisitos fundamentales es necesaria una muy buena estabilidad en la adquisición de datos y un buen control de los parámetros ambientales para evitar que posibles efectos sistemáticos puedan ser tomados por modulación anual de materia oscura. El experimento ANAIS constará de más de cien kilogramos de ioduro de sodio ultrapuro en proceso de fabricación y que serán instalados en los próximos meses en el LSC.
\paragraph{}
Esta tesis ha estado enfocada al diseño, implementación y caracterización de un sistema de adquisición de datos adecuado para el experimento ANAIS teniendo en cuenta los requisitos antes mencionados. Se han estudiado los fotomultiplicadores (PMTs) elegidos para la detección de la luz del centelleo del ioduro de sodio. Además se han desarrollado los algoritmos y protocolos necesarios para hacer el control de calidad de todas la unidades. Se ha descrito el diseño de la electrónica necesaria para la adquisición de datos de los módulos de ioduro y los centelladores plásticos usados como detector de muones junto con el software de adquisición y de análisis de datos. También se ha medido la recogida de luz de tres módulos de Alpha Spectra, los dos primeros de que constó el prototipo ANAIS-25 y un tercero que unido a los dos anteriores formaron ANAIS-37. Por último, se ha hecho un estudio de la estabilidad de los parámetros ambientales y de parámetros cruciales para la adquisición de datos.
\paragraph{}
Los fotomultiplicadores (PMTs) Hamamatsu R12669SEL2 fueron los elegidos para el experimento ANAIS dado que cumplen los requisitos de tener un bajo fondo radioactivo, una alta eficiencia cuántica y un número bajo de cuentas oscuras. Se han presentado la teoría y los métodos utilizados para la caracterización de la respuesta a un solo fotoelectrón (\emph{single electron response}, SER). Además, se han implementado y testado algoritmos para detectar la luz emitida por un LED ultravioleta excitado por pulsos y deconvolucionar la SER de los PMTs. Posteriormente se usan estos algoritmos para determinar la ganancia y la eficiencia cuántica relativa. Se pudieron observar las limitaciones de este método debido a inestabilidades en la electrónica de excitación y se proponen mejoras para los test futuros. Se han medido las cuentas oscuras y su dependencia con la temperatura y con el tiempo posterior a la exposición a la luz ambiente del fotomultiplicador. Se ha propuesto un protocolo a seguir para la caracterización de los fotomultiplicadores y se han presentado los datos de las medidas de las primeras unidades. Estas medidas han consistido en la curva de ganancia frente a voltaje, la relación pico-valle, las cuentas oscuras y la medida de eficiencia cuántica relativa. Todos los PMTs han pasado el control de calidad en dichos parámetros y en cuanto a radiopureza medida en un detector de germanio ultrapuro de bajo fondo en el LSC. Los fotomultiplicadores están actualmente almacenados en el Laboratorio Subterráneo de Canfranc y el control de calidad de las unidades restantes que van a ser usadas en el experimento completo está en proceso.
\paragraph{}
Se han revisado los requerimientos que debe cumplir el sistema electrónico para ANAIS teniendo en cuenta los requisitos del experimento. Se han descrito el diseño de dicho sistema, basado en módulos NIM y VME, y el preamplificador de bajo ruido diseñado específicamente para ANAIS. Se han realizado tests exhaustivos de las partes más críticas de este sistema como el preamplificador, el discriminador de fracción constante (CFD) y la digitalizadora MATACQ teniendo especial cuidado en los aspectos más sensibles para el experimento como la calidad de la señal digitalizada y el bajo umbral de disparo. Además la linea base del experimento se ha caracterizado y mejorado. Se han detectado y resuelto problemas de ruido e interferencias entre señales, algunos de ellos provenientes del incremento en el número de módulos, observadas a distintos niveles como la fuente de alto voltaje o el preamplificador.
\paragraph{}
Se ha diseñado e implementado un sistema de detección de muones y se han presentado los primeros datos preliminares. Se descubrió un importante ritmo de sucesos de ruido en el LSC que hacía imposible el disparo en umbral y se diseñó un sistema de análisis de forma de pulso para descartar ruido en tiempo de análisis. Se han presentado los datos de esta discriminación así como el criterio para ampliar el montaje a los dieciséis plásticos centelleadores de que constará el experimento completo. Se ha medido el ritmo total de muones así como en coincidencia de varias caras, observando una asimetría que sugiere un flujo no simétrico en el ángulo azimutal en consonancia con el perfil asimétrico de la montaña. Se ha observado el efecto de los muones en los cristales de ioduro que podrían afectar al análisis de detección de materia oscura. En particular, se ha podido correlacionar con eventos en los plásticos dos tipos de efectos: eventos a baja energía y episodios de activación debido a interacciones directas de los muones. Este sistema permitirá etiquetar estos eventos para estudiar su influencia y descartarlos de un análisis de materia oscura.
\paragraph{}
Se diseño e implementó un programa de adquisición de datos para el experimento ANAIS. Fue concebido con los requisitos de escalabilidad, estabilidad y rendimiento como guía. Fue implementado en C++ sobre Linux. Se eligió ROOT como formato de salida para facilitar la depuración y el análisis de datos. El programa permite añadir con facilidad nuevas señales y detectores a la adquisición sin recompilar ni el software de adquisición ni el software de análisis guardando la configuración de hardware junto a los datos adquiridos. Además usa un almacenamiento asíncrono que permite absorber las latencias del sistema de ficheros reduciendo el tiempo muerto. Por último, se pueden configurar condiciones evaluadas en tiempo de ejecución que permiten leer y/o guardar solamente los datos relevantes en función de otros valores adquiridos con gran ahorro de espacio en disco y reducción en el tiempo muerto. Hay que hacer notar que el software de adquisición de datos de los ioduros y de los centelleadores plásticos es el mismo ejecutado en dos ordenadores distintos con dos configuraciones distintas demostrando así la versatilidad para la que fue diseñado. Además, se implementó un sistema de monitorización que comprueba el ritmo de adquisición y envía informes diarios para su inspección. Toda esta infraestructura se encuentra actualmente en ejecución y está preparada para el experimento completo.
\paragraph{}
El software de análisis de datos también ha sido preparado para el incremento del número de detectores y testeado con tres detectores. Todos los algoritmos desarrollados a lo largo de la operación de los prototipos anteriores fueron revisados y adaptados a los nuevos formatos de datos y preparados para un gran incremento en el número de señales. Se desarrolló un sistema de configuración del análisis para poder parametrizarlo fácilmente y guardar dicha parametrización junto a los datos analizados. Además del software de adquisición de datos y de análisis, se diseñó un sistema para sincronizar y analizar automáticamente los datos y ejecutar una serie de test de calidad permitiendo enviar avisos de funcionamiento anómalo para poder reaccionar rápidamente y maximizar el tiempo de exposición.
\paragraph{}
El sistema de adquisición fue caracterizado en su conjunto para conocer parámetros relevantes como el tiempo muerto o la eficiencia de trigger hardware. La caracterización del reloj de tiempo real usado en ANAIS fue realizada mediante su verificación con el reloj del PC sincronizado por internet vía NTP (Network Time Protocol). Esta verificación mostró una ligera deriva en el reloj de tiempo real, demostrando no obstante estabilidad y precisión suficientes para los propósitos de ANAIS. Además el sistema NTP se configuró con servidores españoles consiguiendo una sincronización mejor del reloj del PC. Se estudió el comportamiento del tiempo muerto para distintas configuraciones y como resultado el sistema de adquisición se configuró con almacenamiento de datos asíncrono. También se configuró para realizar la adquisición esperando a una IRQ en lugar de haciendo una consulta continua debido a que dicha consulta, a pesar de reducir el tiempo muerto, genera ruido en la digitalización. El mayor tiempo muerto en la espera de la interrupción es absorbido dentro del tiempo muerto de conversión de la digitalizadora (MATACQ) durante el cual se espera dicha interrupción permitiendo además descargar los datos de los otros módulos si la lectura de datos se reordena convenientemente. Se midió el tiempo muerto de los prototipos ANAIS-25 y ANAIS-37 constatando el aumento de tiempo muerto con el aumento de tarjetas digitalizadoras. Se diseñó una estrategia para mantenerlo controlado basada en la descarga y almacenamiento condicionado al disparo del correspondiente detector. Como resultado el tiempo muerto medido para esta configuración es de 2.4~ms por evento. Este valor se verá reducido con la introducción de algunas mejoras en el hardware (tarjeta de comunicaciones con protocolo CONET2 y digitalizadora MATACQ de 14 bits). Teniendo en cuenta también el tiempo de parada de adquisición (que incluye tiempos de calibración, mantenimiento, etc.) a lo largo de los distintos montajes se obtuvo una notable tasa de tiempo vivo a lo largo de ANAIS-25 y ANAIS-37 superior en ambos casos al 95\%. Esta tasa debería ser alcanzable en el experimento final dado que el tiempo muerto previsto es del orden del 2\% y se usarán protocolos similares de calibración y mantenimiento.
\paragraph{}
La eficiencia del trigger hardware también fue caracterizada. Se extrajo la respuesta a un sólo electrón (\emph{single electron response}, SER) con los datos de ANAIS-25 y ANAIS-37. Con los datos de la SER y con las señales de fuentes de calibración se calculó la recogida de luz con un resultado excelente del orden de 14 fotoelectrones por keV. Una vez caracterizada, se computó la eficiencia de trigger de la SER para todos los fotomultiplicadores. Usando estos datos y una ventana de coincidencia de 200 ns, se realizó una simulación para obtener la eficiencia de trigger a baja energía. Posteriormente se compararon estos datos con eventos reales de 0.9 keV de $^{22}Na$ y se encontró una buena concordancia entre ambos resultados. Se determinó que la eficiencia del trigger hardware con eventos de 0.9 keV es superior al 90\%. Además se estudió el impacto de la reducción de la ventana de coincidencia a 100 ns en términos de reducción eventos de ruido (un 35\% menos) y de reducción en la eficiencia de trigger hardware (del orden de un 10\% menos).
\paragraph {}
También han sido estudiados los parámetros ambientales y la estabilidad de parámetros de adquisición como la ganancia, el nivel de trigger o la ventana de coincidencia para los montajes ANAIS-25 y ANAIS-37. El sistema de adquisición de parámetros ambientales y de monitorización (también llamado slow control) toma datos de temperaturas, concentración de radon, flujo de nitrógeno para ventilar el blindaje, presión, humedad, temperatura y voltaje y corriente de los fotomultiplicadores. Los datos de casi cuatro años en el Hall B del Laboratorio Subterráneo de Canfranc han sido analizados y se ha observado una periodicidad estacional en datos de presión, humedad y concentración de radon. Se realizó un ajuste a un seno más una constante y se obtuvo como resultado una modulación con un periodo muy cercano al año. Además se investigó el efecto del aumento en el ritmo de disparo de los prototipos de ANAIS con la falta de flujo de nitrógeno. Este efecto apunta a una excitación en el cristal que emite con posterioridad (con constantes muy largas) fotones que son detectados por la coincidencia de los fotomultiplicadores como coincidencias fortuitas. Todos estos efectos se monitorizarán y mitigarán en lo posible en el experimento completo.
\paragraph{ }
Además de los parámetros ambientales, se ha estudiado la estabilidad de parámetros de adquisición. Una inestabilidad en estos parámetros podría inducir una modulación anual a baja energía y atribuirse erróneamente a materia oscura. La ganancia ha permanecido bastante estable a lo largo de cada uno de los montajes con variaciones del orden del 2\% compatibles en los picos de fondo y de calibración. Dicha calibración permite corregir estas pequeñas variaciones en la ganancia. El análisis del nivel de trigger no ha apreciado variaciones substanciales (por debajo de 0.25 mV) ni con el tiempo ni con la temperatura. La ventana de coincidencia se midió en los datos de fondo aprovechando la digitalización de las dos vías de cada módulo. Se encontró una dependencia con la temperatura en ANAIS-25 (6 ns con 6 ºC de diferencia) que podría tener un impacto significativo en la eficiencia del trigger hardware. Por el contrario en ANAIS-37 el efecto fue muy inferior debido a una mayor estabilidad de la temperatura. Una simulación confirmó que el efecto de una variación de 6 ns podría ser peligrosa usando una ventana de coincidencia de 100~ns en eventos de muy baja energía, donde se espera la modulación anual. Por este motivo una monitorización tanto directa como indirecta de la ventana de coincidencia a lo largo de año es obligatoria y está estudiándose la posibilidad de un sistema de control de temperatura para la electrónica de ANAIS.
\paragraph {}
Como resultado del trabajo recogido en esta tesis se ha construido y caracterizado un sistema de adquisición de datos para ANAIS. Además este sistema ha sido probado con tres detectores, demostrando su versatilidad, escalabilidad, fiabilidad y buen rendimiento que lo hacen estar preparado para la instalación de seis detectores más en los próximos meses. Además se ha estudiado e implementado la monitorización de los parámetros ambientales, que han mostrado un buen comportamiento y se han desarrollado los protocolos necesarios para la monitorización del experimento completo. Estas prestaciones, junto con la buena calidad de los detectores en términos de radiopureza y recolección de luz implican un buen potencial para verificar una posible señal de modulación anual de materia oscura. 
\chapter*{Agradecimientos}
\addcontentsline{toc}{chapter}{Agradecimientos}
\markboth{Agradecimientos}{Agradecimientos}
En primer lugar me gustaría dar las gracias a mis directores de tesis por su guía y apoyo. A José Ángel Villar le agradezco especialmente toda la confianza que puso y ha puesto en mí. A Carlos Pobes que me acercase a este mundo a la vez que toda la paciencia, didáctica y ayuda antes y durante la realización de esta tesis.
\paragraph{ }
También querría agradecer a María Luisa Sarsa todo lo que he podido aprender de su vastísimo conocimiento de la detección directa de materia oscura en todas sus ramificaciones. Este trabajo tampoco habría sido posible sin la sabiduría y dedicación de Julio César Amaré, sin quien nada de lo que aquí se habla de electrónica analógica y filtrado de ruido eléctrico se podría haber llevado a cabo. También agradecer a María Martínez sus conocimientos de los aspectos teóricos y prácticos de la detección directa y su disponibilidad para aclararme dudas. Lo mismo podría decir de Susana Cebrián y su excepcional trato. También Clara Cuesta, con la que comencé aquí ya hace unos años, me introdujo en el experimento ANAIS y me explicó con paciencia todos los detalles. Y por supuesto debo (y quiero) agradecer a Alfonso Ortiz de Solórzano muchísimas cosas: los viajes a Canfranc y las cenas en Jaca, el trabajo con él que nunca es rutina. He podido disfrutar de su dedicación, de su optimismo y de la pasión que pone en lo que hace. Agradezco también la ayuda que me han brindado Eduardo García y Jorge Puimedón cada vez que lo he necesitado, a Patricia Villar por todo su apoyo, especialmente con las medidas de los fotomultiplicadores, a Javier Mena por el soporte y a Ysrael Ortigoza por ser como es.
\paragraph{ }
Por supuesto toda la gente del Laboratorio Subterráneo de Canfranc tiene también mi agradecimiento por su trato. En particular me gustaría dar las gracias a Iulian, Sivia, Alberto y Sergio que son con los que más he trabajado y que me han ayudado también dentro y fuera del Laboratorio. Y al Servicio de Instrumentación Electrónica de la Universidad de Zaragoza, a Pedro y Chema, por su ayuda, consejos y aportaciones y por su excelente disponibilidad.
\paragraph{ }
No me querría olvidar de toda la gente que hacía del café y la comida un sitio distinto: Clara, Asun, David, Héctor, Laura, Carlos, María, Javier, Ángel, Carmen, Paco, Alicia y Diana y de aquellos con los que he compartido (o casi) despacho: Alfredo, Javi, JuanAn, Theopisti y Xavi. También ha sido muy importante para mí el ánimo de mis amigos, especialmente de Emiliano, Rosa, Pedro e Íñigo.
\paragraph{ }
Por último me gustaría acordarme de mi familia, de mis padres por haberme hecho como soy, de todos mis hermanos y familias, de Mario, Miguel, Carolina y Pili, por su constante apoyo, y de Jorge, Elena y Blanca por la comprensión y el ánimo y por todo lo que habéis hecho y dejado de hacer por mí estos últimos tiempos y no podré agradeceros suficiente.

\begingroup
\let\cleardoublepage\relax

\chapter*{Acknowledgments}
I have to thank some other people that have been very helpful for this work. The CAEN staff has been very supportive with every question, problem or issue. I have to thank specially to Gianni Di Maio and Alberto Luchessi for their quick responses and their helpfulness. I'm also very grateful to Marta Trueba and the ATI team for their good treatment.
\paragraph{ }
This work was supported by the Spanish Ministerio de Economía y Competitividad and the European Regional Development Fund (MINECO-FEDER) (FPA2011-23749, FPA2014-55986-P), the Consolider-Ingenio 2010 Programme under grant CPAN CSD2007-00042 and the Gobierno de Aragón (Group in Nuclear and Astroparticle Physics, GIFNA).
\addcontentsline{toc}{chapter}{Acknowledgments}
\markboth{Acknowledgments}{Acknowledgments}
\endgroup
\myblank
\bibliographystyle{ieeetr}
\bibliography{TFM}

\myblank
\appendix


\end{document}